\documentstyle[11pt,epsf]{article}
\begin{document}

%%%%%%%%%%%%%%%%%%%%%
\newcommand{\A}{\u{a}}
\newcommand{\h}{\^{\i}}
%\newcommand{\R}{\c{s}}
%\newcommand{\z}{\c{t}}
%\newcommand{\z}{\c{t}}
%%%%%%%%%%%%%%%%%%%%%%%

\begin{center}
{\bf Arhivele Electronice Los Alamos} 

{\bf http://xxx.lanl.gov/physics/0003106\\}
%%%%%%%%%%%%%%%%%%%%%%%%%%%%%%%%%%%%%%%%%%%%%%%%%%%%%%%%%%%%%%%%%%%%%%%%%%%%%
%Dac\A\ ve\c{t}i \c{s}ti mai mult ve\c{t}i gre\c{s}i mai pu\c{t}in, ve\c{t}i fi 
%mai aten\c{t}i\\
%\c{s}i probabil ve\c{t}i fi mai valoro\c{s}i.\\
%\\
%Acest curs a fost scris \h n Instituto de F\'{\i}sica,\\ 
%Universidad de Guanajuato, Le\'on, Guanajuato, M\'exico. 
%%%%%%%%%%%%%%%%%%%%%%%%%%%%%%%%%%%%%%%%%%%%%%%%%%%%%%%%%%%%%%%%%%%%%%%%%%%%%                                                                                                                                                                                                                                                                                    
\end{center}

\bigskip
\bigskip

\begin{center}
{\huge \bf ELEMENTE DE MECANIC\u{A} CUANTIC\u{A}}\end{center}

%\begin{center} {\large \bf MC I} \end{center}

\bigskip

\begin{center}
{\large \bf HARET C. ROSU}\end{center}
\begin{center} e-mail: rosu@ifug3.ugto.mx\\
fax: 0052-47187611\\
phone: 0052-47183089  \end{center}
%\begin{center} {\bf IFUG (M\'exico) \& IGSS (Rumania)} \end{center}
%Julio de 1998}
%\end{center}

\bigskip
\bigskip

%%%%%%%%%%%%%%
\vskip 2ex
\centerline{
\epsfxsize=280pt
\epsfbox{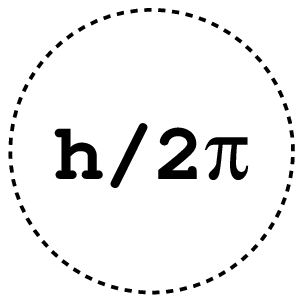}}
\vskip 4ex
%\begin{center}
%{\small{Fig. 1}\\
%}
%\end{center}
%%%%%%%%%%%%%%%%

\vspace{5.5cm}
\begin{center} Pentru to\c{t}i cei atra\c{s}i 
de \c{S}tiin\c{t}ele Fizice\\
\c{s}i se afl\A\ \h n dificilii ani ai \h nceputului de facultate.
\end{center}
\begin{center} .\end{center}
\begin{center} Primul curs de mecanic\A\ cuantic\A\
\^{\i}n rom\^{a}ne\c{s}te  \^{\i}n Internet.\\
Acest curs a fost scris \h n Instituto de F\'{\i}sica,\\ 
Universidad de Guanajuato, Le\'on, Guanajuato, M\'exico.
\end{center}

\begin{center} Copyright \copyright 2000 de c\A tre autor.
%{\cal H}.{\cal C}. {\cal R}{\cal O}{\cal S}{\cal U}.
Orice drept comercial este rezervat.
\end{center}
\vspace{0.2cm}
\centerline{\bf Martie 2000}

\vspace{2cm}

\newpage

\begin{center} English Abstract \end{center}

\bigskip 

\noindent
This is the first graduate course on elementary quantum mechanics in Internet
written in 
Romanian for the benefit of Romanian speaking students (Romania and Moldova). 
It is a translation (with corrections) 
of the Spanish version of the course, which I did at the 
suggestion of Ovidiu Cioroianu, a student of physics in Bucharest.
The topics 
included refer to the postulates of quantum mechanics, one-dimensional barriers 
and wells, angular momentum and spin, WKB method, harmonic oscillator, 
hydrogen atom, quantum scattering, and partial waves.\\

\bigskip

\begin{center} Abstract Rom\^{a}nesc \end{center}

\bigskip

\noindent
Acesta este un curs internetizat de mecanic\A\ cuantic\A\ elementar\A\
pe care l-am tradus cu \h mbun\A t\A \c{t}iri din spaniol\A\ (limba \h n care
l-am predat) la rug\A mintea 
studentului Ovidiu Cioroianu din Bucure\c{s}ti. Este destinat \h n principal
acelor studen\c{t}i care se afl\A\ 
la primele contacte cu aceast\A\ disciplin\A\ de studiu obligatorie, de\c{s}i
ar putea s\A\ fie de un oarecare folos \c{s}i pentru alte categorii de cititori.
Sursele de `inspira\c{t}ie' le-am g\A sit \h n multe din excelentele manuale
de mecanic\A\ cuantic\A\ care au fost publicate de-a lungul anilor.
%\end{center}

%\end{center}

\newpage

%Cuantele de energie au ap\A rut \h n 1900 ca o consecin\c{t}\A\ a 
%lucr\A rilor 
%lui Max Planck (premiul Nobel 1918) asupra problemei radia\c{t}iei de 
%corp negru.

\centerline{{\huge CUPRINS}}

\vspace{0.5cm}

\noindent

{\bf 0. Forward - Cuv\h nt \h nainte} \hfill ... 5%- Haret C. Rosu
\\

{\bf 1. Postulate cuantice} \hfill ... 7
\\

{\bf 2. Bariere \c{s}i gropi rectangulare unidimensionale} \hfill ... 25
%-
\\

{\bf 3. Moment cinetic \c{s}i spin} \hfill ... 47
%- 
\\

{\bf 4. Metoda WKB} \hfill ... 75
%-  
\\

{\bf 5. Oscilatorul armonic} \hfill ... 89
%- 
\\

{\bf 6. Atomul de hidrogen} \hfill ... 111
%- 
\\

{\bf 7. Ciocniri cuantice } \hfill ... 133
%- 
\\

{\bf 8. Unde par\c{t}iale}\hfill ... 147
%-   
\\

Include aproximativ 25 de probleme ilustrative.

\bigskip
\bigskip

\centerline{Unit\A \c{t}ile atomice nerelativiste de spa\c{t}iu \c{s}i timp} 
$$a_H=\hbar ^2/m_ee^2=0.529 \cdot 10^{-8}{\rm cm}$$ 
$$t_{H}=\hbar ^3/m_ee^4=0.242\cdot 10^{-16}{\rm sec}$$ 

\bigskip
\centerline{Unit\A \c{t}ile Planck relativiste de spa\c{t}iu \c{s}i timp} 
$$l_P=\hbar/m_Pc=1.616 \cdot 10^{-33}{\rm cm}$$ 
$$t_{P}=\hbar/m_Pc^2=5.390\cdot 10^{-44}{\rm sec}$$

\newpage

{\sl

\section*{{\huge 0 (E). FORWARD}}  %%%%%%%   0

The energy quanta occured in 1900 in the work of Max Planck (Nobel prize, 1918)
on the black body electromagnetic radiation. Planck's ``quanta of light" have
been used
by Einstein (Nobel prize, 1921) to explain the photoelectric effect, but
the first ``quantization" of a quantity having units of action (the angular
momentum) belongs to Niels Bohr (Nobel Prize, 1922). This
opened the road to the universalization of quanta, since the action is the
basic functional to describe any type of motion. However, only in the
1920's the formalism of quantum mechanics has been developed in a systematic
manner. The remarkable works of that decade contributed in a decisive way
to the rising of quantum mechanics at the level of fundamental theory of the
universe, with successful 
technological applications. Moreover, it is quite probable that many
of the cosmological misteries may be disentangled by means of various
quantization procedures of the gravitational field, advancing our 
understanding of the origins of the universe. On the other hand, in recent years,
there is a strong surge of activity in the information aspect of
quantum mechanics. This aspect, which was generally
ignored in the past,
aims at a very attractive ``quantum computer" technology.

At the philosophical level, the famous paradoxes of quantum mechanics,
which are perfect examples of the difficulties of `quantum' thinking, are
actively pursued ever since they have been first posed. Perhaps the most
famous of them is the EPR paradox (Einstein, Podolsky, Rosen, 1935) on the
existence of {\em elements of physical reality}, or in EPR words:
``If, without in any way disturbing a system, we can predict with certainty
(i.e., with probability equal to unity) the value of a physical quantity, then
there exists an element of physical reality corresponding to this physical
quantity."
Another famous paradox is that of Schr\"odinger's cat which is related to
the fundamental quantum property of entanglement and the way we
understand and detect it.
What one should emphasize is that all these delicate points are the sourse
of many interesting and innovative experiments (such as the so-called 
``teleportation" of quantum states) pushing up the technology.

Here, I present eight elementary topics in nonrelativistic quantum mechanics
from a course in Spanish (``castellano")
on quantum mechanics that I taught in
the Instituto de F\'{\i}sica, Universidad de Guanajuato (IFUG), Le\'on, Mexico,
during the semesters of 1998. 

\hfill Haret C. Rosu

\newpage

\section*{{\huge 0 (R). CUV\^{I}NT \^{I}NAINTE}}  %%%%%%%   0

Cuantele de energie au ap\A rut \h n 1900 ca o consecin\c{t}\A\ a lucr\A rilor 
lui Max Planck (premiul Nobel 1918) asupra problemei radia\c{t}iei de corp negru.
``Cuantele de lumin\u{a}" planckiene au fost folosite de c\A tre Albert Einstein
(premiul Nobel 1921) pentru a explica efectul fotoelectric, dar prima 
``cuantificare" a unei m\A rimi cu unit\A \c{t}i de ac\c{t}iune (momentul cinetic)
se datoreaz\A\ lui Niels Bohr (premiul Nobel 1922). Aceasta a deschis drumul 
universalit\A \c{t}ii cuantelor pentru c\A\ ac\c{t}iunea este func\c{t}ionala
fundamental\A\ pentru descrierea oric\A rui tip de mi\c{s}care. Chiar \c{s}i
\h n aceste condi\c{t}ii, numai anii 1920 se consider\A\ ca 
adev\A ratul \h nceput pentru formalismul cuantic, care a fost capabil s\A\ 
ridice mecanica cuantic\A\ la nivelul unei teorii fundamentale a universului
\c{s}i s\A\ o transforme \h ntr-o surs\A\ de numeroase succese tehnologice. Este
foarte posibil ca multe dintre misteriile cosmologice s\A\ se ascund\A\ \h n 
spatele diferitelor proceduri de cuantificare ale c\h mpului nelinear 
gravita\c{t}ional \c{s}i eventualele progrese \h n aceast\A\ direc\c{t}ie ar
putea contribui la o mai bun\A\ \h n\c{t}elegere a istoriei \c{s}i 
evolu\c{t}iei universului. Pe de alta parte, aspectul informatic al mecanicii
cuantice, care nu a fost mult investigat \h n trecut, cunoa\c{s}te \h n prezent
o perioad\A\ exploziv\A\ de cercet\A ri \h n ideea construirii a\c{s}a-numitelor
``calculatoare cuantice".

\^{I}n domeniul filosofic este de men\c{t}ionat c\A\ \h n mecanica cuantic\A\
exist\A\ paradoxuri faimoase, care \h nc\A\ se men\c{t}in \h n polemic\A\ \c{s}i
care reflect\A\ dificult\A \c{t}ile de logic\A\ pe care le creaz\A\ modul 
de ``g\h ndire cuantic\u{a}" (sau probabilistic\A\ cuantic\A\ ). Unul 
dintre cele mai cunoscute paradoxuri este cel al lui Einstein (care nu a 
acceptat\h n mod total mecanica cuantic\u{a}), Podolsky \c{s}i Rosen (EPR, 1935)
\h n leg\A tur\A\ cu problema dac\A\ exist\A\ sau nu ``elemente adev\A rate de
realitate fizic\u{a}" \h n microcosmosul studiat cu metode cuantice (dup\A\ 
Einstein, mecanica cuantic\A\ interzice existen\c{t}a independent\A\ a actului
de m\A surare de sistemele fizice m\A surate). Alt paradox, de acela\c{s}i rang
de celebritate, este al ``pisicii lui Schr\"odinger". Ceea ce trebuie subliniat
\h n leg\A tur\A\ cu toate aceste puncte teoretice \c{s}i metateoretice delicate
este c\A\ ele genereaz\A\ experimente foarte interesante (cum ar fi, de exemplu,
cele referitoare la a\c{s}a-numita ``teleportare" ale st\A rilor cuantice) care
impulseaz\A\ dezvoltarea tehnologic\A\. 
Ceea ce urmeaz\A\ sunt c\h teva teme introductive \h n mecanica cuantic\A\ 
nerelativist\A\ care au servit ca baz\A\ pentru cursul de graduare \h n mecanica 
cuantic\A\ pe care l-am predat \h n Institutul de Fizic\A\ al 
Universit\A \c{t}ii Statale Guanajuato din Mexic \h n 1998.

\hfill Haret C. Rosu}  

\newpage
%{\scriptsize
%{\small

\section*{\huge 1. POSTULATE CUANTICE} %%%%% 1
%\end{center}
%\author{Martin Gilberto Castro Esparza}
%\date{}
%\maketitle

%\section*{}
Urm\A toarele 6 postulate se pot considera ca baz\A\ pentru teorie \c{s}i
experiment \h n mecanica cuantic\A\ \h n varianta sa cea mai folosit\A\ 
(standard).
\begin{enumerate}
\item[{\bf P1.}-]
Fiec\A\ rei m\A rimi fizice `bine definit\A\ clasic' L \h i corespunde un
operator hermitic $\hat{L}$.
\end{enumerate}
\begin{enumerate}
\item[{\bf P2.}-]
Fiec\A rei st\A ri fizice sta\c{t}ionare \h n care se poate g\A si un sistem
fizic cuantic \h i corespunde o func\c{t}ie de und\A\ normalizat\A\ 
$\psi$ ($\parallel\psi\parallel _{{\cal L}^2}^2=1$).
\end{enumerate}
\begin{enumerate}
\item[{\bf P3.}-]
M\A rimea fizic\A\ L poate s\A\ `ia' experimental numai valorile proprii ale 
$\hat{L}$. De aceea, valorile proprii trebuie s\A\ fie reale, ceea ce are loc
pentru operatori hermitici.
\end{enumerate}
\begin{enumerate}
\item[{\bf P4.}-]
Rezultatul unei m\A sur\A tori pentru determinarea m\A rimii L 
este \h ntotdeauna valoarea medie $\overline{L}$
a operatorului $\hat{L}$
\h n starea $\psi _{n}$, care \h n teorie este elementul de matrice diagonal   

 $<\psi _{n}\mid\hat{L}\mid \psi _{n}>=\overline{L}$.

\end{enumerate}
\begin{enumerate}
\item[{\bf P5.}-]
Elementele de matrice ale operatorilor coordonat\A\ 
\c{s}i moment carteziene
$\widehat{x_{i}}$ \c{s}i $\widehat{p_{k}}$, calculate intre func\c{t}iile de  
und\A\ f \c{s}i g satisfac ecua\c{t}iile de mi\c{s}care Hamilton 
din mecanica clasica \h n forma:\\
$$\frac{d}{dt}<f\mid\widehat{p_{i}}\mid{g}>=-<f\mid\frac{\partial\widehat{H}}
{\partial\widehat{x_{i}}}\mid{g}>$$
$$\frac{d}{dt}<f\mid\widehat{x_{i}}\mid{g}>=<f\mid\frac{\partial\widehat{H}}
{\partial\widehat{p_{i}}}\mid{g}>~,$$

unde $\widehat{H}$ este operatorul hamiltonian, iar derivatele \h n raport cu
operatori se definesc \h n punctul 3 al acestui capitol.
\end{enumerate}
\begin{enumerate}
\item[{\bf P6.}-]
Operatorii $\widehat{p_{i}}$ \c{s}i $\widehat{x_{k}}$ au urm\A torii  
comutatori:
\end{enumerate}
%\begin{center}
%

$$
\qquad \; [\widehat{p_{i}},\widehat{x_{k}}]  =  -i\hbar\delta _{ik},
$$
$$
[\widehat{p_{i}},\widehat{p_{k}}]  =  0, 
$$
$$
[\widehat{x_{i}},\widehat{x_{k}}]  =  0
$$
%\end{eqnarray*}
%\hspace{5mm}

%
\begin{center}
$\hbar=h/2\pi=1.0546\times10^{-27}$ erg.sec.
\end{center}
%\end{enumerate}
%
%%%%%%%%%%%%%%%%%%%%%%%%%%%%%%%%%%%%%%%%%%%%%%%%%%%%%%%%%%%%%%%%%%%%
\begin{enumerate}
\item[1.-]
\underline{Coresponden\c{t}a cu o m\A rime fizic\A\ L care are analog 
clasic $L(x_{i},p_{k})$} 

Aceasta se face substituind $x_{i}$, $p_{k}$ cu $\widehat{x_{i}}$
$\widehat{p_{k}}$. Func\c{t}ia L se presupune c\A\ se poate dezvolta \h n 
serie de puteri pentru orice valoare a argumentelor sale, adic\A\ este 
analitic\A\ . Dac\A\ func\c{t}ia nu con\c{t}ine produse $x_{k}p_{k}$, 
operatorul $\hat{L}$ este hermitic \h n mod automat.\\
Exemplu:
\begin{center}

$T=(\sum_{i}^3p_{i}^2)/2m$ $\longrightarrow$
$\widehat{T}=(\sum_{i}^3\widehat{p}^2)/2m$.\\
\end{center}

Dac\A\ L con\c{t}ine produse mixte $x_{i}p_{i}$ \c{s}i puteri ale acestora, 
$\hat{L}$ nu este hermitic, 
\h n care caz 
L se substituie cu $\hat\Lambda$, partea hermitic\A\ a lui $\hat{L}$ 
($\hat\Lambda$ este un operator autoadjunct).\\
Exemplu:
\begin{center}

$w(x_{i},p_{i})=\sum_{i}p_{i}x_{i}$ $\longrightarrow$             
$\widehat{w}=1/2\sum_{i}^3(\widehat{p_{i}}\widehat{x_{i}}+\widehat
{x_{i}}\widehat{p_{i}})$.\\
\end{center}

Rezult\A\ deasemenea c\A\ timpul nu este un operator 
fiind doar un parametru (care se poate introduce \h n multe feluri). Aceasta 
pentru c\A\ timpul nu depinde de variabilele canonice ci din contr\A\ .

\end{enumerate}
%%%%%%%%%%%%%%%%%%%%%%%%%%%%%%%%%%%%%%%%%%%%%%%%%%%%%%%%%%%%%%
\begin{enumerate}
\item[2.-]
\underline{Probabilitate \h n spectrul discret \c{s}i continuu}

Dac\A\ $\psi_{n}$ este func\c{t}ie proprie a operatorului $\hat{M}$, atunci:\\

$\overline{L}=<n\mid\hat{L}\mid{n}>=<n\mid\lambda_{n}\mid{n}>=
\lambda_{n}<n\mid{n}>=\delta_{nn}\lambda_{n}=\lambda_{n}$.\\

Deasemenea, se poate demonstra c\A\ $\overline{L}^k=(\lambda_{n})^k$.

Dac\A\ func\c{t}ia $\phi$ nu este func\c{t}ie proprie a lui $\hat{L}$ 
se folose\c{s}te dezvoltarea \h n sistem complet de f.p.
ale lui $\hat{L}$ \c{s}i deci:
%
%Fie urm\A toarele defini\c{t}ii:\\
\begin{center}

 $\hat{L}\psi_{n}=\lambda_{n}\psi_{n}$,\hspace{10mm} 
 $\phi=\sum_{n}a_{n}\psi_{n}$\\
\end{center}
combin\h nd aceste dou\A\ rela\c{t}ii  ob\c{t}inem:\\

\begin{center}
 $\hat{L}\phi=\sum_{n}\lambda_{n}a_{n}\psi_{n}$.\\
\end{center}
Putem astfel calcula elementele de matrice
ale operatorului L:\\
\begin{center}
$<\phi\mid\hat{L}\mid{\phi}>=
\sum_{n,m}a_{m}^{\ast}a_{n}\lambda_{n}<m\mid{n}>
=\sum_{m}\mid{a_{m}}\mid^2\lambda_{m}$,\\
\end{center}
ceea ce ne spune c\A\ rezultatul experimentului este $\lambda_{m}$ cu o 
probabilitate $\mid{a_{m}}\mid^2$.\\
Dac\A\ spectrul este discret: de acord cu {\bf P4} \h nseamn\A\ c\A\
$\mid{a_{m}}\mid^2$, deci coeficien\c{t}ii dezvolt\A rii \h ntr-un sistem 
complet de f.p., determin\A\ probabilit\A tile de observare a valorii
proprii $\lambda_{n}$.\\
Dac\A\ spectrul este continuu: folosind urm\A toarea defini\c{t}ie%\\

\begin{center}
$\phi(\tau)=\int{a}(\lambda)\psi{(\tau,\lambda)}d\lambda$,\\
\end{center}
se calculeaz\A\ elementele de matrice pentru spectrul continuu%\\
\begin{center}
$<\phi\mid{\hat{L}}\mid{\phi}>$\\
\end{center}
\begin{center}
$=\int{d}\tau\int{a}^\ast(\lambda)\psi^\ast(\tau,\lambda)d\lambda\int\mu{a}
(\mu)\psi(\tau,\mu)d\mu$\\
\end{center}
\begin{center}
$=\int\int{a}^{\ast} a(\mu)\mu\int\psi^\ast(\tau,\lambda)\psi(tau,\mu)d\lambda
 {d}\mu{d}\tau$\\
\end{center}
\begin{center}
$=\int\int{a}^\ast(\lambda){a}(\mu)\mu\delta(\lambda-\mu){d}\lambda{d}\mu$\\
\end{center}
\begin{center}
$=\int{a}^\ast(\lambda)a(\lambda)\lambda{d}\lambda$\\
\end{center}
\begin{center}
$=\int\mid{a}(\lambda)\mid^2\lambda{d}\lambda$.\\
\end{center}
\^{I}n cazul continuu se spune c\A\ $\mid{a}(\lambda)\mid^2$ este densitatea de 
probabilitate de a observa v.p. $\lambda$ din spectrul continuu. 
Deasemenea, se satisface\\
\begin{center}
 $\overline{L}=<\phi\mid\hat{L}\mid\phi>$.
\end{center}
%\end{enumerate} 

Este comun s\A\ se spun\A\ c\A\ $<\mu\mid \Phi>$ este (reprezentarea lui) 
$\mid \Phi>$ \h n reprezentarea $\mu$, unde $\mid \mu>$ este un vector propriu
al lui $\hat{M}$.
\end{enumerate} 
%%%%%%%%%%%%%%%%%%%%%%%%%%%%%%%%%%%%%%%%%%%%%%%%%%%%%%%%%%%%%%%%%%%%%%%
\newpage
\begin{enumerate}
\item[3.-]
\underline{Defini\c{t}ia unei derivate \h n raport cu un operator}\\
\begin{center}
$\frac{\partial{F(\hat{L})}}{\partial\hat{L}}={\rm lim}_{\epsilon\rightarrow
\infty}
\frac{F(\hat{L}+\epsilon\hat{I})-F(\hat{L})}{\epsilon}.$
\end{center}
\end{enumerate}
 %%%%%%%%%%%%%%%%%%%%%%%%%%%%%%%%%%%%%%%%%%%%%%%%%%%%%%%%%%%%%%%%%%%%%%%
\begin{enumerate}
\item[4.-]\underline{Operatorii de impuls cartezian}

Care este forma concret\A\ a lui $\widehat{p_{1}}$, $\widehat{p_{2}}$ y 
$\widehat{p_{3}}$, dac\A\ argumentele func\c{t}iilor de und\A\ sunt coordonatele 
carteziene $x_{i}$ ?\\
Vom considera urm\A torul comutator:\\
\begin{center}
$[\widehat{p_{i}}, \widehat{x_{i}}^2]= \widehat{p_{i}}\widehat{x_{i}}^2-\widehat{x_{i}}^2
\widehat{p_{i}}$\\
\end{center}
\begin{center}
$= \widehat{p_{i}}\widehat{x_{i}}\widehat{x_{i}}-\widehat{x_{i}}\widehat{p_{i}}
\widehat{x_{i}}+\widehat{x_{i}}\widehat{p_{i}}\widehat{x_{i}}-\widehat{x_{i}}
\widehat{x_{i}}
\widehat{p_{i}}$\\
\end{center}
\begin{center}
$=(\widehat{p_{i}}\widehat{x_{i}}-\widehat{x_{i}}\widehat{p_{i}})\widehat{x_{i}}
+\widehat{x_{i}}(\widehat{p_{i}}\widehat{x_{i}}-\widehat{x_{i}}\widehat{p_{i}})$\\
\end{center}
\begin{center}
$=[\widehat{p_{i}},
\widehat{x_{i}}]\widehat{x_{i}}+\widehat{x_{i}}[\widehat{p_{i}}, \widehat{x_{i}}]$\\
\end{center}
\begin{center}
$=-i\hbar\widehat{x_{i}}-i\hbar\widehat{x_{i}}=-2i\hbar\widehat{x_{i}}$.\\
\end{center}
\^{I}n general, se satisfac:\\

\begin{center}
$\widehat{p_{i}}\widehat{x_{i}}^n-\widehat{x_{i}}^n\widehat{p_{i}}=
-ni\hbar\widehat{x_{i}}^{n-1}.$\\
\end{center}
Atunci, pentru toate func\c{t}iile analitice avem:\\
\begin{center}
$\widehat{p_{i}}\psi(x)-\psi(x)\widehat{p_{i}}=-i\hbar\frac{\partial\psi}
{\partial{x_{i}}}$.\\
\end{center}
Acum, fie $\widehat{p_{i}}\phi=f(x_{1},x_{2},x_{3})$ modul \h n care
ac\c{t}ioneaz\A\ $\widehat{p_{i}}$
asupra lui $\phi(x_{1},x_{2},x_{3})=1$.
Atunci:

 $\widehat{p_{i}}\psi=-i\hbar\frac{\partial\psi}{\partial{x_{1}}}+f_{1}\psi$ 
\c{s}i exist\A\  
rela\c{t}ii analoage pentru $x_{2}$ y $x_{3}$.\\
Din comutatorul $[\widehat{p_{i}},\widehat{p_{k}}]=0$ se ob\c{t}ine 
$\nabla\times\vec{f}=0$,
\c{s}i deci,\\
 $f_{i}=\nabla_{i}F$.\\
Forma cea mai general\A\ a lui $\widehat{p_{i}}$ este
$\widehat{p_{i}}=-i\hbar\frac{\partial}{\partial{x_{i}}}+\frac{\partial{F}}
{\partial{x_{i}}}$, unde F este orice func\c{t}ie.
Func\c{t}ia F se poate elimina folosind o transformare unitar\A\
$\widehat{U}^\dagger=\exp(\frac{i}{\hbar}F)$.\\

\begin{center}
$\widehat{p_{i}}=\widehat{U}^\dagger(-i\hbar\frac{\partial}{\partial{x_{i}}}+
\frac{\partial{F}}{\partial{x_{i}}})\widehat{U}$\\
\end{center}
\begin{center}

$=\exp^{\frac{i}{\hbar}F}(-i\hbar
\frac{\partial}{\partial{x_{i}}}+\frac{\partial{F}}{\partial{x_{i}}})
\exp^{\frac{-i}{\hbar}F}$\\
\end{center}
\begin{center}
$=-i\hbar\frac{\partial}{\partial{x_{i}}}$\\
\end{center}
rezult\h nd  
\hspace{10mm} $\widehat{p_{i}}=-i\hbar\frac{\partial}{\partial{x_{i}}}$ 
$\longrightarrow$ $\widehat{p}=-i\hbar\nabla$.\\

\end{enumerate}
%%%%%%%%%%%%%%%%%%%%%%%%%%%%%%%%%%%%%%%%%%%%%%%%%%%%%%%%%%%%%%%%%%%%%%%%% 5
\begin{enumerate}
\item[5.-]
%%%%%%%%%%%%
\underline{Calculul constantei de normalizare}

Orice func\c{t}ie de und\A\ $\psi(x)$ $\in$ ${\cal L}^2$ de variabil\A\ $x$ se 
poate scrie \h n forma:\\
\begin{center}
$\psi(x)=\int\delta(x-\xi)\psi(\xi)d\xi$\\
\end{center}
\c{s}i se poate considera aceast\A\ expresie ca dezvoltare a lui $\psi$ \h n 
f.p. ale operatorului coordonat\A\ $\hat{x}\delta(x-\xi)=\xi(x-\xi)$.
Atunci, $\mid\psi(x)\mid^2$ este densitatea de probabilitate a coordonatei 
\h n starea $\psi(x)$. De aici rezult\A\ interpretarea normei\\
\begin{center}
$\parallel\psi(x)\parallel^2=\int\mid\psi(x)\mid^2 dx=1$.\\
\end{center}
Intuitiv, aceast\A\ rela\c{t}ie ne spune c\A\ sistemul descris de c\A tre 
func\c{t}ia $\psi(x)$ trebuie s\A\ se g\A seasc\A\
\h ntr-un `loc' pe axa real\u{a}, chiar dac\A\ vom \c{s}ti doar aproximativ 
unde.\\
Func\c{t}iile proprii ale operatorului impuls sunt:\\
$-i\hbar\frac{\partial\psi}{\partial{x_{i}}}=p_{i}\psi$, integr\h nd se
ob\c{t}ine $\psi(x_{i})=A\exp^{\frac{i}{\hbar}p_{i}x_{i}}$, $x$ \c{s}i $p$ 
au spectru continuu \c{s}i deci normalizarea se face cu `func\c{t}ia delta'.\\
Cum se ob\c{t}ine constanta de normalizare ?\\
Se poate ob\c{t}ine utiliz\h nd urm\A toarele transform\A ri Fourier:\\
$f(k)=\int{g(x)}\exp^{-ikx}dx$,\hspace{3mm}$g(x)=\frac{1}{2\pi}\int{f(k)}\exp
^{ikx}dk.$\\
Deasemenea se ob\c{t}ine cu urm\A toarea procedur\A\ :\\
Fie func\c{t}ia de und\A\ nenormalizat\A\ a particulei libere\\
$\phi_{p}(x)=A\exp^{\frac{ipx}{\hbar}}$ \c{s}i formula
\begin{center}
$\delta(x-x^{'})=\frac{1}{2\pi}\int_{-\infty}^{\infty}\exp^{ik(x-x^{'})}dx$\\
\end{center}
Se vede c\A\ \\

\begin{center}
$\int_{-\infty}^{\infty}\phi_{p^{'}}^{\ast}(x)\phi_{p}(x)dx$\\
\end{center}
\begin{center}
$=\int_{-\infty}^{\infty}A^{\ast}\exp^{\frac{-ip^{'}x}{\hbar}}A\exp^{\frac{ipx}
{\hbar}}dx$\\
\end{center}
\begin{center}
$=\int_{-\infty}^{\infty}\mid{A}\mid^2\exp^{\frac{ix(p-p^{'})}{\hbar}}dx$\\
\end{center}
\begin{center}
$=\mid{A}\mid^2\hbar\int_{-\infty}^{\infty}\exp^{\frac{ix(p-p^{'})}{\hbar}}
d\frac{x}{\hbar}$\\
\end{center}
\begin{center}
$=2\pi\hbar\mid{A}\mid^2\delta(p-p^{'})$\\
\end{center}
\c{s}i deci constanta de normalizare este:
\begin{center}              
$A=\frac{1}{\sqrt{2\pi\hbar}}$.\\
\end{center}
Deasemenea rezult\A\ c\A\ f.p. ale operatorului impuls formeaz\A\ un 
sistem complet (\h n sensul cazului continuu) pentru func\c{t}iile de clas\A\  
${\cal L}^2$.\\

\begin{center}
$\psi(x)=\frac{1}{\sqrt{2\pi\hbar}}\int{a(p)}\exp^{\frac{ipx}{\hbar}}dp$\\
\end{center}
\begin{center}
$a(p)=\frac{1}{\sqrt{2\pi\hbar}}\int\psi(x)\exp^{\frac{-ipx}{\hbar}}dx$.\\
\end{center}
Aceste formule stabilesc leg\A tura \h ntre reprezent\A rile x \c{s}i p.

\end{enumerate}
%%%%%%%%%%%%%%%%%%%%%%%%%%%%%%%%%%%%%%%%%%%%%%%%%%%%%%%%%%%%%%%%%%%%%%%
\begin{enumerate}
\item[6.-]
\underline{Reprezentarea p (de impuls)}\\ 

Forma explicit\A\ a 
operatorilor $\hat{p_{i}}$ \c{s}i
$\hat{x_{k}}$ se poate ob\c{t}ine din rela\c{t}iile de comutare, dar \c{s}i 
folosind nucleele\\
\begin{center}

$x(p,\beta)=U^{\dagger}xU=\frac{1}{2\pi\hbar}\int\exp^{\frac{-ipx}{\hbar}}x
\exp^{\frac{i\beta{x}}{\hbar}}dx$\\
\end{center}
\begin{center}
$=\frac{1}{2\pi{\hbar}}\int\exp^{\frac{-ipx}{\hbar}}(-i\hbar\frac{\partial}
{\partial\beta}\exp^{\frac{i\beta{x}}{\hbar}})$.\\
\end{center}
Integrala are forma urm\A toare:
$M(\lambda,\lambda^{'})=\int{U^{\dagger}}(\lambda,x)\widehat{M}U(\lambda^{'},x)
dx$, \c{s}i folosind $\hat{x}f=\int{x}(x,\xi)f(\xi)d\xi$,
ac\c{t}iunea lui $\hat{x}$ asupra lui $a(p)$ $\in$ ${\cal L}^2$ este:\\
\begin{center}
$\hat{x}a(p)=\int{x}(p,\beta)a(\beta)d\beta$\\
\end{center}
\begin{center}
$=\int(\frac{1}{2\pi\hbar}\int
\exp^{\frac{-ipx}{\hbar}}(-i\hbar\frac{\partial}{\partial\beta}\exp^{\frac
{i\beta{x}}{\hbar}})dx)a(\beta)d\beta$\\
\end{center}
\begin{center}
$=\frac{-i}{2\pi}\int\int\exp^{\frac{-ipx}{\hbar}}\frac{\partial}{\partial\beta}
\exp^{\frac{i\beta{x}}{\hbar}}a(\beta)dxd\beta$\\
\end{center}
\begin{center}
$=\frac{-i\hbar}{2\pi}\int\int\exp^{\frac{-ipx}{\hbar}}\frac
{\partial}{\partial\beta}\exp^{\frac{i\beta{x}}{\hbar}}a(\beta)d\frac{x}{\hbar}
d\beta$\\
\end{center}
\begin{center}
$=\frac{-i\hbar}{2\pi}\int\int\exp^{\frac{ix(\beta-p)}{\hbar}}\frac{\partial}
{\partial\beta}a(\beta)d\frac{x}{\hbar}d\beta$\\
\end{center}
\begin{center}
$=-i\hbar\int\frac{\partial{a(p)}}{\partial\beta}\delta(\beta-p)d\beta
=-i\hbar\frac{\partial{a(p)}}
{\partial{p}}$,\\

\end{center}
unde \hspace{15mm}$\delta(\beta-p)=\frac{1}{2\pi}\int\exp^{\frac{ix(\beta-p)}
{\hbar}}d\frac{x}{\hbar}$.\\

Operatorul impuls \h n reprezentarea p se caracterizeaz\A\ prin nucleul:\\

\begin{center}
$p(p,\beta)=\widehat{U}^{\dagger}p\widehat{U}$\\
\end{center}
\begin{center}
$=\frac{1}{2\pi\hbar}\int\exp^{\frac{-ipx}{\hbar}}
(-i\hbar\frac{\partial}{\partial{x}})\exp^{\frac{i\beta{x}}{\hbar}}dx$\\
\end{center}
\begin{center}
$=\frac{1}{2\pi\hbar}\int\exp^{\frac{-ipx}{\hbar}}\beta\exp^{\frac{i\beta{x}}{\hbar}}dx
=\beta\lambda(p-\beta)$ \\
\end{center}
rezult\h nd $\hat{p}a(p)=pa(p)$.\\

Ceea ce se \h nt\h mpl\A\ cu $\hat{x}$ \c{s}i $\hat{p}$ este c\A\
de\c{s}i sunt operatori hermitici pentru toate
f(x) $\in$ $L^2$ nu sunt hermitici exact pentru func\c{t}iile lor proprii.\\
Dac\A\ $\hat{p}a(p)=p_{o}a(p)$ \c{s}i $\hat{x}=\hat{x}^\dagger$ $\hat{p}=
\hat{p}^\dagger$, atunci\\
\begin{center}
$<a\mid\hat{p}\hat{x}\mid{a}>-<a\mid\hat{x}\hat{p}\mid{a}>=-i\hbar<a\mid{a}>$\\
\end{center}
\begin{center}
$p_{o}[<a\mid\hat{x}\mid{a}>-<a\mid\hat{x}\mid{a}>]=-i\hbar<a\mid{a}>$\\
\end{center}
\begin{center}
$p_{o}[<a\mid\hat{x}\mid{a}>-<a\mid\hat{x}\mid{a}>]=0$
\end{center}
Partea st\h ng\A\ este zero, \h n timp ce dreapta este indefinit\A\ , ceea ce este 
o contradic\c{t}ie.
\end{enumerate}
  
%%%%%%%%%%%%%%%%%%%%%%%%%%%%%%%%%%%%%%%%%%%%%%%%%%%%%%%%%%%%%%%%%%%%%%% 7
\begin{enumerate}
\item[7.-]
%%%%%%%%%%%
\underline{Reprezent\A rile Schr\"{o}dinger \c{s}i Heisenberg} \\

Ecua\c{t}iile de mi\c{s}care date prin {\bf P5} au diferite 
interpret\A ri, datorit\A\ faptului c\A\ \h n expresia 
$\frac{d}{dt}<f\mid\hat{L}\mid{f}>$ se poate considera dependen\c{t}a
temporal\A\ ca apar\c{t}in\h nd fie func\c{t}iilor de und\A\ 
fie operatorilor, fie at\h t func\c{t}iilor de und\A\ c\h t \c{s}i
operatorilor. Vom considera numai primele dou\A\ cazuri.\\
\begin{itemize}
\item
Pentru un operator ce depinde de timp $\widehat{O}=\widehat{O(t)}$ avem:\\

\begin{center}
$\hat{p_{i}}=-\frac{\partial\widehat{H}}{\partial\hat{x_{i}}}$,\hspace{5mm}
$\hat{x_{i}}=\frac{\partial\widehat{H}}{\partial\hat{p_{i}}}$\\
\end{center}
\begin{center}
$[\hat{p},f]=\hat{p}f-f\hat{p}=-i\hbar\frac{\partial{f}}{\partial\hat{x_{i}}}$\\
\end{center}
\begin{center}
$[\hat{x},f]=\hat{x}f-f\hat{x}=-i\hbar\frac{\partial{f}}{\partial\hat{p_{i}}}$\\
\end{center}
\c{s}i se ob\c{t}in ecua\c{t}iile de mi\c{s}care Heisenberg:\\

\begin{center}
$\hat{p_{i}}=\frac{-i}{\hbar}[\hat{p},\widehat{H}]$,\hspace{5mm}
$\hat{x_{i}}=\frac{-i}{\hbar}[\hat{x},\widehat{H}]$.
\end{center}

\end{itemize}
\begin{itemize}
\item
Dac\A\ func\c{t}iile de und\A\ depind de timp, \h nc\A\ se poate folosi
$\hat{p_{i}}=\frac{-i}{\hbar}[\hat{p_{i}},\widehat{H}]$, pentru c\A\ este 
o consecin\c{t}\A\ numai a 
rela\c{t}iilor de comutare care nu depind de reprezentare\\

\begin{center}
$\frac{d}{dt}<f\mid\hat{p_{i}}\mid{g}>=\frac{-i}{\hbar}<f\mid[\hat{p},{\widehat{H}}]
\mid{g}>$.\\
\end{center}

Dac\A\ acum $\hat{p_{i}}$ \c{s}i ${\widehat{H}}$ nu depind de timp
\c{s}i se \c{t}ine cont de hermiticitate se ob\c{t}ine:\\

\begin{center}
$(\frac{\partial{f}}{\partial{t}},\hat{p_{i}}g)+(\hat{p_{i}}f,\frac
{\partial{g}}{\partial{t}})$\\
\end{center}
\begin{center}
$=\frac{-i}{\hbar}(f,\hat{p_{i}}\hat{H}g)+\frac
{i}{\hbar}(f,\hat{H}\hat{p_{i}}g)$\\
\end{center}
\begin{center}
$=\frac{-i}{\hbar}(\hat{p}f,\hat{H}g)+\frac{i}{\hbar}(\hat{H}f,\hat{p_{i}}g)$\\
\end{center}

\begin{center}
$(\frac{\partial{f}}{\partial{t}}+\frac{i}{\hbar}\hat{H}f,\hat{p_{i}}g)+
(\hat{p_{i}}f,\frac{\partial{g}}{\partial{t}}-\frac{i}{\hbar}\hat{H}g)=0$\\
\end{center}
Ultima rela\c{t}ie se \h ndepline\c{s}te pentru orice pereche de 
func\c{t}ii $f(x)$ \c{s}i
$g(x)$ \h n momentul ini\c{t}ial dac\A\ fiecare satisface ecua\c{t}ia\\
\begin{center}
$i\hbar\frac{\partial\psi}{\partial{t}}=H\psi$.\\
\end{center}
Aceasta este ecua\c{t}ia Schr\"{o}dinger \c{s}i descrierea sistemului cu
ajutorul operatorilor independen\c{t}i de timp se cunoa\c{s}te ca 
reprezentarea Schr\"{o}dinger.
\end{itemize}
\^{I}n ambele reprezent\A ri evolu\c{t}ia temporal\A\ a sistemului
se caracterizeaz\A\
prin operatorul $\widehat{H}$, care se ob\c{t}ine din func\c{t}ia Hamilton din
mecanica clasic\A\ .\\
Exemplu: $\widehat{H}$ pentru o particul\A\ \h n poten\c{t}ial 
$U(x_{1},x_{2},x_{3})$ este:\\

$\widehat{H}=\frac{\hat{p^2}}{2m}+U(x_{1},x_{2},x_{3})$, \c{s}i \h n
reprezentarea x este:\\
\begin{center}
$\widehat{H}=-\frac{\hbar^{2}}{2m}\nabla ^2_{x}+U(x_{1},x_{2},x_{3})$.
\end{center}         
\end{enumerate} 
%%%%%%%%%%%%%%%%%%%%%%%%%%%%%%%%%%%%%%%%%%%%%%%%%%%%%%%%%%%%%%%%%%%%%  8
\begin{enumerate}
\item[8.-] \underline{Leg\A tura \h ntre reprezent\A rile S \c{s}i H}

{\bf P5} este corect \h n reprezent\A rile Schr\"{o}dinger \c{s}i 
Heisenberg.
De aceea, valoarea medie a oric\A rei observabile coincide \h n cele dou\A\
reprezent\A ri, \c{s}i deci, exist\A\ o transformare unitar\A\ cu care se
poate trece de la o reprezentare la alta.
O astfel de transformare este de forma
$\hat{s}^\dagger=\exp^{\frac{-i\hat{H}t}{\hbar}}$. Pentru a trece la 
reprezentarea Schr\"{o}dinger trebuie folosit\A\ transformata 
Heisenberg $\psi=\hat{s^{\dagger}}f$ cu $f$ \c{s}i $\hat{L}$, \c{s}i
pentru a trece la 
reprezentarea Heisenberg se folose\c{s}te transformarea 
 Schr\"{o}dinger $\hat{\Lambda}=\hat{s^{\dagger}}\hat{L}\hat{s}$ cu $\psi$ 
\c{s}i $\hat{\Lambda}$.
Se poate ob\c{t}ine ecua\c{t}ia Schr\"{o}dinger dup\A\ cum urmeaz\A\ : cum \h n 
transformarea $\psi=\hat{s^{\dagger}}f$ func\c{t}ia $f$ nu depinde de
timp, vom deriva transformarea \h n raport cu timpul pentru a ob\c{t}ine:\\

$\frac{\partial{\psi}}{\partial{t}}=\frac{\partial{s^{\dagger}}}{\partial{t}}f=
\frac{\partial}{\partial{t}}(\exp^{\frac{-i\widehat{H}t}{\hbar}})f=\frac{-i}
{\hbar}\widehat{H}\exp^{\frac{-i\widehat{H}t}{\hbar}}f=\frac{-i}{\hbar}\widehat{H}\hat{s^{\dagger}}f=\frac{-i}{\hbar}\widehat{H}\psi$.\\

\c{s}i deci, avem:\\
\begin{center}
$i\hbar\frac{\partial\psi}{\partial{t}}=\widehat{H}\psi$.\\
\end{center}
\^{I}n continuare ob\c{t}inem ecua\c{t}ia Heisenberg: pun\h nd 
transformarea Schr\"{o}dinger \h n urm\A toarea form\A\ 
$\hat{s}\hat{\Lambda}\hat
{s^{\dagger}}=\hat{L}$ \c{s}i deriv\h nd \h n raport cu timpul se ob\c{t}ine
ecua\c{t}ia Heisenberg\\

\begin{center}
$\frac{\partial\hat{L}}{\partial{t}}=\frac{\partial\hat{s}}{\partial{t}}
\hat{\Lambda}\hat{s^{\dagger}}+\hat{s}\hat{\Lambda}\frac{\partial\hat
{s^{\dagger}}}{\partial{t}}=\frac{i}{\hbar}\widehat{H}\exp^{\frac{i\widehat{H}t}{\hbar}}\hat\Lambda\hat{s^{\dagger}}-\frac{i}{\hbar}\hat{s}\hat\lambda
\exp^{\frac{-i\hat{H}t}{\hbar}}\widehat{H}$\\
\end{center}
\begin{center}
$=\frac{i}{\hbar}(\widehat{H}\hat{s}\hat{\Lambda}
\hat{s^{\dagger}}-\hat{s}\hat{\Lambda}\hat{s^{\dagger}}\widehat{H})=\frac{i}{\hbar}(\widehat{H}\hat{L}-\hat{L}\widehat{H})=\frac{i}{\hbar}[\widehat{H},\hat{L}]$.\\
\end{center}
Prin urmare, avem:\\
\begin{center}
$\frac{\partial\hat{L}}{\partial{t}}=\frac{i}{\hbar}[\widehat{H},\hat{L}]$.\\
\end{center}
Deasemenea, ecua\c{t}ia Heisenberg se poate scrie \h n forma urm\A toare:\\
\begin{center}
$\frac{\partial\hat{L}}{\partial{t}}=\frac{i}{\hbar}\hat{s}[\widehat{H},\hat
{\Lambda}]\hat{s^{\dagger}}$.\\
\end{center}
$\hat{L}$ se cunoa\c{s}te ca integral\A\ de mi\c{s}care dac\A\ $\frac{d}{dt}
<\psi\mid\hat{L}\mid\psi>=0$ \c{s}i se caracterizeaz\A\ prin urm\A torii
comutatori:\\
\begin{center}
$[\widehat{H},\hat{L}]=0$,\hspace{6mm} $[\widehat{H},\hat\Lambda]=0$.

\end{center}
\end{enumerate} 
%%%%%%%%%%%%%%%%%%%%%%%%%%%%%%%%%%%%%%%%%%%%%%%%%%%%%%%%%%%%%%%%%%%%%%%%%  9
\begin{enumerate}
\item[9.-] \underline{St\A ri sta\c{t}ionare}

St\A rile unui sistem descris prin f.p. ale lui $\widehat{H}$ se numesc
st\A ri sta\c{t}ionare ale sistemului, \h n timp ce setul de v.p.  
corespunz\A toare se nume\c{s}te spectrul de energie (spectrul energetic)
al sistemului. \^{I}n astfel de cazuri, ecua\c{t}ia Schr\"{o}dinger este :\\
\begin{center}
$i\hbar\frac{\partial\psi_{n}}{\partial{t}}=E_{n}\psi_{n}=\widehat{H}\psi_{n}$.\\
\end{center}
Solu\c{t}iile sunt de forma: \hspace{11mm}$\psi_{n}(x,t)=\exp^{\frac{-iE_{n}t}{\hbar}}\phi_{n}(x)$.\\
\begin{itemize}
\item
Probabilitatea este urm\A toarea:\\
\begin{center}
$\delta(x)=\mid\psi_{n}(x,t)\mid^2=\mid\exp^{\frac{-iE_{n}t}{\hbar}}\phi_{n}(x)
\mid^2$\\
\end{center}
\begin{center}
$=\exp^{\frac{iE_{n}t}{\hbar}}\exp^{\frac{-iE_{n}t}{\hbar}}\mid\phi_{n}(x)
\mid^2=\mid\phi_{n}(x)\mid^2$.\\
\end{center}
Rezult\A\ c\A\ probabilitatea este constant\A\ \h n timp.
\end{itemize}
\begin{itemize}
\item
\^{I}n st\A rile sta\c{t}ionare, valoarea medie a oric\A rui comutator
 de forma $[\widehat{H},\hat{A}]$ este zero, unde $\hat{A}$ este un operator
arbitrar:\\
\begin{center}
$<n\mid\widehat{H}\hat{A}-\hat{A}\widehat{H}\mid{n}>=<n\mid\widehat{H}\hat{A}\mid{n}>-
<n\mid\hat{A}\widehat{H}\mid{n}>$\\
\end{center}
\begin{center}
$=<n\mid{E_{n}}\hat{A}\mid{n}>-<n\mid\hat{A}E_{n}\mid{n}>$\\
\end{center}
\begin{center}
$=E_{n}<n\mid\hat{A}\mid{n}>-E_{n}<n\mid\hat{A}\mid{n}>=0$.\\
\end{center}
\end{itemize}
\begin{itemize}
\item Teorema de virial (virialului) \h n mecanica cuantic\A\ 
- dac\A\ $\widehat{H}$ este un operator 
hamiltonian al unei particule \h n c\h mpul $U(r)$, folosind\\ 
$\hat{A}=1/2\sum_{i=1}^3(\hat{p_{i}}\hat{x_{i}}-\hat{x_{i}}
\hat{p_{i}})$ se ob\c{t}ine:\\
\begin{center}
$<\psi\mid[\hat{A},\widehat{H}]\mid\psi>=0=<\psi\mid\hat{A}\widehat{H}-\widehat{H}\hat{A}
\mid\psi>$\\
\end{center}
\begin{center}
$=\sum_{i=1}^3<\psi\mid\hat{p_{i}}\hat{x_{i}}\widehat{H}-\widehat{H}\hat{p_{i}}\hat{x_{i}}\mid\psi>$\\
\end{center}
\begin{center}
$=\sum_{i=1}^3<\psi\mid[\widehat{H},\hat{x_{i}}]\hat{p_{i}}+
\hat{x_{i}}[\widehat{H},\hat{p_{i}}]\mid\psi>$.\\
\end{center}
folosind de mai multe ori comutatorii \c{s}i $\hat{p_{i}}
=-i\hbar\nabla _{i}$,
$\hat{H}=\widehat{T}+U(r)$, se poate ob\c{t}ine:\\
\begin{center}
$<\psi\mid[\hat{A},\widehat{H}]\mid\psi>=0$\\
\end{center}
\begin{center}
$=-i\hbar(2<\psi\mid\widehat{T}\mid\psi>-<\psi\mid\vec{r}\cdot\nabla{U(r)}\mid
\psi>)$.\\
\end{center}
Aceasta este teorema virialului. Dac\A\ poten\c{t}ialul este $U(r)=U_{o}r^{n}$, 
atunci se ob\c{t}ine o form\A\ a teoremei de virial ca \h n mecanica clasic\A\ 
cu unica diferen\c{t}\A\ c\A\ se refer\A\ la valori medii\\
\begin{center}
$\overline{T}=\frac{n}{2}\overline{U}$.
\end{center}
\end{itemize}
\begin{itemize}
\item
Pentru un Hamiltonian $\widehat{H}=-\frac{\hbar^2}{2m}\nabla ^{2}+U(r)$ \c{s}i
$[\vec{r},H]=\frac{-i\hbar}{m}\vec{p}$, calcul\h nd elementele de matrice 
se ob\c{t}ine:\\
\begin{center}
$(E_{k}-E_{n})<n\mid\vec{r}\mid{k}>=\frac{i\hbar}{m}<n\mid\hat{p}\mid{k}>$.
\end{center}
\end{itemize}  
\end{enumerate} 
%%%%%%%%%%%%%%%%%%%%%%%%%%%%%%%%%%%%%%%%%%%%%%%%%%%%%%%%%%%%%%%%%%%%%%%% 10
\begin{enumerate}
\item[10.-]\underline{Densitate de curent de probabilitate Schr\"odinger}

Urm\A toarea integral\A\ :\\
\begin{center}
$\int\mid{\psi_{n}}(x)\mid^2dx=1$,\\
\end{center}
este normalizarea unei f.p. din spectrul discret \h n reprezentarea 
de coordonat\A\ , \c{s}i apare ca o condi\c{t}ie asupra mi\c{s}c\A rii
microscopice \h ntr-o
regiune finit\A\ .\\ 
%De accea, st\A rile spectrului discret se numesc st\A ri legate.\\
Pentru f.p. ale spectrului continuu $\psi_{\lambda}(x)$ nu se 
poate da \h n mod direct o interpretare probabilistic\A\ .\\
S\A\ presupunem o func\c{t}ie dat\A\ $\phi$ $\in$ ${\cal L}^2$, 
pe care o scriem ca o combina\c{t}ie
linear\A\ de f.p. \h n continuu:\\
\begin{center}
$\phi=\int{a(\lambda)}\psi_{\lambda}(x)dx.$\\
\end{center}
Se spune c\A\ $\phi$ corespunde unei mi\c{s}c\A ri infinite.\\
\^{I}n multe cazuri, func\c{t}ia $a(\lambda)$ este diferit\A\ de zero numai
\h ntr-o vecin\A tate a unui punct $\lambda=\lambda_{o}$. \^{I}n acest caz 
$\phi$ se cunoa\c{s}te
ca pachet de unde.\\
Vom calcula acum viteza de schimbare a probabilit\A \c{t}ii de a g\A si
sistemul \h n segmentul (volumul) $\Omega$ 
(generalizarea 3D se face trivial \c{s}i se va presupune, chiar 
dac\A\ nota\c{t}ia este 1D).\\
\begin{center}
$P=\int_{\Omega}\mid\psi(x,t)\mid^2dx=\int_{\Omega}\psi^{\ast}(x,t)
\psi(x,t)dx$.\\
\end{center}
Deriv\h nd integrala \h n raport cu timpul g\A sim:\\
\begin{center}
$\frac{dP}{dt}=\int_{\Omega}(\psi\frac{\partial{\psi^{\ast}}}{\partial{t}}+
\psi^{\ast}\frac{\partial{\psi}}{\partial{t}})dx$.\\
\end{center}
Utiliz\h nd ecua\c{t}ia Schr\"{o}dinger \h n integrala din partea dreapt\A\ 
se ob\c{t}ine:\\
\begin{center}
$\frac{dP}{dt}=\frac{i}{\hbar}\int_{\Omega}(\psi\hat{H}\psi^{\ast}-\psi^{\ast}
\hat{H}\psi)dx$.\\
\end{center}
Folosind identitatea $f\nabla ^2{g}-g\nabla ^2{f}=div[(f) grad{(g)}-(g) 
grad{(f)}]$
precum \c{s}i ecua\c{t}ia Schr\"{o}dinger \h n  forma:\\
\begin{center}
$\hat{H}\psi=\frac{\hbar^2}{2m}\nabla ^2{\psi}+V\psi$\\
\end{center}
\c{s}i substituind \h n integral\A\ se ob\c{t}ine:\\

\begin{center}
$\frac{dP}{dt}=\frac{i}{\hbar}\int_{\Omega}[\psi(-\frac{\hbar^2}{2m}\nabla ^2 {\psi
^{\ast}})-\psi^{\ast}(\frac{-\hbar^2}{2m}\nabla ^2{\psi})]dx$\\
\end{center}
\begin{center}
$=-\int_{\Omega}\frac{i\hbar}{2m}(\psi\nabla ^2{\psi^{\ast}}-\psi^{\ast}\nabla ^2
\psi)dx$\\
\end{center}
\begin{center}
$=-\int_{\Omega}div\frac{i\hbar}{2m}(\psi\nabla ^2{\psi^{\ast}}-\psi^{\ast}
\nabla ^2{\psi})dx$.\\
\end{center}
Folosind teorema divergen\c{t}ei pentru a transforma integrala de volum \h n una
de suprafa\c{t}\A\ ob\c{t}inem:\\
\begin{center}
$\frac{dP}{dt}=-\oint\frac{i\hbar}{2m}(\psi\nabla {\psi^{\ast}}-\psi^{\ast}
\nabla {\psi})dx$.\\
\end{center}
M\A rimea $\vec J(\psi)=\frac{i\hbar}{2m}(\psi\nabla{\psi^{\ast}}-
\psi^{\ast}\nabla{\psi})$ se cunoa\c{s}te ca densitate de curent de
probabilitate, pentru care imediat se ob\c{t}ine o ecua\c{t}ie de 
continuitate,\\
\begin{center}
$\frac{d\rho}{dt}+div(\vec J)=0$.\\
\end{center}
\begin{itemize}
\item
Dac\A\ $\psi(x)=AR(x)$, unde R(x) este o func\c{t}ie real\A\ ,
atunci: $\vec J(\psi)=0$.\\
\end{itemize}
\begin{itemize}
\item
Pentru f.p. ale impulsului $\psi(x)=\frac{1}{(2\pi{\hbar})^3/2}
\exp^{\frac{i\vec{p}\vec{x}}{\hbar}}$ se ob\c{t}ine:\\

\begin{center}
$J(\psi)=\frac{i\hbar}{2m}(\frac{1}{(2\pi{\hbar})^3/2}\exp^{\frac
{i\vec{p}\vec{x}}{\hbar}}(\frac{i\vec{p}}{\hbar(2\pi{\hbar})^3/2}\exp
^{\frac{-i\vec{p}\vec{x}}{\hbar}})$\\
\end{center}
\begin{center}
$-(\frac{1}{(2\pi{\hbar})^3/2}\exp^{\frac{-i\vec
{p}\vec{x}}{\hbar}}\frac{i\vec{p}}{\hbar(2\pi{\hbar})^3/2}
\exp^{\frac{i\hbar
\vec{p}\vec{x}}{\hbar}}))$\\
\end{center}
\begin{center}
$=\frac{i\hbar}{2m}(-\frac{2i\vec{p}}{\hbar(2\pi{\hbar})^3})=\frac{\vec{p}}
{m(2\pi{\hbar})^3}$,\\
\end{center}
ceea ce ne indic\A\ o densitatea de curent de probabilitate independent\A\ de 
coordonat\A\ .
\end{itemize}
\end{enumerate} 
%%%%%%%%%%%%%%%%%%%%%%%%%%%%%%%%%%%%%%%%%%%%%%%%%%%%%%%%%%%%%%%%%%%%%%  11
\begin{enumerate}
\item[11.-]\underline{Operator de transport spa\c{t}ial}

Dac\A\ $\widehat{H}$ este invariant la transla\c{t}ii de vector 
arbitrar $\vec{a}$,\\
\begin{center}
$\widehat{H}(\vec{r}+\vec{a})=\widehat{H}\vec{(r)}$~,\\
\end{center}
atunci exist\A\ un $\widehat{T}(\vec{a})$ unitar $\widehat{T}^{\dagger}(\vec
{a})\widehat{H}(\vec{r})\widehat{T}(\vec{a})=\widehat{H}(\vec{r}+\vec{a})$.\\
Din cauza comutativit\A \c{t}ii transla\c{t}iilor
\begin{center}
 $\widehat{T}(\vec{a})\widehat{T}(\vec{b})=
\widehat{T}(\vec{b})\widehat{T}(\vec{a})=\widehat{T}(\vec{a}+\vec{b})$,
\end{center}
rezult\A\ c\A\ $\widehat{T}$ are forma $\widehat{T}=\exp^{i\hat{k}a}$,
unde $\hat{k}=\frac{\hat{p}}{\hbar}$.\\
\^{I}n cazul infinitezimal:
\begin{center}
$\widehat{T}(\delta\vec{a})\widehat{H}\widehat{T}(\delta\vec{a})\approx(\hat{I}+i\hat{k}
\delta\vec{a})\widehat{H}(\hat{I}-i\hat{k}\delta\vec{a})$,
\end{center}
\begin{center}
$\widehat{H}(\vec{r})+i[\hat{K},\widehat{H}]\delta\vec{a}=\widehat{H}(\vec{r})+(\nabla\widehat{H})\delta\vec{a}$.\\
\end{center}
Deasemenea $[\hat{p},\widehat{H}]=0$, unde $\hat{p}$ este o integral\A\ de 
mi\c{s}care.
Sistemul are func\c{t}ii de und\A\ de forma $\psi(\vec{p},\vec{r})=\frac
{1}{(2\pi\hbar)^3/2}\exp^{\frac{i\vec{p}\vec{r}}{\hbar}}$ \c{s}i transformarea 
unitar\A\ face ca 
$\exp^{\frac{i\vec{p}\vec{a}}{\hbar}}\psi(\vec{r})=\psi(\vec{r}+\vec{a})$.
Operatorul de transport spa\c{t}ial $\widehat{T}^\dagger
=\exp^{\frac{-i\vec{p}\vec{a}}
{\hbar}}$ este analogul lui
 $\hat{s}^\dagger=\exp^{\frac{-i\hat{H}t}{\hbar}}$, operatorul de `transport' 
temporal.
\end{enumerate}   
\begin{enumerate}
\item[12.-]\underline{Exemplu: Hamiltonian cristalin}

Dac\A\ $\widehat{H}$ este invariant pentru o transla\c{t}ie discret\A\ 
(de exemplu \h ntr-o re\c{t}ea cristalin\A\ ) 
$\widehat{H}(\vec{r}+\vec{a})=\widehat{H}(\vec{r})$, unde
$\vec{a}=\sum_{i}\vec{a_{i}}n_{i}$, $n_{i}$ $\in$ $N$ \c{s}i $a_{i}$ sunt  
vectorii barici, atunci:
\begin{center}
$\widehat{H}(\vec{r})\psi(\vec{r})=E\psi(\vec{r})$,
\end{center}
\begin{center}
$\widehat{H}(\vec{r}+\vec{a})\psi(\vec{r}+\vec{a})=E\psi(\vec{r}+\vec{a})=\hat{H}
(\vec{r})\psi(\vec{r}+\vec{a})$.
\end{center}
Rezult\A\ c\A\ $\psi(\vec{r})$ \c{s}i $\psi(\vec{r}+\vec{a})$ sunt 
func\c{t}ii de und\A\ pentru aceea\c{s}i v.p. a lui $\widehat{H}$.
Rela\c{t}ia \h ntre $\psi(\vec{r})$ \c{s}i $\psi(\vec{r}+\vec{a})$ se poate
c\A uta \h n forma $\psi(\vec{r}+\vec{a})=\hat{c}(\vec{a})\psi(\vec{r})$ unde
$\hat{c}(\vec{a})$ este o matrice gxg (g este gradul de degenerare al 
nivelului E). Dou\A\ matrici de tip coloan\A\ , $\hat{c}(\vec{a})$ 
 $\hat{c}(\vec{b})$ comut\A\ \c{s}i atunci sunt diagonalizabili 
simultan.\\
\^{I}n plus, pentru elementele diagonale se respect\A\ 
$c_{ii}(\vec{a})c_{ii}(\vec{b})=c_{ii}(\vec{a}+\vec{b})$, i=1,2,....,g, cu
solu\c{t}ii de tipul $c_{ii}(a)=\exp^{ik_{i}a}$. Rezult\A\ c\A\ 
$\psi_{k}(\vec{r})=U_{k}(\vec{r})\exp^{i\vec{k}\vec{a}}$, unde $\vec{k}$
este un vector real arbitrar \c{s}i 
func\c{t}ia $U_{k}(\vec{r})$ este periodic\A\ de
perioad\A\ $\vec{a}$, $U_{k}(\vec{r}+\vec{a})=U_{k}(\vec{r})$.\\
Afirma\c{t}ia c\A\ func\c{t}iile proprii ale unui $\hat{H}$ periodic 
cristalin $\hat{H}(\vec{r}+\vec{a})=\hat{H}(\vec{r})$ se pot scrie 
$\psi_{k}(\vec{r})=U_{k}(\vec{r})\exp{i\vec{k}\vec{a}}$ cu
$U_{k}(\vec{r}+\vec{a})=U_{k}(\vec{r})$ se cunoa\c{s}te ca
 teorema lui Bloch.
\^{I}n cazul continuu, $U_{k}$ trebuie s\A\ fie constant, pentru c\A\ 
o constant\A\ este unica func\c{t}ie periodic\A\ pentru orice 
$\vec{a}$.
Vectorul $\vec{p}=\hbar\vec{k}$ se nume\c{s}te cuasi-impuls 
(prin analogie cu cazul continuu). Vectorul $\vec{k}$ nu este 
determinat \h n mod univoc, 
pentru c\A\ i se poate ad\A uga orice vector $\vec{g}$ pentru care 
$ga=2\pi{n}$ unde n $\in$ N.\\
Vectorul $\vec{g}$ se poate scrie 
$\vec{g}=\sum_{i=1}^{3}\vec{b_{i}}m_{i}$
unde $m_{i}$ sunt numere \h ntregi \c{s}i  $b_{i}$ sunt da\c{t}i de\\
\begin{center}
$\vec{b_{i}}=2\pi\frac{\hat{a_{j}}\times\vec{a_{k}}}{\vec{a_{i}}(\vec{a_{j}}\times\vec
{a_{k}})}$ \\
\end{center}
pentru $i\neq{j}\neq{k}$. $\vec{b_{i}}$ sunt vectorii barici ai 
re\c{t}elei cristaline.\\

\end{enumerate}
\underline{Referin\c{t}e recomandate}

%\noindent 1. Acetatos del Prof. H. Rosu.

\noindent 1. E. Farhi, J. Goldstone, S. Gutmann, ``How probability arises in
quantum mechanics",
Annals of Physics {\bf 192}, 368-382 (1989)

\noindent 2. N.K. Tyagi \h n Am. J. Phys. {\bf 31}, 624 (1963) d\A\ o
demostra\c{t}ie foarte scurt\A\ pentru principiul de incertitudine Heisenberg, 
pe baza c\A ruia se afirm\A\ c\A\ m\A surarea simultan\A\ a doi operatori 
hermitici care nu comut\A\ produce o incertitudine rela\c{t}ionat\A\ cu 
valoarea comutatorului lor.

\noindent 
3. H.N. N\'u\~nez-Y\'epez et al., ``Simple quantum systems in the momentum 
representation", physics/0001030 (Europ. J. Phys., 2000).

\noindent 4. J.C. Garrison, ``Quantum mechanics of periodic systems",
Am. J. Phys. {\bf 67}, 196 (1999).

\noindent 5. F. Gieres, ``Dirac's formalism and mathematical surprises in 
quantum mechanics", quant-ph/9907069 (\h n englez\u{a}); quant-ph/9907070
(\h n francez\u{a}).

\noindent
{\bf 1N. Note}

\noindent
1. Pentru ``crearea mecanicii cuantice...", Werner Heisenberg a fost distins
cu premiul Nobel \h n 1932 (primit \h n  1933). Articolul
``Z\"ur Quantenmechanik. II", [``Asupra mecanicii cuantice.II", 
Zf. f. Physik {\bf 35}, 557-615
(1926) (ajuns la redac\c{t}ie \h n ziua de 16 Noiembrie 
1925) de M. Born, W. Heisenberg \c{s}i
P. Jordan, se cunoa\c{s}te ca ``lucrarea celor trei (oameni)" \c{s}i este
considerat ca cel care a deschis cu adev\A rat vastele orizonturi ale
mecanicii cuantice.

\noindent
2. Pentru ``interpretarea statistic\A\ a func\c{t}iei de und\A\ "
Max Born a primit premiul Nobel \h n 1954.
%\newpage

\bigskip

\section*{{\huge 1P. Probleme}} 
% de mec\'anica cu\'antica:\\}
{\bf Problema 1.1}:
Se consider\A\ doi operatori A \c{s}i B care prin ipotez\A\ comut\u{a}.
\^{I}n acest caz se poate deduce rela\c{t}ia:\\

$e^{A}e^{B}=e^{(A+B)}e^{(1/2[A,B])}$.\hspace{10mm}
%(formula Glauber).\\

\bigskip

{\bf Solu\c{t}ie}

Definim un operator F(t), ca func\c{t}ie de variabil\A\ real\A\ t, prin:
$F(t)=e^{(At)}e^{(Bt)}$.\\
Atunci:
$\frac{dF}{dt}=Ae^{At}e^{Bt}+e^{At}Be^{Bt}=(A+e^{At}Be^{-At})
F(t)$.\\
Acum, aplic\h nd formula $[A,F(B)]=[A,B]F^{'}(B)$, avem \\
$[e^{At},B]=t[A.B]e^{At}$, \c{s}i deci:
$e^{At}B=Be^{At}+t[A,B]e^{At}~,$\\
multiplic\h nd ambele p\A r\c{t}i ale ecua\c{t}iei ultime cu $\exp^{-At}$
\c{s}i substituind \h n prima ecua\c{t}ie, ob\c{t}inem:\\

$\frac{dF}{dt}=(A+B+t[A,B])F(t)$.\\

Operatorii A, B \c{s}i [A,B] comut\A\ prin ipotez\A\ . Deci, putem
integra ecua\c{t}ia diferen\c{t}ial\A\ ca \c{s}i cum 
$A+B$ \c{s}i $[A,B]$ ar fi numere (scalare).\\
Vom avea deci:\\

$F(t)=F(0)e^{(A+B)t+1/2[A,B]t^2}$.\\

Pun\h nd $t=0$, se vede c\A\ $F(0)=1$, \c{s}i :\\

$F(t)=e^{(A+B)t+1/2[A,B]t^2}$.\\

Pun\h nd acum $t=1$, ob\c{t}inem rezultatul dorit.\\

\bigskip

\noindent
{\bf Problema 1.2}:
S\A\ se calculeze comutatorul $[X,D_{x}]$.

\bigskip 

{\bf Solu\c{t}ie}

Calculul se face aplic\h nd comutatorul unei func\c{t}ii
arbitrare $\psi(\vec{r})$:\\
$[X,D_{x}]\psi(\vec{r})=(x\frac{\partial}{\partial{x}}-
\frac{\partial}{\partial{x}}x)\psi(\vec{r})=
x\frac{\partial}{\partial{x}}\psi(\vec{r})-
\frac{\partial}{\partial{x}}[x\psi(\vec{r})]\\
=x\frac{\partial}{\partial{x}}\psi(\vec{r})-
\psi(\vec{r})-x\frac{\partial}{\partial{x}}\psi(\vec{r})=-\psi(\vec{r})$.\\
Cum aceast\A\ rela\c{t}ie se satisface pentru orice $\psi(\vec{r})$, 
se poate deduce c\A\ : $[X,D_{x}]=-1$.         

\bigskip
 
%\newpage
\noindent
{\bf Problema 1.3}:
S\A\ se verifice c\A\ urma de matrice este invariant\A\ la schimb\A ri
de baze ortonormale discrete.

\bigskip

{\bf Solu\c{t}ie}

Suma elementelor diagonale ale unei reprezent\A ri matriceale de un operator
(cuantic) A \h ntr-o baz\A\ arbitrar\A\ nu depinde de baz\A\ .\\
Se va ob\c{t}ine aceast\A\ propietate pentru cazul schimb\A rii 
dintr-o baz\A\ ortonormal\A\ dicret\A\ {$\mid{u_{i}}>$} \h n alta
ortonormal discret\A\ {$\mid{t_{k}}>$}. Avem:\\
$\sum_{i}<u_{i}\mid{A}\mid{u_{i}}>=\sum_{i}<u_{i}
\mid \left(\sum_{k}\mid{t_{k}}><t_{k}\mid \right)A\mid{u_{i}}>$\\

\noindent
(unde s-a folosit rela\c{t}ia de completitudine pentru starea
$t_{k}$). Partea dreapt\A\ este egal\A\ cu:\\

$\sum_{i,j}<u_{i}\mid{t_{k}}><t_{k}\mid{A}\mid{u_{i}}>=\sum_{i,j}
<t_{k}\mid{A}\mid{u_{i}}><u_{i}\mid{t_{k}}>$,\\

\noindent
(este posibil\A\ schimbarea ordinii \h n produsul de dou\A\ numere scalare). 
Astfel, putem \h nlocui 
$\sum_{i}\mid{u_{i}}><u_{i}\mid$ cu unu (rela\c{t}ia de completitudine pentru
st\A rile $\mid{u_{i}}>$), pentru a ob\c{t}ine \h n final:
$$\sum_{i}<u_{i}\mid{A}\mid{u_{i}}>=\sum_{k}<t_{k}\mid{A}\mid{t_{k}}>~.$$
A\c{s}adar, s-a demonstrat proprietatea cerut\A\ de invarian\c{t}\A\ pentru 
urmele matriceale. 

\bigskip

\noindent
{\bf Problema 1.4}:
Dac\A\ pentru operatorul hermitic $N$ exist\A\ operatorii hermitici $L$ \c{s}i
$M$ astfel c\A\ : $[M,N]=0$, $[L,N]=0$, $[M,L]\neq 0$, atunci func\c{t}iile
proprii ale lui $N$ sunt degenerate. 

\bigskip

{\bf Solu\c{t}ie}

Fie $\psi(x;\mu , \nu)$ func\c{t}iile proprii comune ale lui $M$ \c{s}i $N$ 
(fiind de comutator nul sunt observabile simultane). 
Fie $\psi(x;\lambda , \nu)$ func\c{t}iile proprii comune ale lui $L$ \c{s}i $N$ 
(fiind de comutator nul sunt observabile simultane). Parametrii greci indic\A\
valorile proprii ale operatorilor corespunz\A tori. Consider\A m, pentru a 
simplifica, c\A\ $N$ are spectru discret. Atunci:
$$
f(x)=\sum _{\nu}a_{\nu}\psi(x;\mu , \nu)=\sum _{\nu}b_{\nu}
\psi(x;\lambda , \nu)~.
$$
Se calculeaz\A\ acum elementul de matrice $<f|ML|f>$:
$$
<f|ML|f>=\int\sum_{\nu}\mu _{\nu}a_{\nu}\psi ^{*}(x;\mu,\nu)\sum_{\nu ^{'}}
\lambda _{\nu ^{'}}b_{\nu ^{'}}\psi(x;\lambda , \nu ^{'})dx~.
$$
Dac\A\ toate func\c{t}iile proprii ale lui $N$ sunt diferite (nedegenerate)
atunci $<f|ML|f>=\sum _{\nu}\mu _{\nu}a_{\nu}\lambda _{\nu}b_{\nu}$.
Dar acela\c{s}i rezultat se ob\c{t}ine \c{s}i dac\A\ se calculeaz\A\
$<f|LM|f>$ \c{s}i comuatorul ar fi zero. Prin urmare, unele func\c{t}ii 
proprii ale lui $N$ trebuie s\A\ fie degenerate.

%\end{document}

%\newpage
%\vspace{7cm}
%\begin{center} {\huge 2} \end{center}
\newpage
%%%%%%%%%%%%%%%%%%%%%%%%%%%%%%%%%%%%%%%%%%%%%%%%%%%%%%%%%%%%%%%%%%%%
%%%%%%%%%%%%%%%%%%%%%%%%%%%%%%%%%%%%%%%%%%%%%%%%%%%%   Bariere \& Gropi
%%%%%%%%%%%%%%%%%%%%%%%%%%%%%%%%%%%%
%\documentstyle[12pt]{article}
\newcommand{\aple}{\mbox{${}_{\textstyle\sim}^{\textstyle<}$}}
\newcommand{\apge}{\mbox{${}_{\textstyle\sim}^{\textstyle>}$}}
\newcommand{\slsh}[1]{\mbox{$\displaystyle {#1}\!\!\!{/}$}}
\newcommand{\lpr}{\mbox{$ \displaystyle O_L $}}
\newcommand{\rpr}{\mbox{$ \displaystyle O_R $}}
\newcommand{\GeV}{\mbox{$\rm  \, GeV $}}
%\begin{document}
%\date{}
%\title
\section*{\huge 2. BARIERE \c{S}I GROPI RECTANGULARE}      %%%%%%%%%%%%  2
%\end{center}
%\maketitle

%\section*{Comportamentul unei func\c{t}ii de und\A\ sta\c{t}ionare $\psi(x)$}

\subsection*{Regiuni de poten\c{t}ial constant}

\qquad \^{I}n cazul unui poten\c{t}ial rectangular, $V(x)$ este o func\c{t}ie
constant\A\ $V(x)=V$ \h ntr-o regiune oarecare \h n spa\c{t}iu. \^{I}ntr-o 
astfel de regiune, ecua\c{t}ia Schr\"odinger poate fi scris\A\ :
\begin{equation}
\frac{d^2\psi(x)}{dx^2} + \frac{2m}{\hbar^2} (E-V)\psi(x) = 0
\end{equation}
 
Distingem mai multe cazuri:

{\bf (i) $E>V$}

Introducem constanta pozitiv\A\ $k$, definit\A\ prin
\begin{equation}
k = \frac{\sqrt{2m(E-V)}}{\hbar}
\end{equation}

\noindent
Solu\c{t}ia ecua\c{t}iei (1) se poate atunci scrie:
\begin{equation}
\psi(x) = Ae^{ikx} + A'e^{-ikx}
\end{equation}

\noindent
unde $A$ \c{s}i $A'$ sunt constante complexe.

%\newpage
% .................................................

{\bf (ii) $E<V$}

Aceast\A\ condi\c{t}ie corespunde la regiuni de spa\c{t}iu care
ar fi interzise pentru particul\A\ din punctul de vedere al mi\c{s}c\A rii
mecanice clasice. \^{I}n acest caz, introducem constanta
pozitiv\A\ $q$ definit\A\ prin:

\begin{equation}
 q = \frac{\sqrt{2m(V-E)}}{\hbar} 
\end{equation}
\c{s}i solu\c{t}ia lui (1) poate fi scris\A\ :
\begin{equation}
\psi(x) = Be^{q x} + B'e^{-q x}
\end{equation}
unde $B$ \c{s}i $B'$ sunt constante complexe.

{\bf (iii) $E = V$}

\noindent
\^{I}n acest caz special, $\psi(x)$ este o func\c{t}ie linear\A\ de $x$.

\noindent
\subsection*{Comportamentul lui $\psi(x)$ la o
discontinuitate a poten\c{t}ialului}

\qquad S-ar putea crede c\A\ in punctul
$x=x_1$, unde poten\c{t}ialul $V(x)$ este discontinuu,
func\c{t}ia de und\A\ $\psi(x)$ se
comport\A\ mai ciudat, poate c\A\ \h n mod discontinuu, de exemplu. 
Aceasta nu se \h nt\h mpl\A\ : $\psi(x)$ \c{s}i
$\frac{d\psi}{dx}$ sunt continue, \c{s}i numai a doua derivat\A\ prezint\A\ 
discontinuitate \h n $x=x_1$.

\noindent
\subsection*{Viziune general\A\ asupra calculului}

\qquad Procedeul pentru determinarea st\A rii sta\c{t}ionare \h n
poten\c{t}iale rectangulare este deci urm\A torul: \h n toate
regiunile unde $V(x)$ este constant, scriem $\psi(x)$ \h n
oricare dintre cele dou\A\ forme (3) sau (5) \h n func\c{t}ie de 
aplica\c{t}ie; \h n continuare `lipim'
aceste fun\c{t}ii corespunz\A tor cerin\c{t}ei de continuitate pentru $\psi(x)$ 
\c{s}i $\frac{d\psi}{dx}$ \h n punctele unde $V(x)$ este discontinuu.

\noindent
\section*{Examinarea c\h torva cazuri simple}
\qquad S\A\ facem  calculul cantitativ pentru st\A rile
sta\c{t}ionare, conform metodei descrise.

% ========================================================================
\subsection*{Poten\c{t}ial treapt\A\ }

%%%%%%%%%%%%%%
\vskip 2ex
\centerline{
\epsfxsize=280pt
\epsfbox{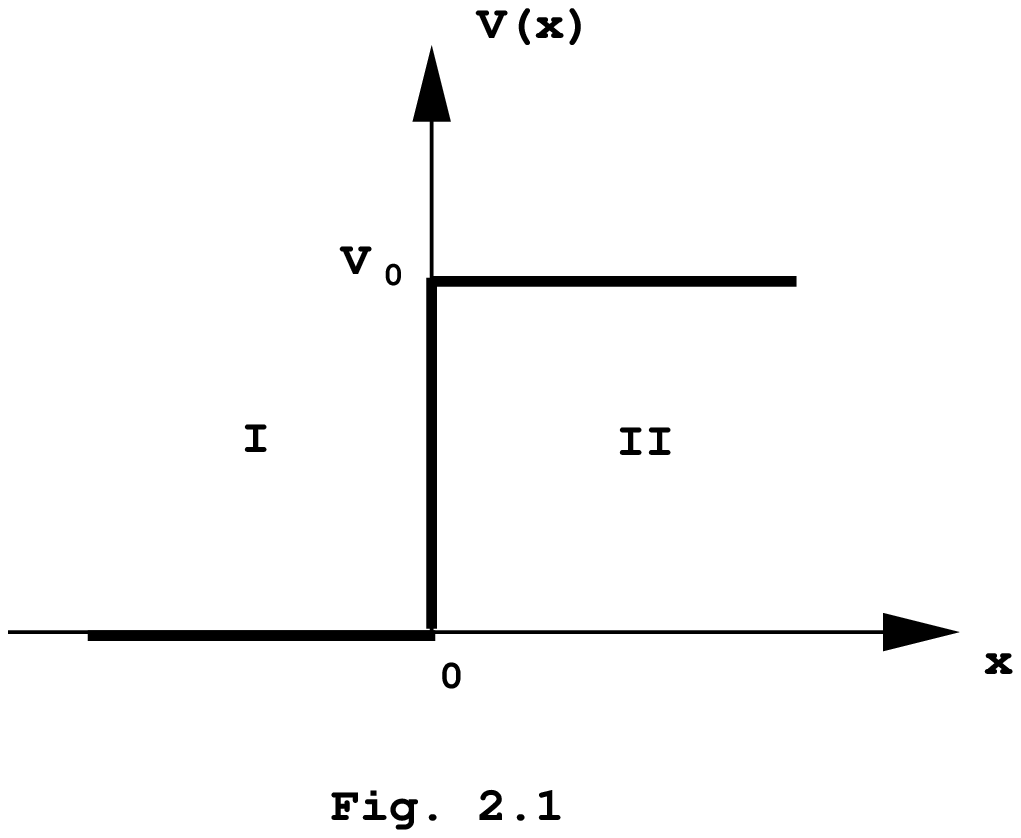}}
\vskip 4ex
%\begin{center}
%{\small{Fig. 1}\\
%}
%\end{center}
%%%%%%%%%%%%%%%%

{\bf a. Cazul $E>V_0$; {\it reflexie par\c{t}ial\A\ }}

S\A\ punem ec. (2) \h n forma:

\begin{eqnarray}
k_1 &=& \frac{\sqrt{2mE}}{\hbar}  \\
\nonumber\\
k_2 &=& \frac{\sqrt{2m(E-V_0)}}{\hbar}
\end{eqnarray}

Solu\c{t}ia ec. (1) are forma din ec. (3) \h n regiunile $I (x<0)$ 
\c{s}i $II (x>0)$:
\begin{eqnarray}
\psi_I &=& A_1e^{ik_1x} + A_1'e^{-ik_1x}  \nonumber
\nonumber\\
\psi_{II} &=& A_2e^{ik_2x} + A_2'e^{-ik_2x} \nonumber
\end{eqnarray}
% .......................................................................
\^{I}n regiunea I ec. (1) ia forma:
\begin{eqnarray}
\psi''(x) + \frac{2mE}{\hbar^2}\psi(x) = \psi''(x) + k^2\psi(x) = 0 \nonumber
\end{eqnarray}

\noindent
iar \h n regiunea II:
\begin{eqnarray}
\psi''(x) - \frac{2m}{\hbar^2} [V_0-E]\phi(x) = \psi''(x) - q^2\psi(x) = 0 \nonumber
\end{eqnarray}
% ......................................................................
Dac\A\ ne limit\A m la cazul unei particule incidente care `vine'
 de la $x=-\infty$, trebuie s\A\ alegem $A_2'=0$ \c{s}i
putem determina raporturile $A_1'/A_1$ \c{s}i $A_2/A_1$. 
Condi\c{t}iile de `lipire' dau atunci:
\begin{itemize}
\item
$\psi_I = \psi_{II}$,\qquad {\mbox \h n} $x=0:$
\begin{equation}
A_1+A_1' = A_2
\end{equation}
\item
$\psi'_I = \psi'_{II}$,\qquad {\mbox \h n} $x=0:$
\begin{equation}
A_1ik_1 - A_1'ik_1 = A_2ik_2
\end{equation}
\end{itemize}

\noindent
Substituind $A_1$ \c{s}i $A_1'$ din (8) \h n (9):
\begin{eqnarray}
A_1' &=& \frac{A_2(k_1 - k_2)}{2k_1} \\
\nonumber\\
A_1 &=& \frac{A_2(k_1 + k_2)}{2k_1}
\end{eqnarray}
\noindent
Egalarea constantei $A_2$ \h n (10) \c{s}i (11) implic\A\ :
\begin{equation}
\frac{A_1'}{A_1} = \frac{k_1 - k_2}{k_1 + k_2}
\end{equation}
\c{s}i din (11) ob\c{t}inem:
\begin{equation}
\frac{A_2}{A_1} = \frac{2k_1}{k_1+k_2}
\end{equation}

\noindent
$\psi(x)$ este o superpozi\c{t}ia de dou\A\ unde. Prima
(termenul \h n $A_1$) corespunde  unei particule incidente, de
moment $p = \hbar k_1$, \h n propagare de la st\h nga la dreapta. A doua
(termenul \h n $A_1'$) corespunde unei particule reflectate,
de impuls $-\hbar k_1$, \h n propagare \h n sens opus. Cum deja am ales
$A_2' = 0$, $\psi_{II}(x)$ con\c{t}ine o singur\A\ und\A\ , care este
asociat\A\ cu o particul\A\ transmis\A\ . (Se va ar\A ta mai departe 
cum este posibil, folosind conceptul de curent de
probabilitate, s\A\ definim coeficientul de transmisie T precum \c{s}i
coeficientul de reflexie R pentru poten\c{t}ialul treapt\A\ ). Ace\c{s}ti
coeficien\c{t}i dau probabilitatea ca o
particul\A\ , sosind de la $x
=-\infty$, ar putea trece de poten\c{t}ialul treapt\A\ \h n $x=0$ sau se 
\h ntoarce. Astfel ob\c{t}inem:
\begin{equation}
R = | \frac{A_1'}{A_1}|^2
\end{equation}

\noindent
iar pentru $T$:
\begin{equation}
T = \frac{k_2}{k_1}| \frac{A_2}{A_1}|^2~.
\end{equation}

\c{T}in\h nd cont de (12) \c{s}i (13), avem:
\begin{eqnarray}
R &=& 1- \frac{4 k_1 k_2}{(k_1 + k_2)^2} \\
\nonumber\\
T &=& \frac{4 k_1 k_2}{(k_1 + k_2)^2}~.
\end{eqnarray}
  
Este u\c{s}or de verificat c\A\ $R+T=1$: 
este deci sigur c\A\ particula va fi transmis\A\ sau reflectat\A\ . 
Contrar predic\c{t}iilor mecanicii clasice, particula incident\A\
are o probabilitate nenul\A\ de a nu se \h ntoarce.

Deasemenea este u\c{s}or de verificat, folosind (6), (7) \c{s}i (17), 
c\A\ dac\A\ $E \gg V_0$ atunci $T \simeq 1$: c\h nd energia particulei este 
suficient de mare \h n compara\c{t}ie cu \h n\A l\c{t}imea treptei, 
pentru particul\A\ este ca \c{s}i cum obstacolul treapt\A\ nu ar exista.

\bigskip

%\newpage
Consider\h nd solu\c{t}ia \h n regiunea I:
\begin{eqnarray}
\psi_I = A_1e^{ik_1x} + A_1'e^{-ik_1x}  \nonumber
\end{eqnarray}

\begin{equation}
j = -\frac{i\hbar}{2m}(\phi^* \bigtriangledown \phi - \phi \bigtriangledown \phi^*)
\end{equation}

cu $A_1 e^{ik_1x}$ \c{s}i conjugata sa $A_1^* e^{-ik_1x}$:
\begin{eqnarray}
j &=& -\frac{i\hbar}{2m}[(A_1^* e^{-ik_1x})(A_1 i k_1 e^{ik_1x})-(A_1 e^{ik_1x})(-A_1^* i k_1 e^{-ik_1x})] \nonumber \\
\nonumber \\
j &=& \frac{\hbar k_1}{m}|A_1|^2 \nonumber
\end{eqnarray}

Acum cu $A'_1 e^{-ik_1x}$ \c{s}i conjugata sa $A_1^* e^{ik_1x}$ rezult\A\ :

\noindent
\begin{eqnarray}
j = -\frac{\hbar k_1}{m}|A'_1|^2~. \nonumber
\end{eqnarray}

Dorim \h n continuare s\A\ verific\A m propor\c{t}ia de curent reflectat
fa\c{t}\A\ de curentul incident (mai precis, dorim s\A\ verific\A m
probabilitatea ca particula s\A\ fie returnat\A\ ):
\begin{eqnarray}
R = \frac{|j(\phi_-)|}{|j(\phi_+)|} = \frac{| -\frac{\hbar k_1}{m}|A'_1|^2|}{| \frac{\hbar k_1}{m}|A_1|^2|} = |\frac{A'_1}{A_1}|^2
\end{eqnarray}

Similar, propor\c{t}ia de transmisie fa\c{t}\A\ de inciden\c{t}\A\
(adic\A\ probabilitatea ca particula s\A\ fie transmis\A\ ) este,
\c{t}in\h nd acum cont de solu\c{t}ia din regiunea II:
\begin{eqnarray}
T = \frac{|\frac{\hbar k_2}{m}|A_2|^2|}{| \frac{\hbar k_1}{m}|A_1|^2|} = \frac{k_2}{k_1}|\frac{A_2}{A_1}|^2
\end{eqnarray}

% ...........................................................................
%\newpage
{\bf a}. Cazul $E<V_0$; {\it reflexie total\A\ }

\^{I}n acest caz avem:
\begin{eqnarray}
k_1 &=& \frac{\sqrt{2mE}}{\hbar}  \\ 
\nonumber\\
q_2 &=& \frac{\sqrt{2m(V_0-E)}}{\hbar}
\end{eqnarray}
\^{I}n regiunea $I (x<0)$, solu\c{t}ia ec. (1) [scris\A\
$\psi(x)'' + k_1^2\psi(x) = 0$] are forma dat\A\ \h n ec. (3):

\begin{equation}
\psi_I = A_1e^{ik_1x} + A_1'e^{-ik_1x} 
\end{equation}
 
\noindent
iar \h n regiunea $II (x>0)$, aceea\c{s}i ec. (1) [acum scris\A\ 
$\psi(x)'' -  q_2^2\psi(x) = 0$] are forma ec. (5):
\begin{equation}
\psi_{II} = B_2e^{q_2x} + B_2'e^{-q_2x}
\end{equation}

\noindent
Pentru ca solu\c{t}ia s\A\ fie men\c{t}inut\A\ finit\A\
c\h nd $x \rightarrow + \infty$, este necesar ca:
\begin{equation}
B_2 = 0
\end{equation}
Condi\c{t}iile de `lipit' \h n $x=0$ dau \h n acest caz:

\begin{itemize}
\item
$\psi _I = \psi_{II}$,\qquad {\mbox \h n} $x=0:$
\begin{equation}
A_1 + A_1' = B_2'
\end{equation}
\item
$\psi'_I = \psi'_{II}$,\qquad {\mbox \h n} $x=0:$
\begin{equation}
A_1 ik_1 - A_1' ik_1 = - B_2' q_2
\end{equation}
\end{itemize}

\noindent
Substituind $A_1$ \c{s}i $A_1'$ din (26) \h n (27):
\begin{eqnarray}
A_1' &=& \frac{B_2'(i k_1 + q_2)}{2i k_1} \\
\nonumber\\
A_1 &=& \frac{B_2'(i k_1 -  q_2)}{2i k1}
\end{eqnarray}
\noindent
Egalarea constantei $B_2'$ \h n (28) \c{s}i (29) duce la:
\begin{equation}
\frac{A_1'}{A_1} = \frac{i k_1 + q_2}{i k_1 - q_2} = \frac{k_1 - iq_2}{k_1 + iq_2}, 
\end{equation}
astfel c\A\ din (29) avem:
\begin{equation}
\frac{B_2'}{A_1} = \frac{2i k_1}{i k_1 - q_2} =\frac{2 k_1}{k_1 - iq_2} 
\end{equation}

\noindent
Coeficientul de reflexie $R$ este deci:
\begin{equation}
R = | \frac{A_1'}{A_1}|^2 = | \frac{k_1 - i q_2}{k_1 + i q_2}|^2 = \frac{k_1^2 + q_2^2}{k_1^2 + q_2^2} = 1   
\end{equation}
\noindent
Ca \h n mecanica clasic\A\ , microparticula este \h ntotdeauna reflectat\A\
(reflexie total\A\ ). Totu\c{s}i, exist\A\ o diferen\c{t}\A\ important\A\ : 
datorit\A\ 
existen\c{t}ei a\c{s}a-numitei unde evanescente $e^{-q_2x}$, particula are o
probabilitate nenul\A\ de a se g\A si `prezent\A\ ' 
\h n regiunea din spa\c{t}iu care este clasic interzis\A\ .
Aceast\A\ probabilitate descre\c{s}te exponen\c{t}ial
cu $x$ \c{s}i ajunge s\A\ fie neglijabil\A\ c\h nd $x$ dep\A \c{s}e\c{s}te
``zona'' $1/q_{2}$ corespunz\A toare undei evanescente. S\A\ not\A m
\c{s}i c\A\ $A_1'/A_1$ este complex. O anumit\A\ diferen\c{t}\A\ de faz\A\ 
apare din cauza
reflexiei, care, fizic, se datoreaz\A\ faptului c\A\ 
particula este ` fr\h nat\u{a}' (\h ncetinit\A\ c\h nd intr\A\ \h n regiunea $x>0$. 
Nu exist\A\ analogie pentru aceasta \h n mecanica clasic\A\ , ci doar \h n 
optica fizic\A\ .

% =============================================================================
\subsection*{Poten\c{t}iale tip barier\A\ }

%%%%%%%%%%%%%%
\vskip 2ex
\centerline{
\epsfxsize=280pt
\epsfbox{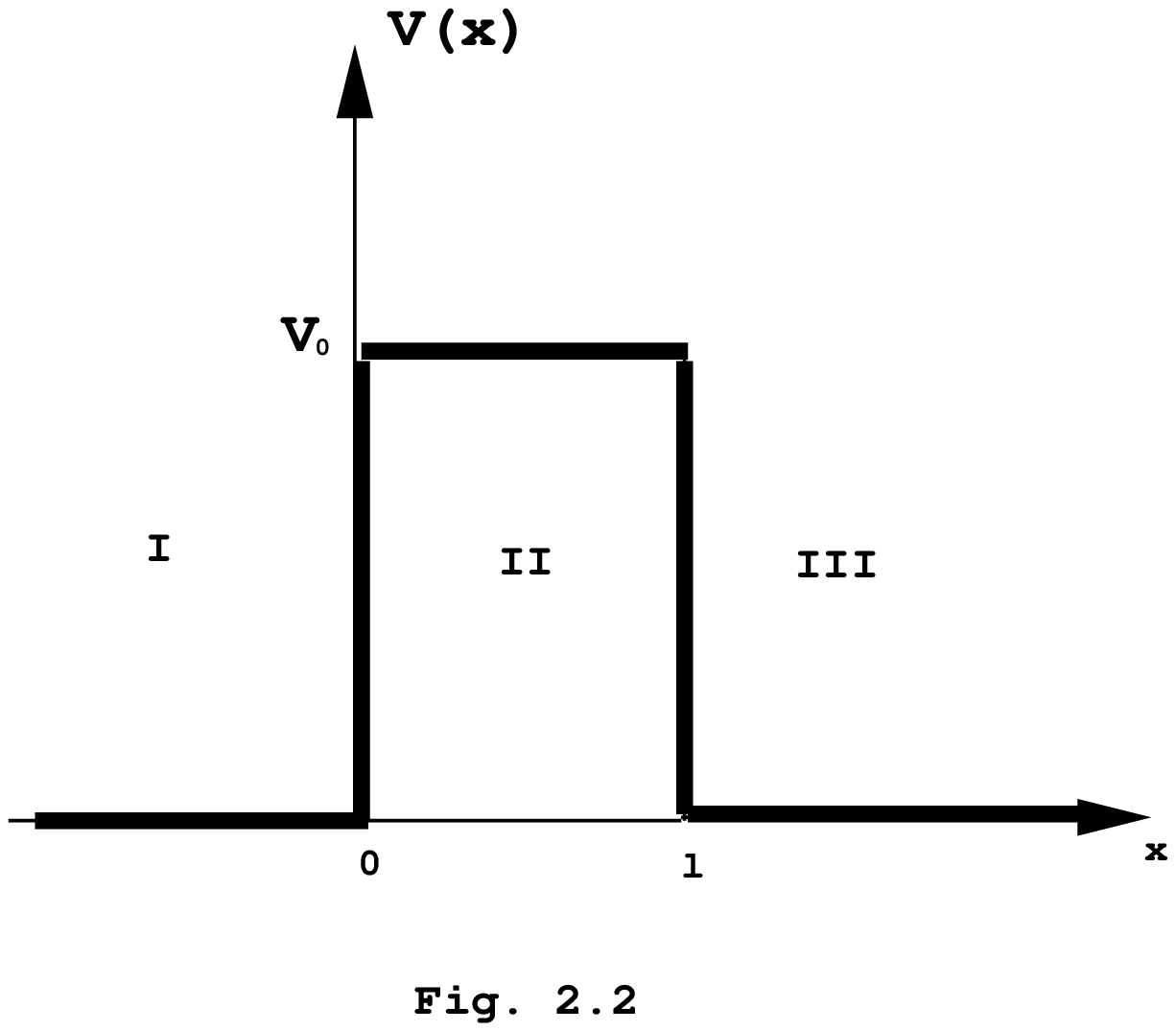}}
\vskip 4ex
%\begin{center}
%{\small{Fig. 1}\\
%}
%\end{center}
%%%%%%%%%%%%%%%%

{\bf a. Cazul $E>V_0$}; {\it rezonan\c{t}e}

S\A\ punem aici ec. (2) \h n forma:
\begin{eqnarray}
k_1 &=& \frac{\sqrt{2mE}}{\hbar}  \\
\nonumber\\
k_2 &=& \frac{\sqrt{2m(E-V_0)}}{\hbar}
\end{eqnarray}

Solu\c{t}ia ec. (1) este ca \h n ec. (3) \h n regiunile 
$I (x<0)$, $II (0<x<a$) \c{s}i $III (x>a):$
\begin{eqnarray}
\psi_I &=& A_1e^{ik_1x} + A_1'e^{-ik_1x} \nonumber
\nonumber\\
\psi_{II} &=& A_2e^{ik_2x} + A_2'e^{-ik_2x}\nonumber
\nonumber\\
\psi_{III} &=& A_3e^{ik_1x} + A_3'e^{-ik_1x}\nonumber   
\end{eqnarray}

Dac\A\ ne limit\A m la cazul unei particule incidente care vine 
de la $x=-\infty$, trebuie s\A\ alegem $A_3'=0$. 
% ......................................................

\begin{itemize}
\item
$\psi_I = \psi_{II}$,\qquad {\mbox \h n} $x=0:$
\begin{equation}
A_1 + A_1' = A_2 + A_2'
\end{equation}
\item
$\psi'_I = \psi'_{II}$,\qquad {\mbox \h n} $x=0:$
\begin{equation}
A_1ik_1 - A_1'ik_1 = A_2ik_2 - A_2'ik_2
\end{equation}
\item
$\psi_{II} = \psi_{III}$,\qquad {\mbox \h n} $x=a:$
\begin{equation}
A_2e^{ik_2a} + A_2'e^{-ik_2a} = A_3e^{ik_1a} 
\end{equation}
\item
$\psi'_{II} = \psi'_{III}$,\qquad {\mbox \h n} $x=a:$ 
\begin{equation}
A_2ik_2e^{ik_2a} - A_2'ik_2e^{-ik_2a} = A_3ik_1e^{ik_1a} 
\end{equation}
\end{itemize}

\noindent
Condi\c{t}iile de continuitate \h n $x=a$ dau $A_2$ \c{s}i $A_2'$ \h n 
func\c{t}ie de $A_3$, \c{s}i cele din $x=0$ dau $A_1$ \c{s}i $A_1'$ \h n 
fun\c{t}ie de $A_2$ \c{s}i $A_2'$ (\c{s}i deci \h n func\c{t}ie de $A_3$). 
Acest procedeu este ar\A tat \h n continuare.

\noindent 
Substituind $A_2'$ din ec. (37) \h n (38):
\begin{equation}
A_2 = \frac{A_3e^{ik_1a}(k_2+k_1)}{2k_2e^{ik_2a}}
\end{equation}

\noindent
Substituind $A_2$ din ec. (37) \h n (38):
\begin{equation}
A_2' = \frac{A_3e^{ik_1a}(k_2-k_1)}{2k_2e^{-ik_2a}}
\end{equation}
\noindent
Substituind $A_1$ din ec. (35) \h n (36):
\begin{equation}
A_1' = \frac{A_2(k_2-k_1)-A_2'(k_2+k_1)}{-2k_1}
\end{equation}

\noindent
Substituind $A_1'$ din ec. (35) \h n (36):
\begin{equation}
A_1 = \frac{A_2(k_2+k_1)-A_2'(k_2-k_1)}{2k_1}
\end{equation}

\noindent
Acum, substituind \h n (41) ecua\c{t}iile (39) \c{s}i (40), avem:
\begin{equation}
A_1' = i \frac{(k_2^2 - k_1^2)}{2 k_1 k_2} (\sin k_2a) e^{ik_1a}A_3
\end{equation}

\noindent
\^{I}n final, substituind \h n (42) ecua\c{t}iile (39) \c{s}i (40):
\begin{equation}
A_1 = [\cos k_2a - i\frac{k_1^2 + k_2^2}{2 k_1 k_2} \sin k_2a] e^{ik_1a}A_3
\end{equation}
$A_1'/A_1$ \c{s}i $A_3/A_1$ [raporturi care se ob\c{t}in egal\h nd
ecua\c{t}iile (43) \c{s}i (44), 
\c{s}i respectiv separ\h nd \h n ec. (44)] ne permit calculul 
coeficientului de reflexie $R$ precum \c{s}i a celui de transmisie $T$ 
pentru acest caz simplu de barier\A\ , fiind ace\c{s}tia 
da\c{t}i de:
\begin{equation}
R = |A_1'/A_1|^2 = \frac{(k_1^2 - k_2^2)^2\sin^2 k_2a}{4k_1^2k_2^2 + (k_1^2-k_2^2)^2 \sin^2 k_2a},
\end{equation}
\begin{equation}
T=|A_3/A_1|^2=\frac{4k_1^2k_2^2}{4k_1^2 k_2^2 + (k_1^2 - k_2^2)^2 \sin^2 k_2a},
\end{equation}

\noindent 
Acum este u\c{s}or de verificar c\A\ $R + T = 1$.

% .......................................................................

{\bf b. Cazul $E<V_0$; {\it efectul tunel}}

\qquad Acum, fie ecua\c{t}iile (2) \c{s}i (4):
\begin{eqnarray}
k_1 &=& \frac{\sqrt{2mE}}{\hbar}  \\
\nonumber\\
q_2 &=& \frac{\sqrt{2m(V_0 - E)}} {\hbar}
\end{eqnarray}

Solu\c{t}ia ec. (1) are forma ec. (3) \h n regiunile $I (x<0)$ \c{s}i $III 
(x>a)$, \h n timp ce \h n regiunea $II (0<x<a$) are forma ec. (5):
\begin{eqnarray}
\psi_I &=& A_1e^{ik_1x} + A_1'e^{-ik_1x}\nonumber
\nonumber\\
\psi_{II} &=& B_2e^{q_2x} + B_2'e^{-q_2x}\nonumber
\nonumber\\
\psi_{III} &=& A_3e^{ik_1x} + A_3'e^{-ik_1x}\nonumber   
\end{eqnarray}

Condi\c{t}iile de `lipit' \h n $x=0$ \c{s}i $x=a$ ne permit calculul
coeficientului de transmisie al barierei. De fapt, nu este necesar a efectua 
\h nc\A\ odata calculul: este suficient de a face substitu\c{t}ia, 
\h n ecua\c{t}ia ob\c{t}inut\A\ \h n primul caz din aceast\A\ sec\c{t}iune
$k_2$ cu $-i q_2$.

\newpage

% ======================================================================
\subsection*{St\A ri legate \h n groap\A\ rectangular\A\ }
{\bf a. Groap\A\ de ad\h ncime finit\A\ }

%%%%%%%%%%%%%%
\vskip 2ex
\centerline{
\epsfxsize=280pt
\epsfbox{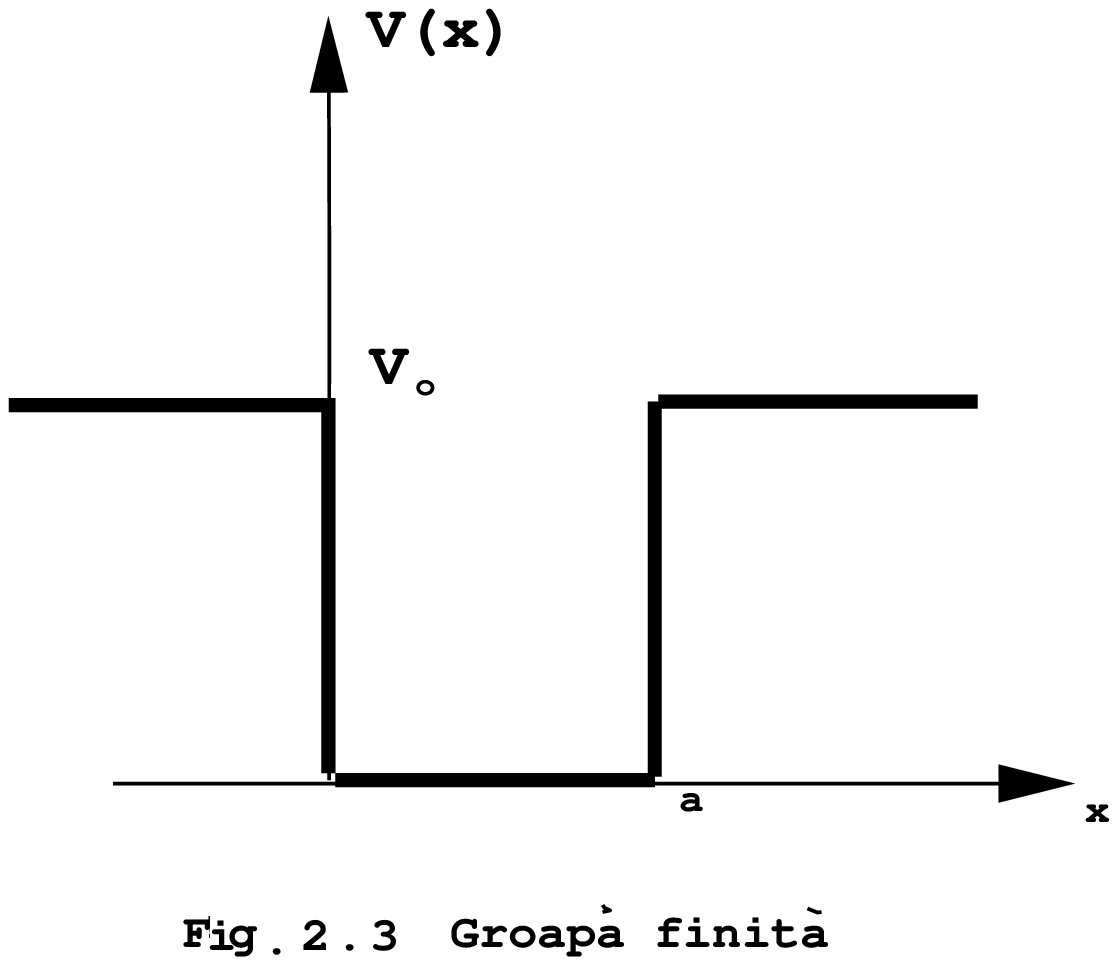}}
\vskip 4ex
%\begin{center}
%{\small{Fig. 1}\\
%}
%\end{center}
%%%%%%%%%%%%%%%%

\^{I}n aceast\A\ parte ne limit\A m la studiul cazului $0<E<V_0$ 
(cazul $E>V_0$ este identic calculului din sec\c{t}iunea precedent\A\ , 
``barier\A\ de poten\c{t}ial''.

Pentru regiunile exterioare I $(x<0)$ \c{s}i III $(x>a)$ folosim ec. (4):

\begin{equation}
q = \frac{\sqrt{2m(V_0-E)}}{\hbar}
\end{equation}

Pentru regiunea central\A\  II $(0<x<a)$ folosim ec. (2):

\begin{equation}
k = \frac{\sqrt{2m(E)}}{\hbar}
\end{equation}

Solu\c{t}ia ec. (1) are forma ec. (5) \h n  regiunile
exterioare \c{s}i \h n forma din ec. (3) \h n regiunea central\A\ :
\begin{eqnarray}
\psi_I &=& B_1e^{q x} + B_1'e^{-q x}\nonumber
\nonumber\\
\psi_{II} &=& A_2e^{ikx} + A_2'e^{-ikx}\nonumber
\nonumber\\
\psi_{III} &=& B_3e^{q x} + B_3'e^{-q x} \nonumber   
\end{eqnarray}

\^{I}n regiunea $(0<x<a)$ ec. (1) are forma:
\begin{equation}
\psi(x)'' + \frac{2mE}{\hbar^2}\psi(x) = \psi(x)'' + k^2\psi(x) = 0
\end{equation}

\noindent
\c{s}i \h n regiunile exterioare:
\begin{equation}
\psi(x)'' - \frac{2m}{\hbar^2} [V_0-E]\phi(x) = \psi(x)'' - q^2\psi(x) = 0
\end{equation}

Pentru c\A\  $\psi$ trebuie s\A\ fie finit\A\ \h n regiunea I, trebuie s\A\
avem:
\begin{equation}
B_1'=0
\end{equation}
Condi\c{t}iile de lipire dau:

$\psi_I = \psi_{II}$,\qquad \h n $x=0:$
\begin{equation}
B_1 = A_2 + A'_2
\end{equation}

$\psi'_I = \psi'_{II}$,\qquad \h n $x=0:$
\begin{equation}
B_1 q = A_2ik - A'_2ik
\end{equation}

$\psi_{II} = \psi_{III}$,\qquad \h n $x=a:$
\begin{equation}
A_2e^{ika} + A'_2e^{-ika} = B_3e^{q a} + B'_3e^{-q a}
\end{equation}

$\psi'_{II} = \psi'_{III}$,\qquad \h n $x=a:$
\begin{equation}
A_2ike^{ika} - A'_2ike^{-ika} = B_3q e^{q a} - B'_3q e^{-q a}
\end{equation}

Substituind constantele $A_2$ \c{s}i $A'_2$ din ec. (54) \h n ec. (55) 
ob\c{t}inem, respectiv:
%                .........................................................
\begin{eqnarray}
A'_2 &=& \frac{B_1(q-ik)}{-2ik}\nonumber
\nonumber\\
A_2 &=& \frac{B_1(q+ik)}{2ik}
\end{eqnarray}

Substituind constanta $A_2$ \c{s}i constanta $A'_2$ din ec. (56) \h n ec. (57) 
ob\c{t}inem, respectiv:
\begin{eqnarray}
B'_3e^{-q a}(ik + q) + B_3e^{q a}(ik-q) + A'_2e^{-ika}(-2ik) &=& 0\nonumber
\nonumber\\
2ikA_2e^{ika} + B'_3e^{-q a}(-ik+q) + B_3E^{q a}(-ik-q) &=& 0
\end{eqnarray}

\noindent
Egal\h nd $B'_3$ din ecua\c{t}iile (59) \c{s}i \c{t}in\h nd cont de
ecua\c{t}iile (58):
\begin{equation}
\frac{B_3}{B_1} = \frac{e^{-q a}}{4ikq}[e^{ika}(q+ik)^2 - e^{-ika}(q - ik)^2]
\end{equation}

\^{I}ns\A\  $\psi(x)$ trebuie s\A\ fie finit\A\ \c{s}i \h n regiunea III. 
Prin urmare, este necesar ca $B_3=0$, \c{s}i deci:
\begin{equation}
[\frac{q - ik}{q + ik}]^2 = \frac{e^{ika}}{e^{-ika}} = e^{2ika}
\end{equation}

Deoarece $q$ \c{s}i $k$ depind de $E$, ec. (1) poate fi satisf\A\ cut\A\
pentru anumite valori ale lui $E$. 
Condi\c{t}ia ca $\psi(x)$ s\A\ fie finit\A\ \h n toate regiunile
spa\c{t}iale impune cuantizarea energiei. \c{S}i mai precis dou\A\ cazuri sunt
posibile:

{\bf (i) dac\A\ :}
\begin{equation}
\frac{q - ik}{q + ik} = - e^{ika}
\end{equation}

\noindent
Egal\h nd \h n ambii membri partea real\A\
\c{s}i cea imaginar\A\ ,respectiv, rezult\A\ :
\begin{equation}
\tan(\frac{ka}{2}) =\frac{q}{k} 
\end{equation}
Pun\h nd:
\begin{equation}
k_0 = \sqrt{\frac{2mV_0}{\hbar}} = \sqrt{k^2 + q^2}
\end{equation}
ob\c{t}inem:
\begin{equation}
\frac{1}{\cos^2(\frac{ka}{2})} = 1 + \tan^2(\frac{ka}{2}) = \frac{k^2 + q^2}{k^2} = (\frac{k_0}{k})^2
\end{equation}

Ec.(63) este astfel echivalent\A\ cu sistemul de ecua\c{t}ii:

% &&&&&&&&&&&&&&&&&&&&&&&&&&&&&&&&&&&&&&&&&&&&&&&&&&&&&&&&&&&&&&&&
% \[
% \left\{
% \begin{array}{ll}
% |\cos(ka/2)|= k/k_0&\mbox{}\\
% \tan(ka/2)>0&\mbox{}
% \end{array}
% \right.
% \]
% &&&&&&&&&&&&&&&&&&&&&&&&&&&&&&&&&&&&&&&&&&&&&&&&&&&&&&&&&&&&&&&
\begin{eqnarray}
|\cos(\frac{ka}{2})| &=& \frac{k}{k_0}
\nonumber\\
\tan(\frac{ka}{2}) &>& 0   
\end{eqnarray}

Nivelele de energie sunt determinate de c\A tre intersec\c{t}ia unei 
linii drepte de \h nclinare $\frac{1}{k_0}$ cu primul set de cosinusoide 
\h ntrerupte \h n figura 2.4. Astfel ob\c{t}inem 
un num\A r de nivele
de energie, ale c\A ror func\c{t}ii de und\A\ sunt pare. Acest lucru devine
mai clar dac\A\ substituim (62) \h n (58) \c{s}i (60). Este u\c{s}or
de verificat c\A\ $B'_3 = B_1$ \c{s}i 
$A_2 = A'_2$, astfel c\A\ $\psi(-x) =\psi(x)$.

{\bf (ii) dac\A\ :}
\begin{equation}
\frac{q - ik}{q + ik} = e^{ika}
\end{equation}
Un calcul de acela\c{s}i tip ne duce la:
\begin{eqnarray}
|\sin(\frac{ka}{2})| &=& \frac{k}{k_0}
\nonumber\\
\tan(\frac{ka}{2}) &<& 0   
\end{eqnarray}
 
Nivelele de energie sunt \h n acest caz determinate de
c\A tre intersec\c{t}ia 
aceleia\c{s}i linii drepte cu al doilea set de cosinusoide \h ntrerupte 
\h n figura 2.4. Nivelele astfel ob\c{t}inute se afl\A\ \h ntre cele
g\A site \h n (i). Se poate ar\A ta u\c{s}or c\A\ func\c{t}iile de und\A\ 
corespunz\A toare sunt impare.

%%%%%%%%%%%%%%
\vskip 2ex
\centerline{
\epsfxsize=280pt
\epsfbox{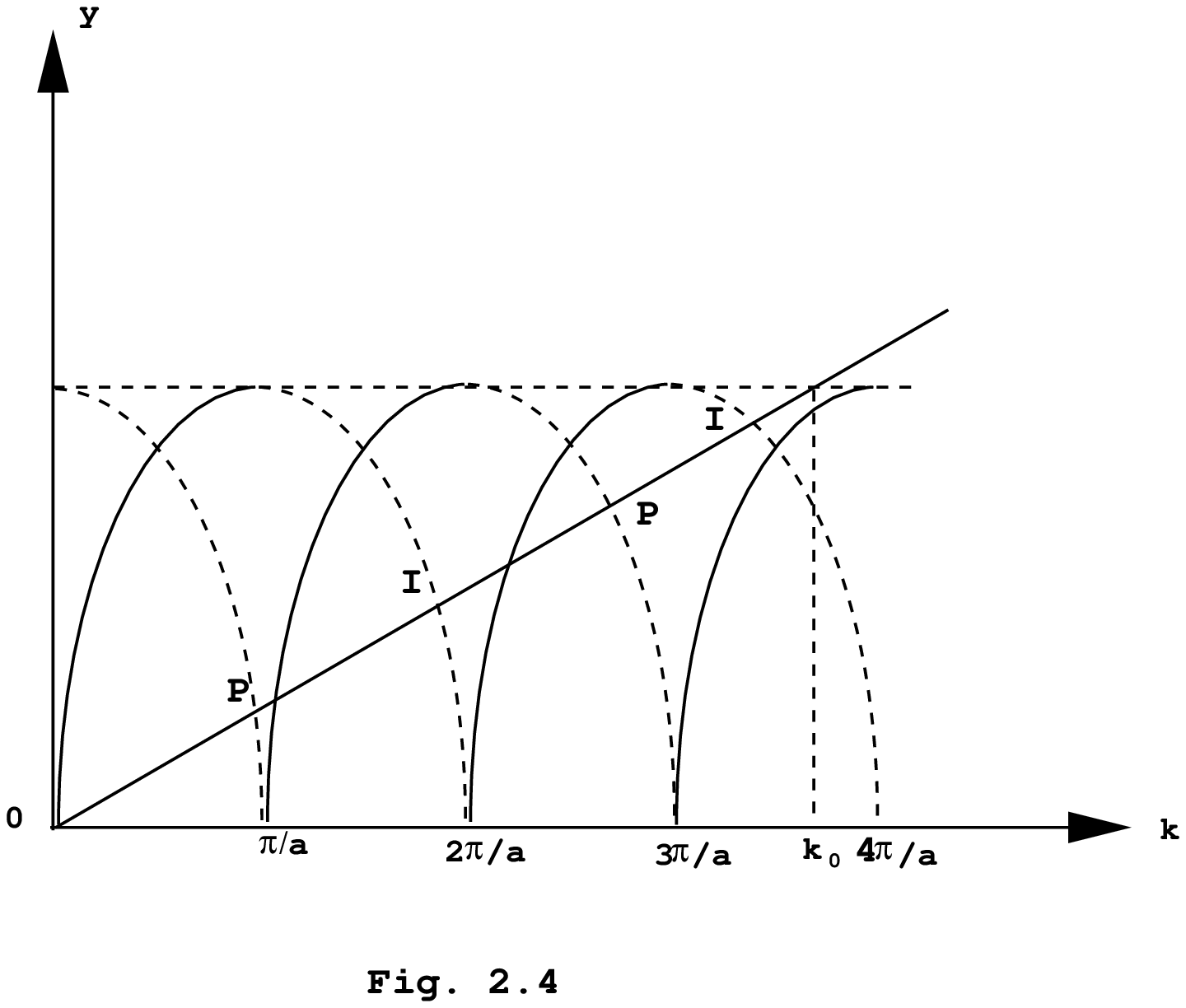}}
\vskip 4ex
%\begin{center}
%{\small{Fig. 1}\\
%}
%\end{center}
%%%%%%%%%%%%%%%%

{\bf b. Groap\A\ de ad\h ncime infinit\A\ }

\noindent
\^{I}n acest caz este convenabil s\A\ se pun\A\ $V(x)$ zero pentru $0<x<a$ 
\c{s}i infinit \h n tot restul axei. 
Pun\h nd:
\begin{equation}
k = \sqrt{\frac{2mE}{\hbar^2}}
\end{equation}
$\psi(x)$ trebuie s\A\ fie zero \h n afara intervalului $[0,a]$, \c{s}i
continu\A\ \h n $x=0$, 
c\h t \c{s}i \h n $x=a$.

\noindent
Acum, pentru $0 \leq x \leq a$:
\begin{equation}
\psi(x) = Ae^{ikx} + A'e^{-ikx}
\end{equation}
Pentru c\A\ $\psi(0)=0$, se poate deduce c\A\ $A' = -A$, ceea ce ne conduce la:
\begin{equation}
\psi(x) = 2iA\sin(kx)
\end{equation}
\^{I}n plus $\psi(a)=0$, astfel c\A\ :
\begin{equation}
k = \frac{n\pi}{a}
\end{equation}
unde $n$ este un \h ntreg pozitiv arbitrar. Dac\A\ normaliz\A m 
func\c{t}ia (71), \c{t}in\h nd cont de (72), atunci ob\c{t}inem func\c{t}iile
de und\A\ sta\c{t}ionare:
\begin{equation}
\psi_n(x) = \sqrt{\frac{2}{a}}\sin(\frac{n\pi x}{a})
\end{equation}
cu energiile:
\begin{equation}
E_n = \frac{n^2\pi^2\hbar^2}{2ma^2}
\end{equation}
Cuantizarea nivelelor de energie este prin urmare extrem de simpl\A\ \h n acest 
caz: energiile sta\c{t}ionare sunt propor\c{t}ionale cu p\A tratele numerelor
naturale.
% ----------------------------------------------------------

\bigskip

%\newpage
\section*{{\huge 2P. Probleme}}

\vspace*{4mm}

\subsection*{Problema 2.1: Poten\c{t}ialul Delta atractiv}

S\A\ presupunem c\A\ avem un poten\c{t}ial de forma:

\begin{eqnarray}
V(x) = -V_0 \delta(x);\qquad  V_0 > 0; \qquad x \in \Re. \nonumber 
\end{eqnarray}
Func\c{t}ia de und\A\ $\psi(x)$ corespunz\A toare se presupune continu\A\ .
 
%{\bf Problem\A\ }.
% 
a) S\A\ se ob\c{t}in\A\ st\A rile legate ($E<0$), dac\A\ exist\A\ , localizate \h n acest
tip de poten\c{t}ial.

b) S\A\ se calculeze dispersia unei unde plane care `cade' pe acest 
poten\c{t}ial \c{s}i s\A\ se ob\c{t}in\A\ {\it  coeficientul de reflexie}
\begin{eqnarray}
R = \frac{|\psi _{refl}|^2}{|\psi _{inc}|^2}|_{x=0} \nonumber
\end{eqnarray}
unde $\psi _{refl}$, $\psi _{inc}$ sunt unda reflectat\A\ \c{s}i respectiv
cea incident\A\ .

\noindent
{\it Sugestie}: Pentru a evalua comportamentul lui $\psi(x)$ \h n x=0, 
se recomand\A\ integrarea ecua\c{t}iei Schr\"odinger \h n intervalul 
($-\varepsilon ,+\varepsilon$), dup\A\ care se aplic\A\ limita 
$\varepsilon$ $\rightarrow$ $0$.

{\bf Solu\c{t}ie.} a) Ecua\c{t}ia Schr\"odinger este:
\begin{equation}
\frac{d^2\psi(x)}{dx^2} + \frac{2m}{\hbar^2} (E+V_0 \delta(x))\psi(x) = 0
\end{equation}
Departe de origine avem o ecua\c{t}ie diferen\c{t}ial\A\ de forma
\begin{equation}
\frac{d^2}{dx^2} \psi (x) = - \frac{2mE}{\hbar^2}\psi(x).
\end{equation}
Func\c{t}iile de und\A\ sunt prin urmare de forma
\begin{equation}
\psi (x) = Ae^{-q x} + Be^{q x} \qquad {\rm pentru} \qquad x>0 \qquad 
{\rm sau} \qquad x<0,
\end{equation}
cu $q = \sqrt{-2mE/ \hbar^2}$ $ \in\Re.$ Cum $|\psi|^2$ trebuie s\A\ fie
integrabil\A\ , nu putem accepta o
parte care s\A\ creasc\A\ exponen\c{t}ial. 
\^{I}n plus func\c{t}ia de und\A\ trebuie s\A\ fie continu\A\ \h n origine. 
Cu aceste condi\c{t}ii,
\begin{eqnarray}
\psi(x) &=& Ae^{q x}; \qquad (x<0), \nonumber
\nonumber\\
\psi(x) &=& Ae^{-q x}; \qquad (x>0).
\end{eqnarray} 
%
%\end{document}
\^{I}ntegr\h nd ecua\c{t}ia Schr\"odinger \h ntre $-\varepsilon$ \c{s}i 
$+\varepsilon$, ob\c{t}inem
\begin{equation}
-\frac{\hbar^2}{2m}[\psi'(\varepsilon)-\psi'(-\varepsilon)] - V_0\psi(0) = E\int^{+\varepsilon} _{-\varepsilon} \psi(x)dx \approx 2\varepsilon E\psi(0)
\end{equation}
Introduc\h nd acum resultatul (78) \c{s}i \c{t}in\h nd cont
de limita $\varepsilon \rightarrow 0$, avem
\begin{equation}
-\frac{\hbar^2}{2m}(-q A-q A) - V_0 A = 0
\end{equation}
sau $E = -m(V_0^2 / 2\hbar^2)$ [$-\frac{V_{0}^2}{4}$ \h n 
unit\A \c{t}i $\frac{\hbar ^2}{2m}$]. \^{I}n mod clar exist\A\ o singur\A\
energie discret\A\ . Constanta de normalizare se g\A se\c{s}te c\A\ este 
$A = \sqrt{mV_0/ \hbar^2}$. Func\c{t}ia de und\A\ a st\A rii legate se
ob\c{t}ine $\psi _{o}=Ae^{V_0|x|/2}$, cu $V_0$ \h n unit\A \c{t}i 
$\frac{\hbar ^2}{2m}$.
% .........
\begin{eqnarray}
\nonumber
\end{eqnarray} 
% .........
b) Func\c{t}ia de und\A\ pentru o und\A\  plan\A\ este dup\A\ cum se \c{s}tie
\begin{equation}
\psi(x) = A e^{ikx}, \qquad k^2 = \frac{2mE}{\hbar^2}~. 
\end{equation}
Se mi\c{s}c\A\ de la st\h nga la dreapta \c{s}i se reflect\A\ 
\h n poten\c{t}ial. Dac\A\ $B$ sau $C$ este amplitudinea undei reflectate 
sau transmise, respectiv, avem
\begin{eqnarray}
\psi(x) &=& Ae^{ikx} + Be^{-ikx}; \qquad (x<0), \nonumber
\nonumber\\
\psi(x) &=& Ce^{ikx}; \qquad \qquad \qquad (x>0).
\end{eqnarray} 

Condi\c{t}iile de continuitate \c{s}i rela\c{t}ia 
$\psi'(\varepsilon)-\psi'(-\varepsilon) = - f\psi(0)$ cu $f = 2mV_0 / \hbar^2$ 
produc
\begin{eqnarray}
A + B &=& C \qquad  \qquad  \qquad B = -\frac{f}{f+2ik}A,     \nonumber
\nonumber\\
ik(C - A + B) &=& -fC \qquad  \qquad  C = \frac{2ik}{f+2ik}A.
\end{eqnarray} 
Coeficientul de reflexie cerut este prin urmare 
\begin{eqnarray}
R = \frac{|\psi_{refl}|^2}{|\psi_{inc}|^2}|_{x=0} = \frac{|B|^2}{|A|^2} = \frac{m^2V_0^2}{m^2V_0^2 + \hbar^4k^2}.
\end{eqnarray}
Dac\A\ poten\c{t}ialul este extrem de puternic 
($V_0 \rightarrow \infty$) se vede c\A\ $R \rightarrow 1$, adic\A\ unda este 
reflectat\A\ \h n totalitate.

{\it Coeficientul de transmisie}, pe de alt\A\ parte, este
\begin{eqnarray}
T = \frac{|\psi_{trans}|^2}{|\psi_{inc}|^2}|_{x=0} = \frac{|C|^2}{|A|^2} 
= \frac{\hbar^4 k^2}{m^2V_0^2 + \hbar^4k^2}.
\end{eqnarray} 
Dac\A\  poten\c{t}ialul este foarte puternic 
($V_0 \rightarrow \infty$) atunci $T \rightarrow 0$, adic\A\ , unda 
transmis\A\ cade rapid de cealalt\A\ parte a poten\c{t}ialului.

Evident, $R + T = 1$ cum era de a\c{s}teptat.

% ...................................................................................
\newpage
\subsection*{Problema 2.2:
Particul\A\ \h ntr-o groap\A\ de poten\c{t}ial finit\A\ 1D }

%{\bf Problema.}
S\A\ se rezolve ecua\c{t}ia Schr\"odinger unidimensional\A\ 
pentru o groap\A\ de poten\c{t}ial finit\A\ descris\A\ prin condi\c{t}iile 
\[
V(x) = \left\{
\begin{array}{ll}
-V_0&\mbox{dac\A\ $|x| \leq a$}\\
0&\mbox{dac\A\ $|x|>a$~.}
\end{array}
\right.
\]

S\A\ se considere numai st\A rile legate ($E<0$).

%%%%%%%%%%%%%%
\vskip 2ex
\centerline{
\epsfxsize=280pt
\epsfbox{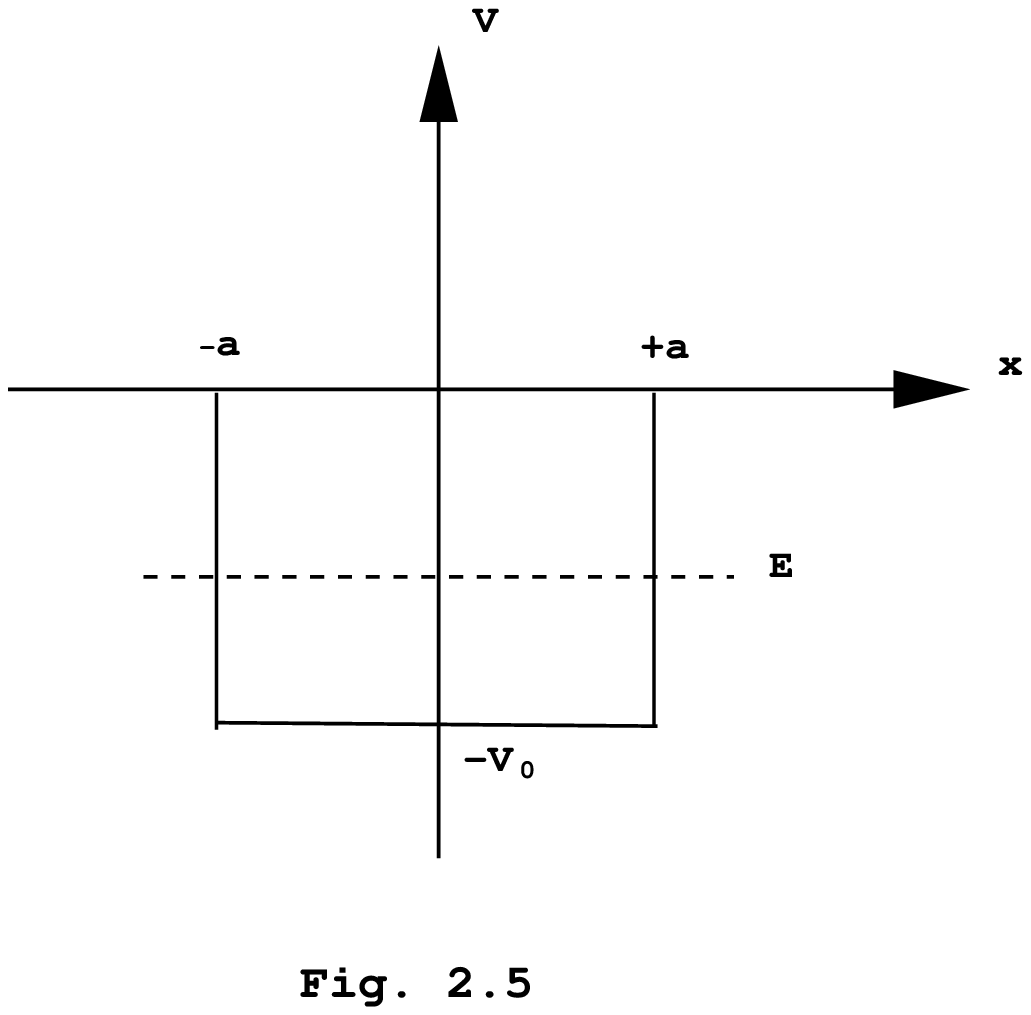}}
\vskip 4ex
%\begin{center}
%{\small{Fig. 1}\\
%}
%\end{center}
%%%%%%%%%%%%%%%%

%\vspace*{75mm}

\noindent
{\bf Solu\c{t}ie.} 

a) Func\c{t}ia de und\A\ pentru $|x|<a$ \c{s}i $|x|>a$.
%%%%%%%%%%%%%%%%%%%%%%%%%%%%%%%%%%%%%%%%%%%%%%%%%%%%%%%% 

Ecua\c{t}ia Schr\"odinger corespunz\A toare este
\begin{equation}
-\frac{\hbar^2}{2m}\psi ^{''}(x) + V(x)\psi(x) = E\psi(x)~.
\end{equation}
Definim
\begin{equation}
q^2 = -\frac{2mE}{\hbar^2}, \qquad  k^2 = \frac{2m(E+V_0)}{\hbar^2}
\end{equation}
%
%\newpage

\noindent
\c{s}i ob\c{t}inem:
\begin{eqnarray}
{\rm 1)~ pentru ~~~x<-a:}\qquad \psi ^{''}_{1}(x) - q^2 \psi _1 &=& 0,\ 
\psi _1 = A_1e^{q x} + B_1e^{-q x};\nonumber
\nonumber\\
{\rm 2)~ pentru ~-a\leq x\leq a:}~\psi ^{''}_{2}(x) + k^2 \psi _2 &=& 0, 
\ \psi _2 = A_2 \cos(kx) + B_2 \sin(kx); \nonumber
\nonumber\\
{\rm 3)~ pentru ~~~x>a:~~}\qquad \psi ^{''}_{3}(x) - q^2 \psi _3 &=& 0,\ 
\psi _3 = B_3 e^{q x} + B_3 e^{-q x}. \nonumber   
\end{eqnarray}

b) Formularea condi\c{t}iilor de frontier\A\ . 
%%%%%%%%%%%%%%%%%%%%%%%%%%%%%%%%%%%%%%%%%%%%%%%

\noindent
Normalizarea 
st\A rilor legate cere ca solu\c{t}ia s\A\ fie zero la infinit. 
Aceasta \h nseamn\A\ c\A\ $B_1=A_3=0$. \^{I}n plus, $\psi(x)$ trebuie s\A\
fie continuu diferen\c{t}iabil\A\ . Toate solu\c{t}iile
particulare sunt fixate \h n a\c{s}a fel \h nc\h t $\psi$ 
precum \c{s}i prima sa derivat\A\ $\psi'$ sunt continue \h n 
acea valoare a lui x corespunz\h nd frontierei \h ntre 
zona interioar\A\ \c{s}i cea exterioar\A\ . 
A doua derivat\A\ $\psi''$ con\c{t}ine saltul (discontinuitatea)
impus de c\A tre poten\c{t}ialul particular de tip `cutie' al acestei 
ecua\c{t}ii Schr\"odinger. Toate acestea ne conduc la
\begin{eqnarray}
\psi_1(-a) &=& \psi_2(-a),\qquad  \psi_2(a) = \psi_3(a), \nonumber
\nonumber\\
\psi'_1(-a) &=& \psi'_2(-a),\qquad  \psi'_2(a) = \psi'_3(a).
\end{eqnarray} 

c) Ecua\c{t}iile de valori proprii. 
%%%%%%%%%%%%%%%%%%%%%%%%%%%%%%%%%%%%

Din (88) ob\c{t}inem patru ecua\c{t}ii
lineare \c{s}i omogene pentru coeficien\c{t}ii $A_1$, $A_2$, $B_2$ \c{s}i $B_3$:
\begin{eqnarray}
A_1 e^{-qa} &=& A_2\cos(ka) - B_2\sin(ka), \nonumber
\nonumber\\
qA_1 e^{-qa} &=& A_2k\sin(ka) + B_2k\cos(ka),  \nonumber
\nonumber\\
B_3 e^{-qa} &=& A_2\cos(ka)  + B_2\sin(ka), \nonumber
\nonumber\\
-qB_3 e^{-qa} &=& -A_2k\sin(ka) + B_2k\cos(ka).
\end{eqnarray} 

%\newpage

\noindent
Adun\h nd \c{s}i sc\A z\h nd ob\c{t}inem un sistem de ecua\c{t}ii mai u\c{s}or
de rezolvat:
\begin{eqnarray}
 (A_1+B_3) e^{-qa} &=& 2A_2\cos(ka) \nonumber
\nonumber\\
q(A_1+B_3) e^{-qa} &=& 2A_2k\sin(ka) \nonumber
\nonumber\\
(A_1-B_3)  e^{-qa} &=& -2B_2\sin(ka) \nonumber
\nonumber\\
q(A_1-B_3) e^{-qa} &=&  2B_2k\cos(ka).
\end{eqnarray} 
Dac\A\ se presupune c\A\ $A_1+B_3 \neq 0$ \c{s}i $A_2 \neq 0$, primele dou\A\ 
ecua\c{t}ii dau
\begin{equation}
q = k\tan(ka).
\end{equation}
care pus \h n ultimele dou\A\ d\A\
\begin{equation}
A_1 = B_3; \qquad B_2 = 0.
\end{equation}
De aici, ca rezultat, ob\c{t}inem o solu\c{t}ie simetric\A\ ,
$\psi(x) = \psi(-x)$, sau de {\it paritate pozitiv\A\ }.

Un calcul practic identic ne duce pentru $A_1 - B_3 \neq 0$ \c{s}i 
pentru $B_2 \neq 0$
la
\begin{equation}
q = -k\cot(ka) \qquad y \qquad A_1 = -B_3; \qquad A_2 = 0. 
\end{equation}
Func\c{t}ia de und\A\ astfel ob\c{t}inut\A\ este antisimetric\A\ , 
corespunz\h nd unei parit\A \c{t}i {\it negative}.

d) Solu\c{t}ie calitativ\A\ a problemei de valori proprii.
%%%%%%%%%%%%%%%%%%%%%%%%%%%%%%%%%%%%%%%%%%%%%%%%%%%%%%%%%% 

Ecua\c{t}ia care leag\A\ 
$q$ \c{s}i $k$, pe care am ob\c{t}inut-o deja, d\A\ condi\c{t}ii pentru
autovalorile de energie. Folosind forma scurt\A\
\begin{equation}
\xi = ka, \qquad \eta = qa,
\end{equation}
ob\c{t}inem din defini\c{t}ia (87)
\begin{equation}
\xi^2 + \eta^2 = \frac{2mV_0a^2}{\hbar^2} = r^2.
\end{equation}
Pe de alt\A\ parte, folosind (91) \c{s}i (93) ob\c{t}inem ecua\c{t}iile
\begin{eqnarray}
\eta = \xi \tan(\xi), \qquad \eta = -\xi\cot(\xi). \nonumber
\end{eqnarray}
Astfel autovalorile de energie c\A utate pot fi ob\c{t}inute
construind intersec\c{t}ia acestor dou\A\ curbe cu cercul
definit de (95), \h n planul $\xi$-$\eta$ (vezi figura 2.6).

%%%%%%%%%%%%%%
\vskip 2ex
\centerline{
\epsfxsize=280pt
\epsfbox{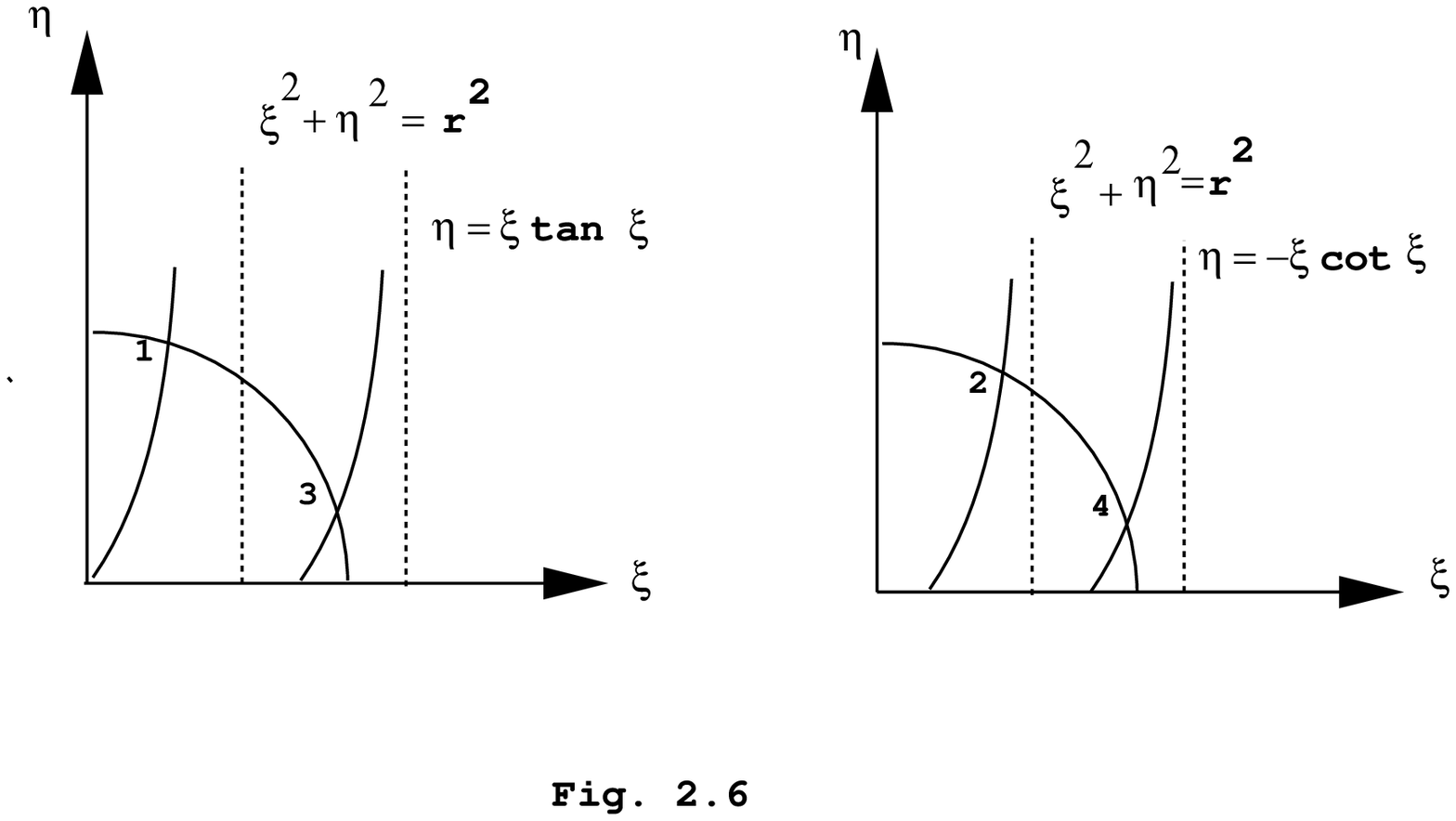}}
\vskip 4ex
%\begin{center}
%{\small{Fig. 1}\\
%}
%\end{center}
%%%%%%%%%%%%%%%%

%\vspace*{70mm}

Cel pu\c{t}in o solu\c{t}ie exist\A\ pentru valori arbitrare ale
parametrului $V_0$, \h n cazul parit\A \c{t}ii pozitive, pentru c\A\ 
func\c{t}ia tangent\A\ intersecteaz\A\ originea. Pentru paritatea negativ\A\ , 
raza cercului trebuie s\A\ fie mai mare dec\h t o anumit\A\ valoare 
minim\A\ astfel c\A\ cele dou\A\ curbe s\A\ se poat\A\ 
intersecta. Poten\c{t}ialul trebuie s\A\ aib\A\ o anumit\A\ 
ad\h ncime rela\c{t}ionat\A\ cu o scal\A\ spa\c{t}ial\A\ dat\A\ 
$a$ \c{s}i o scal\A\ de mas\A\ dat\A\ $m$,
pentru a permite o solu\c{t}ie de paritate negativ\A\ . Num\A rul de nivele
de energie cre\c{s}te cu $V_0$, $a$ \c{s}i masa $m$. Pentru cazul \h n care
$mVa^2 \rightarrow \infty$, intersec\c{t}iile se afl\A\ din
\begin{eqnarray}
\tan(ka) &=& \infty \qquad \longrightarrow \qquad  ka=\frac{2n-1}{2}\pi, \nonumber
\nonumber\\
-\cot(ka) &=& \infty \qquad \longrightarrow \qquad ka = n \pi, 
\end{eqnarray} 
unde $n=1,2,3, \, \ldots $

\noindent
sau combin\h nd:
\begin{equation}
k(2a) = n \pi.
\end{equation}
Pentru spectrul de energie acest lucru \h nseamn\A\ c\A\
\begin{equation}
E_n = \frac{\hbar^2}{2m}(\frac{n \pi}{2a})^2 - V_0.
\end{equation}
L\A rgind groapa de poten\c{t}ial \c{s}i/sau masa particulei $m$, 
diferen\c{t}a \h ntre dou\A\  autovalori de energie vecine va descre\c{s}te. 
Nivelul cel mai de jos ($n=1$) nu este localizat \h n $-V_0$, ci un pic
mai sus. Aceast\A\  diferen\c{t}\A\ se nume\c{s}te {\it energia de punct zero}.

e) Formele func\c{t}iilor de und\A\ se arat\A\ \h n figura 2.7 pentru 
solu\c{t}iile discutate . 

%%%%%%%%%%%%%%
\vskip 2ex
\centerline{
\epsfxsize=280pt
\epsfbox{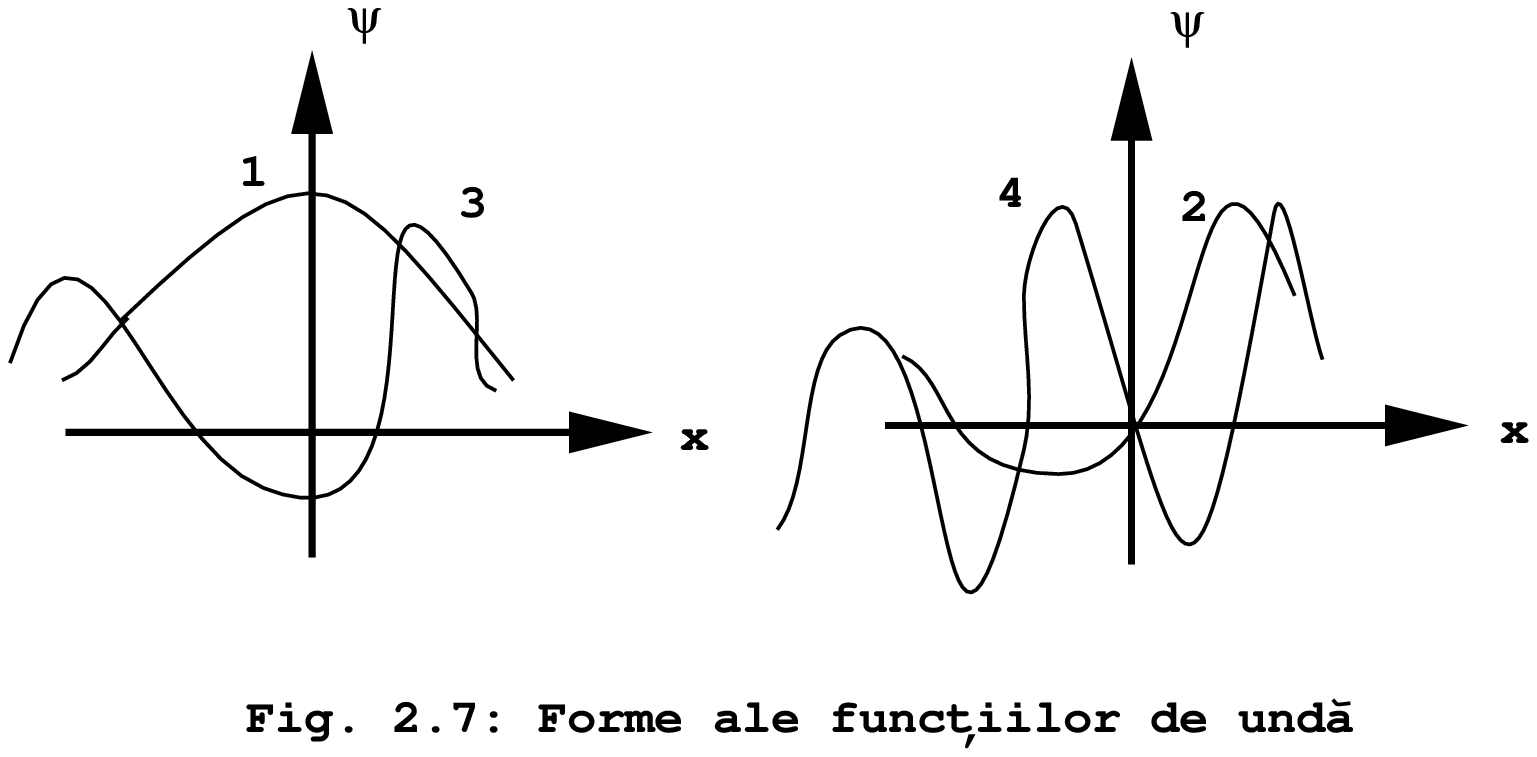}}
\vskip 4ex
%\begin{center}
%{\small{Fig. 1}\\
%}
%\end{center}
%%%%%%%%%%%%%%%%

%\end{document}

%\newpage

\subsection*{Problema 2.3:
Particul\A\ \h n groap\A\ rectangular\A\ 1D infinit\A\ }

%{\bf Problema.}
S\A\ se rezolve ecua\c{t}ia Schr\"odinger unidimensional\A\ 
pentru o particul\A\ care se afl\A\ intr-o groap\A\ de
poten\c{t}ial infinit\A\ descris\A\ de:
\[
V(x) = \left\{
\begin{array}{ll}
0&\mbox{pentru $x'<x<x'+2a$}\\
\infty&\mbox{pentru $x'\geq x~~{\rm o}~~x\geq x'+2a$.}
\end{array}
\right.
\]
Solu\c{t}ia \h n form\A\ general\A\ este
\begin{equation}
\psi(x) = A\sin(kx) + B\cos(kx)
\end{equation}
unde
\begin{equation}
k = \sqrt{\frac{2mE}{\hbar^2}}~.
\end{equation}
Cum $\psi$ trebuie s\A\ \h ndeplineasc\A\ $\psi(x') = \psi(x'+2a) = 0$, 
se ob\c{t}ine:
\begin{eqnarray}
A~\sin(kx')~~~ +~~~ B~\cos(kx') = 0 \\
A\sin[k(x'+2a)] + B\cos[k(x'+2a)] = 0~.
\end{eqnarray} 
Multiplic\h nd (101) cu $\sin[k(x'+2a)]$ \c{s}i (102) cu $\sin(kx')$ \c{s}i
\h n continuare sc\A z\h nd ultimul rezultat din primul ob\c{t}inem:
\begin{equation}
B[~~\cos(kx') \sin[k(x'+2a)] - \cos[k(x'+2a)]\sin(kx')~~] = 0~.
\end{equation}
sau prin intermediul unei identit\A \c{t}i trigonometrice:
\begin{equation}
B \sin(2ak) = 0
\end{equation}
Multiplic\h nd (101) cu $\cos[k(x'+2a)]$ \c{s}i 
sc\A z\h nd (102) multiplicat\A\ cu $\cos(kx')$, se ob\c{t}ine:
\begin{equation}
A[~~\sin(kx') \cos[k(x'+2a)] - \sin[k(x'+2a)]\cos(kx')~~] = 0~,
\end{equation}
sau folosind aceea\c{s}i identitate trigonometric\A\ :
\begin{equation}
A \sin[k(-2ak)] = A \sin[k(2ak)] =  0~.
\end{equation}

Cum nu se \c{t}ine cont de solu\c{t}ia trivial\A\ $\psi=0$, 
atunci pe baza ecua\c{t}iilor
(104) \c{s}i (106) trebuie ca $\sin(2ak)=0$, care are loc numai dac\A\ 
$2ak = n \pi$, cu $n$ un \h ntreg. Conform celor anterioare $k=n \pi/2a$ \c{s}i
cum $k^2=2mE/\hbar^2$ atunci rezult\A\ c\A\ autovalorile sunt date de c\A tre
expresia:
\begin{equation}
E = \frac{\hbar^2\pi^2n^2}{8a^2m}~.
\end{equation}
Energia este cuantizat\A\ pentru c\A\ numai pentru fiecare
$k_n = n\pi/2a$ le corespunde o
energie $E_n=[n^2/2m][\pi\hbar/2a]^2$.

Solu\c{t}ia \h n forma general\A\ :
\begin{equation}
\psi_n = A\sin(\frac{n\pi x}{2a}) + B\cos(\frac{n\pi x}{2a}).
\end{equation}
se poate normaliza:
\begin{equation}
1 = \int_{x'} ^{x'+2a} \psi \psi^* dx = a(A^2+B^2)
\end{equation}
ceea ce ne conduce la:
\begin{equation}
A = \pm \sqrt{1/a - B^2}~.
\end{equation}
Substituind aceast\A\ valoare a lui $A$ \h n (101) se ob\c{t}ine:
\begin{equation}
B = \mp\frac{1}{\sqrt{a}}\sin(\frac{n\pi x'}{2a})~.
\end{equation}
Substituind acum aceasta valoare a lui $B$ \h n (110) se ob\c{t}ine:
\begin{equation}
A = \pm\frac{1}{\sqrt{a}}\cos(\frac{n\pi x'}{2a})
\end{equation}
Folosind semnele superioare pentru $A$ \c{s}i $B$, 
prin substituirea valorilor acestora \h n
(108) se ob\c{t}ine:
\begin{equation}
\psi_n =\frac{1}{\sqrt{a}}\sin(\frac{n\pi}{2a})(x-x')~.
\end{equation}
Utiliz\h nd semnele inferioare pentru A \c{s}i B se ob\c{t}ine:
\begin{equation}
\psi_n =-\frac{1}{\sqrt{a}}\sin(\frac{n\pi}{2a})(x-x').
\end{equation}

%\end{document}
\newpage
%%%%%%%%%%%%%%%%%%%%%%%%%%%%%%%%%%%%%%%%%%%%%%%%%%%%%%%%%%%%%%%%%%%%%%%%%
%%%%%%%%%%%%%%%%%%%%%%%%%%%%%%%%%%%%%%%%%%%%%%%%%%%%%%%%%%  Moment cinetic
%%%%%%%%%%%%%%%%%%%%%%%%%%%%%%%%%%%%%%%%%%%%%%%%%%%%%%%
%\documentstyle[12pt]{article}
%\newcommand{\aple}{\mbox{${}_{\textstyle\sim}^{\textstyle<}$}}
%\newcommand{\apge}{\mbox{${}_{\textstyle\sim}^{\textstyle>}$}}
%\newcommand{\slsh}[1]{\mbox{$\displaystyle {#1}\!\!\!{/}$}}
%\newcommand{\lpr}{\mbox{$ \displaystyle O_L $}}
%\newcommand{\rpr}{\mbox{$ \displaystyle O_R $}}
%\newcommand{\GeV}{\mbox{$\rm  \, GeV $}}
%\baselineskip 25.1pt plus 0.2pt minus 0.1pt
%\begin{document}
%\baselineskip 20pt plus 0.2pt minus 0.1pt
%%%%%%%%%%%%%%%%%%%%%%%%%%%%%%%%%%%%%%%%%%%%%%%%%%%%%%%%%%%%%%
%\title
%\setcounter{equation}\\
%\section*
\begin{center}
{\huge 3. MOMENT CINETIC \c{S}I SPIN }
\end{center}
%\setcounter{equation}
%\author{Teodoro C\'ordova Fraga}
%\date{}
%\maketitle

%\vspace*{-20pt}
\section*{Introducere}
Din {\it Mecanica Clasic\u{a}} se \c{s}tie c\A\ , {\it momentul cinetic} 
$\bf{l}$ pentru particulele macroscopice este dat de
\setcounter{equation}{0}
\begin{equation}
{\bf l=r} \times {\bf p},
\label{1}
\end{equation}
unde $\bf r$ \c{s}i $\bf p$ sunt respectiv vectorul de pozi\c{t}ie \c{s}i
impulsul lineal.

Totu\c{s}i, \h n {\it Mecanica Cuantic\A\ }, exist\A\ 
operatori de tip moment cinetic 
(OTMC) care nu obligatoriu se exprim\A\ (numai) prin operatorii de 
coordonat\A\ $\hat{x}_j$ \c{s}i impuls $\hat{p}_k$
ac\c{t}ion\h nd (numai) asupra func\c{t}iilor proprii de coordonate. 
Prin urmare,
este foarte important s\A\ se stabileasc\A\ , mai \h nt\h i de toate, 
rela\c{t}ii de comutare generale
pentru componentele OTMC. 

\^{I}n {\it Mecanica Cuantic\A\ }
$\bf l$ se reprezint\A\ prin operatorul 
\begin{equation}
{\bf l}=-i\hbar {\bf r} \times \nabla,
\label{2}
\end{equation}
ale c\A rui componente sunt operatori care satisfac urm\A toarele reguli de 
comutare

\begin{equation}
[l_x,l_y]=il_z, \qquad  [l_y,l_z]=il_x, \qquad  [l_z,l_x]=il_y,
\label{3}
\end{equation}
\c{s}i \h n plus, fiecare dintre ele comut\A\ cu p\A tratul 
momentului cinetic, adic\A\

\begin{equation}
l^2=l^2_x+l^2_y+l^2_z, \qquad [l_i,l^2]=0, \qquad i=1,2,3.
\label{4}
\end{equation}
Aceste rela\c{t}ii, \h n afar\A\ de a fi corecte pentru momentul cinetic, 
se satisfac
pentru importanta clas\A\ OTMC a operatorilor de spin, care sunt 
lipsi\c{t}i de analogi \h n {\it mecanica clasic\A\ }.

\noindent
Aceste rela\c{t}ii de comutare sunt bazice pentru a ob\c{t}ine spectrul
operatorilor men\c{t}iona\c{t}i, 
precum \c{s}i reprezent\A rile lor diferen\c{t}iale.

\section*{Momentul cinetic orbital}
Pentru un punct oarecare al unui spa\c{t}iu fix (SF), se poate introduce o
func\c{t}ie $\psi(x,y,z)$, pentru care, s\A\ consider\A m dou\A\ sisteme 
carteziene $\Sigma$ \c{s}i $\Sigma '$, unde $\Sigma '$ se ob\c{t}ine
prin rota\c{t}ia axei $z$ a lui $\Sigma$.

\^{I}n cazul general un SF se refer\A\ la un sistem de coordonate 
diferite~de~$\Sigma$~\c{s}i~${\Sigma}'$.

Acum, s\A\ compar\A m valorile lui $\psi$ \h n dou\A\ puncte ale SF
cu acelea\c{s}i coordonate (x,y,z) \h n $\Sigma$ \c{s}i ${\Sigma}'$, ceea ce 
este echivalent cu rota\c{t}ia vectorial\A\
\begin{equation}
\psi(x',y',z') = R \psi(x,y,z)
\label{5}
\end{equation}
unde $R$ este matricea de rota\c{t}ie \h n {\sc R}$^3$
\begin{equation}
\left(\begin{array}{c}
x' \\ y' \\ z' 
\end{array}\right)
=
\left(\begin{array}{ccc}
\cos \phi & -\sin \phi & 0 \\
\sin \phi &  \cos \phi & 0 \\
0         &  0         & z 
\end{array}\right)
\left(\begin{array}{c}
x \\ y \\ z
\end{array}\right),
\label{6}
\end{equation}
atunci
\begin{equation}
R \psi (x,y,z) =
\psi(x \cos \phi -y \sin \phi ,
     x \sin \phi  +y \cos \phi , z).
\label{7}
\end{equation}

Pe de alt\A\ parte este important de amintit c\A\ func\c{t}iile de 
und\A\ nu depind de sistemul de coordonate \c{s}i c\A\ trasformarea 
la rota\c{t}ii
\h n cadrul SF se face cu ajutorul operatorilor unitari \c{s}i deci pentru a
stabili forma operatorului unitar $U^\dagger(\phi)$ care trece
$\psi$ \h n $\psi '$, se consider\A\ o rota\c{t}ie infinitezimal\A\ $d\phi$,
men\c{t}in\h nd numai termenii lineari \h n $d\phi$ atunci c\h nd se
face dezvoltarea \h n serie Taylor a lui $\psi '$ \h n jurul
punctului~$x$
\begin{eqnarray}\vspace*{-20pt}
\psi(x',y',z') & \approx &  \psi(x+yd\phi, xd\phi+y, z), \nonumber\\
	       & \approx & \psi(x,y,z) 
+ d\phi\left(y \frac{\partial \psi}{\partial x} 
      - x \frac{\partial \psi}{\partial y}\right), \nonumber\\
               & \approx & (1-id\phi l_z)\psi(x,y,z),
\label{8} 
\end{eqnarray}

\noindent
unde am folosit nota\c{t}ia\footnote{Demostra\c{t}ia lui (8) 
se prezint\A\ ca problema 3.1}
\begin{equation}
l_z = \hbar^{-1}(\hat{x}\hat{p}_y -\hat{y}\hat{p}_x ),
\label{9}
\end{equation}
care, dup\A\ cum se va vedea mai t\h rziu, corespunde operatorului de 
proiec\c{t}ie \h n~$z$ al momentului cinetic de acord cu defini\c{t}ia 
din (2) \c{s}i ad\A ug\h nd factorul~$\hbar ^{-1}$, astfel c\A\  
rota\c{t}iile de unghi $\phi$
finit se pot reprezenta ca exponen\c{t}iale de tipul:
\begin{equation}
\psi(x', y', z) = e^{il_z \phi}\psi(x,y,z),
\label{10}
\end{equation}
unde
\begin{equation}
\hat{U}^\dagger(\phi)=e^{il_z\phi}.
\label{11}
\end{equation}
Pentru a reafirma  conceptul de rota\c{t}ie, s\A\ \h l consider\A m
\h ntr-un tratament mai general cu ajutorul operatorului
 vectorial $\hat{\vec{A}}$ care ac\c{t}ioneaz\A\ asupra lui $\psi$, 
presupun\h nd c\A\ $\hat{A}_x$,  $\hat{A}_y$ $\hat{A}_z$ au aceea\c{s}i form\A\
\h n $\Sigma$ \c{s}i $\Sigma '$, adic\A\ , valorile medii ale lui
$\hat{\vec{A}}$ calculate \h n $\Sigma$ \c{s}i $\Sigma '$ trebuie s\A\ fie
egale
c\h nd se v\A d din SF:
\begin{eqnarray}
&&\int \psi^*(\vec{r}')
(\hat{A}_x\hat{\imath}' + \hat{A}_y\hat{\jmath}' + \hat{A}_z\hat{k}')
\psi^*(\vec{r}')\,d\vec{r} \nonumber\\
&& \qquad =\int \psi^*(\vec{r})
(\hat{A}_x\hat{\imath} + \hat{A}_y\hat{\jmath} + \hat{A}_z\hat{k})
\psi^*(\vec{r})\,d\vec{r},
\label{12}
\end{eqnarray}
unde

\begin{equation}
\hat{\imath}' = \hat{\imath}\cos\phi + \hat{\jmath}\sin\phi, \qquad
\hat{\jmath}' = \hat{\imath}\sin\phi + \hat{\jmath}\cos\phi, \qquad
\hat{k}' = \hat{k}.
\label{13}
\end{equation}

Prin urmare, dac\A\ vom combina (10), (12) \c{s}i (13) ob\c{t}inem
\begin{eqnarray}
e^{il_z\phi} \hat{A}_x e^{-il_z\phi}&=& \hat{A}_x\cos\phi -\hat{A}_y\sin\phi,
\nonumber\\
e^{il_z\phi} \hat{A}_y e^{-il_z\phi}&=& \hat{A}_x\sin\phi -\hat{A}_y\cos\phi,
\nonumber\\
e^{il_z\phi} \hat{A}_z e^{-il_z\phi}&=& \hat{A}_z.
\label{14}
\end{eqnarray}

Din nou consider\h nd rota\c{t}ii infinitezimale \c{s}i
dezvolt\h nd  p\A r\c{t}ile din st\h nga \h n
(14) se pot determina rela\c{t}iile de 
comutare ale lui $\hat{A}_x$, $\hat{A}_y$ \c{s}i $\hat{A}_z$ cu $\hat{l}_z$
\begin{equation}
[l_z,A_x]=iA_y, \qquad  [l_z,A_y]=-iA_x, \qquad  [l_z,A_z]= 0,
\label{15}
\end{equation}
\c{s}i \h n acela\c{s}i mod pentru $l_x$ \c{s}i $l_y$. 
\\

Condi\c{t}iile bazice pentru a ob\c{t}ine aceste rela\c{t}ii
de comutare sunt

\begin{itemize}

\item[$\star$]
FP se transform\A\ ca \h n (7) c\h nd $\Sigma \rightarrow \Sigma '$.

\item[$\star$]
Componentele $\hat{A}_x$, $\hat{A}_y$, $\hat{A}_z$ au aceea\c{s}i form\A\ \h n
$\Sigma$ \c{s}i $\Sigma '$.

\item[$\star$] 
Vectorii valorilor medii ale lui $\hat{A}$ \h n $\Sigma$ \c{s}i $\Sigma '$
coincid (sunt egale) pentru un observator din SF.

\end{itemize}

Deasemenea se poate folosi alt\A\ reprezentare \h n care $\psi(x,y,z)$
nu se schimb\A\ c\h nd $\Sigma \rightarrow \Sigma '$ \c{s}i operatorii 
vectoriali se 
transform\A\ ca orice vectori. Pentru a trece la o astfel de 
representare c\h nd $\phi$
se rote\c{s}te \h n jurul lui $z$ se folose\c{s}te operatorul $\hat{U}(\phi)$, 
adic\A\

\begin{equation}
e^{il_z\phi} \psi'(x,y,z) = \psi(x,y,z),
\label{16}
\end{equation}
\c{s}i deci
\begin{equation}
e^{-il_z\phi} \hat{\vec{A}} e^{il_z\phi} = \hat{\vec{A}}.
\label{17}
\end{equation}
Utiliz\h nd rela\c{t}iile date \h n (14) ob\c{t}inem
\begin{eqnarray}
\hat{A}_x' & = & \hat{A}_x\cos\phi + \hat{A}_y\sin\phi 
             =   e^{-il_z\phi} \hat{A}_x e^{il_z\phi}, \nonumber\\ 
\hat{A}_y' & = & -\hat{A}_x\sin\phi + \hat{A}_y\cos\phi 
             =  e^{-il_z\phi} \hat{A}_y e^{il_z\phi}, \nonumber\\ 
\hat{A}_z' & = & e^{-il_z\phi} \hat{A}_z e^{il_z\phi}.
\label{18}
\end{eqnarray}

Pentru c\A\ transform\A rile noii reprezent\A ri se fac cu ajutorul 
operatorilor unitari, rela\c{t}iile de comutare nu se schimb\A\ .

\subsection*{Observa\c{t}ii}

\begin{itemize}

\item[$\star$]
Operatorul $\hat{A}^2$ este invariant la rota\c{t}ii, adic\A\
\begin{equation}
e^{-il_z\phi} \hat{A}^2 e^{il_z\phi} = \hat{A}'^2 = \hat{A}^2~.
\label{19}
\end{equation}

\item[$\star$]
Rezult\A\ c\A\ 
\begin{equation}
[\hat{l}_i, \hat{A}^2] = 0~.
\label{20}
\end{equation}

\item[$\star$]
Dac\A\ operatorul Hamiltonian este de forma
\begin{equation}
\hat{H} = \frac{1}{2m}\hat{p}^2 + U(|\vec{r}|),
\label{21}
\end{equation}
atunci se men\c{t}ine invariant la rota\c{t}ii produse \h n oricare ax\A\
care trece prin originea de coordonate
\begin{equation}
[\hat{l}_i, \hat{H}] = 0~,
\label{22}
\end{equation}
unde $\hat{l}_i$ sunt integrale de mi\c{s}care.
\end{itemize}

\subsection*{Defini\c{t}ie}
Dac\A\ $\hat{A}_i$ sunt componentele unui operator vectorial care 
ac\c{t}ioneaz\A\ asupra unei func\c{t}ii de und\A\ dependent\A\ numai de 
coordonate \c{s}i dac\A\ exist\A\
operatori $\hat{l}_i$ care satisfac urm\A toarele
rela\c{t}ii de comutare:
\begin{equation}
[\hat{l}_i, \hat{A}_j] = i\varepsilon_{ijk}\hat{A}_k, \qquad
[\hat{l}_i, \hat{l}_j] = i\varepsilon_{ijk}\hat{l}_k~,
\label{23}
\end{equation}

\noindent
atunci, $\hat{l}_i$ se numesc {\it componentele operatorului
moment cinetic (orbital)} \c{s}i putem s\A\ deducem pe baza lui (20) \c{s}i 
(23) c\A\ 
\begin{equation}
[\hat{l}_i, \hat{l}^2]=0.
\label{24}
\end{equation}
 
Prin urmare, cele trei componente operatoriale asociate cu componentele 
unui moment cinetic
clasic arbitrar satisfac rela\c{t}ii de comutare de tipul (23), (24).
Mai mult, se poate ar\A ta c\A\ aceste rela\c{t}ii conduc la 
propriet\A \c{t}i geometrice specifice ale rota\c{t}iilor \h ntr-un spa\c{t}iu
euclidean tridimensional. Aceasta are loc dac\A\ adopt\A m un punct de 
vedere mai
general \c{s}i definim un operator moment unghiular 
$\bf J$ (nu mai punem simbolul de operator)
ca orice set de trei observabile $J_x$, $J_y$ \c{s}i
$J_z$ care \h ndeplinesc rela\c{t}iile de comutare:
\begin{equation}
[J_i, J_j] = i\varepsilon_{ijk}J_k.
\label{25}
\end{equation}

\^{I}n plus, s\A\ introducem operatorul 
\begin{equation}
{\bf J}^2 = J^2_x + J^2_y + J^2_z,
\label{26}
\end{equation}
p\A tratul scalar al momentului unghiular $\bf J$. Acest operator este 
hermitic pentru c\A\ $J_x$, $J_y$ \c{s}i $J_z$ sunt hermitici
%\c{s}i presupun\h nd c\A\ este o observabil\A\ ,
\c{s}i se arat\A\ u\c{s}or c\A\ $\bf J^2$ comut\A\ cu cele trei componente
ale lui $\bf J$
\begin{equation}
[{\bf J}^2, J_{i}]=0.
\label{27}
\end{equation}

Deoarece $\bf J^2$ comut\A\ cu fiecare dintre componente rezult\A\ c\A\ exist\A\
un sistem complet de FP, respectiv
\begin{equation}
{\bf J^2}\psi_{\gamma \mu} = f(\gamma ^2)\psi_{\gamma \mu}, \qquad
    J_i\psi_{\gamma \mu} = g(\mu)    \psi_{\gamma \mu},
\label{28}
\end{equation}
unde, a\c{s}a cum se va ar\A ta \h n continuare, FP-urile depind de doi subindici,
care se vor determina odat\A\ cu forma 
func\c{t}iilor $f(\gamma)$ \c{s}i $g(\mu)$. 
Operatorii $J_i$ \c{s}i $J_k$ $(i \neq k)$ nu comut\A\ (\h ntre ei), adic\A\ , 
nu au FP \h n comun. Din motive at\h t fizice c\h t \c{s}i matematice, suntem
interesa\c{t}i s\A\ determin\A m func\c{t}iile proprii comune ale lui 
${\bf J^2}$ \c{s}i $J_{z}$, adic\A\ vom lua $i=z$ \h n (28).

\^{I}n loc de a folosi componentele $J_x$ \c{s}i $J_y$ ale momentului
unghiular $\bf J$,
este mai convenabil s\A\ se introduc\A\ urm\A toarele combina\c{t}ii lineare
\begin{equation}
J_+ = J_x + iJ_y, \qquad J_{-} = J_x - iJ_y.
\label{29}
\end{equation}
Ace\c{s}ti operatori nu sunt hermitici, a\c{s}a cum sunt 
operatorii $a$ \c{s}i $a^\dagger$
ai oscilatorului armonic (vezi capitolul 5), sunt numai adjunc\c{t}i.
Se pot deduce u\c{s}or urm\A toarele propriet\A \c{t}i
\begin{equation}
[J_z,   J_{\pm}] = \pm J_{\pm}, \qquad 
[J_{+}, J_{-}]   =     2J_z, 
\label{30}
\end{equation}
\begin{equation}
[J^2, J_{+}]   =  [J^2, J_{-}] = [J^2, J_{z}] = 0.
\label{31}
\end{equation}
\begin{equation}
J_z(J_{\pm}\psi_{\gamma\mu}) = \{J_{\pm}J_z + [J_z, J_{\pm}]\}\psi_{\gamma\mu}
=(\mu \pm 1) (J_{\pm} \psi_{\gamma\mu}).
\label{32}
\end{equation}

Prin urmare $J_{\pm}\psi_{\gamma\mu}$ sunt FP ale lui $J_z$
corespunz\A toare autovalorilor $\mu \pm 1$, adic\A\ func\c{t}iile 
respective sunt identice p\h n\A\
la factorii consta\c{t}i (de determinat) $\alpha_\mu$ \c{s}i $\beta_\mu$:
\begin{eqnarray}
J_{+} \psi_{\gamma\mu - 1} &=& \alpha_\mu \psi_{\gamma\mu}, \nonumber\\
J_{-}\psi_{\gamma\mu}         &=& \beta_\mu \psi_{\gamma\mu-1}.
\label{33}
\end{eqnarray}
Pe de alt\A\ parte
\begin{equation}
\alpha^*_{\mu} = (J_{+}\psi_{\gamma \mu -1} , \psi_{\gamma \mu}) 
               = (\psi_{\gamma \mu -1} J_{-} \psi_{\gamma \mu})= \beta_\mu,
\label{34}
\end{equation}
astfel c\A\ , lu\h nd o faz\A\ de tipul $e^{ia}$ (cu $a$ real) pentru 
func\c{t}ia
$\psi_{\gamma \mu}$ se poate face $\alpha_\mu$ real \c{s}i egal cu $\beta _\mu$, 
ceea ce \h nseamn\A\
\begin{equation}
J_{+}\psi_{\gamma , \mu - 1} = \alpha \mu \psi_{\gamma \mu},
J_{-}\psi_{\gamma \mu    } = \alpha \mu \psi_{\gamma , \mu - 1},
\label{35}
\end{equation}
\c{s}i deci
\begin{eqnarray}
\gamma &=& (\psi_{\gamma \mu}, [J_x^2 + J_y^2 + J_z^2] \psi_{\gamma \mu})
= \mu^2 + a + b, \nonumber\\
a & = & (\psi_{\gamma \mu}, J_{x}^{2} \psi_{\gamma \mu}) = 
       (J_x\psi_{\gamma \mu}, J_x \psi_{\gamma \mu}) \geq 0, \nonumber\\
b & = &  (\psi_{\gamma \mu}, J_{y}^{2} \psi_{\gamma \mu}) = 
        (J_y\psi_{\gamma \mu}, J_y \psi_{\gamma \mu}) \geq 0.
\label{36}
\end{eqnarray}

Constanta de normalizare a oric\A rei func\c{t}ii de und\A\ nu este negativ\A\ ,
ceea ce implic\A\
\begin{equation}
\gamma \geq \mu^2,
\label{37}
\end{equation}
pentru $\gamma$ fix, valoarea lui $\mu$ are limite at\h t superioar\A\ c\h t
\c{s}i inferioar\A\ (adic\A\ ia valori \h ntr-un interval finit). 

Fie $\Lambda$ \c{s}i $\lambda$ aceste limite (superioar\A \c{s}i
inferiar\A\ de $\mu$) pentru un $\gamma$ dat
\begin{equation}
J_{+}\psi_{\gamma \Lambda} = 0, \qquad J_{-}\psi_{\gamma \lambda} = 0.
\label{38}
\end{equation}

Utiliz\h nd urm\A toarele identit\A \c{t}i operatoriale
\begin{eqnarray}
J_{-}J_{+} &=& {\bf J^2} - J^2_z + J_z = {\bf J^2} - J_z(J_z-1),\nonumber\\
J_{+}J_{-} &=& {\bf J^2} - J^2_z + J_z = {\bf J^2} - J_z(J_z+1),
\label{39}
\end{eqnarray}
ac\c{t}ion\h nd asupra lui $\psi _ {\gamma \Lambda}$ \c{s}i 
$\psi_{\gamma \lambda}$ se ob\c{t}ine
\begin{eqnarray}
\gamma - \Lambda^2 - \Lambda &=& 0, \nonumber \\
\gamma - \lambda^2 + \lambda &=& 0, \nonumber \\
(\lambda - \lambda + 1) (\lambda + \lambda) &=& 0.
\label{40}
\end{eqnarray}

\^{I}n plus 
\begin{equation}
\Lambda \geq \lambda \rightarrow \Lambda = -\lambda = J \rightarrow
\gamma = J(J+1).
\label{41}
\end{equation} 

Pentru un $\gamma$ dat (fix) proiec\c{t}ia momentului $\mu$ ia
$2J+1$ valori care difer\A\ printr-o unitate \h ntre $J$ \c{s}i $-J$. 
De aceea,
diferen\c{t}a $\Lambda -\lambda = 2J$ trebuie s\A\ fie un num\A r \h ntreg
\c{s}i deci 
autovalorile lui $J_z$ numerotate prin $m$ sunt \h ntregi
\begin{equation}
m=k, \qquad k=0, \pm 1, \pm 2, \, \ldots \, ,
\label{42}
\end{equation}
sau semi\h ntregi
\begin{equation}
m=k + {1 \over 2}, \qquad k=0, \pm 1, \pm 2, \, \ldots \, .
\label{43}
\end{equation}

O stare de $\gamma = J(J+1)$ dat\A\ indic\A\ o degenerare de grad $g=2J+1$
fa\c{t}\A\ de autovalorile $m$ (aceasta pentru c\A\
$J_i,~J_k$ comut\A\ cu $J^2$ dar nu comut\A\ \h ntre ele).
 
Prin ``stare de moment unghiular $J$''  se \h n\c{t}elege \h n majoritatea
cazurilor, o stare cu $\gamma = J(J+1)$ care are valoarea maxim\A\ a
momentului proiectat, respectiv $J$. Este comun s\A\ se noteze
aceste st\A ri cu $\psi_{jm}$ sau de ket Dirac $|jm\rangle$.

S\A\ ob\c{t}inem acum elementele de matrice ale lui $J_x,~J_y$ 
\h n reprezentarea \h n care
$J^2$ \c{s}i $J_z$ sunt diagonale. \^{I}n acest caz, din (35) \c{s}i (39) se
ob\c{t}ine c\A\
\begin{eqnarray}
J_{-}J_{+} \psi_{jm-1} = \alpha_mJ_{-}\psi_{jm} = \alpha_m\psi_{jm-1},
\nonumber\\
J(J+1)-(m-1)^2-(m-1) = \alpha_m^2,\nonumber\\
\alpha_m = \sqrt{(J+m)(J-m+1)}.
\label{44}
\end{eqnarray}

Combin\h nd (44) cu (35) se ob\c{t}ine
\begin{equation}
J_+\psi_{jm-1} = \sqrt{(J+m)(J-m+1)}\psi_{jm},
\label{45}
\end{equation}
rezult\A\ c\A\ elementul de matrice al lui $J_{+}$ este
\begin{equation}
\langle jm | J_+ | jm-1 \rangle = \sqrt{(J+m)(J-m+1)} \delta_{nm},
\label{46}
\end{equation}
\c{s}i analog
\begin{equation}
\langle jn | J_{-} | jm \rangle = -\sqrt{(J+m)(J-m+1)} \delta_{nm-1}~.
\label{47}
\end{equation}
\^{I}n sf\h r\c{s}it, din defini\c{t}iile (29) pentru $J_{+},\ J_{-}$ se 
ob\c{t}ine u\c{s}or 
\begin{eqnarray}
\langle jm | J_x | jm-1 \rangle &=& {1 \over 2}\sqrt{(J+m)(J-m+1)} ,
\nonumber\\
\langle jm | J_y | jm-1 \rangle &=& {-i \over 2}\sqrt{(J+m)(J-m+1)}~.
\label{48}
\end{eqnarray}

\subsection*{Concluzii par\c{t}iale}
\begin{itemize}
\item[$\alpha$] \underline{{\it Propriet\A \c{t}i ale autovalorilor lui  
$\bf J$ \c{s}i $J_z$}}\\
Dac\A\ $j(j+1)\hbar^2$ \c{s}i $m\hbar$ sunt autovalori ale lui $\bf J$ \c{s}i $J_z$
asociate cu autovectorii $|kjm\rangle$, atunci $j$ \c{s}i $m$ satisfac
inegalitatea 
\[
-j \leq m \leq j.
\]

\item[$\beta$] \underline{{\it Propriet\A \c{t}i ale vectorului 
$J_{-}|kjm\rangle$}}\\
Fie $|kjm\rangle$ un eigenvector al lui $\bf J^2$ \c{s}i $J_{z}$ 
cu autovalorile $j(j+1)\hbar^2$ \c{s}i $m\hbar$
\begin{itemize}
\item{(i)}
Dac\A\ $m=-j$ atunci $J_{-}|kj-j\rangle=0$.
\item{(ii)}
Dac\A\ $m>-j$ atunci $J_{-}|kjm \rangle$ este un vector propriu nenul al lui 
$J^2$ \c{s}i $J_z$ cu autovalorile $j(j+1)\hbar^2$ \c{s}i $(m-1)\hbar$.
\end{itemize}

\item[$\gamma$] \underline{{\it Propriet\A \c{t}i ale vectorului 
$J_+|kjm\rangle$}}\\
Fie $|kjm\rangle$ un vector (ket) propriu al lui $\bf J^2$ \c{s}i $J_z$ cu
autovalorile $j(j+1)\hbar$ \c{s}i $m\hbar$
\begin{itemize}
\item[$\star$]
Dac\A\ $m=j$ atunci $J_+|kjm\rangle =0.$
\item[$\star$]
Dac\A\ $m<j$ atunci $J_+|kjm\rangle$ este un vector nenul al lui $\bf J^2$ 
\c{s}i $J_z$ cu autovalorile $j(j+1)\hbar^2$ \c{s}i $(m+1)\hbar$
\end{itemize}

\item[$\delta$] \underline{{\it Consecin\c{t}e ale propriet\A \c{t}ilor 
anterioare}}
\begin{eqnarray}
J_z|kjm\rangle &=& m\hbar|kjm\rangle,\nonumber\\
J_+|kjm\rangle &=& m\hbar\sqrt{j(j+1) - m(m+1)}|kjm+1\rangle, \nonumber\\
J_-|kjm\rangle &=& m\hbar\sqrt{j(j+1) - m(m-1)}|kjm+1\rangle. \nonumber
\end{eqnarray}
\end{itemize}

\section*{Aplica\c{t}ii ale momentului cinetic orbital}
Am considerat p\h n\A\ acum propriet\A \c{t}ile momentului cinetic
derivate numai pe baza rela\c{t}iilor de comutare. S\A\ ne \h ntoarcem la
momentul cinetic $\bf l$ al unei particule f\A r\A\ rota\c{t}ie intrinsec\A\ 
\c{s}i s\A\ vedem cum se aplic\A\ teoria din sec\c{t}iunea anterioar\A\
la cazul particular important
\begin{equation}
[\hat{l}_i, \hat{p}_j] = i\varepsilon_{ijk}\hat{p}_k.
\label{49}
\end{equation} 
Mai \h nt\h i, $\hat{l}_z$ \c{s}i $\hat{p}_j$ au un sistem comun de func\c{t}ii
proprii. Pe de alt\A\ parte, Hamiltonianul unei particule libere
\[
\hat{H} = \left(\frac{\hat{\vec{p}}}{\sqrt{2m}}\right)^2,
\]
fiind p\A tratul unui operator vectorial are acela\c{s}i
sistem complet de FP cu $\hat{l^2}$ \c{s}i $\hat{l}_z$. \^{I}n plus, aceasta implic\A\ 
c\A\ particula liber\A\ se poate g\A si \h ntr-o stare cu $E$, $l$, $m$
bine determinate.

%%%%%%%%%%%%%%%%%%%%%%%%%%%%%%%%%%%%%%%%%%%%%%%%%%%%%%%%%%%%%%%%%%%%%
\subsection*{Valori proprii \c{s}i func\c{t}ii proprii ale lui $\bf l^2$ \c{s}i 
$\bf L_{z}$}
Este mai convenabil s\A\ se lucreze cu coordonate sferice (sau polare),
pentru c\A\ , a\c{s}a cum vom vedea, diferi\c{t}i
 operatori de moment cinetic ac\c{t}ioneaz\A\ numai asupra variabilelor
unghiulare $\theta,\ \phi$ \c{s}i nu \c{s}i asupra variabilei $r$.
\^{I}n loc de a caracteriza vectorul $r$ prin componentele carteziene
$x,\ y,\ z$ s\A\ determin\A m punctul corespunz\A tor $M$ de raz\A\ vectoare
$\bf r$ prin coordonatele sferice tridimensionale
\begin{equation}
x=r\cos\phi\sin\theta, \qquad  
y=r\sin\phi\sin\theta, \qquad
z=r\cos\theta,
\label{50}
\end{equation}
cu
\[
r \geq 0, \qquad 
0 \leq \theta \leq \pi, \qquad 
0 \leq \phi   \leq 2\pi.
\]

Fie $\Phi(r, \theta,\phi)$ \c{s}i $\Phi'(r, \theta,\phi)$ func\c{t}iile de 
und\A\ ale unei 
particule \h n $\Sigma$ \c{s}i $\Sigma'$ \h n care rota\c{t}ia 
infinitezimal\A\ este dat\A\  prin ~$\delta\alpha$ \h n jurul lui~$z$
\begin{eqnarray}
\Phi'(r, \theta,\phi) &=& \Phi(r, \theta,\phi+\delta\alpha),\nonumber\\
&=& \Phi(r, \theta,\phi) + \delta\alpha\frac{\partial \Phi}{\partial\phi},
\label{51}
\end{eqnarray}
sau
\begin{equation}
\Phi'(r, \theta,\phi) = (1+i\hat{l}_z\delta\alpha)\Phi(r, \theta,\phi).
\label{52}
\end{equation}

Rezult\A\ c\A\
\begin{equation}
\frac{\partial \Phi}{\partial \phi} = i\hat{l_z}\Phi, \qquad 
\hat{l}_z = -i{\partial \over \partial \phi}.
\label{53}
\end{equation}

Pentru o rota\c{t}ie infinitezimal\A\ \h n $x$
\begin{eqnarray}
\Phi'(r, \theta,\phi) &=& \Phi+\delta\alpha
\left(\frac{\partial \Phi}{\partial \theta}
      \frac{\partial \theta}{\partial \alpha} +
      \frac{\partial \Phi}{\partial \theta}
      \frac{\partial \phi}{\partial \alpha}
\right), \nonumber\\
&=& (1+i\hat{l}_x\delta\alpha)\Phi(r, \theta,\phi),
\label{54}
\end{eqnarray}
\h ns\A\ \h ntr-o astfel de rota\c{t}ie
\begin{equation}
z' = z+y\delta\alpha; \qquad
z' = z+y\delta\alpha; \qquad x' = x
\label{55}
\end{equation}
\c{s}i din (50) se ob\c{t}ine
\begin{eqnarray}
r\cos(\theta + d\theta) &=& 
r\cos\theta + r\sin\theta\sin\phi\delta\alpha, \nonumber\\
r\sin\phi\sin(\theta + d\theta) &=& 
r\sin\theta\sin\phi + r\sin\theta\sin\phi -r\cos\theta\delta\alpha,
\label{56}
\end{eqnarray}
adic\A\
\begin{equation}
\sin\theta d\theta = \sin\theta\sin\phi \, \delta\alpha \rightarrow
{d\theta \over d\alpha} = -\sin\phi,
\label{57}
\end{equation}
\c{s}i
\begin{eqnarray}
\cos\theta\sin\phi \, d\theta + \sin\theta\cos\phi \, d\phi 
&=& -\cos\theta \, \delta\alpha,\nonumber\\
\cos\phi\sin\theta{d\phi \over d\alpha} &=& -\cos\theta - 
\cos\theta\sin\phi{d\theta \over d\alpha}~.
\label{58}
\end{eqnarray}
Substituind (57) \h n (56) duce la 
\begin{equation}
\frac{d\phi}{d\alpha} = -\cot\theta\cos\phi~.
\label{59}
\end{equation}
Cu (56) \c{s}i (58)  substituite \h n (51) \c{s}i compar\h nd p\A r\c{t}ile
din dreapta din (51) se ob\c{t}ine
\begin{equation}
\hat{l}_x = i\left(
\sin\phi {\partial \over \partial \theta} + 
\cot \theta\cos\phi {\partial \over \partial \phi}
\right).
\label{60}
\end{equation}

\^{I}n cazul rota\c{t}iei \h n $y$, rezultatul este similar
\begin{equation}
\hat{l}_y = i\left(
-\cos\phi {\partial \over \partial \theta} + 
\cot \theta\sin\phi {\partial \over \partial \phi}
\right).
\label{61}
\end{equation}

Folosind $\hat{l}_x, \ \hat{l}_y$ se pot ob\c{t}ine deasemenea  
$\hat{l}_\pm, \ \hat{l}^2$
\begin{eqnarray}
\hat{l}_\pm &=& \exp\left[\pm i\phi
\left(
\pm{\partial \over \partial\theta} + i\cot\theta{\partial \over \partial\phi}
\right)\right], \nonumber\\
\hat{l}^2 &=& \hat{l}_{-}\hat{l}_{+} + \hat{l}^2 + \hat{l}_z, \nonumber\\
&=&-\left[
{1 \over \sin^2\theta}
{\partial^2 \over \partial \phi^2} +
{1 \over \sin\theta}
{\partial \over \partial \theta}
\bigg(
\sin\theta{\partial \over \partial\theta} 
\bigg)\right].
\label{62}
\end{eqnarray}
Din (62) se vede c\A\ $\hat{l}^2$ este identic p\h n\A\ la o constant\A\ cu
partea unghiular\A\ a operatorului Laplace la o raz\A\ fix\A\
\begin{equation}
\nabla^2 f = {1 \over r^2}{\partial \over \partial r}
\left( r^2{\partial f \over \partial r} \right) + 
{1 \over r^2}
\left[
{1 \over \sin\theta}
{\partial \over \partial\theta} 
\left(
\sin\theta{\partial f \over \partial\theta}
\right) + 
{1 \over \sin^2\theta}{\partial^2 \over \partial \phi^2}
\right].
\label{63}
\end{equation}

\subsection*{Func\c{t}ii proprii ale lui $l_z$}
%%%%%%%%%%%%%%%%%%%%%%%%%%%%%%%%%%%%%%%%%%%%%%%
\begin{eqnarray}
\hat{l}_z\Phi_m = m\Phi = -i{\partial \Phi_m \over \partial \phi},
\nonumber\\
\Phi_m = {1 \over \sqrt{2\pi}}e^{im\phi}.
\label{64}
\end{eqnarray}

\subsection*{Condi\c{t}ii de hermiticitate ale lui $\hat{l}_z$}
%%%%%%%%%%%%%%%%%%%%%%%%%%%%%%%%%%%%%%%%%%%%%%%%%%%%%%%%%%%%%%%
\begin{equation}
\int_0^{2\pi} f^*\hat{l}_zg\,d\phi = 
\left( \int_0^{2\pi} g^*\hat{l}_zf\,d\phi \right)^* +
f^*g(2\pi) - f^*g(0).
\label{65}
\end{equation}

Rezult\A\ c\A\ $\hat{l}_z$ este hermitic \h n clasa de func\c{t}ii pentru care
\begin{equation}
f^*g(2\pi) = f^*g(0).
\label{66}
\end{equation}

Func\c{t}iile proprii $\Phi_m$ ale lui $\hat{l}_z$ apar\c{t}in clasei integrabile
${\cal L}^2(0, 2\pi)$ \c{s}i \h ndeplinesc (66), 
cum se \h nt\h mpl\A\ pentru orice func\c{t}ie care se poate dezvolta 
\h n $\Phi_m(\phi)$
\begin{eqnarray}
F(\phi) &=& \sum^k a_ke^{ik\phi}, \qquad k = 0, \pm 1, \pm 2, \, \ldots \, ,
\nonumber\\
G(\phi) &=& \sum^k b_ke^{ik\phi},\qquad  
k = \pm 1/2,\pm 3/2, \pm 5/2 \, \ldots \, ,
\label{67}
\end{eqnarray}
cu $k$ numai \h ntregi sau numai semi\h ntregi, dar nu pentru 
combina\c{t}ii de $F(\phi)$ \c{s}i $G(\phi)$.

Alegerea corect\A\ a lui $m$ se bazeaz\A\ \h n FP 
comune ale lui $\hat{l}_z$ \c{s}i $\hat{l}^2$.

%%%%%%%%%%%%%%%%%%%%%%%%%%%%%%%%%%%%%%%%%%%%%%%%%%%%%%%%%%%%%%%%
\subsection*{Armonice sferice}
\^{I}n reprezentarea $\{\bf \vec{r}\}$, func\c{t}iile proprii asociate
cu autovalorile $l(l+1)\hbar^2$, ale lui $\bf l^2$ \c{s}i $m\hbar$ 
ale lui $l_z$ sunt solu\c{t}ii ale ecua\c{t}iilor diferen\c{t}iale par\c{t}iale
\begin{eqnarray}
-\left({\partial^2 \over \partial \theta^2} + {1 \over \tan\theta}
{\partial \over \partial \theta} + {1 \over \sin^2\theta} 
{\partial^2 \over \partial \phi^2}\right)\psi(r, \theta, \phi) &=&
l(l+1)\hbar ^2\psi(r, \theta, \phi), \nonumber\\
-i{\partial \over \partial \phi}\psi(r, \theta, \phi) &=& 
m\hbar\psi(r, \theta, \phi). 
\label{68}
\end{eqnarray}

Consider\h nd c\A\ rezultatele generale ob\c{t}inute sunt aplicabile la cazul 
momentului cinetic, \c{s}tim c\A\ $l$ este un \h ntreg sau semi\h ntreg \c{s}i
c\A\ pentru $l$ fixa\c{t}i $m$ poate lua numai valorile 
\[
-l, -l+1, \, \dots \, ,l-1, l.
\]

\^{I}n (68), $r$ nu apare \h n operatorul diferen\c{t}ial, a\c{s}a c\A\
se poate 
considera ca un parametru \c{s}i se poate \c{t}ine cont numai de dependen\c{t}a
\h n $\theta,\ \phi$ a lui $\psi$. Se folose\c{s}e nota\c{t}ia 
$Y_{lm}(\theta, \phi)$ pentru o astfel de func\c{t}ie proprie comun\A\ a lui 
$\bf l^2$ \c{s}i $l_z$, corespunz\A toare 
autovalorilor $l(l+1)\hbar^2, m\hbar$ \c{s}i se nume\c{s}te armonic\A\ 
sferic\A\ .
\begin{eqnarray}
{\bf l^2}Y_{lm}(\theta, \phi) &=& l(l+1)\hbar^2Y_{lm}(\theta,\phi), \nonumber\\
 l_z Y_{lm}(\theta, \phi)     &=& m\hbar Y_{lm}(\theta,\phi).
\label{69}
\end{eqnarray}

Pentru a fi c\h t mai riguro\c{s}i, trebuie s\A\ introducem un indice 
adi\c{t}ional cu scopul de a distinge \h ntre diferite solu\c{t}ii ale lui (69), 
care corespund acelea\c{s}i perechi de valori $l,\ m$. \^{I}ntr-adev\A r, 
dup\A\ cum se va vedea mai departe, 
aceste ecua\c{t}ii au o solu\c{t}ie unic\A\ (p\h n\A\ la un factor 
constant) pentru fiecare pereche de valori permise ale lui $l,\ m$; aceasta 
pentru c\A\ subindicii $l,\ m$ sunt suficien\c{t}i \h n contextul respectiv.
%Ec.\ (69) di\'o a $\theta,\ \phi$ dependencia de las eigenfunci\'ones
%de $\bf L^2$ y $l_z$. 
Solu\c{t}iile $Y_{lm}(\theta,\, \phi)$ au fost g\A site prin metoda separ\A rii
variabilelor \h n coordonate sferice (vezi capitolul {\em Atomul de hidrogen})
%\footnote{Demonstra\c{t}ia \h n problema 3.4}
%
\begin{equation}
\psi_{lm}(r, \theta,\phi) = f(r)\psi_{lm}(\theta,\phi),
\label{70}
\end{equation}
unde $f(r)$ este o func\c{t}ie de $r$, care apare ca o constant\A\ de
integrare din punctul de vedere al ecua\c{t}iilor diferen\c{t}iale
par\c{t}iale din (68). Faptul c\A\ $f(r)$ este arbitrar\A\ arat\A\ c\A\ 
$\bf L^2$ \c{s}i $l_z$ nu formeaz\A\ un set
complet de observabile\footnote{Prin defini\c{t}ie, operatorul 
hermitic A este o observabil\A\ dac\A\ acest sistem ortogonal de vectori
reprezint\A\ o baz\A\ \h n spa\c{t}iul configura\c{t}iilor de st\A ri.} 
\h n spa\c{t}iul 
$\varepsilon_r$\footnote{Fiecare stare cuantic\A\ a particulei este
caracterizat\A\ printr-o stare vectorial\A\
apar\c{t}in\h nd unui spa\c{t}iu vectorial abstract
$\varepsilon_r$.} al func\c{t}iilor de $\vec{r}$ (sau de $r, \theta, \phi$).

Cu obiectivul de a normaliza $\psi_{lm}(r, \theta, \phi)$, este convenabil s\A\
se normalizeze $Y_{lm}(\theta, \phi) $ \c{s}i $f(r)$ \h n mod separat: 
%(a\c{s}a cum se arat\A\ ).
%\^{i}n acest caz, trebuie s\A\ lu\A m un element diferen\c{t}ial 
%de unghi solid 
%
\begin{eqnarray}
\int_0^{2\pi}d\phi\int_0^{\pi}\sin\theta|\psi_{lm}(\theta, \phi)|^2d\theta 
&=& 1, 
\label{71} \\
\int_0^\infty r^2|f(r)|^2dr &=& 1.
\label{72}
\end{eqnarray}

%%%%%%%%%%%%%%%%%%%%%%%%%%%%%%%%%%%%%%%%%%%%%%%
\subsection*{Valorile perechii $l,\ m$}

\noindent
($\alpha$):  {\it $l,\ m$ trebuie s\A\ fie \h ntregi}\\
Folosind $l_z = {\hbar \over i}{\partial \over \partial \phi}$, putem scrie
(69) dup\A\ cum urmeaz\A\ 
\begin{equation}
{\hbar \over i}{\partial \over \partial \phi}Y_{lm}(\theta, \phi) = m\hbar 
Y_{lm}(\theta, \phi),
\label{73}
\end{equation}
care arat\A\ c\A\  
\begin{equation}
Y_{lm}(\theta, \phi) = F_{lm}(\theta, \phi)e^{im\phi}.
\label{74}
\end{equation}

Dac\A\ permitem ca $0\leq \phi < 2\pi$, atunci putem acoperi tot spa\c{t}iul 
pentru c\A\ func\c{t}ia trebuie s\A\ fie continu\A\ \h n orice zon\A\ , astfel 
c\A\
\begin{equation}
Y_{lm}(\theta, \phi=0) = Y_{lm}(\theta, \phi=2\pi),
\label{75}
\end{equation}
ceea ce implic\A\ 
\begin{equation}
e^{im\pi} = 1.
\label{76}
\end{equation}

A\c{s}a cum s-a v\A zut, $m$ este un \h ntreg sau semi\h ntreg; \h n cazul 
aplica\c{t}iei la momentul cinetic orbital, $m$ trebuie s\A\ fie un \h ntreg. 
($e^{2im\pi}$
ar fi $-1$ dac\A\ $m$ ar fi semi\h ntreg).

%%%%%%
\noindent
($\beta$): Pentru o valoare dat\A\ a lui $l$, toate armonicele sferice $Y_{lm}$
corespunz\A toare se pot ob\c{t}ine prin metode algebrice folosind
%%%%%%%%%%%%
\begin{equation}
l_+Y_{ll}(\theta, \phi)=0,
\label{77}
\end{equation}
care combinat\A\ cu ec.~(62) pentru $l_+$ duce la
\begin{equation}
\left({d \over d\theta} - l\cot\theta \right) F_{ll}(\theta) = 0.
\label{78}
\end{equation}
Aceast\A\ ecua\c{t}ie poate fi integrat\A\ imediat dac\A\ not\A m rela\c{t}ia
\begin{equation}
\cot\theta d\theta = \frac{d(\sin\theta)}{\sin\theta}~.
\label{79}
\end{equation}
Solu\c{t}ia sa general\A\ este

\begin{equation}
F_{ll}=c_l(\sin\theta)^l,
\label{80}
\end{equation}
unde $c_l$ este o constant\A\ de normalizare.

Rezult\A\ c\A\ pentru orice valoare pozitiv\A\ sau zero a lui $l$, exist\A\ o 
func\c{t}ie $Y_{ll}(\theta, \phi)$ care este (cu un factor constant asociat)
\begin{equation}
Y^{ll}(\theta, \phi) = c_l(\sin\theta)^le^{il\phi}.
\label{81}
\end{equation}

Folosind ac\c{t}iunea lui $l_{-}$ \h n mod repetat se poate construi tot setul
de func\c{t}ii
$Y_{ll-1}(\theta, \phi), \, \dots \, , Y_{l0}(\theta, \phi),$  
$\dots\, , Y_{l-l}(\theta, \phi)$. \^{I}n continuare, vedem modul \h n care se 
corespund aceste func\c{t}ii cu
perechea de autovalori $l(l+1)\hbar, m\hbar $ (unde $l$ este un \h ntreg 
pozitiv arbitrar sau zero \c{s}i $m$ este alt \h ntreg
astfel c\A\ $l\leq m \leq l$ ). Pe baza lui
(78) ajungem la concluzia c\A\ orice func\c{t}ie proprie
$Y_{lm}(\theta, \phi)$, poate fi ob\c{t}inut\A\ \h n mod neambiguu din $Y_{ll}$. 

%%%%%%%%%%%%%%%%%%%%%%%%%%%%%%%%%%%%%%%%%%%%%%%%%%%%%%%%%%%%%%%%%
\subsection*{Propriet\A \c{t}i ale armonicelor sferice}

\noindent
$\alpha\ $ {\it Rela\c{t}ii de recuren\c{t}\A\ }\\
Urm\h nd rezultatele generale avem
\begin{equation}
l_\pm Y_{lm}(\theta,\phi)=\hbar\sqrt{l(l+1)-m(m\pm 1)}Y_{lm\pm1}(\theta, \phi).
\label{82}
\end{equation}
Folosind expresia (62) pentru 
$l_\pm $ \c{s}i faptul c\A\ $Y_{lm}(\theta,\phi)$
este produsul \h ntre o func\c{t}ie dependent\A\ numai de $\theta$ \c{s}i 
$e^{\pm i\phi}$, ob\c{t}inem
\begin{equation}
e^{\pm i\phi}\left(\frac{\partial}{\partial \theta} - m\cot\theta \right)
Y_{lm}(\theta,\phi) = \sqrt{l(l+1)-m(m\pm 1)}Y_{lm\pm 1}(\theta, \phi)
\label{83}
\end{equation}

\noindent
$\beta \ $ {\it Ortonormare \c{s}i rela\c{t}ii de completitudine}\\
Ecua\c{t}ia (68) determin\A\ numai armonicele sferice p\h n\A\ la un factor 
constant. Acum vom alege acest factor astfel ca s\A\ avem ortonormalizarea 
func\c{t}iilor 
$Y_{lm}(\theta, \phi)$ (ca func\c{t}ii de variabilele 
unghiulare $\theta,\ \phi$)
\begin{equation}
\int^{2\pi}_0d\phi\int^\pi_0\sin\theta\,d\theta Y^*_{lm}(\theta, \phi)
Y_{lm}(\theta, \phi) = \delta_{l'l}\delta_{m'm}.
\label{84}
\end{equation}
\^{I}n plus, orice func\c{t}ie continu\A\ 
de $\theta,\ \phi$ pot fi expresate cu ajutorul armonicelor sferice, adic\A\
\begin{equation}
f(\theta, \phi) = \sum^\infty_{l=0}\sum^l_{m= -l}c_{lm}Y_{lm}(\theta,\phi),
\label{85}
\end{equation}
unde
\begin{equation}
c_{lm}=\int^{2\pi}_0d\phi\int^\pi_0\sin\theta\,d\theta\, Y^*_{lm}(\theta, \phi)
f(\theta, \phi).
\label{86}
\end{equation}

Rezultatele (85), (86) sunt consecin\c{t}a definirii armonicele sferice ca 
baz\A\ ortonormal\A\ complet\A\ \h n spa\c{t}iul
$\varepsilon_{\Omega}$ a func\c{t}iilor de $\theta,\ \phi$. 
Rela\c{t}ia de completitudine este
\begin{eqnarray}
 \sum^\infty_{l=0}\sum^l_{m=l}Y_{lm}(\theta,\phi)Y^*_{lm}(\theta',\phi)
&=&\delta(\cos\theta-\cos\theta' )\delta(\phi, \phi),\nonumber\\
&=&\frac{1}{\sin\theta}\delta(\theta-\theta')\delta(\phi, \phi).
\label{87} 
\end{eqnarray}

\noindent
`Func\c{t}ia' $\delta(\cos\theta-\cos\theta' )$ apare 
%\c{s}i nu $\delta(\theta-\theta')$  
%care intr\A\ \h n partea dreapt\A\ a rela\c{t}iei de completitudine 
pentru c\A\
integrala peste variabila $\theta$ se efectueaz\A\ folosind
elementul diferen\c{t}ial $\sin\theta\,d\theta = -d(\cos\theta)$.

%%%%%%%%%%%%%%%%%%%%%%%%%%%%%%%%%%%%%%%%%%%%%%%%%%%%%%%%%%%%%%%%%%%%%%%%%%%%
\subsection*{Operatorul de paritate $\cal P$ \h n cazul armonicelor sferice}
Comportamentul lui ${\cal P}$ \h n trei dimensiuni este destul de 
asem\A n\A tor celui \h ntr-o singur\A\ dimensiune, respectiv, 
c\h nd se aplic\A\ unei func\c{t}ii
$\psi(x,y,z)$ de coordonate carteziene \h i modific\A\ numai
semnul fiec\A reia dintre coordonate
\begin{equation}
{\cal P}\psi(x,y,z) = \psi(-x,-y,-z).
\label{88}
\end{equation}
$\cal P$ are propriet\A \c{t}ile unui operator hermitic \c{s}i \h n plus
este un operator unitar precum \c{s}i proiector deoarece
${\cal P}^2$ este un operator identitate
\begin{eqnarray}
\langle \psi(\bf{r}) | {\cal P} | \psi(\bf{r}) \rangle =
\langle \psi(\bf{r}) | \psi(\bf{-r}) \rangle  = 
\langle \psi(\bf{-r}) | \psi(\bf{r}) \rangle,
%%\delta({\bf{r} + \bf{-r'} }), 
\nonumber\\
{\cal P}^2 |\bf{r}\rangle = {\cal P}({\cal P}|\bf{r}\rangle  = 
\cal{P}|-\bf{r}\rangle  = |\bf{r}\rangle
\label{89}
\end{eqnarray}
\c{s}i deci
\begin{equation}
{\cal P}^2 = \hat{1}, 
\label{90}
\end{equation}
pentru care valorile proprii sunt $P=\pm1$. 
FP-urile se numesc pare dac\A\ $P = 1$ \c{s}i impare dac\A\
$P=-1$. \^{I}n mecanica cuantic\A\ nerelativist\A\ , operatorul $\hat{H}$
pentru un sistem conservativ este invariant la transform\A ri unitare discrete
\begin{equation}
{\cal P}\hat{H}{\cal P} = {\cal P}^{-1}\hat{H}{\cal P} = \hat{H}.
\label{91}
\end{equation}
Astfel $\hat{H}$ comut\A\ cu $\cal P$ \c{s}i deci paritatea st\A rii
este constant\A\ de mi\c{s}care. Deasemenea $\cal P$ comut\A\ cu operatorii
$\hat{l}$ \c{s}i $\hat{l}_\pm$
\begin{equation}
[{\cal P}, \hat{l}_i] = 0, \qquad [{\cal P}, \hat{l}_\pm]=0.
\label{92}
\end{equation}
%
%Dac\A\ $\hat{H}$ este par \c{s}i consider\A m o FP $|\Phi_n \rangle$ 
%care are 
%paritate bine definit\A\ $({\cal P}|\Phi_n \rangle )$, necolinear\A\ cu $|\psi_n \rangle$,
%se ha encontrado y puede inferirse que el eigenvalor correspondiente es
%degenerado con un grado de degeneraci\'on $n^2$, 
%din faptul c\A\ $\cal P$ comut\A\
%cu $\hat{H}$ deducem c\A\ $({\cal P}|\Phi_n \rangle )$ este un vector propriu
%al lui $\hat{H}$
%cu aceea\c{s}i valoare proprie ca \c{s}i $|\Phi_n \rangle )$. 
Dac\A\ $\psi$ este FP a tripletei 
${\cal P},\ \hat{l}$ \c{s}i $\hat{l}_z$, din (92) rezult\A\ c\A\ 
parit\A \c{t}ile st\A rilor care difer\A\  numai \h n $\hat{l}_z$ coincid. 
\^{I}n felul acesta se identific\A\  
paritatea unei particule de moment unghiular $\hat{l}$.

\^{I}n coordonate sferice, pentru acest operator vom considera urm\A toarea
substitu\c{t}ie
\begin{equation}
r\rightarrow r, \qquad \theta \rightarrow \pi-\theta \qquad
\phi \rightarrow \pi+\phi.
\label{93}
\end{equation}

\noindent
Prin urmare, dac\A\ folosim o baz\A\ standard \h n spa\c{t}iul func\c{t}iilor 
de und\A\ a unei particule f\A r\A\ `rota\c{t}ie proprie', partea radial\A\ 
a func\c{t}iilor $\psi_{klm}(\vec{r})$ din baz\A\
nu este modificat\A\ de c\A tre operatorul paritate. 
Doar armonicele sferice se schimb\A\ .
Transform\A rile (93) se traduc trigonometric \h n: 
\begin{equation}
\sin(\pi-\theta) \rightarrow \sin\theta, \qquad \cos(\pi-\theta) \rightarrow
-\cos\theta \qquad e^{im(\pi +\phi} \rightarrow (-1)^me^{im\phi}
\label{94}
\end{equation}
\c{s}i \h n aceste condi\c{t}ii, func\c{t}ia $Y_{ll}(\theta, \phi)$ 
se transform\A\ \h n
\begin{equation}
Y_{ll}(\phi - \theta, \pi + \phi) = (-1)^l Y_{ll}(\theta, \phi)~.
\label{95}
\end{equation}
Din (95) rezult\A\ c\A\ paritatea lui $Y_{ll}$ este $(-1)^l$.
Pe de alt\A\ parte, $l_{-}$ (ca \c{s}i $l_{+}$ este invariant la 
transform\A rile: 
\begin{equation}
{\partial \over \partial(\pi-\theta)} \rightarrow -{\partial 
\over \partial\theta},
\quad {\partial \over \partial(\pi+\phi)} 
\rightarrow {\partial \over \partial\phi}\quad
e^{i(\pi+\phi)}\rightarrow -e^{i\phi}\quad
{\rm cot}(\pi-\theta)\rightarrow -{\rm cot}\theta~.
\label{96}
\end{equation}
Cu alte cuvinte $l_{\pm}$ sunt pari. Prin urmare, deducem c\A\ paritatea 
oric\A rei armonice sferice este
$(-1)^l$, adic\A\ este invariant\A\ la schimb\A ri azimutale: 
\begin{equation}
Y_{lm}(\phi - \theta, \pi + \phi) = (-1)^l Y_{lm}(\theta, \phi).
\label{97}
\end{equation}
A\c{s}adar, armonicele sferice sunt func\c{t}ii de paritate 
bine definit\A\ , independent\A\ de $m$, par\A\ dac\A\ $l$ este par \c{s}i
impar\A\ dac\A\ $l$ este impar.

\section*{Operatorul de spin}
%%%%%%%%%%%%%%%%%%%%%%%%%%%%%
Unele particule, posed\A\ nu numai moment cinetic \h n raport cu axe exterioare
ci \c{s}i un  
{\it moment propriu\/}, care se cunoa\c{s}te ca {\it spin\/} \c{s}i pe care
\h l vom nota cu $\hat{S}$. Acest operator nu este rela\c{t}ionat cu 
rota\c{t}ii normale fa\c{t}\A\ de axe `reale' \h n spa\c{t}iu, 
dar respect\A\ rela\c{t}ii de comutare
de acela\c{s}i tip ca ale momentului cinetic orbital, respectiv
\begin{equation}
[\hat{S}_i,\hat{S}_j]= i\varepsilon_{ijk}\hat{S}_k~,
\label{98}
\end{equation}
odat\A\ cu urm\A toarele propriet\A \c{t}i:

\begin{itemize}

\item[(1).]
Pentru spin, se satisfac toate formulele de la (23) p\h n\A\ la (48) ale
momentului cinetic. %care sunt similare cu (98)

\item[(2).] 
Spectrul proiec\c{t}iilor de spin, este o secven\c{t}\A\ de numere \h ntregi
sau semi\h ntregi care difer\A\ printr-o unitate.

\item[(3).]
Valorile proprii ale lui $\hat{S}^2$ sunt 
\begin{equation}
\hat{S}^2\psi _{s}=S(S+1)\psi_s.
\label{99}
\end{equation}

\item[(4).]
Pentru un $S$ dat, componentele $S_z$ pot s\A\ ia numai $2S+1$ valori, 
de la $-S$ la $+S$.
 
\item[(5).]
FP-urile particulelor cu spin, pe l\h ng\A\ dependen\c{t}a de $\vec{r}$ 
\c{s}i/sau
$\vec{p}$, depind de o variabil\A\ discret\A\ (proprie spinului)~$\sigma$,
care denot\A\ proiec\c{t}ia spinului pe axa $z$.

\item[(6).]
FP-urile $\psi(\vec{r}, \sigma)$ ale unei particule cu spin se pot 
dezvolta \h n FP-uri cu proiec\c{t}ii date ale spinului $S_z$, adic\A\
\begin{equation}
\psi(\vec{r}, \sigma) = \sum_{\sigma = -S}^S \psi_\sigma(\vec{r})\chi(\sigma),
\label{100}
\end{equation}
unde $\psi_\sigma(\vec{r})$ este partea orbital\A\ \c{s}i $\chi(\sigma)$ este 
partea spinorial\A\ .

\item[(7).]
Func\c{t}iile de spin (spinorii) $\chi(\sigma_i)$ sunt ortogonale pentru
orice pereche $\sigma_i \ne \sigma_k$. Func\c{t}iile 
$\psi_\sigma(\vec{r})\chi(\sigma)$ din suma (100) sunt componentele unei FP 
a unei particule cu spin.

\item[(8).]
Func\c{t}ia $\psi_\sigma(\vec{r})$ se nume\c{s}te parte orbital\A\ 
a spinorului, sau mai scurt orbital.

\item[(9)]
Normalizarea spinorilor se face \h n felul urm\A tor
\begin{equation}
\sum_{\sigma = -S}^S ||\psi_\sigma(\vec{r})|| = 1.
\label{101}
\end{equation}

\end{itemize}

Rela\c{t}iile de comutare permit s\A\ se stabileasc\A\ forma concret\A\ a
operatorilor (matricelor) de spin care ac\c{t}ioneaz\A\ \h n spa\c{t}iul 
FP-urilor operatorului proiec\c{t}iei spinului.

Multe particule `elementare', cum ar fi electronul, neutronul, protonul, 
etc.\ au spin $1/2$ \c{s}i de aceea proiec\c{t}ia spinului lor ia numai dou\A\
valori, respectiv $S_z = \pm 1/2$ (\h n unit\A \c{t}i $\hbar$). Fac parte din
clasa fermionilor, datorit\A\ statisticii pe care o prezint\A\ c\h nd 
formeaz\A\ sisteme de multe corpuri.

Pe de alt\A\ parte, matricele $S_x,\ S_y,\ S_z$ \h n spa\c{t}iul FP-urilor lui
$\hat{S}^2,\ \hat{S}_z$ sunt
\begin{eqnarray}
S_x = {1 \over 2}\left(\begin{array}{cc}
 0 & 1  \\
 1 & 0   
\end{array}\right), \qquad
&&S_y = {1 \over 2}\left(\begin{array}{cc}
 0 & -i  \\
 i &  0   
\end{array}\right), \nonumber\\
\nonumber\\
S_z = {1 \over 2}\left(\begin{array}{cc}
 1 &  0  \\
 0 & -1  
\end{array}\right), \qquad
&&S^2 = {3 \over 4}\left(\begin{array}{cc}
 1 &  0  \\
 0 &  1  
\end{array}\right).
\label{102}
\end{eqnarray}

\subsection*{Defini\c{t}ia matricelor Pauli}
Matricele
\begin{equation}
\sigma_i = 2S_i
\label{103}
\end{equation}
se numesc matricele (lui) Pauli. Sunt matrici hermitice \c{s}i 
au aceea\c{s}i ecua\c{t}ie caracteristic\A\
\begin{equation}
\lambda^2 - 1 = 0,
\label{104}
\end{equation}
\c{s}i prin urmare, autovalorile lui $\sigma_x,\ \sigma_y$ \c{s}i  $\sigma_z$ 
sunt
\begin{equation}
\lambda = \pm 1.
\label{105}
\end{equation}
%
%De aceea, sunt consistente cu faptul c\A\ $S_x,\ S_y$ \c{s}i $S_z$ sunt egali 
%cu $ \pm 1$. 
Algebra acestor matrici este urm\A toarea: 
\begin{equation}
\sigma_i^2 = \hat{I}, \qquad  \sigma_k\sigma_j = -\sigma_j\sigma_k
= i\sigma_z,\qquad \sigma_j\sigma_k = i\sum_l \varepsilon_{jkl}\sigma_l. +
\delta_{jk}I~.
\label{106}
\end{equation}

\noindent
\^{I}n cazul \h n care sistemul cu spin are simetrie sferic\A\ 
\begin{equation}
\psi_1(r, +\textstyle{1\over 2}), \qquad
\psi_1(r, -\textstyle{1\over 2})~,
\label{107}
\end{equation}
sunt solu\c{t}ii diferite datorit\A\ proiec\c{t}iilor $S_z$ diferite.
Valoarea probabilit\A \c{t}ii uneia sau alteia dintre proiec\c{t}ii este 
determinat\A\ de modulul p\A trat $||\psi_{1}||^2$ sau $||\psi_{2}||^2$
\h n a\c{s}a fel c\A\
\begin{equation}
||\psi_1||^2 + ||\psi_2||^2 = 1.
\label{109}
\end{equation}
Cum  FP ale lui $S_z$ are dou\A\ componente, atunci
\begin{equation}
\chi_1= \left(\begin{array}{c}  1  \\ 0  
\end{array}\right), \qquad
\chi_2= \left(\begin{array}{c}  0  \\ 1  
\end{array}\right),
\label{109}
\end{equation}
astfel c\A\ FP a unei particule de spin $1/2$ se poate scrie
ca o matrice de o column\A\
\begin{equation}
\psi = \psi_1\chi_1 + \psi_2\chi_2 = 
\left(\begin{array}{c}  \psi_1 \\ \psi_2 \end{array}\right).
\label{110}
\end{equation}

\noindent
\^{I}n continuare, orbitalii vor fi substitui\c{t}i prin numere datorit\A\
faptului c\A\ ne intereseaz\A\ numai partea de spin.

\section*{Transform\A rile la  rota\c{t}ii ale spinorilor}
%%%%%%%%%%%%%%%%%%%%%%%%%%%%%%%%%%%%%%%%%%%%%%%%%%%%%%%%%%
Fie $\psi$ func\c{t}ia de und\A\ a unui sistem cu spin \h n $\Sigma$. 
S\A\ determin\A m 
probabilitatea proiec\c{t}iei spinului \h ntr-o direc\c{t}ie arbitrar\A\ 
\h n spa\c{t}iul tridimensional (3D) care se poate alege ca ax\A\ $z'$ a lui
$\Sigma'$.
Cum am v\A zut deja pentru cazul momentului cinetic, exist\A\ dou\A\ metode 
de abordare \c{s}i solu\c{t}ionare a acestei probleme:

\begin{itemize}

\item[($\alpha$)]
$\psi$ nu se schimb\A\ c\h nd $\Sigma \rightarrow \Sigma'$ \c{s}i operatorul
$\hat{\Lambda}$ se transform\A\ ca un vector. Trebuie s\A\ g\A sim  FP-urile  
proiec\c{t}iilor $S'_z$ \c{s}i s\A\ dezvolt\A m $\psi$ \h n aceste FP-uri. 
P\A tratele modulelor coeficien\c{t}ilor dau rezultatul
\begin{eqnarray}
\hat{S}_x' = \hat{S}_x\cos\phi + \hat{S}_y\sin\phi &=& 
e^{-il\phi}\hat{S}_x e^{il\phi},\nonumber\\
\hat{S}_y' = -\hat{S}_x\sin\phi + \hat{S}_y\cos\phi &=& 
e^{-il\phi}\hat{S}_y e^{il\phi},\nonumber\\
\hat{S}_z' = -\hat{S}_z = e^{il\phi}\hat{S}_z,
\label{111}
\end{eqnarray}
cu rota\c{t}ii infinitezimale \c{s}i din rela\c{t}iile de comutare pentru spin
se poate g\A si
\begin{equation}
\hat{l} =\hat{S}_z,
\label{112}
\end{equation}
unde $\hat{l}$ este generatorul infinitezimal.

\item[($\beta$)]
A doua reprezentare este:\\
\noindent
$\hat{S}$ nu se schimb\A\ c\h nd $\Sigma \rightarrow \Sigma'$ \c{s}i 
componentele lui $\psi$ se schimb\A\ .
Transformarea la aceast\A\ reprezentare se face cu o 
transformare unitar\A\ de forma
\begin{eqnarray}
\hat{V}^\dagger\hat{S}'\hat{V} &=& \hat{\Lambda}, \nonumber\\
\left(\begin{array}{c}  \psi_1'  \\  \psi_2'  \end{array}\right) &=&
\hat{V}^\dagger
\left(\begin{array}{c}  \psi_1  \\  \psi_2  \end{array}\right)~. 
\label{113}
\end{eqnarray}
Pe baza lui (111) \c{s}i (113) rezult\A\ c\A\  
\begin{eqnarray}
\hat{V}^\dagger e^{-i\hat{S}_z\phi}\hat{S} e^{i\hat{S}_z\phi}
\hat{V}&=&\hat{S},\nonumber\\
\hat{V}^\dagger &=& e^{i\hat{S}_z\phi},
\label{114}
\end{eqnarray}
\c{s}i din (114) se ob\c{t}ine
\begin{equation}
\left(\begin{array}{c}  \psi_1'  \\  \psi_2'  \end{array}\right) =
e^{i\hat{S}_z\phi}
\left(\begin{array}{c}  \psi_1  \\  \psi_2  \end{array}\right)~. 
\label{115}
\end{equation}
Folosind forma concret\A\ a lui $\hat{S}_z$ \c{s}i propriet\A \c{t}ile
matricelor 
Pauli se ob\c{t}ine forma concret\A\ $\hat{V}^\dagger_z$, astfel c\A\
\begin{equation}
\hat{V}^\dagger_z(\phi) = \left(\begin{array}{cc}  
e^{{i \over 2}\phi}  & 0 \\  
0 &  e^{{-i \over 2}\phi}  \end{array}\right).
\label{116}
\end{equation}

\end{itemize}

\section*{Un rezultat al lui Euler}
%%%%%%%%%%%%%%%%%%%%%%%%%%%%%%%%%%%
Se poate ajunge la orice sistem de referin\c{t}\A\ $\Sigma'$ de orientare 
arbitrar\A\ fa\c{t}\A\ de 
$\Sigma$ prin numai trei rota\c{t}ii, prima de unghi $\phi$
\h n jurul axei $z$, urm\A toarea rota\c{t}ie de unghi $\theta$   
\h n jurul noii axe  de coordonate $x'$ \c{s}i ultima de unghi $\psi_a$
\h n jurul lui $z'$. Acest rezultat important apar\c{t}ine lui 
Euler.

Parametrii $(\varphi, \theta, \psi_a)$ se numesc unghiurile (lui) Euler
\begin{equation}
\hat{V}^\dagger(\varphi, \theta, \psi_a) = 
\hat{V}^\dagger_{z'}(\psi_a)\hat{V}^\dagger_{x'}(\theta)
\hat{V}^\dagger_{z}(\varphi).
\label{117}
\end{equation}

Matricele $\hat{V}^\dagger_z$ sunt de forma (116), \h n timp ce  
$\hat{V}^\dagger_x$ este de forma
\begin{equation}
\hat{V}^\dagger_x(\varphi) = \left(\begin{array}{cc}  
\cos{\theta \over 2}  & i\sin{\theta \over 2} \\  
i\sin{\theta \over 2} &  \cos{\theta \over 2}  
\end{array}\right),
\label{118}
\end{equation}
astfel c\A\  
\begin{equation}
\hat{V}^\dagger(\varphi, \theta, \psi_a) = 
\left(\begin{array}{cc}  
e^{i{\varphi + \psi_a \over 2}}\cos{\theta \over 2}   & 
ie^{i{\psi_a - \varphi \over 2}}\sin{\theta \over 2}  \\  
ie^{i{\varphi - \psi_a \over 2}}\sin{\theta \over 2}  &
e^{-i{\varphi + \psi_a \over 2}}\cos{\theta \over 2} 
\end{array}\right).
\label{119}
\end{equation}

Rezult\A\ deci c\A\ prin rota\c{t}ia lui $\Sigma$, componentele func\c{t}iei
spinoriale se transform\A\ dup\A\ cum urmeaz\A\
\begin{eqnarray}
\psi'_1 &=& \psi_1 e^{i{\varphi + \psi_a \over 2}}\cos{\theta \over 2} +
       i\psi_2 e^{i{\psi_a - \varphi \over 2}}\sin{\theta \over 2},\nonumber\\ 
\psi'_2 &=&i\psi_1 e^{i{\varphi - \psi_a \over 2}}\sin{\theta \over 2} +
\psi_2 e^{-i{\varphi + \psi_a \over 2}}\cos{\theta \over 2}.
\label{120}
\end{eqnarray}
Din (120) se poate vedea c\A\ unei rota\c{t}ii \h n $E_3$ \h i corespunde o 
transformare linear\A\ \h n $E_2$, spa\c{t}iul euclidean bidimensional, 
rela\c{t}ionat\A\ cu cele dou\A\ componente ale func\c{t}iei spinoriale. 
Rota\c{t}ia \h n
$E_3$ nu implic\A\ o rota\c{t}ie \h n $E_2$, ceea ce \h nseamn\A\
\begin{equation}
\langle \Phi' | \psi' \rangle = 
\langle \Phi | \psi \rangle = 
\Phi^*_1\psi_1 + \Phi^*_2\psi_2.
\label{121}
\end{equation}

Din (119) se ob\c{t}ine c\A\ (121) nu se satisface, totu\c{s}i exist\A\
o invarian\c{t}\A\ \h n
transform\A rile (119) \h n spa\c{t}iul $E_2$ al func\c{t}iilor spinoriale,
\begin{equation}
\{\Phi | \psi \}= 
\psi_1\Phi_2 - \psi_2\Phi_1.
\label{122}
\end{equation} 

Transform\A rile lineare care men\c{t}in invariante astfel de forme
bilineare se numesc binare.

O m\A rime fizic\A\ cu dou\A\ componente pentru care o 
rota\c{t}ie
a sistemului de coordonate este o transformare binar\A\ se nume\c{s}te 
{\it spin de ordinul \h nt\h i } sau pe scurt {\it spin}.

%%%%%%%%%%%%%%%%%%%%%%%%%%%%%%%%%%%%%%%%%%%%%%%%%%%%%%%%%%%%%%%%%%%%%%%%%%
\subsection*{ Spinorii
%\hspace*{1pt}
unui sistem de doi fermioni}
Func\c{t}iile proprii ale lui $_i\hat{s}^2\ _i\hat{s}_z$, cu $i = 1,2$ au
urm\A toarea form\A\
\begin{equation}
i| + \rangle = \left(\begin{array}{c}
1 \\ 0
\end{array}\right)_i, \qquad 
i| - \rangle = \left(\begin{array}{c}
0 \\ 1
\end{array}\right)_i.
\label{123}
\end{equation}

Un operator foarte folosit \h ntr-un sistem de doi 
fermioni este spinul total
\begin{equation}
\hat{S} = _1\hat{S} + _2\hat{S}
\label{124}
\end{equation}

Spinorii lui $\hat{s}^2\ \hat{s}_z $
sunt ket-uri $|\hat{S}, \sigma  \rangle$, care sunt combina\c{t}ii 
lineare ale spinorilor $_i\hat{s}^2\ _i\hat{s}_z$
\begin{eqnarray}
| + +  \rangle = 
\left(\begin{array}{c} 1 \\ 0 \end{array}\right)_1
\left(\begin{array}{c} 1 \\ 0 \end{array}\right)_1, && \qquad 
| + -  \rangle = 
\left(\begin{array}{c} 1 \\ 0 \end{array}\right)_1
\left(\begin{array}{c} 0 \\ 1 \end{array}\right)_2, \nonumber\\
| - +  \rangle = 
\left(\begin{array}{c} 0 \\ 1 \end{array}\right)_2
\left(\begin{array}{c} 1 \\ 0 \end{array}\right)_1, && \qquad 
| - -  \rangle = 
\left(\begin{array}{c} 0 \\ 1 \end{array}\right)_2
\left(\begin{array}{c} 0 \\ 1 \end{array}\right)_2.
\label{125}
\end{eqnarray}

Func\c{t}iile spinoriale din (125) se consider\A\ ortonormalizate. 
\^{I}n $E_n$ ket-ul 
$| ++ \rangle$ este de  $S_z = 1$ \c{s}i \h n acela\c{s}i timp este
func\c{t}ie proprie a operatorului 
\begin{equation}
\hat{S} = _1\hat{s}^2 + 2(_1\hat{s})(_2\hat{s})+_2\hat{s}^2.
\label{126}
\end{equation}
Dup\A\ cum se poate vedea din 
\begin{eqnarray}
\label{127}
\hspace*{-30pt}
\hat{S}^2 &=& | + +  \rangle = \textstyle{3 \over 2} | + +  \rangle +
2(_1\hat{s}_x \cdot _2\hat{s}_x + _1\hat{s}_y \cdot _2\hat{s}_y +
  _1\hat{s}_z \cdot _2\hat{s}_z)| + +  \rangle , \hspace*{-30pt}\\
\label{128}\hspace*{-30pt}
\hat{S}^2 &=& | + +  \rangle = 2 | + +  \rangle  =1(1+1) | + +  \rangle .
\hspace*{-30pt}
\end{eqnarray} 

Dac\A\ se introduce operatorul 
\begin{equation}
\hat{S}_{-} = _1\hat{s}_{-} + _2\hat{s}_{-},
\label{129}
\end{equation}
se ob\c{t}ine c\A\
\begin{equation}
[\hat{S}_{-} ,\hat{S}^2] = 0.
\label{130}
\end{equation}
Atunci $(\hat{S}_{-})^k|1,1\rangle$ se poate scrie \h n func\c{t}ie de 
FP-urile operatorului $\hat{S}^2$, respectiv
\begin{equation}
\hat{S}_{-}|1,1\rangle = \hat{S}_{-}|+ +\rangle =
\sqrt{2}|+ -\rangle + \sqrt{2}|- +\rangle.
\label{131}
\end{equation}
Rezult\A\ c\A\ $S_z = 0$ \h n starea $\hat{S}_{-}|1,1\rangle $. 
Pe de alt\A\ parte,
din condi\c{t}ia de normalizare avem
\begin{eqnarray}
|1,0\rangle  = \textstyle{1\over\sqrt{2}}(|+ -\rangle + |- +\rangle)\\
\label{132}
\hat{S}_{-}|1,0\rangle =|- -\rangle + |- -\rangle = \alpha|1, -1\rangle.
\label{133} 
\end{eqnarray}

Din condi\c{t}ia de normalizare 
\begin{equation}
|1, -1\rangle  = |-, -\rangle. 
\label{134}
\end{equation}

Exist\A\ \h nc\A\ o singur\A\ combina\c{t}ie linear independent\A\  
de func\c{t}ii de tip (125)
diferit\A\ de $|1, 1\rangle,\ |1, 0\rangle$ y $|1, -1\rangle$, respectiv
\begin{eqnarray}
\psi_4 = \textstyle{1\over\sqrt{2}}(|+ -\rangle - |- +\rangle), \label{135}\\
\hat{S}_z \psi_4 = 0, \qquad \hat{S}^2 \psi_4.
\label{136}
\end{eqnarray}
Prin urmare
\begin{equation}
\psi_4 =|0, 0\rangle.
 \label{137}
\end{equation}
$\psi_4$ descrie starea unui sistem de doi fermioni cu spinul total
egal zero. Acest tip de stare se nume\c{s}te {\it singlet\/}. Pe de alt\A\ 
parte,
starea a doi fermioni de spin total egal unu se poate numi {\it triplet\/}
av\h nd un grad de degenerare $g=3$.

\section*{Moment unghiular total}
%%%%%%%%%%%%%%%%%%%%%%%%%%%%%%%%%
Momentul unghiular total este un operator care se
 introduce ca suma momentului unghiular orbital \c{s}i de spin, respectiv
\begin{equation}
\hat{J} = \hat{l}+\hat{S},
\label{138}
\end{equation}
unde $\hat{l}$ \c{s}i $\hat{S}$, a\c{s}a cum am v\A zut, 
ac\c{t}ioneaz\A\ \h n spa\c{t}ii
diferite, dar p\A tratele lui $\hat{l}$ \c{s}i $\hat{S}$ comut\A\ cu
$\hat{J}$, adic\A\
\begin{equation}
[\hat{J}_i, \hat{J}_j] = i\varepsilon_{ijk}\hat{J}_k, \qquad
[\hat{J}_i, \hat{l}^2] = 0, \qquad [\hat{J}_i, \hat{S}^2] = 0,
\label{139}
\end{equation}
Din (139) rezult\A\ c\A\ $\hat{l}^2$ \c{s}i $\hat{S}^2$ au un sistem comun 
de FP-uri cu $\hat{J}^2$ \c{s}i $\hat{J}_z$.

S\A\ determin\A m spectrul proiec\c{t}iilor lui $\hat{J}_z$ pentru un fermion.
Starea de proiec\c{t}ie de $\hat{J}_z$ maxim se poate scrie 
\begin{eqnarray}
\bar{\psi} &=& \psi_{ll}
\left(\begin{array}{c} 1 \\ 0 \end{array}\right) =
 |l,l,+ \rangle \\
\label{140}
\hat{\jmath}_z\psi &=& (l + \textstyle{1 \over 2})\bar{\psi}, \rightarrow
j= l + \textstyle{1 \over 2}.
\label{141}
\end{eqnarray}

Introducem operatorul $\hat{J}_{-}$ definit prin
\begin{equation}
\hat{J}_{-}=\hat{l}_{-}+\hat{S}_{-} = \hat{l}_{-}+
\left(\begin{array}{cc} 0 & 0 \\ 1 & 0 \end{array}\right).
\label{142}
\end{equation}

Pe baza normaliz\A rii $\alpha = \sqrt{(J+M)(J-M+1)}$ se ob\c{t}ine
\begin{equation}
\hat{J}_{-}\psi_{ll}\left(\begin{array}{c} 1 \\ 0 \end{array}\right) =
\sqrt{2l}|l,l-1, +\rangle + |l,l-1, -\rangle,
\label{143}
\end{equation}
astfel c\A\ valoarea proiec\c{t}iei lui $\hat{j}_{-}$ \h n 
$\hat{j}_{-}\bar{\psi}$ va fi
\begin{equation}
\hat{\jmath}_z = (l-1) + \textstyle{1 \over 2} = l - \textstyle{1 \over 2}~.
\label{144}
\end{equation}
Rezult\A\ c\A\ $\hat{\jmath}_{-}$ mic\c{s}oreaz\A\ cu o unitate ac\c{t}iunea lui
$\hat{J}_z$.

\^{I}n cazul general avem
\begin{equation}
\hat{\jmath}_{-}^k = \hat{l}_{-}^k + k\hat{l}_{-}^{k-1}\hat{S}_{-}~.
\label{145}
\end{equation}
Se observ\A\ c\A\ (145) se ob\c{t}ine din dezvoltarea binomial\A\ 
consider\h nd \h n plus c\A\ 
$\hat{s}^2_{-}$  \c{s}i toate puterile superioare ale lui $\hat{s}$ sunt zero.
\begin{equation}
\hat{\jmath}_{-}^k |l,l,+\rangle = \hat{l}_{-}^k |l,l,+\rangle + 
k\hat{l}_{-}^{k-1} |l,l,-\rangle.
\label{146}
\end{equation}

\c{S}tim c\A\ 
\[
(\hat{l}_{-})^k\psi_{l,l} =
\textstyle{\sqrt{\frac{k!(2l)!}{(2l-k)!}}\psi_{l,l-k}}
\]
\c{s}i folosind-o se ob\c{t}ine
\begin{equation}
\hat{\jmath}_{-}^k | l,l,+\rangle = 
\textstyle{\sqrt{\frac{k!(2l)!}{(2l-k)!}}}| l,l-k,+\rangle +
\textstyle{\sqrt{\frac{(k+1)!(2l)!}{(2l-k+1)!}}} k| l,l-k+1,-\rangle~.
\label{147}
\end{equation}
Not\A m acum $m = l-k$
\begin{equation}
\hat{\jmath}_{-}^{l-m} | l,l,+\rangle = 
\textstyle{\sqrt{\frac{(l-m)!(2l)!}{(l+m)!}}}| l,m,+\rangle +
\textstyle{\sqrt{\frac{(l-m-1)!(2l)!}{(2l+m+1)!}}} (l-m)| l,m+1,-\rangle.
\label{148}
\end{equation}

\noindent
Valorile proprii ale proiec\c{t}iei  momentului unghiular total sunt date de 
secven\c{t}a de numere care difer\A\ printr-o unitate de la 
$j=l+{1\over 2}$ p\h n\A\ la $j=l-{1\over 2}$.
Toate aceste st\A ri apar\c{t}in acelea\c{s}i func\c{t}ii proprii a lui 
$\hat{J}$
ca \c{s}i $|l,l,+\rangle$ pentru c\A\ $[\hat{J}_{-},\hat{J}^2]=0$
\begin{eqnarray}
\hat{J}^2|l,l,+\rangle &=& 
(\hat{l}^2 + 2\hat{l}\hat{S} + \hat{S}^2)|l,l,+\rangle, \nonumber\\ &=& 
[l(l+1) + 2l\textstyle{1\over 2} + \textstyle{3\over 4}]|l,l,+\rangle
\label{149}
\end{eqnarray}
unde $j(j+1) = (l+{1 \over 2})(l + {3 \over 2})$.

\^{I}n partea dreapt\A\ a lui (149) o contribu\c{t}ie diferit\A\ de zero d\A\
numai 
$j=\hat{l}_z\hat{S}_z$. Atunci FP-urile ob\c{t}inute corespund perechii 
$j=l+{1 \over 2}$, $m_j=m+{1 \over 2}$ \c{s}i sunt de forma
\begin{equation} 
 |l+{1 \over 2}, m+{1 \over 2} \rangle = 
\sqrt{l+m+1 \over 2l+1}|l, m, + \rangle +
\sqrt{l-m \over 2l+1}|l, m+1, - \rangle.
\label{150}
\end{equation}

Num\A rul total de st\A ri linear independente este
\begin{equation}
N=(2l+1)(2s + 1) = 4l+2,
\label{151}
\end{equation}
din care \h n (150) s-au construit (2j+1)=2l+3. Restul de 
$2l-1$ func\c{t}ii proprii se pot ob\c{t}ine din condi\c{t}ia de 
ortonormalizare:
\begin{equation}
|l-\textstyle{1 \over 2}, m-\textstyle{1 \over 2} \rangle = 
\sqrt{l-m \over  2l+1} | l, m, + \rangle -
\sqrt{l+m+1 \over  2l+1} | l, m+1, - \rangle.
\label{152}
\end{equation}

Dac\A\ dou\A\ subsisteme sunt \h n interac\c{t}iune \h n a\c{s}a fel \h nc\h t
fiecare moment unghiular  $\hat{j}_i$ se conserv\A\ , atunci FP-urile 
operatorului moment unghiular total
\begin{equation}
\hat{J} =\hat{\jmath}_1 + \hat{\jmath}_2,
\label{153}
\end{equation}
se pot ob\c{t}ine printr-o procedur\A\ asem\A n\A toare
celei anterioare. Pentru valori proprii fixe ale lui  
$\hat{\jmath}_1$ \c{s}i  $\hat{\jmath}_2$ exist\A\ 
$(2j_1+1)(2j_2+1)$
FP-uri ortonormalizate ale proiec\c{t}iei momentului unghiular total 
$\hat{J}_z$, iar cea care 
corespunde valorii maxime a proiec\c{t}iei $\hat{J}_z$, adic\A\  
$M_J = j_1 + j_2$,
se poate construi \h n mod unic \c{s}i prin urmare $J = j_1 + j_2$ este 
valoarea maxim\A\ a momentului unghiular total al sistemului. Aplic\h nd
operatorul
$\hat{J} = \hat{\jmath}_1 + \hat{\jmath}_2$ \h n mod repetat func\c{t}iei
\begin{equation}
|j_1 + j_2, j_1 + j_2, j_1 + j_2 \rangle = |j_1 , j_1\rangle \cdot
|j_2 , j_2\rangle,
\label{154}
\end{equation}
se pot ob\c{t}ine toate cele $2(j_1 + j_2) + 1$ FP ale 
lui
$\hat{J} = j_1 + j_2$ cu diferi\c{t}i $M$: 
\[
-(j_1 + j_2) \leq M \leq (j_1 + j_2).
\]
De exemplu, FP pentru $M = j_1 + j_2-1$ este:
\begin{equation}
|j_1 + j_2, j_1 + j_2-1, j_1 , j_2 \rangle =
\sqrt{j_1 \over j_1 + j_2}|j_1,j_1-1, j_2, j_2 \rangle +
\sqrt{j_2 \over j_1 + j_2}|j_1,j_1, j_2, j_2-1 \rangle .
\label{155}
\end{equation}

Aplic\h nd \h n continuare de mai multe ori operatorul $\hat{J}_{-}$ se pot
ob\c{t}ine
cele $2(j_1 + j_2-1) -1$ func\c{t}ii ale lui $J = j_1 + j_2-1$.

Se poate demonstra c\A\ 
\[
|j_1 - j_2| \leq J \leq j_1 + j_2
\]
astfel c\A\
\begin{equation}
\sum_{{\rm min} \ J}^{{\rm max} \ J}(2J +1) = (2J_1 +1)(2J_2 +1)
\label{156}
\end{equation}
\c{s}i deci
\begin{equation}
|J,M,j_1,j_2\rangle = 
\sum_{m_1+m_2 = M} (j_{1} m_{1}j_{2}m_{2}|J M) | j_1, m_1, j_2, m_2 \rangle~,
\label{156}
\end{equation}
unde coeficien\c{t}ii $(j_{1} m_{1}j_{2}m_{2}|J M)$ determin\A\ contribu\c{t}ia
diferitelor func\c{t}ii $| j_1, m_1, j_2, m_2 \rangle$ \h n func\c{t}iile 
proprii ale lui $\hat{J^2}$, $\hat{J_{z}}$ de valori proprii $J(J+1)$, $M$ 
\c{s}i sunt numi\c{t}i coeficien\c{t}ii Clebsch-Gordan.

\bigskip

\noindent
\underline{Referin\c{t}e}: 

%1. Acetatos del Prof. H. Rosu

\noindent
%Referin\c{t}e bibliografice:

\noindent
1. H.A. Buchdahl, ``Remark concerning the eigenvalues of orbital angular
momentum",

\noindent
Am. J. Phys. {\bf 30}, 829-831 (1962)

\newpage

\noindent
{\bf 3N. Not\u{a}}:
 1. Operatorul corespunz\A tor vectorului Runge-Lenz din problema Kepler
clasic\A\ se scrie 
$$
\hat{\vec{A}}=\frac{{\bf \hat{r}}}{r}+\frac{1}{2}\Bigg[(\hat{l}\times \hat{p})-
(\hat{p}\times \hat{l})\Bigg]~.
$$
unde s-au folosit unit\A \c{t}i atomice \c{s}i s-a 
considerat $Z=1$ (atomul de hidrogen).
Acest operator comut\A\ cu Hamiltonianul atomului de hidrogen
$\hat{H}=\frac{\hat{p^2}}{2}-\frac{1}{r}$, adic\A\ este integral\A\ de 
mi\c{s}care cuantic\A\ . Componentele sale au 
comutatori de tipul $[A_i,A_j]=-2i\epsilon _{ijk}l_{k}\cdot H$, iar
comutatorii componentelor Runge-Lenz 
cu componentele momentului cinetic sunt de tipul
$[l_i,A_j]=i\epsilon _{ijk}A_k$, adic\A\ respect\A\ condi\c{t}iile (23).
De demonstrat toate aceste rela\c{t}ii poate fi un exerci\c{t}iu util.

%\newpage
\section*{{\huge 3P. Probleme}}
\subsection*{Problema 3.1}
S\A\ se arate c\A\ orice operator de transla\c{t}ie, 
pentru care $\psi (y+a) = T_{a}\psi(y)$, se poate scrie 
ca un operator exponen\c{t}ial \c{s}i s\A\ se aplice acest rezultat pentru 
$y=\vec{r}$ \c{s}i pentru rota\c{t}ia finit\A\ $\alpha$ \h n jurul axei $z$.\\
\noindent
{\bf Solu\c{t}ie}\\
Demonstra\c{t}ia se ob\c{t}ine dezvolt\h nd $\psi(y+a))$ \h n serie Taylor 
\h n vecin\A tatea infinitezimal\A\ a punctului $x$, adic\A\ \h n puteri ale
lui $a$
$$
\psi(y+a)=\sum _{n=0}^{\infty}\frac{a^{n}}{n!}\frac{d^{n}}{dx^{n}}\psi(x)
$$
Observ\A m c\A\
$$
\sum _{n=0}^{\infty}\frac{a^{n}\frac{d^{n}}{dx^{n}}}{n!}=e^{a\frac{d}{dx}}
$$
\c{s}i deci $T_{a}=e^{a\frac{d}{dx}}$ \h n cazul 1D. \^{I}n 3D, $y=\vec{r}$
\c{s}i $a\rightarrow \vec{a}$. Rezultatul este 
$T_{\vec{a}}=e^{\vec{a}\cdot 
\vec{\nabla}}
$.

Pentru rota\c{t}ia finit\A\ $\alpha$ \h n jurul lui $z$ avem $y=\phi$ \c{s}i
$a=\alpha$. Rezult\A\
$$
T_{\alpha}=R_{\alpha}=e^{\alpha \frac{d}{d\phi}}~.
$$

O alt\A\ form\A\ exponen\c{t}ial\A\ a rota\c{t}iei \h n $z$ este cea \h n
func\c{t}ie de operatorul moment cinetic a\c{s}a cum s-a comentat \h n acest
capitol. Fie $x' = x+dx$, \c{s}i consider\h nd numai primul ordin al seriei
Taylor 
\begin{eqnarray}
\psi(x',y',z') &=& \psi(x,y,z) + (x'-x)\frac{\partial}{\partial x'}
\psi(x',y',z')\bigg|_{\vec{r'} = \vec{r}}  \nonumber \\
&& + (y'-y)\frac{\partial}{\partial y'} \psi(x',y',z')
\bigg|_{\vec{r'} = \vec{r}} \nonumber \\ 
&& + (z'-z)\frac{\partial}{\partial z'} \psi(x',y',z')
\bigg|_{\vec{r'} = \vec{r}}\, . \nonumber 
\end{eqnarray}

\c{T}in\h nd cont de faptul c\A\
\begin{eqnarray}
\frac{\partial}{\partial x'_i}\psi(\vec{r}')\bigg|_{\vec{r}'} &=& 
\frac{\partial}{\partial x_i}\psi(\vec{r}), \nonumber\\ 
x' = x -y d \phi, \qquad y' &=& y + xd \phi, \qquad z' = z, \nonumber
\end{eqnarray}
se poate reduce seria din trei dimensiuni la numai dou\A\  
\begin{eqnarray}
\psi(\vec{r}') &=& \psi(\vec{r}) + 
  (x-yd\phi-x)\frac{\partial\psi(\vec{r})}{\partial x} 
+ (y+xd\phi -y)\frac{\partial\psi(\vec{r})}{\partial y'}, \nonumber\\
&=& \psi(\vec{r})-y d\phi \frac{\partial\psi(\vec{r})}{\partial x}  
	         +x d\phi x\frac{\partial\psi(\vec{r})}{\partial y},\nonumber\\ 
&=& \left[1 - d\phi\left(-x\frac{\partial}{\partial y} 
            + y\frac{\partial}{\partial x}\right)\right]\psi(\vec{r})~.\nonumber
\end{eqnarray}
Cum $i\hat{l}_z = 
 \left(x\frac{\partial}{\partial y} - y\frac{\partial}{\partial x}\right)$
rezult\A\ c\A\
$
R = \left[1 - d\phi\left(x\frac{\partial}{\partial y} 
            - y\frac{\partial}{\partial x}\right)\right]~.
$
Continu\h nd \h n ordinul doi se poate ar\A ta c\A\ se ob\c{t}ine
$\textstyle{1 \over 2!}(i\hat{l}_zd\phi)^2$ \c{s}i a\c{s}a mai departe.
Prin urmare, $R$ poate fi scris ca exponen\c{t}ial\A\
\[
R=e^{i\hat{l}_zd\phi}.
\]

%\newpage
\subsection*{Problema 3.2}
S\A\ se arate c\A\ pe baza expresiilor date \h n (14) se poate ajunge la 
(15). \\

\bigskip

{\bf{Solu\c{t}ie}}\\
S\A\ consider\A m numai termenii lineari \h n dezvoltarea \h n serie
Taylor (rota\c{t}ii infinitezimale)
\[
e^{i\hat{l}_zd\phi} = 1 + i\hat{l}_zd\phi + 
\textstyle{1 \over 2!}(i\hat{l}_zd\phi)^2 + \ldots\, , \nonumber
\]
\c{s}i deci
\begin{eqnarray}
(1 + i\hat{l}_zd\phi)\hat{A}_x(1-i\hat{l}_zd\phi) &=& 
\hat{A}_x -\hat{A}_xd\phi,\nonumber\\
(\hat{A}_x + i\hat{l}_zd\phi\hat{A}_x)(1-i\hat{l}_zd\phi) &=& 
\hat{A}_x -\hat{A}_xd\phi,\nonumber\\
\hat{A}_x -\hat{A}_xi\hat{l}_zd\phi+ i\hat{l}_zd\phi\hat{A}_x +
\hat{l}_zd\phi\hat{A}_x \hat{l}_zd\phi &=&\hat{A}_x -\hat{A}_xd\phi,\nonumber\\
i(\hat{l}_z\hat{A}_x -\hat{A}_x\hat{l}_z)d\phi &=&-\hat{A}_yd\phi.\nonumber
\end{eqnarray} 
Ajungem u\c{s}or la concluzia c\A\
\[
[ \hat{l}_z, \hat{A}_x]  = i\hat{A}_y~. \nonumber
\]
Deasemenea, $[ \hat{l}_z, \hat{A}_y]  = i\hat{A}_x$ se ob\c{t}ine din:
\begin{eqnarray}
(1 + i\hat{l}_zd\phi)\hat{A}_y(1-i\hat{l}_zd\phi) &=& 
\hat{A}_xd\phi -\hat{A}_y,\nonumber\\
(\hat{A}_y + i\hat{l}_zd\phi\hat{A}_y)(1-i\hat{l}_zd\phi) &=& 
\hat{A}_xd\phi -\hat{A}_y,\nonumber\\
\hat{A}_y -\hat{A}_yi\hat{l}_zd\phi+ i\hat{l}_zd\phi\hat{A}_y +
\hat{l}_zd\phi\hat{A}_y \hat{l}_zd\phi &=&\hat{A}_xd\phi -\hat{A}_y,\nonumber\\
i(\hat{l}_z\hat{A}_y -\hat{A}_y\hat{l}_z)d\phi &=&-\hat{A}_xd\phi.\nonumber
\end{eqnarray} 

\subsection*{Problema 3.3}
S\A\ se determine operatorul $\frac{d\hat{\sigma}_{x}}{dt}$ pe baza 
Hamiltonianului unui electron cu spin aflat \h ntr-un c\h mp magnetic de
induc\c{t}ie $\vec{B}$.

\bigskip

\noindent
{\bf Solu\c{t}ie}\\
Hamiltonianul \h n acest caz este 
$\hat{H}(\hat{{\bf p}},\hat{{\bf r}},\hat{{\bf \sigma}})=
\hat{H}(\hat{{\bf p}},\hat{{\bf r}})+\hat{{\bf \sigma}}\cdot \vec{{\bf B}}$, 
unde ultimul termen este Hamiltonianul Zeeman pt. electron.
Cum $\hat{\sigma}_{x}$ comut\A\ cu impulsurile \c{s}i coordonatele, aplicarea
ecua\c{t}iei de mi\c{s}care Heisenberg conduce la:
$$  
\frac{d\hat{\sigma}_{x}}{dt}=\frac{i}{\hbar}[\hat{H},\hat{\sigma}_{x}]=
-\frac{i}{\hbar}\frac{e\hbar}{2m_e}((\hat{\sigma}_{y}B_y+\hat{\sigma}_{z}B_z)
\hat{\sigma}_{x}-\hat{\sigma}_{x}(\hat{\sigma}_{y}B_y+
\hat{\sigma}_{z}B_z))
$$
Folosind $[\sigma _{x},\sigma _y]=i\sigma _z$, rezult\A\ :
$$
\frac{d\hat{\sigma}_{x}}{dt}=
\frac{e}{m_e}(\hat{\sigma}_{y}B_{z}-\hat{\sigma}_{z}B_{y})
=\frac{e}{m_e}(\vec{\sigma}\times \vec{B})_{x}~.
$$

%\newpage
%\subsection*{Problema 3.3}
%S\A\ se determine precesia spinului unui electron 
%\h ntr-un c\h mp magnetic omogen.\\

%\bigskip

%\noindent
%{\bf Solu\c{t}ie}\\
%Electronul se mi\c{s}c\A\ circular \h n jurul unei axe de c\h mp
%magnetic omogen \c{s}i uniform cu frecven\c{t}a
%\[
%\omega= 2\omega_L = {-eB \over mc}.
%\]
%Acesta este un rezultat binecunoscut care este consecin\c{t}a echilibr\A rii
%for\c{t}ei Lorentz cu for\c{t}a centrifug\A\ 
%
%\[
%{eBv \over c} = my\omega^2,
%\]
%\c{s}i atunci
%\[
%\omega = - {eB \over mc}.
%\]
%Astfel
%\[
%\omega_L = {eB \over 2mc}
%\]
%care se cunoa\c{s}te ca {\it frecven\c{t}a Larmor}.

%\newpage
%\subsection*{Problema 3.4}
%
%S\A\ se rezolve ec.\ Laplace folosind coordonate sferice.\\

%\bigskip

%{\bf Solu\c{t}ie}\\
%Folosind separarea de func\c{t}ii \c{s}i variabile 
%\[
%U(r, \theta, \phi) = R(r)\Theta(\theta, \phi),
%\]
%se vede c\A\
%\[
%{r \over R(r)} {\partial^2 \over \partial r^2} [rR(r)] = {1 \over \Theta} 
%{\bf L^2}\Theta = l(l+1)
%\]

\newpage
%%%%%%%%%%%%%%%%%%%%%%%%%%%%%%%%%%%%%%%%%%%%%%%%%%%%%%%%%%%%%%%%%%%%%%
%%%%%%%%%%%%%%%%%%%%%%%%%%%%%%%%%%%%%%%%%%%%%%%%%%%   W K B
%%%%%%%%%%%%%%%%%%%%%%%%%%%%%%%%%%%%%%%
%\documentclass[a4paper,12pt]{article}
%\usepackage{latexsym}
%\pagestyle{empty}
%\begin{document}
%\begin{center}
%\setcounter{equation}
%%%%%%%%%%%%%%%%%%%%%%%%%%%%%%%%%%%%%%%%
%\documentclass[a4paper,12pt]{article}
%\usepackage{latexsym}
%\pagestyle{empty}
%\begin{document}
\begin{center}
{\huge{4. METODA WKB}}
%\section*{4. METODA WKB}
\end{center}
\setcounter{equation}{0}
\hspace{0.6cm} Pentru a studia poten\c{t}iale
mai realiste dec\h t cele de  
{\em bariere \c{s}i gropi rectangulare}, este necesar de multe ori
s\A\ se foloseasc\A\
metode care s\A\ permit\A\ 
rezolvarea ecua\c{t}iei Schr\"odinger pentru clase generale de 
poten\c{t}iale \c{s}i care s\A\ fie o bun\A\ aproximare a solu\c{t}iilor 
exacte. 

Scopul diferitelor metode de aproximare este s\A\ ofere  
solu\c{t}ii suficient de bune \c{s}i simple, care s\A\ permit\A\ \h n acest
fel \h n\c{t}elegerea comportamentului sistemului \h n form\A\ cuasianalitic\A\ .

\^{I}n cadrul mecanicii cuantice, una dintre cele mai vechi \c{s}i eficiente
metode a fost dezvoltat\A\ \h n mod aproape simultan de c\A tre  
{\em G. Wentzel, H. A. Kramers \c{s}i L.
Bri\-llouin} \h n 1926, de la al c\A ror nume deriv\A\ acronimul 
{\em WKB} sub care este cunoscut\A\ aceast\A\ metod\A\ (corect este {\em JWKB},
vezi nota 4N).
%\textit{\textbf{WKB}}.

Este important de men\c{t}ionat c\A\ metoda WKB se aplic\A\ mai ales  
ecua\c{t}iilor Schr\"odinger 1D \c{s}i exist\A\ 
dificult\A \c{t}i serioase \h n generaliz\A rile la dimensiuni superioare.

Pentru a rezolva ecua\c{t}ia Schr\"odinger
\begin{equation}
-\frac{\hbar^2}{2m}\frac{d^2\psi}{dy^2}+u(y)\psi=E\psi
\end{equation}
presupunem c\A\ poten\c{t}ialul are forma:
\begin{equation}
u(y)=u_0f\Big(\frac{y}{a}\Big)
\end{equation}

\noindent
\c{s}i facem schimbul de variabil\A\ :
\begin{equation}
\xi^2=\frac{\hbar^2}{2mu_0a^2}
\end{equation}
\begin{equation}
\eta=\frac{E}{u_0}
\end{equation}
\begin{equation}
x=\frac{y}{a}~.
\end{equation}
Din ecua\c{t}ia $(5)$ ob\c{t}inem:
\begin{equation}
\frac{d}{dx}=
\frac{dy}{dx}\frac{d}{dy}=
a\frac{d}{dy}
\end{equation}
\begin{equation}
\frac{d^2}{dx^2}
=\frac{d}{dx}\Big(a\frac{d}{dy}\Big)
=\Big(a\frac{d}{dx}\Big)\Big(a\frac{d}{dx}
\Big)=a^2\frac{d^2}{dy^2}
\end{equation}
\c{s}i ec. Schr\"odinger se scrie:
\begin{equation}
-\xi^2\frac{d^2\psi}{dx^2}+f(x)\psi=\eta\psi~.
\end{equation}
Multiplic\h nd cu $-1/\xi^2$ \c{s}i definind $r(x)=\eta-f(x)$, este posibil 
s\A\ o scriem \h n forma:
\begin{equation}
\frac{d^2\psi}{dx^2}+\frac{1}{\xi^2}r(x)\psi=0~.
\end{equation}
Pentru a rezolva (9) se propune urm\A toarea solu\c{t}ie:
\begin{equation}
\psi(x)=\exp\Bigg[\frac{i}{\xi}\int_a^x{q(x)dx}\Bigg]~.
\end{equation}

%\^{I}n general: 
%$\int_a^xq(x)dx=Q(x)\vert_a^x=
%Q(x)-Q(a)\ni\frac{\partial{Q(x)}}{\partial{x}}=\frac{dQ(x)}{dx}=q(x)$ conform 
%teoremei fundamentale a calculului.

A\c{s}adar:
%\begin{displaymath}
$$
\frac{d^2{\psi}}{dx^2}
=\frac{d}{dx}\bigg(\frac{d\psi}{x}\bigg)
=\frac{d}{dx}\Bigg\{\frac{i}{\xi}q(x)
\exp\Bigg[{\frac{i}{\xi}\int_a^xq(x)dx\Bigg]}\Bigg\}
$$
%\end{displaymath}
%\begin{displaymath}
$$
\Longrightarrow\frac{d^2\psi}{dx^2}
=\frac{i}{\xi}\Bigg\{\frac{i}{\xi}q^2(x)\exp\Bigg[\frac{i}{\xi}
\int_a^xq(x)dx\Bigg]+\frac{\partial{q(x)}}{\partial{x}}
\exp\Bigg[\frac{i}{\xi}\int_a^xq(x)dx\Bigg]\Bigg\}~.
$$
%\end{displaymath}
Factoriz\h nd $\psi$ avem:
\begin{equation}
\frac{d^2\psi}{dx^2}=\Bigg[-\frac{1}
{\xi^2}q^2(x)+\frac{i}{\xi}\frac{dq(x)}{dx}\Bigg]\psi~.
\end{equation}

\noindent
Neglij\h nd pe moment dependen\c{t}a \h n $x$, ecua\c{t}ia Schr\"odinger 
se poate scrie :
\begin{equation}
\Bigg[-\frac{1}{\xi^2}q^2+\frac{i}{\xi}\frac{\partial{q}}{\partial{x}}
+\frac{1}{\xi^2}r\Bigg]\psi=0
\end{equation}

\c{s}i cum \h n general $\psi\neq0$, avem:
\begin{equation}
i\xi\frac{dq}{dx}+r-q^2=0~,
\end{equation}
care este o ecua\c{t}ie diferen\c{t}ial\A\ linear\A\ de tip Riccati, 
a c\A rei solu\c{t}ii se caut\A\
\h n forma unei serii de puteri ale lui $\xi$ cu presupunerea c\A\ $\xi$ este
foarte mic. 

Mai exact seria se propune de forma:
\begin{equation}
q(x)=\sum^\infty_{n=0}(-i\xi)^nq_n(x)~.
\end{equation}
Substituind-o \h n Riccati ob\c{t}inem:
\begin{equation}
i\xi\sum_{n=0}^\infty(-i\xi)^n\frac{dq_n}{dx}+r(x)-
\sum_{\mu=0}^\infty(-i\xi)^{\mu}q_{\mu}
\sum_{\nu=0}^\infty(-i\xi)^{\nu}q_{\nu}=0~.
\end{equation}
Printr-o rearanjare a termenilor ob\c{t}inem:
\begin{equation}
\sum_{n=0}^\infty(-1)^n(i\xi)^{n+1}\frac{dq_n}{dx}+r(x)-
\sum_{\mu=0}^\infty\sum_{\nu=0}^\infty(-i\xi)^{\mu+\nu}q_{\mu}q_{\nu}=0~.
\end{equation}
Seriile duble au urm\A toarea proprietate:
\begin{displaymath}
\sum_{\mu=0}^\infty\sum_{\nu=0}^\infty{a_{\mu\nu}}=\sum_{n=0}^\infty
\sum_{m=0}^n{a_{m,n-m}}~,
\end{displaymath}
unde: $\mu=n-m\quad ,\nu=m$~.\\\\
\^{I}n acest fel:
\begin{equation}
\sum_{n=0}^\infty(-1)^n(i\xi)^{n+1}\frac{dq_n}{dx}+r(x)-
\sum_{n=0}^\infty\sum_{m=0}^n(-i\xi)^{n-m+m}q_{m}q_{n-m}=0~.
\end{equation}

S\A\ vedem explicit c\h \c{t}iva termeni \h n fiecare din seriile din
ecua\c{t}ia (17): 
\begin{equation}
\sum_{n=0}^\infty(-1)^n(i\xi)^{n+1}\frac{dq_n}{dx}=i\xi
\frac{dq_0}{dx}+\xi^2\frac{dq_1}{dx}-i\xi^3\frac{dq_2}{dx}+\dots
\end{equation}
\begin{equation}
\sum_{n=0}^\infty\sum_{m=0}^n(-i\xi)^{n}q_{m}q_{n-m}=q^2_0-i2{\xi}q_0q_1+\dots
\end{equation}
Pentru ca ambele serii s\A\ con\c{t}in\A\ pe $i\xi$ \h n primul termen 
trebuie s\A\ le scriem:
$$
%\begin{displaymath}
\sum_{n=1}^\infty(-1)^{n-1}(i\xi)^n\frac{dq_{n-1}}{dx}+r(x)-q_0^2-
\sum_{n=1}^\infty\sum_{m=0}^n(-i\xi)^nq_mq{n-m}=0~,
%\end{displaymath}
$$
ceea ce ne conduce la:
\begin{equation}
\sum_{n=1}^\infty\Bigg[-(-i\xi)^n\frac{dq_{n-1}}{dx}-
\sum_{m=0}^n(-i\xi)^nq_mq_{n-m}\Bigg]+\Bigg[r(x)-q_0^2\Bigg]=0~.
\end{equation}

Pentru ca aceast\A\ ecua\c{t}ie s\A\ se satisfac\A\ avem condi\c{t}iile:
\begin{equation}
r(x)-q_0^2=0 \quad\Rightarrow\quad q_0=\pm\sqrt{r(x)}
\end{equation}
$$
%\begin{displaymath}
-(-i\xi)^n\frac{dq_{n-1}}{dx}-\sum_{m=0}^n(-i\xi)^nq_mq_{n-m}=0 \quad 
%\end{displaymath}
$$
\begin{equation}
\Rightarrow\quad\quad\frac{dq_{n-1}}{dx}=-\sum_{m=0}^{n}q_mq_{n-m}
\quad\quad{n\geq1}~.
\end{equation}
Aceasta ultima este o rela\c{t}ie de recuren\c{t}\A\ care apare \h n metoda WKB.
Este momentul s\A\ amintim c\A\ am definit 
$r(x)=\eta-f(x),\quad\eta=\frac{E}{u_0}\quad\&\quad{f(x)=\frac{u}{u_0}}$ 
\c{s}i cu ajutorul ec. $(21)$ ob\c{t}inem:
\begin{equation}
q_0=\pm\sqrt{\eta-f(x)}=\pm\sqrt{\frac{E}{u_0}-\frac{u}{u_0}}=
\pm\sqrt{\frac{2m(E-u)}{2mu_0}}~,
\end{equation}
care ne indic\A\ natura clasic\A\
a impulsului WKB 
a particulei de energie $E$ \h n poten\c{t}ialul $u$ \c{s}i \h n unit\A \c{t}i 
$\sqrt{2mu _0}$. Astfel:
$$
%\begin{displaymath}
q_0=p(x)=\sqrt{\eta-f(x)}$$ \hspace{1mm} {\bf nu este un operator}.
%\end{displaymath}
%$$
Dac\A\ aproxim\A m p\h n\A\ la ordinul doi, ob\c{t}inem:
$$
%\begin{displaymath}
q(x)=q_0-i{\xi}q_1-\xi^2q_2
%\end{displaymath}
$$
\c{s}i folosind rela\c{t}ia de recuren\c{t}\A\ WKB (22)
calcul\A m $q_1$ \c{s}i 
$q_2$:
$$
%\begin{displaymath}
\frac{dq_0}{dx}=-2q_0q_1
\quad \Rightarrow \quad 
q_1=-\frac{1}{2}\frac{\frac{dq_0}{dx}}{q_0}=
-\frac{1}{2}\frac{d}{dx}(\ln\vert{q_0}\vert)
%\end{displaymath}
$$
\begin{equation}
\Rightarrow \quad q_1=-\frac{1}{2}\frac{d}{dx}(\ln\vert p(x)\vert)
\end{equation}
\begin{equation}
\frac{dq_1}{dx}=-2q_0q_2-q_1^2 \quad\Rightarrow\quad 
q_2=-\frac{\frac{dq_1}{dx}-q_1^2}{2q_0}~.
\end{equation}

Din ecua\c{t}ia $(24)$, ne d\A m seama c\A\ m\A rimea $q_1$ este panta cu 
semnul schimbat a lui $\ln\vert q_0\vert$; c\h nd $q_0$ este foarte mic, 
$q_1\ll0\quad\Rightarrow\quad -\xi{q_1}\gg0$ \c{s}i prin urmare 
seria diverge. Pentru a evita acest lucru se impune urm\A toarea
{\bf condi\c{t}ie WKB}:
$$
%\begin{displaymath}
\vert q_0\vert\gg\vert -\xi{q_1}\vert=\xi\vert{q_1}\vert~.$$
%\hspace{1mm}

Este important de observat c\A\ aceast\A\ condi\c{t}ie WKB nu se satisface 
pentru acele puncte $x_k$ unde:
$$
%\begin{displaymath}
q_0(x_k)=p(x_k)=0~.
%\end{displaymath}
$$
Dar $q_0=p=\sqrt{\frac{2m(E-u)}{2mu_0}}$ \c{s}i deci ecua\c{t}ia 
precedent\A\ ne conduce la:
\begin{equation}
E=u(x_k)~.
\end{equation}

\^{I}n mecanica clasic\A\ punctele $x_k$ care satisfac  
(26) se numesc {\bf puncte de \h ntoarcere} pentru c\A\ \h n ele are loc 
schimbarea sensului de mi\c{s}care a particulei macroscopice.

\^{I}n baza acestor argumente, putem s\A\ spunem despre $q_0$ c\A\ este
o solu\c{t}ie 
clasic\A\ a problemei examinate \c{s}i c\A\ m\A rimile $q_1$ \& $q_2$ 
sunt respectiv prima \c{s}i a doua corec\c{t}ie cuantic\A\ \h n problema WKB.

Pentru a ob\c{t}ine func\c{t}iile de und\A\ vom considera numai solu\c{t}ia 
clasic\A\ \c{s}i prima corec\c{t}ie cuantic\A\ a problemei pe care le substituim 
\h n forma WKB a lui $\psi$:
$$
%\begin{displaymath}
\psi=\exp\Bigg[\frac{i}{\xi}\int_a^x{q(x)dx}\Bigg]=
\exp\Bigg[\frac{i}{\xi}\int_a^x(q_0-i\xi{q_1})dx\Bigg]
%\end{displaymath}
$$
$$
%\begin{displaymath}
\Rightarrow\quad\psi=\exp\Bigg(\frac{i}{\xi}\int_a^xq_0dx\Bigg)\cdot
\exp\Bigg(\int_a^xq_1dx\Bigg)~.
%\end{displaymath}
$$

Pentru al doilea factor avem:
$$
%\begin{displaymath}
\exp\Bigg(\int_a^xq_1dx\Bigg)=\exp\Bigg[-\frac{1}{2}
\int_a^x\frac{d}{dx}(\ln \vert p(x)\vert)dx\Bigg]=
%\end{displaymath}
$$
$$
%\begin{displaymath}
\quad\quad\quad\quad\quad\quad\quad=\exp\Bigg[-\frac{1}{2}(\ln\vert 
p(x)\vert)\Big{\vert}_a^x\Bigg]=\frac{A}{\sqrt{p(x)}}~, 
%\end{displaymath}
$$
cu $A$ o constant\A\ , \h n timp ce pentru primul factor se ob\c{t}ine:
$$
%\begin{displaymath}
\exp\Bigg(\frac{i}{\xi}\int_a^xq_0dx\Bigg)=\exp\Bigg[\pm\frac{i}{\xi}
\int_a^xp(x)dx\Bigg]~.
%\end{displaymath}
$$
Astfel putem scrie $\psi$ \h n urm\A toarea form\A\ :
\begin{equation}
\psi^{\pm}=\frac{1}{\sqrt{p(x)}}\exp\Bigg[\pm\frac{i}{\xi}\int_a^xp(x)dx\Bigg]~,
\end{equation}
care sunt cunoscute ca %\textbf
{\bf solu\c{t}ii WKB ale ecua\c{t}iei Schr\"odinger 
unidimensionale}.
Solu\c{t}ia general\A\ WKB \h n regiunea \h n care 
condi\c{t}ia WKB se satisface se scrie:
\begin{equation}
\psi=a_+\psi^++a_-\psi^-~.
\end{equation}

\noindent
A\c{s}a cum deja s-a men\c{t}ionat, nu exist\A\ solu\c{t}ie WKB \h n punctele
de \h ntoarcere, ceea ce ridic\A\ problema modului \h n care se face trecerea
de la $\psi(x<x_k)$ la
$\psi(x>x_k)$. Rezolvarea acestei dificult\A \c{t}i se face prin introducerea 
formulelor de conexiune WKB.

%pero antes veamos como hacer una estimaci\'on
%del error que se comete al resolver una ecuaci\'on diferencial
%ordinaria por el m\'etodo aproximativo WKB. %\end{document}
\bigskip

%%%%%%%%%%%%%%%%%%%%%%%%%%%%%%%%%%%%%%%%%%
%\documentclass[a4paper,12pt]{article}
%\usepackage{latexsym}
%\pagestyle{empty}
%\begin{document}
%\begin{center}
%\centerline{\huge
\subsection*{Formulele de Conexiune}
%\end{center}

\hspace{0.6cm}Deja s-a v\A zut c\A\ solu\c{t}iile WKB sunt singulare 
\h n punctele de \h ntoarcere clasic\A\ ; totu\c{s}i aceste solu\c{t}ii sunt
corecte la st\h nga \c{s}i la dreapta acestor puncte  
$x_k$. Ne \h ntreb\A m deci cum schimb\A m $\psi(x<x_k)$ \h n
$\psi(x>x_k)$ \h n aceste puncte. R\A spunsul este dat de a\c{s}a numitele
formule de conexiune.

Din teoria ecua\c{t}iilor diferen\c{t}iale ordinare \c{s}i pe baza analizei de
func\c{t}ii de variabil\A\ complex\A\  
se poate demonstra c\A\ formulele de conexiune exist\A\ \c{s}i c\A\ sunt 
urm\A toarele: 
$$
%\begin{displaymath}
\psi_1(x)=
\frac{1}{\left[-r(x)\right]^{\frac{1}{4}}}
\exp\left(-\int_x^{x_k}\sqrt{-r(x)}dx\right)\rightarrow 
%\end{displaymath}
$$
\begin{equation}
\rightarrow\frac{2}{\left[r(x)\right]^{\frac{1}{4}}}
\cos\left(\int_{x_k}^x\sqrt{r(x)}dx-\frac{\pi}{4}\right)~,
\end{equation}
unde $\psi_1(x)$ are numai comportament atenuant exponen\c{t}ial pentru 
$x<x_k$. Prima formul\A\ de conexiune arat\A\ c\A\ 
func\c{t}ia $\psi(x)$, care la st\h nga punctului de \h ntoarcere se 
comport\A\ atenuant exponen\c{t}ial, trece \h n dreapta acelui punct
\h ntr-o cosinusoid\A\ de faz\A\ $\phi=\frac{\pi}{4}$ \c{s}i de amplitudine
dubl\A\ fa\c{t}\A\ de amplitudinea exponen\c{t}ialei.

\^{I}n cazul unei func\c{t}ii $\psi(x)$ mai generale, respectiv o
func\c{t}ie care s\A\ aib\A\ un comportament exponen\c{t}ial cresc\A tor
\c{s}i atenuant, formula de conexiune corespunz\A toare este:
$$
%\begin{displaymath}
\sin\left(\phi+\frac{\pi}{4}\right)
\frac{1}{\left[-r(x)\right]^{\frac{1}{4}}}
\exp\left(\int_x^{x_k}\sqrt{-r(x)}dx\right)\leftarrow
%\end{displaymath}
$$
\begin{equation}
\leftarrow\frac{1}{\left[r(x)\right]^{\frac{1}{4}}}
\cos\left(\int_{x_k}^x\sqrt{r(x)}dx+\phi\right)~,
\end{equation}
cu condi\c{t}ia ca $\phi$ s\A\ nu ia o valoare prea apropiat\A\ de
$-\frac{\pi}{4}$. Motivul este c\A\ dac\A\ $\phi=-\frac{\pi}{4}$, 
func\c{t}ia sinus se anuleaz\A\ . Aceast\A\ a doua  
formul\A\ de conexiune signific\A\ c\A\ o func\c{t}ie care se 
comport\A\ ca o cosinusoid\A\ la dreapta unui punct de \h ntoarcere 
trece \h n partea sa st\h nga ca o exponen\c{t}ial\A\ cresc\A toare 
cu amplitudinea modulat\A\ de c\A tre o sinusoid\A\ .

Pentru a studia detaliile procedurii de ob\c{t}inere a  formulelor de 
conexiune se poate consulta expunerea din cartea %\textbf
{\em Mathematical Methods of Physics} de %\textbf
{\em J. Mathews \& R.L. Walker.}\\
%\textbf
\bigskip
%\newpage
%\documentclass[a4paper,12pt]{article}
%\usepackage{latexsym}
%\pagestyle{empty}
%\begin{document}
%\begin{center}
%\centerline{\huge
%%%%%%%%%%%%%%%%%%%%%%%%%%%%%%%%%%%%%%%%%%%%%%%%%%%%%%%%%%%%%%%%%
\subsection*{Estimarea erorii introduse de aproxima\c{t}ia WKB}

\hspace{0.6cm}Am g\A sit solu\c{t}ia ecua\c{t}iei  
Schr\"odinger \h n orice regiune unde se satisface  
condi\c{t}ia WKB. Totu\c{s}i, solu\c{t}iile WKB diverg \h n  
punctele de \h ntoarcere a\c{s}a cum am semnalat. 
Vom analiza, de\c{s}i superficial, aceast\A\ problematic\A\ cu scopul de a 
propune  
%\textbf
{\em formulele de conexiune} \h ntr-o vecin\A tate
redus\A\ a punctelor de \h ntoarcere.

S\A\ presupunem c\A\ $x=x_k$ este un punct de \h ntoarcere, unde avem: 
$q_0(x_k)=p(x_k)=0\quad\Rightarrow\quad 
E=u(x_k)$. La st\h nga lui $x_k$, adic\A\ \h n semidreapta 
$x<x_k$, vom presupune c\A\  
$E<u(x)$, astfel c\A\ \h n aceast\A\ regiune solu\c{t}ia WKB este:
$$
%\begin{displaymath}
\psi(x)=\frac{a}{\left[\frac{u(x)-E}{u_0}\right]^\frac{1}{4}}\exp\left(-\frac{1}{\xi}\int_x^{x_k}\sqrt{\frac{u(x)-E}{u_0}}dx\right)\quad+
%\end{displaymath}
$$
\begin{equation}
\quad\quad+\quad\frac{b}{\left[\frac{u(x)-E}{u_0}\right]^\frac{1}{4}}
\exp\left(\frac{1}{\xi}\int_x^{x_k}\sqrt{\frac{u(x)-E}{u_0}}dx\right)~.
\end{equation}
\^{I}n acela\c{s}i mod, la dreapta lui $x_k$ ( \h n semidreapta $x>x_k$ ) 
presupunem $E>u(x)$. \^{I}n consecin\c{t}\A\ 
solu\c{t}ia WKB \h n aceast\A\ zon\A\ este:
$$
%\begin{displaymath}
\psi(x)=\frac{c}{\left[\frac{E-u(x)}{u_0}\right]^\frac{1}{4}}
\exp\left(\frac{i}{\xi}\int_{x_k}^x\sqrt{\frac{E-u(x)}{u_0}}dx\right)\quad+
%\end{displaymath}
$$
\begin{equation}
\quad\quad\quad\quad+
\quad\frac{d}{\left[\frac{E-u(x)}{u_0}\right]^\frac{1}{4}}
\exp\left(-\frac{i}{\xi}\int_{x_k}^x\sqrt{\frac{E-u(x)}{u_0}}dx\right)~.
\end{equation}

Dac\A\ $\psi(x)$ este o func\c{t}ie real\A\ , 
va avea aceast\A\ proprietate at\h t la dreapta c\h t \c{s}i la st\h nga lui 
$x_k$. Vom numi acest fapt %\textbf
{\it ``condi\c{t}ia de realitate''}, care \h nseamn\A\ c\A\ dac\A\
$a,b\in\Re$, atunci $c=d^*$.

Problema noastr\A\ este de a conecta aproxima\c{t}iile din cele dou\A\ laturi
ale lui $x_k$ pentru ca ele s\A\ se refere la aceea\c{s}i solu\c{t}ie. 
Aceasta \h nseamn\A\ a g\A si 
$c$ \c{s}i $d$ dac\A\ se cunosc $a$ \c{s}i $b$, precum \c{s}i viceversa. 
Pentru a efectua aceast\A\  
conexiune, trebuie s\A\ utiliz\A m o solu\c{t}ie aproximat\A\ , care s\A\ fie
corect\A\ de-a lungul unui drum care leag\A\ regiunile din cele dou\A\ 
laturi ale lui $x_k$, unde solu\c{t}iile WKB s\A\ fie deasemenea corecte. 

Cel mai comun este s\A\ se recurg\A\ la o metod\A\ propus\A\ de c\A tre %\textbf
{\em Zwann} \c{s}i %\textbf
{\em Kemble} care 
const\A\ \h n a ie\c{s}i de pe axa real\A\ \h n vecin\A tatea lui $x_k$, 
pe un contur \h n jurul lui $x_k$ \h n  
planul complex. Pe acest contur se consider\A\ c\A\ 
solu\c{t}iile WKB continu\A\ s\A\ fie corecte.
\^{I}n aceast\A\ prezentare vom folosi aceast\A\ metod\A\ , dar numai cu scopul 
de a ob\c{t}ine un mijloc de a estima erorile \h n  
aproxima\c{t}ia WKB.

Estimarea erorilor este \h ntotdeauna important\A\ pentru
solu\c{t}iile aproximate prin diferite metode \c{s}i \h n plus \h n cazul WKB
aproxima\c{t}ia se face pe intervale mari ale axei reale, ceea ce poate duce la   
acumularea erorilor \c{s}i la eventuale artefacte datorate \c{s}ifturilor de 
faz\A\ astfel introduse. 

S\A\ definim 
{\em func\c{t}iile WKB asociate} \h n felul urm\A tor:
\begin{equation}
W_{\pm}=\frac{1}{\left[\frac{E-u(x)}{u_0}\right]^\frac{1}{4}}
\exp\left(\pm\frac{i}{\xi}\int_{x_k}^x\sqrt{\frac{E-u(x)}{u_0}}dx\right)~,
\end{equation}
pe care le consider\A m ca func\c{t}ii de variabil\A\ complex\A\ .  
Vom folosi t\A ieturi pentru a evita discontinuit\A \c{t}ile 
din zerourile 
lui $r(x)=\frac{E-u(x)}{u_0}$. Aceste func\c{t}ii satisfac o 
ecua\c{t}ie diferen\c{t}ial\A\ care se poate ob\c{t}ine 
diferen\c{t}iindu-le \h n raport cu 
$x$, conduc\h nd la:
$$
%\begin{displaymath}
W_{\pm}'=\left(\pm\frac{i}{\xi}\sqrt{r}-\frac{1}{4}\frac{r'}{r}\right)W_{\pm}
%\end{displaymath}
$$
\begin{equation}
W_{\pm}''+\left[\frac{r}{\xi^2}+\frac{1}{4}\frac{r''}{r}-\frac{5}{16}
\left(\frac{r'}{r}\right)^2\right]W_{\pm}=0~.
\end{equation}
S\A\ not\A m:
\begin{equation}
s(x)=\frac{1}{4}\frac{r''}{r}-\frac{5}{16}\left(\frac{r'}{r}\right)^2~,
\end{equation}
atunci $W_{\pm}$ sunt solu\c{t}ii exacte ale ecua\c{t}iei:
\begin{equation}
W_{\pm}''+\left[\frac{1}{\xi^2}r(x)+s(x)\right]W_{\pm}=0~,
\end{equation}
dar satisfac numai aproximativ ecua\c{t}ia 
Schr\"odinger, care este regular\A\ \h n $x=x_k$ \h n timp ce  
ecua\c{t}ia pentru func\c{t}iile WKB asociate este singular\A\ 
\h n acel punct.

Vom defini func\c{t}iile $\alpha_{\pm}(x)$ care s\A\ satisfac\A\
urm\A toarele dou\A\ rela\c{t}ii: 
\begin{equation}
\psi(x)=\alpha_+(x)W_+(x)+\alpha_-(x)W_-(x)
\end{equation}
\begin{equation}
\psi'(x)=\alpha_+(x)W_+'(x)+\alpha_-(x)W_-'(x)~,
\end{equation}
unde $\psi(x)$ este solu\c{t}ie a ecua\c{t}iei Schr\"odinger. 
Rezolv\h nd ecua\c{t}iile anterioare pentru $\alpha_{\pm}$ 
avem:
$$
%\begin{displaymath}
\alpha_+=\frac{\psi W_-'-\psi'W_-}{W_+W_-'-W_+'W_-}
\qquad\qquad\alpha_-=-\frac{\psi W_+'-\psi'W_+}{W_+W_-'-W_+'W_-}~,
%\end{displaymath}
$$
unde num\A r\A torul este exact {\bf Wronskianul} lui $W_+$ \c{s}i $W_-$. Nu
este dificil de demonstrat c\A\ acesta ia valoarea  
$-\frac{2}{\xi}i$, astfel c\A\ $\alpha_{\pm}$ se simplific\A\ 
la forma:
\begin{equation}
\alpha_+=\frac{\xi}{2}i\left(\psi W_-'-\psi'W_-\right)
\end{equation}
\begin{equation}
\alpha_-=\frac{-\xi}{2}i\left(\psi W_+'-\psi'W_+\right)~.
\end{equation}
Efectu\h nd derivata \h n $x$ \h n ecua\c{t}iile $(39)$ \c{s}i $(40)$, avem:
\begin{equation}
\frac{d\alpha_{\pm}}{dx}=\frac{\xi}{2}i\left(\psi'W_{\mp}'+\psi W_{\mp}''
-\psi''W_{\mp}-\psi'W_{\mp}'\right)~.
\end{equation}
\^{I}n paranteze, primul \c{s}i al patrulea termen se anuleaz\A\ ; 
amintim c\A\ :
$$
%\begin{displaymath}
\psi''+\frac{1}{\xi^2}r(x)\psi=0\quad\&\quad 
W_{\pm}''+\left[\frac{1}{\xi^2}r(x)+s(x)\right]W_{\pm}=0
%\end{displaymath}
$$
\c{s}i deci putem scrie ecua\c{t}ia $(41)$ \h n forma:
$$
%\begin{displaymath}
\frac{d\alpha_{\pm}}{dx}=\frac{\xi}{2}i\left[-\psi\left(\frac{r}{\xi^2}+s\right)W_{\mp}+\frac{r}{\xi^2}\psi W_{\mp}\right]
%\end{displaymath}
$$
\begin{equation}
\frac{d\alpha_{\pm}}{dx}=\mp\frac{\xi}{2}is(x)\psi(x)W_{\mp}(x)~,
\end{equation}
care \h n baza ecua\c{t}iilor $(33)$ \c{s}i $(37)$ devine:
\begin{equation}
\frac{d\alpha_{\pm}}{dx}=
\mp\frac{\xi}{2}i\frac{s(x)}{\left[r(x)\right]^\frac{1}{2}}
\left[\alpha_{\pm}+\alpha_{\mp}\exp\left(\mp\frac{2}{\xi}i
\int_{x_k}^x\sqrt{r(x)}dx\right)\right]~.
\end{equation}

Ecua\c{t}iile $(42)$ \c{s}i $(43)$ se folosesc pentru %\textbf
{\em a estima eroarea care se 
comite \h n aproxima\c{t}ia WKB \h n cazul unidimensional.} 

Motivul pentru care $\frac{d\alpha_{\pm}}{dx}$ se poate considera ca m\A sur\A\
a erorii WKB este c\A\ \h n 
ecua\c{t}iile $(31)$ \c{s}i $(32)$ constantele $a$, $b$ \c{s}i $c$, $d$, 
respectiv, ne dau numai solu\c{t}ii $\psi$ aproximative, \h n timp ce 
func\c{t}iile $\alpha_{\pm}$ introduse \h n
ecua\c{t}iile $(37)$ \c{s}i $(38)$ produc solu\c{t}ii $\psi$ exacte. Din punct
de vedere geometric derivata d\A\ panta dreptei 
tangente la aceste func\c{t}ii \c{s}i indic\A\ m\A sura \h n care 
$\alpha_{\pm}$ deviaz\A\ fa\c{t}\A\ de constantele $a$, $b$, $c$ \c{s}i $d$.
%\end{document}

\bigskip

\noindent
\underline{{\bf 4N. Not\u{a}}}: Articolele (J)WKB originale sunt 
urm\A toarele:\\

\noindent
G. Wentzel, ``Eine Verallgemeinerung der Wellenmechanik", [``O generalizare a
mecanicii ondulatorii"],

\noindent
Zeitschrift f\"ur Physik {\bf 38}, 518-529 (1926) [primit 18 June 1926]\\

\noindent
L. Brillouin, ``La m\'ecanique ondulatoire de Schr\"odinger: une m\'ethode
g\'en\'erale de resolution par approximations successives",
[``Mecanica ondulatorie a lui Schr\"odinger: o metod\A\ general\A\ de rezolvare
prin aproxim\A ri succesive"],

\noindent
Comptes Rendus Acad. Sci. Paris {\bf 183}, 24-26 (1926) [primit 5 July 1926]\\

\noindent
H.A. Kramers, ``Wellenmechanik und halbzahlige Quantisierung",
[``Mecanica ondulatorie \c{s}i cuantizarea semi\h ntreag\A\ "],

\noindent
Zf. Physik {\bf 39}, 828-840 (1926) [primit 9 Sept. 1926]\\

\noindent
H. Jeffreys, ``On certain approx. solutions of linear diff. eqs. of the
second order",
[``Asupra unor solu\c{t}ii aproximative a ecua\c{t}iilor diferen\c{t}iale 
lineare de ordinul doi"],

\noindent
Proc. Lond. Math. Soc. {\bf 23}, 428-436 (1925)

\bigskip

%\newpage
\centerline{{\huge 4P.  Probleme}}
%%%%%%%%%%%%%%%%%%%%%%%%%%%%%%%%%%%%%%%

{\em Problema 4.1}\\

S\A\ se foloseasc\A\ metoda WKB pentru
o particul\A\ de energie $E$ care se mi\c{s}c\A\ \h ntr-un 
poten\c{t}ial $u(x)$ de forma ar\A tat\A\ \h n figura 4.1.

%%%%%%%%%%%%%%
\vskip 2ex
\centerline{
\epsfxsize=280pt
\epsfbox{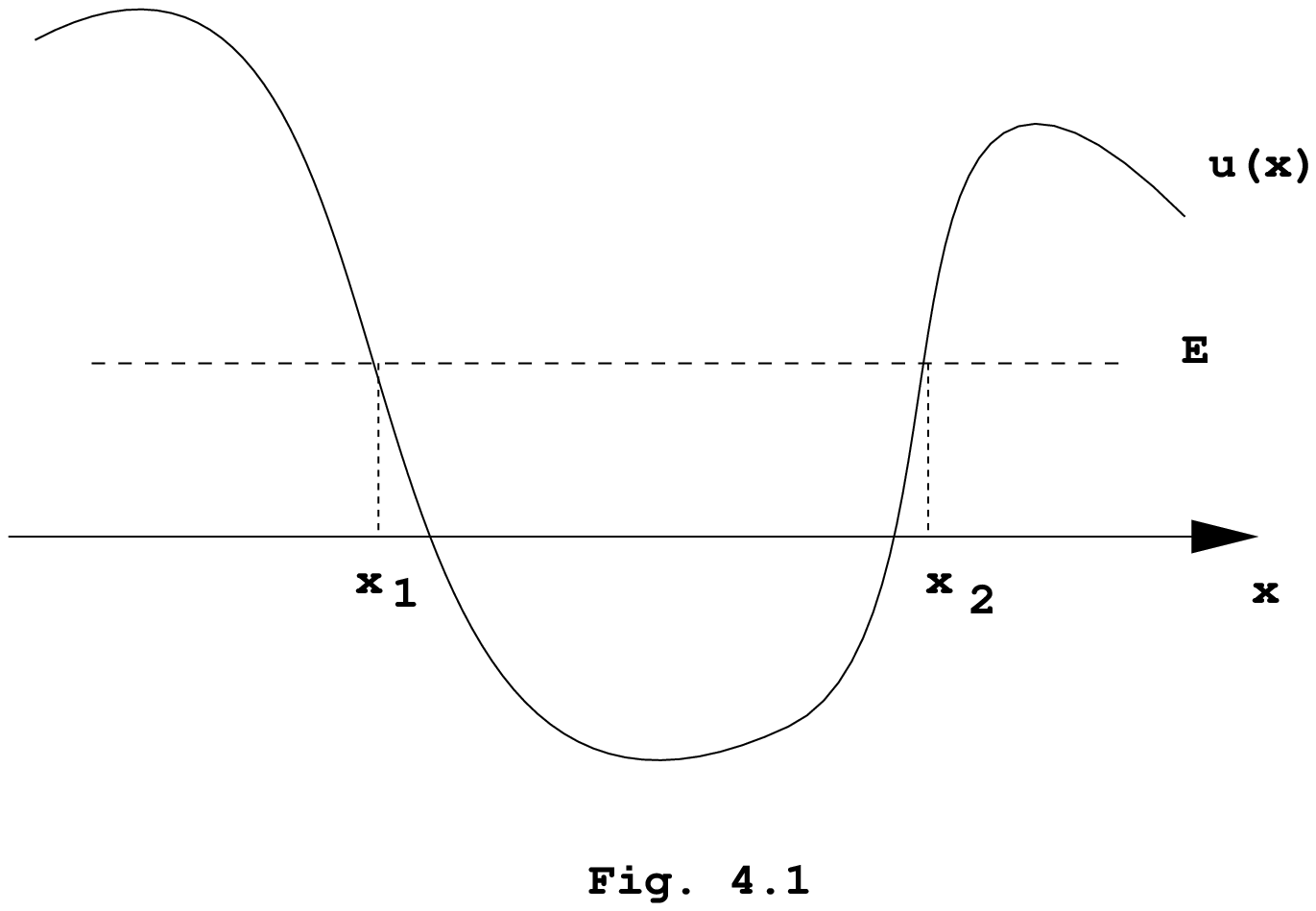}}
\vskip 4ex
%\begin{center}
%{\small{Fig. 1}\\
%}
%\end{center}
%%%%%%%%%%%%%%%%

{\bf Solu\c{t}ie}

Ecua\c{t}ia  Schr\"odinger corespunz\A toare este:
\begin{equation}
\frac{d^2\psi}{dx^2}+\frac{2m}{\hbar^2}\left[E-u(x)\right]\psi=0~.
\end{equation}

Dup\A\ cum putem vedea:
$$%\begin{displaymath}
r(x)=\frac{2m}{\hbar^2}\left[E-u(x)\right]\qquad
\left\{
\begin{array}{ll}
\mbox{este pozitiv\A\ pentru $a<x<b$}\\
\mbox{este negativ\A\ pentru $x<a, x>b$.}
\end{array}
\right.
$$%\end{displaymath}

Dac\A\ $\psi(x)$ corespunde zonei \h n care $x<a$, la trecerea \h n
intervalul $a<x<b$, formula de conexiune este dat\A\ de ecua\c{t}ia $(29)$ 
\c{s}i ne spune c\A\ :
\begin{equation}
\psi(x)\approx\frac{A}{\left[E-u\right]^{\frac{1}{4}}}\cos\left(\int_a^x
\sqrt{\frac{2m}{\hbar^2}(E-u)}dx-\frac{\pi}{4}\right)
\end{equation}
unde $A$ este o constant\A\ arbitrar\A\ .

C\h nd $\psi(x)$ corespunde zonei $x>b$, la trecerea \h n intervalul $a<x<b$ 
avem \h n mod similar: 
\begin{equation}
\psi(x)\approx-\frac{B}{\left[E-u\right]^{\frac{1}{4}}}\cos\left(\int_x^b
\sqrt{\frac{2m}{\hbar^2}(E-u)}dx-\frac{\pi}{4}\right)~,
\end{equation}
unde $B$ este o constant\A\ arbitrar\A\ . Motivul pentru care
formula de conexiune este din nou ecua\c{t}ia (29), se \h n\c{t}elege
examin\h nd ce se \h nt\h mpl\A\ c\h nd 
particula ajunge la al doilea punct clasic de \h ntoarcere $x=b$. Acesta
produce inversia direc\c{t}iei de mi\c{s}care \c{s}i atunci particula apare
ca venind de la dreapta spre st\h nga. Cu alte cuvinte, ne g\A sim \h n
prima situa\c{t}ie (de la st\h nga la dreapta)
numai c\A\ v\A zut\A\ \h ntr-o oglind\A\ \h n punctul $x=a$.

Aceste dou\A\ expresii trebuie s\A\ fie acelea\c{s}i independent de 
constantele $A$ \c{s}i $B$, astfel c\A\ :
$$
%\begin{displaymath}
\cos\left(\int_a^x\sqrt{\frac{2m}{\hbar^2}(E-u)}dx-\frac{\pi}{4}\right)
=-\cos\left(\int_x^b\sqrt{\frac{2m}{\hbar^2}(E-u)}dx-\frac{\pi}{4}\right)\\
$$%\end{displaymath}
\begin{equation}
\Rightarrow\cos\left(\int_a^x\sqrt{\frac{2m}{\hbar^2}(E-u)}dx-\frac{\pi}{4}
\right)+\cos\left(\int_x^b\sqrt{\frac{2m}{\hbar^2}(E-u)}dx-
\frac{\pi}{4}\right)=0~.
\end{equation}
Amintind c\A\ :
$$
%\begin{displaymath}
\cos A+\cos B= 2\cos\left(\frac{A+B}{2}\right)\cos\left(\frac{A-B}{2}\right)~,
$$%\end{displaymath}
ecua\c{t}ia $(47)$ se scrie:
$$%\begin{displaymath}
2\cos\left[\frac{1}{2}\left(
\int_a^x\sqrt{\frac{2m}{\hbar^2}(E-u)}dx-\frac{\pi}{4}
+\int_x^b\sqrt{\frac{2m}{\hbar^2}(E-u)}dx-\frac{\pi}{4}\right)\right]\cdot
$$%\end{displaymath}
\begin{equation}
\cdot\cos\left[\frac{1}{2}\left(\int_a^x\sqrt{\frac{2m}{\hbar^2}(E-u)}dx
-\frac{\pi}{4}-\int_x^b\sqrt{\frac{2m}{\hbar^2}(E-u)}dx
+\frac{\pi}{4}\right)\right]=0~,
\end{equation}
ceea ce implic\A\ pentru argumentele acestor cosinusoide c\A\ sunt
multipli \h ntregi  
de $\frac{\pi}{2}$; argumentul celui de-a doua cosinusoide nu ne duce 
la nici un rezultat netrivial, a\c{s}a c\A\ ne fix\A m aten\c{t}ia numai asupra
argumentului primei cosinusoide, 
care este esen\c{t}ial pentru ob\c{t}inerea unui rezultat important:
$$%\begin{displaymath}
\frac{1}{2}\left(\int_a^x\sqrt{\frac{2m}{\hbar^2}(E-u)}dx
-\frac{\pi}{4}+\int_x^b\sqrt{\frac{2m}{\hbar^2}(E-u)}dx
-\frac{\pi}{4}\right)=\frac{n}{2}\pi\quad\mbox{pt. n impar}
$$%\end{displaymath}
$$%\begin{displaymath}
\Rightarrow\quad\quad\int_a^b\sqrt{\frac{2m}{\hbar^2}(E-u)}dx
-\frac{\pi}{2}=n\pi
$$%\end{displaymath}
$$%\begin{displaymath}
\Rightarrow\quad\quad\int_a^b\sqrt{\frac{2m}{\hbar^2}(E-u)}dx=
(n+\frac{1}{2})\pi
$$%\end{displaymath}
\begin{equation}
\Rightarrow\quad\quad\int_a^b\sqrt{2m(E-u)}dx=(n+\frac{1}{2})\pi\hbar~.
\end{equation}

\noindent
Acest rezultat este foarte similar regulilor de cuantizare  %\textbf
{\em Bohr - Sommerfeld}.

Amintim c\A\ postulatul lui Bohr stabile\c{s}te c\A\ momentul unghiular al unui
electron care se mi\c{s}c\A\ pe o `orbit\A\ permis\A\ ' \h n jurul
nucleului atomic este cuantizat\A\ \c{s}i valoarea sa este: $L=n\hbar$, 
$n=1,2,3,\dots$. Amintim deasemenea c\A\ regulile de cuantizare  
Wilson - Sommerfeld stabilesc c\A\ orice coordonat\A\ a unui sistem 
fizic care variaz\A\ periodic \h n timp trebuie s\A\ satisfac\A\ 
condi\c{t}ia cuantic\A\ : $\oint p_qdq=n_q h$; unde $q$ este o 
coordonat\A\ periodic\A\ , $p_q$ este impulsul asociat acesteia, $n_q$ este un 
num\A r \h ntreg \c{s}i $h$ este constanta lui Planck. Se vede c\A\ 
rezultatul ob\c{t}inut \h n aproxima\c{t}ia WKB este cu adev\A rat 
foarte asem\A n\A tor.\\
%\textbf

\bigskip
%\newpage

{\em Problema 4.2}\\

S\A\ se estimeze eroarea care se comite \h n solu\c{t}ia WKB \h ntr-un 
punct $x_1\neq x_k$, cu $x_k$ un punct clasic de \h ntoarcere, pentru
ecua\c{t}ia diferen\c{t}ial\A\ $y''+xy=0$. {\em Solu\c{t}ia acestei 
probleme este important\A\ \h n studiul c\h mpurilor uniforme, a\c{s}a cum sunt 
cele gravita\c{t}ionale sau cele electrice produse
de pl\A ci plane \h nc\A rcate.}\\

\bigskip

{\bf Solu\c{t}ie:}\\
Pentru aceast\A\ ecua\c{t}ie diferen\c{t}ial\A\ avem:
$$%\begin{displaymath}
\xi=1,\qquad r(x)=x\qquad\&\qquad s(x)=-\frac{5}{16}x^{-2}~.
$$%\end{displaymath}
$r(x)=x$ are un singur zero \h n $x_k=0$, astfel c\A\ pentru $x\gg0$: 
\begin{equation} 
W_{\pm}=x^{-\frac{1}{4}}
\exp\left(\pm i\int_0^x\sqrt{x}dx\right)
=x^{-\frac{1}{4}}\exp\left(\pm\frac{2}{3}ix^{\frac{3}{2}}\right)~.
\end{equation}
Deriv\h nd $W_{\pm}$ p\h n\A\ la a doua derivat\A\ \h n $x$, ne d\A m seama
c\A\ se  satisface urm\A toarea ecua\c{t}ie diferen\c{t}ial\A\ :
\begin{equation}
W_{\pm}''+(x-\frac{5}{16}x^{-2})W_{\pm}=0~.
\end{equation}

Solu\c{t}ia exact\A\ $y(x)$ a acestei ecua\c{t}ii diferen\c{t}iale o scriem 
ca o combina\c{t}ie linear\A\ de $W_{\pm}$, a\c{s}a cum s-a indicat 
\h n sec\c{t}iunea corespondent\A\  
estim\A rii erorii \h n aproxima\c{t}ia WKB;  
amintim c\A\ s-a propus o combina\c{t}ie linear\A\ de
forma:
$$%\begin{displaymath}
y(x)=\alpha_+(x)W_+(x)+\alpha_-(x)W_-(x) 
$$%\end{displaymath}

Pentru $x$ mari, o solu\c{t}ie general\A\ a ecua\c{t}iei noastre 
diferen\c{t}iale se poate scrie \h n aproxima\c{t}ia WKB \h n forma:
\begin{equation}
y(x)=Ax^{-\frac{1}{4}}
\cos\left(\frac{2}{3}x^{\frac{3}{2}}
+\delta\right)\qquad \mbox{c\h nd} \quad x\rightarrow\infty~,
\end{equation}
astfel c\A\ $\alpha_+\rightarrow\frac{A}{2}e^{i\delta}$ \c{s}i 
$\alpha_-\rightarrow\frac{A}{2}e^{-i\delta}$ pentru 
$x\rightarrow\infty$. Vrem s\A\ calcul\A m eroarea datorat\A\ acestei 
solu\c{t}ii WKB. O m\A sur\A\ a acestei erori este devierea lui 
$\alpha_+$ \c{s}i a lui $\alpha_-$ fa\c{t}\A\ de constantele $A$. Pentru aceasta
folosim ecua\c{t}ia:
$$%\begin{displaymath}
\frac{d\alpha_{\pm}}{dx}
=\mp\frac{\xi}{2}i\frac{s(x)}{\sqrt{r(x)}}
\left[\alpha_{\pm}+\alpha_{\mp}
\exp\left(\mp2i\int_{x_k}^x\sqrt{r(x)}dx\right)\right]
$$%\end{displaymath}
\c{s}i efectu\h nd substitu\c{t}iile corespunz\A toare avem:
\begin{equation}
\frac{d\alpha_{\pm}}{dx}
=\mp\frac{i}{2}\left(-\frac{5}{16}x^{-2}\right)
x^{-\frac{1}{2}}\left[\frac{A}{2}e^{\pm i\delta}
+\frac{A}{2}e^{\mp i\delta}\exp\left(\mp 2i\frac{2}{3}
x^{\frac{3}{2}}\right)\right]~.
\end{equation}
\c{S}tim c\A\ $\Delta\alpha_{\pm}$ reprezint\A\ schimb\A rile pe care le 
prezint\A\  
$\alpha_{\pm}$ c\h nd $x$ variaz\A\ \h ntre $x_1$ \c{s}i $\infty$, 
ceea ce ne permite 
calculul prin intermediul lui: 
$$%\begin{displaymath}
\frac{\Delta\alpha_{\pm}}{A/2}
=\frac{2}{A}\int_{x_1}^\infty\frac{d\alpha_{\pm}}{dx}dx
=\qquad\qquad\qquad\qquad\qquad\qquad\qquad\qquad\quad
$$%\end{displaymath}
\begin{equation}
=\pm i\frac{5}{32}e^{\pm i\delta}\left[\frac{2}{3}x_1^{-\frac{3}{2}}+e^{\mp 
2i\delta}\int_{x_1}^\infty x^{-\frac{5}{2}}\exp\left(\mp 
i\frac{4}{3}x^\frac{3}{2}\right)dx\right]~. 
\end{equation}
Al doilea termen din paranteze este mai pu\c{t}in important dec\h t 
primul pentru c\A\ exponen\c{t}iala complex\A\ oscileaz\A\ 
\h ntre $1$ \c{s}i $-1$ \c{s}i deci $x^{-\frac{5}{2}}<x^{-\frac{3}{2}}$. 
Prin urmare:
\begin{equation}
\frac{\Delta\alpha_{\pm}}{A/2}\approx\pm\frac{5}{48}ie^{\pm 
i\delta}x_1^{-\frac{3}{2}} \end{equation}
\c{s}i cum putem vedea eroarea care se introduce este \h ntr-adev\A r 
mic\A\ dac\A\ tinem cont \c{s}i de faptul c\A\ 
exponen\c{t}iala complex\A\ oscileaz\A\ \h ntre $-1$ \c{s}i $1$, iar 
$x_1^{-\frac{3}{2}}$ este deasemenea mic.\\
%\textbf

%{\em Problema 4.3}\\

%Dece ecua\c{t}ia diferen\c{t}ial\A\ pentru care func\c{t}iile WKB asociate sunt
%solu\c{t}ii, difer\A\ de ecua\c{t}ia Schr\"odinger pe care o satisfac 
%func\c{t}iile WKB prin includerea func\c{t}iei $s(x)$, dac\A\ 
%func\c{t}iile WKB \c{s}i asociate WKB au aceea\c{s}i form\A\ ?\\

%{\bf Justificare:}\\

%Amintim c\A\ \h n procesul de ob\c{t}inere a solu\c{t}iilor WKB, \h nt\h lnim o
%ecua\c{t}ie diferen\c{t}ial\A\ de tip Riccati; pentru aceasta am  
%propus o solu\c{t}ie \h n forma unei serii de puteri \h n $-i\xi$, respectiv 
%$q(x)=\sum_{n=0}^\infty(-i\xi)^nq_n(x)$. Dar amintim deasemenea c\A\ am 
%men\c{t}inut numai primii doi termeni (aproxima\c{t}ie de ordinul doi). Prin 
%urmare func\c{t}iile $\psi^{\pm}(x)$ satisfac  
%ecua\c{t}ia Schr\"odinger %\%textbf
%{\em numai aproximativ}. Pe de alta parte,
%se propun func\c{t}iile WKB asociate $W_{\pm}$, ca 
%func\c{t}ii care au form\A\ func\c{t}iilor 
%$\psi^{\pm}$; pentru a ob\c{t}ine ecua\c{t}ia diferen\c{t}ial\A\ pe care
%o satisfac trebuie doar s\A\ le deriv\A m;  
%ecua\c{t}ia diferen\c{t}ial\A\ este prin urmare satisf\A cut\A\ {\em exact} 
%de c\A tre aceste func\c{t}ii \c{s}i \h n plus func\c{t}ia s(x) se introduce
%\h n mod natural pentru a ar\A ta c\h t de mult  %\textbf
%{\em ``deviaz\A\ ''} 
%func\c{t}iile $\psi^{\pm}$ de la solu\c{t}ia exact\A\ a ecua\c{t}iei 
%Schr\"odinger 1D .

%\end{document}

% References

% R. Landauer, ``Path concepts in HJ theory", AJP {\bf }, 363-367 (1952)

%\end{document}

\newpage
%%%%%%%%%%%%%%%%%%%%%%%%%%%%%%%%%%%%%%%%%%%%%%%%%%%%%%%%%%%%%%%%%%%%%%%%%
%%%%%%%%%%%%%%%%%%%%%%%%%%%%%%%%%%%%%%%%%%%%%%%%%%%%   Oscilator armonic
%%%%%%%%%%%%%%%%%%%%%%%%%%%%%%%%%%%%%
%\documentstyle[12pt]{article}
%%%%%%%%%%%%%%%%%%%%%%%%%%%%%%%%%%%%%%%%%%%%%%
%\baselineskip 24.1pt plus 0.2pt minus 0.1pt
%%%%%%%%%%%%%%%%%%%%%%%%%%%%%%%%%%%%%%%%%%%%%%%%
%\newcommand{\oa}{oscilador arm\'onico\hspace{0.2cm}}
\newcommand{\bc}{\begin{center}}
\newcommand{\ec}{\end{center}}
\newcommand{\ii}{\'{\i}}
\newcommand{\be}{\begin{equation}}
\newcommand{\ee}{\end{equation}}
\newcommand{\dd}{\dagger}
\newcommand{\ad}{a^{\dd}}
\newcommand{\m}{\mid}
%\begin{document}

%\setcounter{equation}
%\baselineskip 24.1pt plus 0.2pt minus 0.1pt
%%%%%%%%%%%%%%%%%%%%%%%%%%%%%%%%%%%%%%%%%%%%%%%%%%%
\section*{{\huge 5. OSCILATORUL  ARMONIC (OA)}}
%\author{ Jos\'e Torres Arenas}
%\date {}
%\maketitle
%\setcounter{equation}
\section*{Solu\c{t}ia ecua\c{t}iei Schr\"odinger pentru OA}
\setcounter{equation}{0}
Oscilatorul armonic (OA) poate fi considerat ca o paradigm\A\
a Fizicii. Utilitatea sa apare \h n marea majoritate a domeniilor, de la 
fizica clasic\A\ p\h n\A\ la electrodinamica
cuantic\A\ \c{s}i teorii ale obiectelor colapsate gravita\c{t}ional.\\
Din mecanica clasic\A\ \c{s}tim c\A\ multe poten\c{t}iale complicate pot fi
aproximate \h n vecin\A tatea pozi\c{t}iilor de echilibru printr-un 
poten\c{t}ial OA
\be
V(x) \sim \frac{1}{2}V^{\prime\prime}(a)(x-a)^2~.
\ee

Acesta este un caz unidimensional. Pentru acest caz, func\c{t}ia 
Hamiltonian\A\ clasic\A\ a unei particule de
mas\A\ {\em m}, oscil\h nd cu frecven\c{t}a $\omega$ are urm\A toarea form\A\ :
\be
H=\frac{p^2}{2m}+\frac{1}{2}m\omega^2x^2
\ee
\c{s}i Hamiltonianul cuantic corespunz\A tor \h n spa\c{t}iul de 
configura\c{t}ii este :
\be
\hat{H}=\frac{1}{2m}(-i\hbar\frac{d}{dx})^2+\frac{1}{2}m\omega^2x^2
\ee
\be
\hat{H}=-\frac{\hbar^2}{2m}\frac{d^2}{dx^2}+\frac{1}{2}m\omega^2x^2~.
\ee

Dat faptul c\A\ poten\c{t}ialul este independent de timp,
FP $\Psi_n$ \c{s}i autovalorile $E_n$ se determin\A\ cu ajutorul
ecua\c{t}iei Schr\"odinger independent\A\ de timp :
\be
\hat{H}\Psi_n=E_n\Psi_n~.
\ee

Consider\h nd Hamiltonianul pentru OA , ecua\c{t}ia Schr\"odinger 
pentru acest caz este :
\be
\frac{d^2\Psi}{dx^2}+\Bigg[\frac{2mE}{\hbar^2}
-\frac{m^2\omega^2}{\hbar^2}x^2\Bigg]\Psi=0~.
\ee

Am suprimat subindicii lui $E$ \c{s}i $\Psi$ pentru c\A\ nu au nici o 
importan\c{t}\A\ aici.
Definind:
\be
k^2=\frac{2mE}{\hbar^2}
\ee
\be
\lambda=\frac{m\omega}{\hbar}~,
\ee

\noindent
ecua\c{t}ia Schr\"odinger devine:
\be
\frac{d^2\Psi}{dx^2}+[k^2-\lambda^2x^2]\Psi=0~,
\ee

\noindent
cunoscut\A\ ca ecua\c{t}ia diferen\c{t}ial\A\ Weber \h n matematic\A\ . \\
Vom face \h n continuare transformarea:
\be
y=\lambda x^2~.
\ee

\^{I}n general, cu schimbul de variabil\A\
de la $x$ la $y$ , operatorii diferen\c{t}iali iau forma:
\be
\frac{d}{dx}=\frac{dy}{dx}\frac{d}{dy}
\ee
\be
\frac{d^2}{dx^2}=\frac{d}{dx}(\frac{dy}{dx}\frac{d}{dy})
=\frac{d^2y}{dx^2}\frac{d}{dy}+(\frac{dy}{dx})^2\frac{d^2}{dy^2}~.
\ee

Aplic\h nd aceast\A\ regul\A\ evident\A\ transform\A rii propuse ob\c{t}inem
urm\A toarea 
ecua\c{t}ie diferen\c{t}ial\A\ \h n variabila $y$ :
\be
y\frac{d^2\Psi}{dy^2}+\frac{1}{2}\frac{d\Psi}{dy}+[\frac{k^2}{4\lambda}
-\frac{1}{4}y]\Psi=0~,
\ee

\noindent
sau, definind :
\be
\kappa=\frac{k^2}{2\lambda}=\frac{\bar k^2}{2m\omega}=\frac{E}{\hbar\omega}~,
\ee
ob\c{t}inem:
\be
y\frac{d^2\Psi}{dy^2}+\frac{1}{2}\frac{d\Psi}{dy}
+[\frac{\kappa}{2}-\frac{1}{4}y]\Psi=0~.
\ee

S\A\ trecem la rezolvarea acestei ecua\c{t}ii, efectu\h nd mai \h nt\h i
analiza sa asimptotic\A\ \h n
limita $y\rightarrow\infty$. Pentru aceasta. se rescrie
ecua\c{t}ia anterioar\A\ \h n forma :
\be
\frac{d^2\Psi}{dy^2}+\frac{1}{2y}\frac{d\Psi}{dy}
+[\frac{\kappa}{2y}-\frac{1}{4}]\Psi=0~.
\ee

Observ\A m c\A\ \h n
limita $y\rightarrow\infty$ ecua\c{t}ia se comport\A\ astfel:
\be
\frac{d^2\Psi_{\infty}}{dy^2}-\frac{1}{4}\Psi_{\infty}=0~.
\ee

Aceast\A\ ecua\c{t}ie are ca solu\c{t}ie:
\be
\Psi_{\infty}(y)=A\exp{\frac{y}{2}}+B\exp{\frac{-y}{2}}~.
\ee

Elimin\A m $\exp{\frac{y}{2}}$ lu\h nd $A=0$ pentru c\A\
diverge \h n limita $y\rightarrow\infty$ \c{s}i r\A m\h nem
cu exponen\c{t}iala atenuat\A\ . Putem sugera
acum c\A\ $\Psi$ are urm\A toarea forma:
\be
\Psi(y)=\exp{\frac{-y}{2}}\psi(y)~.
\ee

Substituind-o \h n ecua\c{t}ia diferen\c{t}ial\A\ pentru $y$ ( ec. $15$) 
se ob\c{t}ine:
\be
y\frac{d^2\psi}{dy^2}
+(\frac{1}{2}-y)\frac{d\psi}{dy}+(\frac{\kappa}{2}-\frac{1}{4})\psi=0~.
\ee

Ceea ce am ob\c{t}inut este ecua\c{t}ia hipergeometric\A\ confluent\A\
\footnote{Deasemenea cunoscut\A\ ca ecua\c{t}ia diferen\c{t}ial\A\ Kummer.} :
\be
z\frac{d^2y}{dz^2}+(c-z)\frac{dy}{dz}-ay=0~.
\ee

Solu\c{t}ia general\A\ a acestei ecua\c{t}ii este :
\be
y(z)=A \hspace{.2cm} _1F_1(a;c,z)+
B \hspace{.2cm} z^{1-c} \hspace{.1cm}  _1F_1(a-c+1;2-c,z)~,
\ee

unde func\c{t}ia hipergeometric\A\ confluent\A\ este definit\A\ prin :
\be
_1F_1(a;c,z)=\sum_{n=0}^{\infty}\frac{(a)_n x^n}{(c)_n n!}~.
\ee

Compar\h nd acum ecua\c{t}ia noastr\u{a}, cu ecua\c{t}ia hipergeometric\A\ 
confluent\u{a}, se observ\A\ c\A\ solu\c{t}ia general\A\ a primei este :
\be
\psi(y)=A\hspace{.2cm} _1F_1(a;\frac{1}{2},y)+
B \hspace{.2cm} y^{\frac{1}{2}}
\hspace{.2cm} _1F_1(a+\frac{1}{2};\frac{3}{2},y)
\ee

unde
\be
a=-(\frac{\kappa}{2}-\frac{1}{4})~.
\ee

Dac\A\ men\c{t}inem aceste solu\c{t}ii \h n forma \h n care se prezint\A\ , 
condi\c{t}ia de
normalizare pentru func\c{t}ia de und\A\ nu se satisface, 
pentru c\A\ din comportamentul asimptotic al func\c{t}iei hipergeometrice 
confluente \footnote{ Comportamentul asimptotic
pentru $\mid x \mid\rightarrow \infty$ este:
\bc
$_1F_1(a;c,z)\rightarrow
\frac{\Gamma(c)}{\Gamma(c-a)}e^{-ia\pi}x^{-a}
+\frac{\Gamma(c)}{\Gamma(a)}e^{x}x^{a-c}~.$
\ec
} rezult\A\
( consider\h nd numai comportamentul dominant exponen\c{t}ial ) :
\be
\Psi(y)=e^{\frac{-y}{2}}\psi(y)\rightarrow
\hspace{.3cm}const. \hspace{.2cm} e^{\frac{y}{2}}y^{a-\frac{1}{2}}~.
\ee

Aceast\A\ ultim\A\ aproxima\c{t}ie ne duce la o divergen\c{t}\A\ \h n
integrala de normalizare, care
fizic este inacceptabil\A\ . Ceea ce se face \h n acest caz, este s\A\ se 
impun\A\ 
condi\c{t}ia de terminare a seriei \footnote{Condi\c{t}ia de 
truncare a seriei pentru func\c{t}ia hipergeometric\A\ 
confluent\A\ $_1F_1(a;c,z)$ este $a=-n$, cu $n$ un \h ntreg 
nenegativ ( adic\A\ , include zero ).} , adic\A\ , seria are numai un num\A r
finit de termeni fiind deci un polinom de grad $n$.\\
{\em Observ\A m astfel c\A\ faptul de a cere ca integrala de 
normalizare s\A\ fie finit\A\ (dup\A\ cum \c{s}tim condi\c{t}ie obligatorie
pentru semnifica\c{t}ia fizic\A\ \h n termeni de probabilit\A \c{t}i), 
ne conduce la truncarea seriei, fapt care \h n acela\c{s}i timp 
produce cuantizarea energiei.}\\
Consider\A m \h n continuare cele dou\A\ cazuri posibile :

$1)\hspace{.4cm} a=-n \hspace{.3cm}$ \c{s}i $ B=0$
\be
\frac{\kappa}{2}-\frac{1}{4}=n~.
\ee

FP-urile sunt date de:
\be
\Psi_n(x)=D_n \exp{\frac{-\lambda x^2}{2}}
\hspace{.1cm} _1F_1(-n;\frac{1}{2},\lambda x^2)
\ee

\c{s}i energia este:
\be
E_n=\hbar\omega(2n+\frac{1}{2})~.
\ee

$2)\hspace{.4cm} a+\frac{1}{2}=-n \hspace{.3cm}$ \c{s}i $A=0$
\be
\frac{\kappa}{2}-\frac{1}{4}=n+\frac{1}{2}~.
\ee

FP-urile sunt acum:
\be
\Psi_n(x)=D_n \exp{\frac{-\lambda x^2}{2}}
\hspace{.2cm}x \hspace{.2cm}_1F_1(-n;\frac{3}{2},\lambda x^2)~,
\ee

iar energiile sta\c{t}ionare sunt:
\be
E_n=\hbar\omega[(2n+1)+\frac{1}{2}]~. 
\ee

Polinoamele ob\c{t}inute \h n urma acestei trunc\A ri a seriei hipergeometrice
se numesc polinoame Hermite \c{s}i se pot scrie ca urm\A toarele
func\c{t}ii hipergeometrice :
\be
H_{2n}(\eta)=(-1)^n \frac{(2n)!}{n!}
\hspace{.2cm} _1F_1(-n;\frac{1}{2},\eta^2)
\ee
\be
H_{2n-1}(\eta)=(-1)^n \frac{2(2n+1)!}{n!}
\hspace{.2cm}\eta \hspace{.2cm} _1F_1(-n;\frac{3}{2},\eta^2)~.
\ee

Putem acum combina rezultatele
ob\c{t}inute ( pentru c\A\ unele ne dau valorile pare \c{s}i altele pe cele
impare ) \h ntr-o singur\A\ expresie pentru autovalori \c{s}i
func\c{t}ii proprii :
\be
\Psi_n (x)=D_n \exp{ \frac{-\lambda x^2}{2}} H_n (\sqrt{\lambda}x)
\ee
\be
E_n =(n+\frac{1}{2})\hbar\omega \hspace{1cm}n=0,1,2~\ldots
\ee

Spectrul de energie al OA 
este echidistant, adic\A\ , exist\A\ aceea\c{s}i 
diferen\c{t}\A\ $\hbar \omega$ \h ntre oricare dou\A\ nivele. Alt\A\
observa\c{t}ie pe care o putem face, este \h n leg\A tur\A\ cu
valoarea minim\A\ de energie pe care o are oscilatorul; poate \h n mod
surprinz\A tor este diferit\A\ de zero; acesta se consider\A\ un rezultat
pur cuantic, pentru c\A\ dispare dac\A\ $\hbar\rightarrow 0$. 
Se cunoa\c{s}te
ca {\em energia de punct zero} \c{s}i faptul c\A\ este diferit\A\ de
zero , este o caracteristic\A\ a tuturor poten\c{t}ialelor
confinante .\\

Constanta de normalizare poate fi calculat\A\ u\c{s}or \c{s}i are valoarea:
\be
D_n = \Bigg[ \sqrt{\frac{\lambda}{\pi}}\frac{1}{2^n n!}\Bigg]^{\frac{1}{2}}~.
\ee

Prin urmare se ob\c{t}in func\c{t}iile proprii normalizate
ale OA unidimensional :
\be
\Psi_n (x)= \Bigg[ \sqrt{\frac{\lambda}{\pi}}\frac{1}{2^n n!}\Bigg]^{\frac{1}{2}}
\hspace{.2cm} \exp( \frac{-\lambda x^2}{2})
\hspace{.2cm} H_n( \sqrt{\lambda} x)~.
\ee

%\newpage

%%%%%%%%%%%%%%%%%%%%%%%%%%%%%%%%%%%%%%%%%%%%%%%%%%%%%%%%%%
\section*{Operatori de creare $\hat{a}^{\dagger}$ \c{s}i 
anihilare $\hat{a}$}

Exist\A\ o alt\A\ form\A\ de a trata oscilatorul armonic fa\c{t}\A\ de cea 
conven\c{t}ional\A\ de a rezolva ecua\c{t}ia Schr\"odinger. Este vorba de 
metoda algebric\A\ sau metoda operatorilor de scar\A\ , o
metod\A\ foarte eficient\A\ care se poate aplica cu mult succes
a multe probleme de
mecanic\A\ cuantic\A\ de spectru discret.\\
Definim doi operatori nehermitici $a$ \c{s}i $a^{\dd}$ :
\be
a=\sqrt{\frac{m\omega}{2\hbar}}(x+\frac{ip}{m\omega})
\ee
\be
a^{\dd}=\sqrt{\frac{m\omega}{2\hbar}}(x-\frac{ip}{m\omega})~.
\ee

Ace\c{s}ti operatori sunt cunoscu\c{t}i
ca \hspace{.1cm} {\em operator de anihilare}
\hspace{.1cm}  \c{s}i \hspace{.1cm} {\em operator de creare},
\hspace{.1cm}  respectiv  (justificarea acestor denumiri se va vedea mai departe,
de\c{s}i se poate spune c\A\ vine din teoria cuantic\A\ a c\h mpurilor ).\\
S\A\ calcul\A m acum comutatorul acestor doi operatori:
\be
[a,a^{\dd}]=\frac{m\omega}{2\hbar}[x
+\frac{ip}{m\omega},x-\frac{ip}{m\omega}]=\frac{1}{2\hbar}(-i[x,p]+i[p,x])=1~,
\ee

unde am folosit comutatorul:
\be
[x,p]=i\hbar~.
\ee

Prin urmare operatorii de creare \c{s}i anihilare nu comut\A\ , satisf\A c\h nd 
rela\c{t}iile de comutare :
\be
[a,a^{\dd}]=1~.
\ee

S\A\ definim deasemenea importantul operator de num\A r $\hat{N}$:
\be
\hat{N}=\ad a~.
\ee

Acest\hspace{.1cm} operator este
hermitic\hspace{.1cm} dup\A\ cum se poate demonstra
u\c{s}or\hspace{.1cm} folosind $(AB)^{\dd}=B^{\dd}A^{\dd}$ :
\be
\hat{N}^{\dd}=(\ad a)^{\dd}=\ad (\ad)^{\dd}=\ad a=\hat{N}~.
\ee

Consider\h nd acum c\A\ :
\be
\ad a =\frac{m\omega}{2\hbar}(x^2+\frac{p^2}{m^2\omega^2})+\frac{i}{2\hbar}[x,p]=\frac{\hat{H}}{\hbar\omega}-\frac{1}{2}
\ee

\noindent
observ\A m c\A\ Hamiltonianul se scrie \h ntr-o form\A\ 
simpl\A\ \h n func\c{t}ie de operatorul de num\A r :
\be
\hat{H}=\hbar\omega(\hat{N}+\frac{1}{2})~.
\ee

Operatorul de num\A r are acest nume datorit\A\ faptului c\A\ autovalorile sale
sunt exact subindicii func\c{t}iei de und\A\ asupra c\A reia ac\c{t}ioneaz\A\ :
\be
\hat{N}\m n>=n\m n>~,
\ee

\noindent unde am folosit nota\c{t}ia:
\be
\m \Psi_n> \hspace{.2cm}= \hspace{.2cm}\m n>~.
\ee

Aplic\h nd acest fapt lui $(47)$ avem :
\be
\hat{H}\m n>=\hbar\omega(n+\frac{1}{2})\m n>~.
\ee

Dar \c{s}tim din ecua\c{t}ia Schr\"odinger c\A\ $\hat{H}\m n>=E\m n>$ pe baza 
c\A reia rezult\A\ c\A\ autovalorile energetice sunt date de :
\be
E_n=\hbar\omega(n+\frac{1}{2})~.
\ee

Acest rezultat este identic (cum \c{s}i trebuia s\A\ fie ) cu rezultatul 
$(36)$.\\
\^{I}n continuare s\A\ ar\A t\A m de ce 
operatorii $a$ \c{s}i $\ad$ au numele pe care le au. Pentru aceasta s\A\
calcul\A m comutatorii:
\be
[\hat{N},a]=[\ad a,a]=\ad[a,a]+[\ad,a]a=-a~,
\ee

rezultate care se ob\c{t}in din $[a,a]=0$ \c{s}i $(43)$.
Similar, s\A\ calcul\A m:
\be
[\hat{N},\ad]=[\ad a,\ad]=\ad[a,\ad]+[\ad,\ad]a=\ad~.
\ee
Cu ace\c{s}ti doi comutatori putem s\A\ scriem:
\begin{eqnarray}
\hat{N}(\ad \m n>)&=&([\hat{N},\ad]+\ad\hat{N})\m n>\nonumber\\
&=&(\ad+\ad\hat{N})\m n>\\
&=&\ad(1+n)\m n>=(n+1)\ad\m n>~.\nonumber
\end{eqnarray}

Cu un procedeu similar se ob\c{t}ine deasemenea:
\be
\hat{N}(a\m n>)=([\hat{N},a]+a\hat{N})\m n>=(n-1)a\m n>~.
\ee
Expresia $(54)$ implic\A\ c\A\ se poate considera
ket-ul $\ad \m n>$ ca eigenket al operatorului de num\A r , unde 
autovaloarea increment\A\ cu unu, adic\A\ , a fost produs\A\
o cuant\A\ de energie prin ac\c{t}iunea lui
$\ad$ asupra ket-ului. Aceasta explic\A\ \hspace{.1cm} numele de operator de
creare (crea\c{t}ie).
\hspace{.3cm} Comentarii\hspace{.1cm} urm\h nd\hspace{.1cm} 
aceea\c{s}i\hspace{.1cm} linie\hspace{.1cm} de\hspace{.1cm} 
ra\c{t}ionament acela\c{s}i tip de concluzie rezult\A\ pentru operatorul 
$a$, ceea ce \h i d\A\ numele de 
operator de\hspace{.1cm}
anihilare ( o cuant\A\ \hspace{.1cm} de
energie  este\hspace{.1cm} eliminat\A\ \hspace{.1cm}
c\h nd ac\c{t}ioneaz\A\ \hspace{.1cm} acest\hspace{.1cm} 
operator ).\\
Ecua\c{t}ia $(54)$ deasemenea implic\A\ propor\c{t}ionalitatea
ket-urilor $\ad\m n>$ \c{s}i $\m n+1>$:
\be
\ad\m n>=c\m n+1>~,
\ee

unde $c$ este o constant\A\ care trebuie determinat\A\ . Consider\h nd 
\h n plus c\A\ :
\be
(\ad\m n>)^{\dd}=<n\m a=c^*<n+1\m~,
\ee
putem realiza urm\A torul calcul:
\be
<n\m a( \ad\m n>)=c^*<n+1\m (c\m n+1>)
\ee
\be
<n\m a\ad \m n>=c^*c<n+1\m n+1>
\ee
\be
<n\m a\ad \m n>=\m c\m^2~.
\ee

Dar din rela\c{t}ia de comutare pentru operatorii $a$ \c{s}i $\ad$ :
\be
[a,\ad]=a\ad-\ad a=a\ad-\hat{N}=1~,
\ee

avem c\A\ :
\be
a\ad=\hat{N}+1
\ee

Substituind \h n $(60)$:
\be
<n\m \hat{N}+1\m n>=<n\m n>+<n\m \hat{N}\m n>=n+1=\m c\m^2~.
\ee

Cer\h nd $c$ s\A\ fie real \c{s}i pozitiv ( prin conven\c{t}ie ), ob\c{t}inem
urm\A toarea valoare:
\be
c=\sqrt{n+1}~.
\ee

Cu aceasta avem rela\c{t}ia:
\be
\ad \m n>=\sqrt{n+1}\m n+1>~.
\ee

Urm\h nd acela\c{s}i procedeu se poate ajunge la o rela\c{t}ie pentru operatorul
de anihilare :
\be
a\m n>=\sqrt{n}\m n-1>~.
\ee

S\A\ ar\A t\A m acum c\A\ valorile lui $n$ trebuie s\A\ fie \h ntregi 
nenegativi. 
Pentru aceasta, recurgem la cerin\c{t}a de pozitivitate a normei, aplic\h nd-o 
\h n special vectorului de stare $a\m n>$. Aceast\A\ condi\c{t}ie ne spune
c\A\ produsul interior (intern) al acestui vector cu adjunctul s\A u
($ (a\m n>)^\dd=<n\m \ad$) 
trebuie s\A\ fie mai mare sau egal\A\ cu zero :
\be
( <n\m \ad)\cdot(a\m n>)\geq 0~.
\ee

Dar aceast\A\ rela\c{t}ie nu este dec\h t :
\be
<n\m \ad a\m n>=<n\m \hat{N}\m n>=n \geq 0~.
\ee

Prin urmare $n$ nu poate fi negativ \c{s}i trebuie s\A\ fie \h ntreg pentru c\A\
dac\A\ nu ar fi prin aplicarea consecutiv\A\ a operatorului de anihilare ne-ar
duce la valori negative ale lui $n$, ceea ce este \h n contradic\c{t}ie cu ce
s-a spus anterior.\\
Este posibil s\A\ se exprime starea $n$ $(\m n>)$ direct \h n func\c{t}ie de 
starea baz\A\ $(\m 0>)$ folosind operatorul de creare. S\A\ vedem cum
se face aceast\A\ itera\c{t}ie important\A\ :

\begin{eqnarray}
 \m 1>=\ad \m 0> \\
 \m 2>=[\frac{\ad}{\sqrt{2}}]\m 1>=[\frac{(\ad)^2}{\sqrt{2!}}]\m 0> \\
 \m 3>=[\frac{\ad}{\sqrt{3}}]\m 2>=[ \frac{ (\ad)^3}{\sqrt{3!}}]\m 0> 
\end{eqnarray}
%\begin{center}
\vdots
%\end{center}
\begin{eqnarray}
 \m n>=[ \frac{ (\ad)^n}{\sqrt{n!}}]\m 0>
\end{eqnarray}

Putem deasemenea aplica aceast\A\ metod\A\ pentru a ob\c{t}ine FP-urile
\h n spa\c{t}iul configura\c{t}iilor. Pentru a realiza acest lucru, vom pleca
din starea baz\A\ :
\be
a\m 0>=0~.
\ee
\^{I}n reprezentarea $x$ avem:
\be
\hat{ a}  
\Psi_0(x)=\sqrt{\frac{m\omega}{2\hbar}} (x+\frac{ip}{m\omega}) \Psi_0(x)=0~.
\ee
Amintindu-ne forma pe care o ia operatorul impuls \h n reprezentarea $x$, 
putem ajunge la o ecua\c{t}ie diferen\c{t}ial\A\ pentru func\c{t}ia de und\A\
a st\A rii fundamentale; s\A\ introducem deasemenea urm\A toarea 
defini\c{t}ie $x_0=\sqrt{\frac{\hbar}{m\omega}}$ , cu care avem :
\be
(x+x_0^2\frac{d}{dx})\Psi_0=0
\ee
Aceast\A\ ecua\c{t}ie se poate rezolva u\c{s}or, 
\c{s}i prin normalizare ( integrala sa de la $-\infty$ la $\infty$ 
trebuie s\A\ fie pus\A\ egal\A\ cu unu ), ajungem la func\c{t}ia de und\A\
a st\A rii fundamentale :
\be
\Psi_0(x)=(\frac{1}{\sqrt{ \sqrt{\pi}x_0}})e^{ -\frac{1}{2}(\frac{x}{x_0})^2}
\ee
Restul de FP, care descriu st\A rile excitate ale OA , se pot ob\c{t}ine
folosind operatorul de crea\c{t}ie. Procedeul este urm\A torul:
\begin{eqnarray}
\Psi_1=\ad \Psi_0 =(\frac{1}{\sqrt{2}x_0})(x-x_0^2\frac{d}{dx})\Psi_0\\
\Psi_2=\frac{1}{\sqrt{2}}(\ad)^2\Psi_0=\frac{1}{\sqrt{2!}}(\frac{1}
{\sqrt{2}x_0})^2(x-x_0^2\frac{d}{dx})^2\Psi_0~.
\end{eqnarray}
Continu\h nd, se poate ar\A ta prin induc\c{t}ie c\A\ :
\be
\Psi_n=\frac{1}{\sqrt{ \sqrt{\pi}2^nn!}}\hspace{.2cm}
\frac{1}{x_0^{n+\frac{1}{2}}}
\hspace{.2cm}(x-x_0^2\frac{d}{dx})^n
\hspace{.2cm}e^{-\frac{1}{2}(\frac{x}{x_0})^2}~.
\ee

%\newpage
%\\

\section*{Evolu\c{t}ia temporal\A\ a oscilatorului}

\^{I}n aceast\A\ sec\c{t}iune vom ilustra prin intermediul OA modul 
\h n care se lucreaz\A\ cu  
reprezentarea Heisenberg \h n care st\A rile 
sunt fixate \h n timp \c{s}i se permite evolu\c{t}ia temporal\A\
a operatorilor. 
Vom considera operatorii ca func\c{t}ii de timp \c{s}i vom ob\c{t}ine \h n mod 
concret cum  
evolu\c{t}ioneaz\A\ operatorii de pozi\c{t}ie , impuls, $a$ \c{s}i $\ad$ \h n
timp pentru cazul OA.
Ecua\c{t}iile de mi\c{s}care Heisenberg pentru $p$ \c{s}i $x$ sunt :
\begin{eqnarray}
\frac{d\hat{p}}{dt}&=&-\frac{\partial}{\partial\hat{x}}V({\bf \hat{x})}\\
\nonumber\\
\frac{d\hat{x}}{dt}&=&\frac{\hat{p}}{m}~.
\end{eqnarray}

De aici rezult\A\ c\A\ ecua\c{t}iile de mi\c{s}care 
pentru $x$ y $p$ \h n cazul OA sunt:
\begin{eqnarray}
\frac{d\hat{p}}{dt}&=&-m\omega^2\hat{x}\\
\nonumber\\
\frac{d\hat{x}}{dt}&=&\frac{\hat{p}}{m}~.
\end{eqnarray}

Prin urmare, dispunem de o pereche de ecua\c{t}ii cuplate , 
care sunt echivalente unei perechi de ecua\c{t}ii pentru operatorii de 
crea\c{t}ie \c{s}i anihilare, care \h ns\A\ nu sunt cuplate. \^{I}n mod 
explicit :
\begin{eqnarray}
\frac{d a}{dt}&=&\sqrt{\frac{m\omega}{2\hbar}}\frac{d}{dt}(\hat{x}+\frac{i\hat{p}}{m\omega})\\
\nonumber\\
\frac{da}{dt}&=&\sqrt{\frac{m\omega}{2\hbar}}(\frac{d\hat{x}}{dt}+
\frac{i}{m\omega}\frac{d\hat{p}}{dt})~.
\end{eqnarray}
 
Substituind $(82)$ \c{s}i $(83)$ \h n $(85)$ :
\be
\frac{da}{dt}=\sqrt{\frac{m\omega}{2\hbar}}(\frac{\hat{p}}{m}-
i\omega\hat{x})=-i\omega a~.
\ee
Similar, se poate ob\c{t}ine o ecua\c{t}ie diferen\c{t}ial\A\ pentru
operatorul de crea\c{t}ie :
\be
\frac{d\ad}{dt}=i\omega\ad
\ee
Ecua\c{t}iile diferen\c{t}iale pe care le-am ob\c{t}inut pentru evolu\c{t}ia
temporal\A\ a operatorilor de crea\c{t}ie \c{s}i anihilare , pot fi
integrate imediat, d\h nd evolu\c{t}ia explicit\A\ a acestor operatori:
\begin{eqnarray}
a(t)&=&a(0)e^{-i\omega t}\\
\ad (t)&=&\ad (0)e^{i\omega t}~.
\end{eqnarray}

Se poate remarca din aceste rezultate \c{s}i din ecua\c{t}iile $(44)$ \c{s}i
$(47)$ c\A\ at\h t Hamiltonianul ca \c{s}i operatorul de num\A r , 
nu depind de timp, a\c{s}a cum era de a\c{s}teptat.\\
Cu cele dou\A\ rezultate anterioare , putem s\A\ ob\c{t}inem
operatorii de pozi\c{t}ie \c{s}i impuls ca func\c{t}ii de timp, 
pentru c\A\ sunt da\c{t}i \h n func\c{t}ie de operatorii de crea\c{t}ie \c{s}i
anihilare:
\begin{eqnarray}
\hat{x}&=&\sqrt{\frac{\hbar}{2m\omega}}(a+\ad)\\
\hat{p}&=&i\sqrt{ \frac{m\hbar\omega}{2}}(\ad-a)~.
\end{eqnarray}

Substituind-i se ob\c{t}ine:
\begin{eqnarray}
\hat{x}(t)&=&\hat{x}(0)\cos{\omega t}+\frac{\hat{p}(0)}{m\omega}\sin{\omega t}\\
\nonumber\\
\hat{p}(t)&=&-m\omega\hat{x}(0)\sin{\omega t}+\hat{p}(0)\cos{\omega t}~.
\end{eqnarray}

Evolu\c{t}ia temporal\A\ a acestor operatori este aceea\c{s}i
ca \h n cazul ecua\c{t}iilor clasice de mi\c{s}care.\\
Astfel, am ar\A tat forma explicit\A\ de evolu\c{t}ie 
a patru operatori bazici \h n cazul OA , ar\A t\h nd modul \h n care se 
lucreaz\A\ \h n reprezentarea Heisenberg.

%\newpage

\section*{OA tridimensional}

La \h nceputul analizei noastre a OA cuantic am f\A cut comentarii
\h n leg\A tur\A\ cu importan\c{t}a pentru fizic\A\ a OA . 
Dac\A\ vom considera un analog
tridimensional, ar trebui s\A\ consider\A m o dezvoltare Taylor \h n trei
variabile\footnote{Este posibil s\A\ se exprime dezvoltarea Taylor \h n jurul
lui ${\bf r_{0}}$ ca un 
operator exponen\c{t}ial :
\bc
$e^{[ (x-x_o)+(y-y_o)+(z-z_o)]
(\frac{\partial}{\partial x}+\frac{\partial}{\partial y}+
\frac{\partial}{\partial z})} \hspace{.1cm} f({\bf r_o})~.$\\
\ec }
re\c{t}in\h nd termeni numai p\h n\A\ \h n ordinul doi inclusiv, 
ceea ce ob\c{t}inem este o 
form\A\ cuadratic\A\ (\h n cazul cel mai general). Problema de rezolvat \h n
aceast\A\ aproxima\c{t}ie nu este chiar at\h t de simpl\A\ cum ar p\A rea
dintr-o prim\A\ examinare a poten\c{t}ialului corespunz\A tor :
%\bc

\be
V(x,y,z)=ax^2+by^2+cz^2+dxy+exz+fyz~.
\ee

%\ec

Exist\A\ \h ns\A\ multe sisteme care posed\A\ simetrie sferic\A\ sau
pentru care aproxima\c{t}ia acestei simetrii este satisf\A c\A toare. 
\^{I}n acest caz:
%\bc

\be
V(x,y,z)=K(x^2+y^2+z^2)~,
\ee

%\ec
\noindent
ceea ce este echivalent cu a spune c\A\ derivatele par\c{t}iale
secunde ( nemixte ) iau toate aceea\c{s}i valoare ( \h n cazul 
anterior reprezentate prin $K$). Putem ad\A uga c\A\
aceast\A\ este o bun\A\ aproxima\c{t}ie \h n cazul \h n care valorile
derivatelor par\c{t}iale secunde
mixte sunt mici \h n compara\c{t}ie cu cele nemixte.\\
C\h nd se satisfac aceste condi\c{t}ii \c{s}i poten\c{t}ialul este
dat de $(95)$ spunem c\A\ sistemul este un 
{\em OA tridimensional sferic simetric}.\\
Hamiltonianul pentru acest caz este de forma:
%\bc

\be
\hat{H}=\frac{-\hbar^2}{2m}\bigtriangledown^2 + \frac{m\omega^2}{2}r^2~,
\ee

%\ec
\noindent
unde Laplaceanul este dat \h n coordonate sferice \c{s}i $r$ este 
variabila sferic\A\ radial\A\ .\\
Fiind vorba de un poten\c{t}ial independent de timp, 
energia se conserv\A\ ; \h n plus dat\A\ simetria sferic\A\ ,  
momentul cinetic deasemenea se conserv\A\ . Avem deci dou\A\
m\A rimi conservate, ceea ce ne permite s\A\ spunem c\A\
fiec\A reia \h i corespunde un num\A r cuantic. 
Putem s\A\ presupunem c\A\ func\c{t}iile de und\A\ depind de dou\A\ numere 
cuantice (de\c{s}i \h n acest caz vom vedea c\A\ apare \h nc\A\ unul ). Cu 
aceste comentarii, ecua\c{t}ia de interes este :
%\bc

\be
\hat{H}\Psi_{nl}=E_{nl}\Psi_{nl}~.
\ee

%\ec

Laplaceanul \h n coordonate sferice este :
%\bc

\be
\bigtriangledown^2
=\frac{\partial^2}{\partial r^2}+\frac{2}{r}\frac{\partial}{\partial r}
-\frac{\hat{L}^2}{\hbar^2r^2}
\ee

%\ec
\c{s}i rezult\A\ din faptul cunoscut :
%\bc

\be
\hat{L}^2=-\hbar^2[ \frac{1}{\sin{\theta}}
\frac{\partial}{\partial\theta}
( \sin{\theta}\frac{\partial}{\partial\theta})
+\frac{1}{\sin{\theta}^2}\frac{\partial^2}{\partial\varphi^2}]~.
\ee

%\ec

{\em Func\c{t}iile proprii ale lui $\hat{L}^2$ sunt armonicele sferice}, 
respectiv:
%\bc

\be
\hat{L}^2Y_{lm_{l}}(\theta,\varphi)=-\hbar^2l(l+1)Y_{lm_{l}}(\theta,\varphi)
\ee

%\ec
Faptul c\A\ armonicele sferice
poart\A\ num\A rul cuantic $m_{l}$ face ca acesta s\A\ fie introdus \h n
func\c{t}ia de und\A\ $\Psi_{nlm_{l}}$.\\
Pentru a realiza separarea variabilelor \c{s}i func\c{t}iilor se propune
substitu\c{t}ia:
%\bc

\be
\Psi_{nlm_{l}}(r, \theta,\varphi)=\frac{R_{nl}(r)}{r}
Y_{lm_{l}}(\theta,\varphi)~.
\ee

%\ec
Odat\A\ introdus\A\ \h n ecua\c{t}ia Schr\"odinger va separa 
partea spa\c{t}ial\A\ de cea unghiular\A\ ; ultima se identific\A\ cu 
un operator propor\c{t}ional cu operatorul moment cinetic p\A trat, pentru care
func\c{t}iile proprii sunt armonicele sferice, \h n timp ce \h n partea
spa\c{t}ial\A\ ob\c{t}inem ecua\c{t}ia :
%\bc

\be
R_{nl}^{\prime\prime}+(\frac{2mE_{nl}}{\hbar^2}
-\frac{m^2\omega^2}{\hbar^2}r^2-\frac{l(l+1)}{r^2})R_{nl}(r)=0~.
\ee

%\ec

Folosind defini\c{t}iile $(7)$ \c{s}i $(8)$ , ecua\c{t}ia anterioar\A\ ia
exact forma lui $(9)$, cu excep\c{t}ia termenului de moment
unghiular, care \h n mod comun se cunoa\c{s}te ca 
{\em barier\A\ de moment unghiular}.
%\bc

\be
R_{nl}^{\prime\prime}+(k^2-\lambda^2r^2-\frac{l(l+1)}{r^2})R_{nl}=0~.
\ee

%\ec

\noindent
Pentru a rezolva aceast\A\ ecua\c{t}ie ,
vom pleca dela analiza sa asimptotic\A\ . Dac\A\ vom
considera mai \h nt\h i $r\rightarrow\infty$, observ\A m c\A\
termenul de moment unghiular este neglijabil, astfel c\A\ \h n aceast\A\ 
limit\A\
comportamentul asimptotic este identic aceluia a lui $(9)$, ceea ce ne conduce
la:
%\bc

\be
R_{nl}(r)\sim\exp{\frac{-\lambda r^2}{2}}\hspace{2cm}\mbox{\h n}
\hspace{.3cm}\lim\hspace{.1cm}r\rightarrow\infty~.
\ee

%\ec

Dac\A\ studiem acum comportamentul \h n jurul lui zero, vedem c\A\ 
termenul dominant este cel de moment unghiular,
adic\A\ , ecua\c{t}ia
diferen\c{t}ial\A\ $(102)$ se converte \h n aceast\A\ limit\A\ \h n :
%\bc

\be
R_{nl}^{\prime\prime}-\frac{l(l+1)}{r^2}R_{nl}=0~.
\ee

%\ec

Aceasta este o ecua\c{t}ie diferen\c{t}ial\A\ de tip
Euler \footnote{O ecua\c{t}ie de tip Euler este de forma :
%\bc

\[x^n y^{(n)}(x)+x^{n-1} y^{(n-1)}(x)+\cdots+x y^{\prime}(x)+y(x)=0~.\]

%\ec
Solu\c{t}iile ei sunt de tipul $x^{\alpha}$, care se substituie \c{s}i se
g\A se\c{s}te un polinom \h n $\alpha$.} , a c\A rei rezolvare duce la
dou\A\ solu\c{t}ii independente:
%\bc
\be
R_{nl}(r)\sim \hspace{.2cm}r^{l+1}\hspace{.2cm}\mbox{sau}
\hspace{.4cm}r^{-l}\hspace{2cm}\mbox{\h n}\hspace{.4cm}\lim\hspace{.1cm}r
\rightarrow 0~.
\ee
%\ec

Argumentele anterioare ne conduc la a propune substitu\c{t}ia :
%\bc
\be
R_{nl}(r)=r^{l+1}\exp{\frac{-\lambda r^2}{2}}\phi(r)~.
\ee
%\ec

S-ar putea deasemenea face \c{s}i urm\A toarea substitu\c{t}ie:
%\bc
\be
R_{nl}(r)=r^{-l}\exp{\frac{-\lambda r^2}{2}}v(r)~,
\ee
%\ec
care \h ns\A\ ne conduce la acelea\c{s}i solu\c{t}ii
ca \c{s}i $(107)$ ( de ar\A tat acest lucru este un bun exerci\c{t}iu).
Substituind $(107)$ \h n $(103)$ , se ob\c{t}ine urm\A toarea 
ecua\c{t}ie diferen\c{t}ial\A\ pentru $\phi$ :
%\bc
\be
\phi^{\prime\prime}+2(\frac{l+1}{r}-\lambda r)\phi^{\prime}
-[ \lambda (2l+3)-k^2]\phi=0~.
\ee
%\ec

Cu schimbarea de variabil\A\ $w=\lambda r^2$, ob\c{t}inem:
%\bc
\be
w\phi^{\prime\prime}+(l+\frac{3}{2}-w)\phi^{\prime}-[ \frac{1}{2}(l+
\frac{3}{2})-\frac{\kappa}{2}]\phi=0~,
\ee
%\ec
unde am introdus $\kappa =\frac {k^2}{2\lambda}=\frac{E}{\hbar\omega}$. 
Am ajuns din nou la o ecua\c{t}ie diferen\c{t}ial\A\ de tip hipergeometric 
confluent\A\ cu solu\c{t}iile ( a se vedea $(21)$ \c{s}i $(22)$):
%\bc
\be
\phi(r)=A\hspace{.2cm}_1F_1[\frac{1}{2}(l+\frac{3}{2}-\kappa);l+\frac{3}{2},
\lambda r^2]+B\hspace{.2cm}r^{-(2l+1)}
\hspace{.3cm}_1F_1[\frac{1}{2}(-l+\frac{1}{2}-\kappa);-l+\frac{1}{2},
\lambda r^2]~.
\ee
%\ec
 
A doua solu\c{t}ie particular\A\ nu poate fi normalizat\A\ , 
pentru c\A\ diverge puternic \h n zero, ceea ce oblig\A\ a lua $B=0$, deci :
%\bc
\be
\phi(r)=A\hspace{.2cm}_1F_1[\frac{1}{2}(l+\frac{3}{2}-\kappa);l+\frac{3}{2},
\lambda r^2]~.
\ee
%\ec
Folosind acelea\c{s}i argumente ca \h n cazul OA unidimensional,
respectiv, a impune ca solu\c{t}iile s\A\ fie regulare \h n infinit, 
\h nseamn\A\ condi\c{t}ia de truncare a seriei, ceea ce implic\A\ din nou
cuantizarea energiei . Truncarea este \h n acest caz:
\be
\frac{1}{2}(l+\frac{3}{2}-\kappa)=-n~,
\ee
\noindent unde introduc\h nd explicit $\kappa$, ob\c{t}inem spectrul de
energie :
\be
E_{nl}=\hbar\omega(2n+l+\frac{3}{2})~.
\ee

Putem observa c\A\ pentru OA tridimensional sferic simetric exist\A\ 
o energie de punct zero 
$\frac{3}{2}\hbar\omega$.\\
Func\c{t}iile proprii nenormalizate sunt:
\be
\Psi_{nlm}(r,\theta,\varphi)
=r^{l}e^{\frac{-\lambda r^2}{2}}\hspace{.2cm}_1F_1(-n;l
+\frac{3}{2},\lambda r^2)\hspace{.1cm}Y_{lm}(\theta,\varphi)~.
\ee

%\newpage

\section*{{\huge 5P. Probleme}}

\subsection*{Problema 5.1}

{\bf S\A\ se determine autovalorile \c{s}i func\c{t}iile proprii
ale OA \h n spa\c{t}iul impulsurilor}.

Hamiltonianul cuantic de OA este:
\[
\hat{H}=\frac{\hat{p}^2}{2m}+\frac{1}{2}m\omega^2\hat{x}^2~.
\]
\^{I}n spa\c{t}iul impulsurilor, operatorii  $\hat{x}$  
\c{s}i $\hat{p}$ au urm\A toarea form\A\ :
\[
\hat{p}\rightarrow\hspace{.2cm}p
\]
\[
\hat{x}\rightarrow\hspace{.2cm}i\hbar\frac{\partial}{\partial p}~.
\]
Prin urmare Hamiltonianul cuantic OA \h n reprezentarea de impuls este :
\[
\hat{H}=\frac{p^2}{2m}-\frac{1}{2}m\omega^2\hbar^2\frac{d^2}{dp^2}~.
\]
Avem de rezolvat problema de autovalori 
( ceea ce \h nseamn\A\ de ob\c{t}inut func\c{t}iile proprii \c{s}i
autovalorile) dat\A\ prin $(5)$ , 
care cu Hamiltonianul anterior, este urm\A toarea ecua\c{t}ie 
diferen\c{t}ial\A\ :
\be
\frac{d^2\Psi(p)}{dp^2}+( \frac{2E}{m\hbar^2\omega^2}-
\frac{p^2}{m^2\hbar^2\omega^2})\Psi(p)=0~.
\ee
Se poate observa c\A\ aceast\A\ ecua\c{t}ie diferen\c{t}ial\A\ , este 
identic\A\ , p\h n\A\ la constante, cu ecua\c{t}ia diferen\c{t}ial\A\ 
din spa\c{t}iul configura\c{t}iilor ( ec. $(6)$ ). 
Pentru a exemplifica o alt\A\ form\A\ de a o rezolva, nu vom urma exact
acela\c{s}i drum.\\
Definim doi parametri, analogi celor din $(7)$ \c{s}i $(8)$:
\be
k^2=\frac{2E}{m\hbar^2\omega^2} \hspace{1cm}\lambda=\frac{1}{m\hbar\omega}~.
\ee
Cu aceste defini\c{t}ii, ajungem la ecua\c{t}ia diferen\c{t}ial\A\ $(9)$\c{s}i
deci solu\c{t}ia c\A utat\A\ ( \h n urma efectu\A rii analizei asimptotice ) 
este de forma:
\be
\Psi(y)=e^{-\frac{1}{2}y}\phi (y)~,
\ee
unde $y$ este dat de $y=\lambda p^2$ \c{s}i $\lambda$ este definit \h n $(117)$.
Substituind $(118)$ \h n $(116)$ , av\h nd grij\A\ s\A\ punem $(118)$ \h n 
variabila $p$ . Se ob\c{t}ine astfel o ecua\c{t}ie diferen\c{t}ial\A\ pentru
 $ \phi$ :
\be
\frac{d^2\phi(p)}{dp^2}-2\lambda p\frac{d\phi (p)}{dp}+(k^2-\lambda)\phi (p)=0~.
\ee
Vom face acum schimbul de variabil\A\ $u=\sqrt{\lambda}p$ , 
care ne conduce la ecua\c{t}ia Hermite :
\be
\frac{d^2\phi (u)}{du^2}-2u\frac{d\phi (u)}{du}+2n\phi(u)=0~,
\ee
cu $n$ un \h ntreg nenegativ , \c{s}i unde am pus :
\[
\frac{k^2}{\lambda}-1=2n~.
\]
De aici \hspace{.1cm}  \c{s}i din defini\c{t}iile date \h n $(117)$  
rezult\A\ c\A\ autovalorile de energie sunt date de :
\[
E_n=\hbar\omega(n+\frac{1}{2})~.
\]
Solu\c{t}iile pentru $(120)$ sunt polinoamele Hermite $\phi(u)=H_n(u)$ \c{s}i
func\c{t}iile proprii nenormalizate sunt :
\[
\Psi(p)=A e^{-\frac{\lambda}{2}p^2}H_n(\sqrt{\lambda}p)~.
\]

\vspace{1mm}
%\newpage

\subsection*{Problema 5.2}

{\bf S\A\ se demonstreze c\A\ polinoamele Hermite pot fi expresate
\h n urm\A toarea reprezentare integral\A\ :
\be
H_n(x)=\frac{2^n}{\sqrt{\pi}}\int_{-\infty}^{\infty} (x+iy)^n e^{-y^2}dy~.
\ee
}

Aceast\A\ reprezentare a polinoamelor Hermite nu este foarte uzual\A\ , 
de\c{s}i se poate dovedi util\A\ \h n unele cazuri. Ceea ce vom face 
pentru a realiza demonstra\c{t}ia , este s\A\ dezvolt\A m integrala 
\c{s}i s\A\ demonstr\A m c\A\ ceea ce se ob\c{t}ine este identic cu 
reprezentarea \h n serie a polinoamelor Hermite pentru care avem :
\be
\sum_{k=0}^{[\frac{n}{2}]} \frac{ (-1)^k n!}{(n-2k)!k!}(2x)^{n-2k}~,
\ee

unde simbolul $[c]$ unde se termin\A\ seria signific\A\ cel mai mare \h ntreg 
mai mic sau egal cu $c$.\\
Primul lucru pe care \h l vom face este s\A\ dezvolt\A m binomul 
din integral\A\ folosind binecunoscuta teorem\A\ a binomului:
\[
(x+y)^n = \sum_{m=0}^n \frac{n!}{(n-m)!m!}x^{n-m}y^m~.
\]
Astfel:
\be
(x+iy)^n= \sum_{m=0}^n \frac{n!}{(n-m)!m!}i^mx^{n-m}y^m~,
\ee

care substituit \h n integral\A\ duce la:
\be
\frac{2^n}{\sqrt{\pi}}\sum_{m=0}^n \frac{n!}{(n-m)!m!}i^m x^{n-m}
\int_{-\infty}^{\infty} y^m e^{-y^2}dy~.
\ee

Din forma expresiei din integral\A\ putem s\A\ vedem c\A\ este diferit\A\ de 
zero c\h nd $m$ este par , \h n cazul impar integrala se anuleaz\A\ din 
motive evidente. Folosind nota\c{t}ia par\A\ $m=2k$ avem:

\be
\frac{2^n}{\sqrt{\pi}}
\sum_{k=0}^{[\frac{n}{2}]}\frac{n!}{(n-2k)!(2k)!}i^{2k}x^{n-2k}
\hspace{.2cm}2\int_{0}^{\infty} y^{2k}e^{-y^2}dy~.
\ee

Cu schimbul de variabil\A\ $u=y^2$, integrala devine o func\c{t}ie gamma :
\be
\frac{2^n}{\sqrt{\pi}}\sum_{k=0}^{[\frac{n}{2}]}
\frac{n!}{(n-2k)!(2k)!}i^{2k}x^{n-2k}\int_{0}^{\infty}u^{k-\frac{1}{2}}e^{-u}du~,
\ee

\noindent
respectiv $\Gamma(k+\frac{1}{2})$ , care \h n plus se poate exprima
prin factoriali ( desigur pentru $k$ \h ntreg ) :
\[
\Gamma(k+\frac{1}{2})=\frac{(2k)!}{2^{2k}k!}\sqrt{\pi}~.
\]
Substituind aceast\A\ expresie \h n sum\A\ \c{s}i folosind
faptul c\A\ $i^{2k}=(-1)^k$ se ob\c{t}ine

\be
\sum_{k=0}^{[\frac{n}{2}]} \frac{ (-1)^k n!}{(n-2k)!k!}(2x)^{n-2k}~,
\ee

\noindent
care este identic cu $(122)$ , ceea ce completeaz\A\ demonstra\c{t}ia.

\vspace{1mm}
%\newpage

\subsection*{Problema 5.3}

{\bf S\A\ se arate c\A\ rela\c{t}ia de incertitudine 
Heisenberg se satisface efectu\h nd calculul cu func\c{t}iile proprii ale OA}~.

Trebuie s\A\ ar\A t\A m c\A\ pentru oricare $\Psi_n$ se satisface:
\be
<(\Delta p)^2(\Delta x)^2>\hspace{.3cm}\geq \frac{\hbar^2}{4}~,
\ee
unde nota\c{t}ia $<>$ \h nseamn\A\ valoare medie.\\
Vom calcula \h n mod separat $<(\Delta p)^2>$ \c{s}i $<(\Delta x)^2>$ , 
unde fiecare din aceste expresii este :

\[<(\Delta p)^2>=< (p-<p>)^2 >=< p^2 - 2p<p>+<p>^2>=<p^2>-<p>^2~,\]

\[<(\Delta x)^2>=< (x-<x>)^2 >=< x^2 - 2x<x>+<x>^2>=<x^2>-<x>^2~. \]

Mai \h nt\h i vom ar\A ta c\A\ at\h t media lui $x$ c\h t \c{s}i a lui $p$ 
sunt zero. Pentru media lui $x$ avem:

\[<x>=\int_{-\infty}^{\infty} x[\Psi_n(x)]^2 dx~.\]

Aceast\A\ integral\A\ se anuleaz\A\ datorit\A\ imparit\A \c{t}tii expresiei
de integrat, care este manifest\A\ . 
%Un argument mai lung este c\A\ $[\Psi_n(x)]^2$ este o
%func\c{t}ie par\A\ , dac\A\ \c{t}inem cont c\A\ paritatea este dat\A\ 
%de partea polinomial\A\ ( exponen\c{t}iala implicat\A\ fiind o func\c{t}ie
%par\A\ ). Polinoamele Hermite au paritate bine definit\A\ , ceea ce conduce
%la dou\A\ cazuri, $n$ par sau impar. 
%Dac\A\ $n$ este par $[\Psi_n(x)]^2$ este deasemenea. 
%Dac\A\ $n$ este impar, $H_n(-x)=(-1)^n H_n(x)$ , imediat se vede c\A\
%p\A tratul este par. 
%Hemos mostrado que $[\Psi_n(x)]^2$ es una funci\'on par para $n$ cualquiera, 
%por tanto al multiplicarla por $x$ se vuelve impar, de manera que la integral 
%se anula. 
Rezult\A\ deci c\A\ :
\be
<x>=0~.
\ee

Acelea\c{s}i argumente sunt corecte pentru media lui $p$ , 
dac\A\ efectu\A m calculul \h n spa\c{t}iul impulsurilor, respectiv cu ajutorul
func\c{t}iilor ob\c{t}inute \h n problema 1 . Este suficient s\A\ vedem c\A\
forma fun\c{t}ional\A\ este acea\c{s}i (se schimb\A\ doar simbolul). 
Deci:
\be
<p>=0~.
\ee

S\A\ calcul\A m acum media lui $x^2$. Vom folosi teorema virialului 
\footnote{ Amintim c\A\ teorema virialului \h n mecanica cuantic\A\ 
afirm\A\ c\A\ :
\[ 2<T>=<{\bf r}\cdot\bigtriangledown V({\bf r})>~.\]
Pentru un poten\c{t}ial de forma $V=\lambda x^n$ se satisface:
\[
2<T>=n<V>~,
\]
unde $T$ reprezint\A\ energia cinetic\A\ \c{s}i V este
energia poten\c{t}ial\A\ .}. Observ\A m mai \h nt\h i c\A\ :
\[
<V>=\frac{1}{2}m\omega^2 <x^2>~.
\]
Prin urmare este posibil\A\ rela\c{t}ionarea mediei lui $x^2$ direct cu 
media poten\c{t}ialului \h n acest caz (\c{s}i deci folosirea teoremei
virialului).
\be
<x^2>=\frac{2}{m\omega^2}<V>~.
\ee

Avem nevoie deasemenea de media energiei totale :
\[
<H>=<T>+<V>~,
\]

pentru care din nou se poate folosi teorema virialului ( pentru $n=2$ ) :
\be
<H>=2<V>~.
\ee

Astfel, se ob\c{t}ine:
\be
<x^2>=\frac{<H>}{m\omega^2}=\frac{\hbar\omega(n+\frac{1}{2})}{m\omega^2}
\ee

\be
<x^2>=\frac{\hbar}{m\omega}( n+\frac{1}{2})~.
\ee

Similar, media lui $p^2$ se poate calcula  explicit:
\be
<p^2>=2m<\frac{p^2}{2m}>=2m<T>=m<H>=m\hbar\omega(n+\frac{1}{2})~.
\ee

Cu $(133)$ \c{s}i $(135)$ avem:
\be
<(\Delta p)^2(\Delta x)^2>=(n+\frac{1}{2})^2\hbar^2~.
\ee

Pe baza acestui rezultat ajungem la concluzia c\A\
\h n st\A rile sta\c{t}ionare ale OA, care practic nu au fost folosite \h n mod 
direct, rela\c{t}ia de incertitudine Heisenberg se satisface \c{s}i are 
valoarea minim\A\ pentru starea fundamental\A\ $n=0$.

\vspace{1mm}
%\newpage

\subsection*{Problema 5.4}

{\bf S\A\ se ob\c{t}in\A\ elementele de
matrice ale operatorilor $a$, $\ad$, $\hat{x}$ \c{s}i $\hat{p}$}.

S\A\ g\A sim mai \h nt\h i elementele de matrice pentru operatorii de 
crea\c{t}ie \c{s}i anihilare, care sunt de mult ajutor pentru
a ob\c{t}ine elemente de matrice pentru restul operatorilor.\\
Vom folosi rela\c{t}iile $(65)$ \c{s}i $(66)$, care duc la:
\be
<m \m a\m n>=\sqrt{n}<m\m n-1>=\sqrt{n}\delta_{m,n-1}~.
\ee

\^{I}n mod similar pentru operatorul de crea\c{t}ie avem rezultatul:
\be
<m\m \ad \m n>=\sqrt{n+1}<m\m n+1>=\sqrt{n+1}\delta_{m,n+1}~.
\ee

S\A\ trecem acum la calculul elementelor de matrice ale operatorului
de pozi\c{t}ie. Pentru al efectua , s\A\ exprim\A m operatorul de
posi\c{t}ie \h n func\c{t}ie de operatorii de crea\c{t}ie \c{s}i
anihilare. Folosind defini\c{t}iile $(39)$ \c{s}i $(40)$ , se demonstreaz\A\
imediat c\A\ operatorul de posi\c{t}ie este dat de :
\be
\hat{x}=\sqrt{ \frac{\hbar}{2m\omega}}(a+\ad)~.
\ee

Folosind acest rezultat, elementele de matrice ale operatorului $\hat{x}$ 
pot fi calculate \h n manier\A\ imediat\A\ :
\begin{eqnarray}
<m\m \hat{x}\m n>&=&<m\m \sqrt{ \frac{\hbar}{2m\omega}}(a+\ad)\m n>\nonumber\\
&=&\sqrt{ \frac{\hbar}{2m\omega}}[\sqrt{n}\delta_{m,n-1}+
\sqrt{n+1}\delta_{m,n+1}]~.
\end{eqnarray}
Urm\h nd acela\c{s}i procedeu putem calcula elementele de matrice ale 
operatorului impuls, consider\h nd c\A\ $\hat{p}$ este dat 
\h n func\c{t}ie de operatorii de crea\c{t}ie \c{s}i anihilare \h n forma :
\be
\hat{p}= i\sqrt{ \frac{m\hbar\omega}{2}}(\ad -a)~,
\ee

ceea ce ne conduce la:
\be
<m\m\hat{p}\m n>= i\sqrt{ \frac{m\hbar\omega}{2}}[\sqrt{n+1}\delta_{m,n+1}
-\sqrt{n}\delta_{m,n-1}]~.
\ee

Se poate vedea u\c{s}urin\c{t}a cu care se pot face calculele dac\A\ se 
folosesc elementele de matrice ale operatorilor de crea\c{t}ie \c{s}i anihilare. 
Finaliz\A m cu o observa\c{t}ie \h n leg\A tur\A\ cu nediagonalitatea 
elementelor de matrice ob\c{t}inute. Aceasta este de a\c{s}teptat 
datorit\A\ faptului c\A\ reprezentarea folosit\A\ este cea a 
operatorului de num\A r \c{s}i nici unul dintre cei patru operatori
nu comut\A\ cu el.

\vspace{1mm}
%\newpage

\subsection*{Problema 5.5}

{\bf S\A\ se g\A seasc\A\ valorile medii ale lui
$\hat{x}^2$ \c{s}i $\hat{p}^2$ pentru OA unidimensional \c{s}i s\A\ se
foloseasc\A\ acestea pentru calculul 
valorilor medii (de a\c{s}teptare) ale energiei cinetice \c{s}i  celei
poten\c{t}iale. S\A\ se compare acest ultim rezultat cu teorema virialului.}

Mai \h nt\h i s\A\ ob\c{t}inem valoarea medie a lui $\hat{x}^2$. Pentru aceasta
recurgem la expresia $(139)$, care ne conduce la :
\be
\hat{x}^2 = \frac{\hbar}{2m\omega} (a^2 + (\ad) ^2 +\ad a+a\ad)~.
\ee
Se aminte\c{s}te c\A\ operatorii de crea\c{t}ie \c{s}i anihilare nu comut\A\
\h ntre ei . Av\h nd $(143)$ putem calcula valoarea medie a lui $\hat{x}^2$ :
\begin{eqnarray}
<\hat{x}^2>&=&<n\m \hat{x}^2\m n>\nonumber\\
&=&\frac{\hbar}{2m\omega}[ \sqrt{n(n-1)}\delta_{n,n-2}+\sqrt{(n+1)(n+2)}
\delta_{n,n+2}\nonumber\\
&+& \hspace{.2cm}n\hspace{.1cm}\delta_{n,n}\hspace{.2cm}+
\hspace{.2cm}(n+1)\hspace{.1cm}\delta_{n,n}]~,
\end{eqnarray}

ceea ce arat\A\ c\A\ :
\be
<\hat{x}^2>=<n\m \hat{x}^2\m n> = \frac{\hbar}{2m\omega}(2n+1)~.
\ee
Pentru calcularea valorii medii a lui $\hat{p}^2$ folosim $(141)$ pentru a 
exprima acest operator \h n func\c{t}ie de operatorii de crea\c{t}ie \c{s}i
anihilare:
\be
\hat{p}^2 = -\frac{m\hbar\omega}{2}(a^2+(\ad)^2-a\ad-\ad a)~,
\ee
ceea ce ne conduce la:
\be
<\hat{p}^2>=<n\m \hat{p}^2\m n>=\frac{m\hbar\omega}{2}(2n+1)~.
\ee
Ultimul rezultat ne d\A\ practic media energiei cinetice :
\be
<\hat{T}>=<\frac{\hat{p}^2}{2m}>=\frac{1}{2m}<\hat{p}^2>
=\frac{\hbar\omega}{4}(2n+1)~.
\ee

Valoarea medie a energiei poten\c{t}iale este:
\be
<\hat{V}>=<\frac{1}{2}m\omega^2 \hat{x}^2>=\frac{1}{2}m\omega^2 <\hat{x}^2>
=\frac{\hbar\omega}{4}(2n+1)~,
\ee
unde s-a folosit $(145)$.

Observ\A m c\A\ aceste valori medii coincid pentru orice $n$, 
ceea ce este \h n conformitate cu teorema 
virialului, care ne spune c\A\ pentru un poten\c{t}ial cuadratic ca cel de OA,
valorile medii ale energiei cinetice \c{s}i poten\c{t}iale trebuie s\A\
coincid\A\ \c{s}i deci s\A\ fie jum\A tate din valoarea medie a energiei
totale.

\vspace{1mm}

\newpage
%%%%%%%%%%%%%%%%%%%%%%%%%%%%%%%%%%%%%%%%%%%%%%%%%%%%%%%%%%% Hydrogen Atom
%%%%%%%%%%%%%%%%%%%%%%%%%%%%%%%%%%%%%%%%%%%%%%%%%%%%%%%%%%%%%%%%%%%%%%
%%%%%%%%%%%%%%%%%%%%%%%%%%%%%%%%%%%%%%%%%%%%%%%%%%%%%%%%%%%%%%%%%%%
%\documentstyle[aps,preprint,tighten]{revtex}
%\begin{document}
%\draft
\def\bi{bigskip}
\def\noi{noindent}
\def\ii{\'{\i}}
\begin{center}{\huge 6. ATOMUL DE HIDROGEN}
\end{center}
%\author{Edgar Alvarado Anell}
%\address{Universidad de Guanajuato,
%Guanajuato; M\'exico.}
%\maketitle
%\begin{abstract}
%\begin{center}
\section*{Introducere} %la mecanica cuantic\A\ pentru hidrogen}
\^{I}n acest capitol vom studia atomul de hidrogen, rezolv\h nd ecua\c{t}ia 
Schr\"odinger independent\A\ de timp cu un poten\c{t}ial produs de  
dou\A\ particule \h nc\A rcate electric cum este cazul electronului \c{s}i
protonului, cu  
Laplaceanul \h n coordonate sferice. Din punct de vedere matematic, se va folosi  
metoda separ\A rii de variabile, d\h nd o interpretare fizic\A\ 
func\c{t}iei de und\A\ ca solu\c{t}ie a ecua\c{t}iei Schr\"odinger 
pentru acest caz important, odat\A\ cu interpretarea 
numerelor cuantice \c{s}i a densit\A \c{t}ilor de probabilitate.\\
%\end{center}
%\end{abstract}
%%%%%%%%%%%%%%%%%%%%%%%%%%%%%%%%%%%%%%%%%%%%%%%%%%%%%%%%%%%%%%%%%%%
%\setcounter{equation}
%\section*{Introducere la mecanica cuantic\A\ pentru hidrogen}
\setcounter{equation}{0}
Scala spa\c{t}ial\A\ foarte mic\A\ a atomului de hidrogen   
intr\A\ \h n domeniul de aplicabilitate al mecanicii cuantice, pentru care
fenomenele atomice au fost o arie de verificare \c{s}i interpretare a 
rezultatelor \h nc\A\ de la
bun \h nceput. Cum mecanica cuantic\A\ d\A\ , \h ntre altele, 
rela\c{t}ii \h ntre
m\A rimile observabile \c{s}i cum principiul de incertitudine modific\A\
radical defini\c{t}ia teoretic\A\ a unei ``observabile" este important 
s\A\ \h n\c{t}elegem
\h n mod c\h t mai clar no\c{t}iunea cuantic\A\ de observabil\A\ \h n c\h mpul
atomic. De acord cu principiul de incertitudine, pozi\c{t}ia \c{s}i
impulsul unei particule nu se pot m\A sura simultan sub o anumit\A\ precizie
impus\A\ de comutatorii cuantici.
De fapt, m\A rimile asupra c\A rora mecanica cuantic\A\ d\A\ rezultate \c{s}i pe
care le rela\c{t}ioneaz\A\ sunt \h ntotdeauna 
probabilit\A \c{t}i. \^{I}n loc de a afirma, de exemplu, c\A\ raza orbitei 
electronului \h ntr-o stare fundamental\A\ a atomului 
de hidrogen este \h ntotdeauna $5.3 \times 10^{-11}$ m, mecanica
cuantic\A\ afirm\A\ c\A\ aceasta este doar raza medie; dac\A\ efectu\A m  un
experiment adecuat, vom ob\c{t}ine exact ca \h n cazul experimentelor cu 
detectori macroscopici pe probe macroscopice diferite valori aleatorii
dar a c\A ror medie va fi $5.3 \times 10^{-11}$ m. A\c{s}adar, din punctul de 
vedere al erorilor experimentale nu exist\A\ nici o diferen\c{t}\A\ 
fa\c{t}\A\ de fizica clasic\A\ . 

%%%%%%%%%%%%%%%%%%%%%%%%%%%%%%%%%%%%%%%%%%%%%%%%%%%%%%
%\section*{Ecua\c{t}ia de und\A\ }
Dup\A\ cum se \c{s}tie, pentru calculul valorilor medii \h n mecanica 
cuantic\A\ este necesar\A\ o
func\c{t}ie de und\A\ corespunz\A toare
$\Psi$. De\c{s}i $\Psi$ nu are o interpretare fizic\A\ direct\A\ ,
modulul p\A trat $\mid \Psi \mid^{2}$ calculat \h ntr-un punct arbitrar din 
spa\c{t}iu \c{s}i la un moment dat este propor\c{t}ional cu probabilitatea de a
g\A si particula \h ntr-o vecin\A tate infinitezimal\A\ a acelui punct
 acel loc \c{s}i la momentul dat. Scopul 
mecanicii cuantice este determinarea lui $\Psi$ pentru o microparticul\A\ 
\h n diferite condi\c{t}ii experimentale. 

\^{I}nainte de a trece la calculul efectiv al lui $\Psi$ pentru cazul 
electronului hidrogenic, trebuie s\A\ 
stabilim unele rechizite generale (care trebuie s\A\ se respecte \h n orice
situa\c{t}ie). \^{I}n primul r\h nd,
pentru c\A\ $\mid \Psi \mid^{2}$ este propor\c{t}ional cu probabilitatea P de 
a g\A si particula descris\A\ prin $\Psi$, integrala $\mid \Psi 
\mid^{2}$ pe tot spa\c{t}iul trebuie s\A\ fie finit\A\ , pentru ca 
\h ntr-adev\A r particula s\A\ poat\A\ fi localizat\A\ . Deasemenea, dac\A\
\begin{equation} %1
\int_{-\infty}^{\infty} \mid \Psi \mid^{2} dV = 0
\end{equation}
particula nu exist\A\ , iar dac\A\ integrala este $\infty$
nu putem avea semnifica\c{t}ie fizic\A\ ; $\mid \Psi \mid^{2}$ nu poate fi
negativ\A\ sau 
complex\A\ din simple motive matematice, astfel c\A\ unica
posibilitate r\A m\h ne ca integrala s\A\ fie finit\A\  pentru a avea o
descriere acceptabil\A\ a unei particule reale.
\^{I}n general, este convenabil de a identifica $\mid \Psi \mid^{2}$ cu 
probabilitatea P de a g\A si particula descris\A\ de c\A tre $\Psi$ \c{s}i nu
doar simpla propor\c{t}ionalitate cu P. Pentru ca $\mid \Psi \mid^{2}$ s\A\ fie 
egal\A\ cu P se impune
\begin{equation} %2
\int_{-\infty}^{\infty}\mid \Psi \mid^{2} dV = 1~,
\end{equation}
pentru c\A\ 
\begin{equation} %3
\int_{-\infty}^{\infty}{\rm P} dV = 1
\end{equation}
este afirma\c{t}ia matematic\A\ a faptului c\A\ particula exist\A\ \h ntr-un
loc din spa\c{t}iu la orice moment.
O func\c{t}ie care respect\A\ ec. 2 se spune c\A\ este normalizat\A\ . 
Pe l\h ng\A\ aceast\A\ condi\c{t}ie fundamental\A\ , $\Psi$ trebuie s\A\ aib\A\
o valoare unic\A\ , 
pentru c\A\ P are o singur\A\ valoare \h ntr-un loc \c{s}i la un moment 
determinat. O alt\A\ condi\c{t}ie pe care $\Psi$ trebuie s\A\ o satisfac\A\
este c\A\ at\h t ea c\h t \c{s}i  
derivatele sale par\c{t}iale $\frac{\partial \Psi}{\partial x}$, $\frac{\partial 
\Psi}{\partial y}$, $\frac{\partial \Psi}{\partial z}$ trebuie s\A\ fie
continue \h n orice punct arbitrar.

Ecua\c{t}ia Schr\"odinger este considerat\A\  ecua\c{t}ia fundamental\A\ a 
mecanicii cuantice \h n acela\c{s}i sens \h n care legea for\c{t}ei este 
ecua\c{t}ia fundamental\A\ a mecanicii newtoniene cu deosebirea important\A\
c\A\ este o ecua\c{t}ie de und\A\ pentru $\Psi$.

Odat\A\ ce energia poten\c{t}ial\A\ este cunoscut\A\ , se poate 
rezolva ecua\c{t}ia Schr\"odinger 
pentru func\c{t}ia de und\A\ $\Psi$ a particulei, a c\A rei 
densitate de probabilitate $\mid \Psi \mid^{2}$ se poate determina pentru 
$x,y,z,t$.
\^{I}n multe situa\c{t}ii, energia poten\c{t}ial\A\ a unei particule nu 
depinde explicit de timp; for\c{t}ele care ac\c{t}ioneaz\A\ asupra ei 
se schimb\A\ \h n func\c{t}ie numai de
posi\c{t}ia particulei. 
\^{I}n aceste condi\c{t}ii, ecua\c{t}ia Schr\"odinger se 
poate simplifica elimin\h nd tot ce se refer\A\ la $t$. S\A\ not\A m c\A\
se poate scrie func\c{t}ia de und\A\ unidimensional\A\ a unei particule libere
\h n forma
\begin{eqnarray} %22
\Psi(x,t) & = &  Ae^{(-i/\hbar)(Et - px)} \nonumber \\
& = & Ae^{-(iE/\hbar)t}e^{(ip/\hbar)x} \nonumber\\
& = & \psi(x) e^{-(iE/\hbar)t}~.
\end{eqnarray}
$\Psi(x,t)$ este produsul \h ntre o func\c{t}ie dependent\A\ de timp 
$e^{-(iE/\hbar)t}$ \c{s}i una sta\c{t}ionar\A\ , dependent\A\ 
numai de pozi\c{t}ie $\psi(x)$. 
%Varia\c{t}iile \h n timp de todas las funciones de 
%part\ii culas, sobre las que act\'uan fuerzas estacionarias, tienen la 
%misma forma que las de una part\ii cula libre. 
%Substituind $\Psi$ din 
%ec. 21 \h n ec. Schr\"odinger dependent\A\ de timp, 
%se ob\c{t}ine 
%\begin{equation} %23
%E \psi e^{-(iE/\hbar)t} = -\frac{\hbar^{2}}{2m}e^{-(iE/\hbar)t}
%\frac{\partial^{2} \psi}{\partial x^{2}} + V \psi e^{-(iE/\hbar)t}
%\end{equation}
%\c{s}i elimin\h nd factorul exponen\c{t}ial comun,
%\begin{equation} %24
%\frac{\partial^{2} \psi}{\partial x^{2}} + \frac{2m}{\hbar^{2}}(E-V)\psi = 0~,
%\end{equation}
%care este ec. Schr\"odinger pentru o stare sta\c{t}ionar\A\ . \^{I}n 
%trei dimensiuni, avem 
%\begin{equation} %25
%\frac{\partial^{2} \psi}{\partial x^{2}} + \frac{\partial^{2} 
%\psi}{\partial y^{2}} + \frac{\partial^{2} \psi}{\partial z^{2}} + 
%\frac{2m}{\hbar^{2}}(E-V)\psi = 0~. 
%\end{equation}

\^{I}n cazul general \h ns\A\ , ecua\c{t}ia Schr\"odinger pentru o 
stare sta\c{t}ionar\A\ 
se poate rezolva numai pentru anumite valori ale energiei E. 
Nu este vorba de dificult\A \c{t}i matematice, ci de un aspect fundamental. 
``A rezolva" ecua\c{t}ia Schr\"odinger pentru un sistem dat 
\h nseamn\A\ a ob\c{t}ine o func\c{t}ie de und\A\ $\psi$ care nu numai
c\A\ satisface 
ecua\c{t}ia \c{s}i condi\c{t}iile de frontier\A\ impuse, ci este o 
func\c{t}ie de und\A\ acceptabil\A\ , respectiv, 
func\c{t}ia \c{s}i derivata sa s\A\ fie continue, finite \c{s}i 
univoce. Astfel, cuantizarea energiei apare \h n 
mecanica ondulatorie ca un element teoretic natural, iar \h n practic\A\ ca  
un fenomen universal, caracteristic tuturor  
sistemelor microscopice stabile .

%%%%%%%%%%%%%%%%%%%%%%%%%%%%%%%%%%%%%%%%%%%%%%%%%%%%%%%%%%%%%%%%%%%%%
\section*{Ecua\c{t}ia Schr\"odinger pentru atomul de hidrogen}
\^{I}n continuare, vom aplica ecua\c{t}ia Schr\"odinger  
atomului de hidrogen, despre care se \c{s}tia pe baza experimentelor
Rutherford c\A\ este format dintr-un proton, particul\A\ 
cu sarcina electric\A\ +$e$ \c{s}i un electron de sarcin\A\ -$e$ \c{s}i care
fiind de   
1836 de ori mai u\c{s}or dec\h t protonul este cu mult mai mobil. 

Dac\A\ interac\c{t}iunea \h ntre dou\A\ particule este de tipul $u(r)=u ( \mid 
\vec r_{1} - \vec r_{2} \mid)$, problema de mi\c{s}care  
%part\ii culas en mec\'anica cu\'antica y tambien en mec\'anica cl\'asica 
se reduce at\h t clasic c\h t \c{s}i cuantic la mi\c{s}carea unei singure 
particule \h n c\h mpul de simetrie 
sferic\A\ . \^{I}ntr-adev\A r Lagrangeanul:
\begin{equation} %26
L = \frac{1}{2} m_{1} \dot { \vec r_{1}^{2}} + \frac{1}{2} m_{2} \dot {\vec
r_{2}^{2} } - u ( \mid \vec r_{1} - \vec r_{2} \mid)
\end{equation}
se transform\A\ folosind:
\begin{equation} %27
\vec r = \vec r_{1} - \vec r_{2}
\end{equation}
\c{s}i
\begin{equation} %28
\vec R = \frac{m_{1} \vec r_{1} + m_{2} \vec r_{2}}{m_{1} + m_{2}}~,
\end{equation}
\h n Lagrangeanul:
\begin{equation} %29
L = \frac{1}{2} M \dot { \vec R^{2}} + \frac{1}{2} \mu \dot {\vec
r^{2} } - u (r)~,
\end{equation}
unde
\begin{equation} %30
M = m_{1} + m_{2}
\end{equation}
\c{s}i
\begin{equation} %31
\mu =\frac{m_{1} m_{2}}{m_{1} + m_{2}}~.
\end{equation}

Pe de alt\A\ parte, introducerea impulsului se face cu formulele  
Lagrange
\begin{equation} %32
\vec P = \frac{\partial L}{\partial \dot { \vec R}} = M \dot { \vec R}
\end{equation}
\c{s}i
\begin{equation} %33
\vec p = \frac{\partial L}{\partial \dot { \vec r}} = m \dot { \vec r}~,
\end{equation}
ceea ce permite scrierea func\c{t}iei Hamilton clasice 
\begin{equation} %34
H = \frac{P^{2}}{2M} + \frac{p^{2}}{2m} + u(r)~.
\end{equation}

Astfel, se poate ob\c{t}ine operatorul hamiltonian pentru problema 
corespunz\A toare cuantic\A\ cu comutatori de tipul
\begin{equation} %35
[P_{i},P_{k}] = -i \hbar \delta_{ik}
\end{equation}
\c{s}i
\begin{equation} %36
[p_{i},p_{k}] = -i \hbar \delta_{ik}~.
\end{equation}
Ace\c{s}ti comutatori implic\A\ un operator Hamiltonian de forma
\begin{equation} %37
\hat H = -\frac{\hbar^{2}}{2M}\nabla_{R}^{2} - 
\frac{\hbar^{2}}{2m}\nabla_{r}^{2} + u(r)~,
\end{equation}
care este fundamental pentru studiul atomului de hidrogen cu ajutorul
ecua\c{t}iei Schr\"odinger \h n forma sta\c{t}ionar\A\
\begin{equation} %38
\hat H \psi = E \psi ~,
\end{equation}
%lo cual es una forma muy practica de escribirla, pero lo mas importante 
%hasta ahora escrito en \'esta secci\'on es que se ha tratado al sistema 
%formado por el prot\'on y el electr\'on como un sistema cl\'asico con 
%part\ii culas de masa no despreciable, como lo demuestran las ecs. 24-29, 
%ya que no se estan tomando en cuenta velocidades cercanas a la de la luz, 
ceea ce presupune c\A\ nu se iau \h n considerare efecte relativiste (viteze
apropriate de cele ale luminii \h n vid).

%Ec. Schr\"odinger pentru electronul atomic \h n trei 
%dimensiuni este ec. 21. Se folose\c{s}te ec. Schr\"odinger 
%independent\A\ de timp datorit\A\ faptului c\A\ poten\c{t}ialul $V$ depinde
%numai de $r$ \c{s}i nu de timp. 

Energia poten\c{t}ial\A\ $u(r)$ este cea electrostatic\A\ 
%energ\ii a potencial electrost\'atica de una carga -$e$ a una distancia 
%$r$ de otra carga +$e$, es
\begin{equation} %39
u = -\frac{e^{2}}{4\pi \epsilon _{0} r}
\end{equation}

%Deoarece $V$ este o func\c{t}ie de $r$ \h n loc de $x,y,z$, nu putem 
%substitui ec. 39 direct \h n ec. 21. 
Exist\A\ dou\A\ 
posibilit\A \c{t}i: prima, de a exprima $u$ \h n func\c{t}ie de coordonatele
carteziene 
$x,y,z$ substituind $r$ prin $\sqrt{x^{2}+y^{2}+z^{2}}$, a doua, de a exprima
ecua\c{t}ia Schr\"odinger \h n func\c{t}ie de coordonatele polare 
sferice $r,\theta,\phi$. \^{I}n virtutea simetriei sferice a 
situa\c{t}iei fizice, vom trata ultimul caz pentru c\A\ problema matematic\A\ se 
simplific\A\ considerabil.

Prin urmare, \h n coordonate polare sferice, ecua\c{t}ia 
Schr\"odinger este
%%%%%%%%%%%%%%%%%%%%%%%%%%%%%
\begin{equation} %40
\frac{1}{r^{2}} \frac{\partial}{\partial r}\left(r^{2} \frac{\partial 
\psi}{\partial r}
\right) + \frac{1}{r^{2} \sin\theta} \frac{\partial}{\partial 
\theta} \left(\sin\theta \frac{\partial \psi}{\partial \theta}\right) + 
\frac{1}{r^{2}\sin^{2}\theta} \frac{\partial^{2} \psi}{\partial \phi^{2}} 
+ \frac{2m}{\hbar^{2}}(E - u)\psi = 0
\end{equation}
%%%%%%%%%%%%%%%%%%%%%%%
Substituind (18) \c{s}i multiplic\h nd toat\A\ 
ecua\c{t}ia cu $r^{2}\sin^{2}\theta$, se ob\c{t}ine
%%%%%%%%%%%%%%%%%%%%%%%%%%%%%%%%%%
\begin{equation} %41
\sin^{2}\theta \frac{\partial}{\partial r}\left(r^{2}
\frac{\partial \psi}{\partial r}\right) + \sin\theta \frac{\partial}{\partial
\theta}\left(\sin\theta \frac{\partial \psi}{\partial \theta}\right) +
\frac{\partial^{2} \psi}{\partial \phi^{2}} +
\frac{2mr^{2}\sin^{2}\theta}{\hbar^{2}} \left(\frac{e^{2}}{4\pi 
\epsilon_{0}r} + E\right)\psi = 0~. 
\end{equation}
%%%%%%%%%%%%%%%%%%%%%%%%%%%%% 
Aceast\A\ ecua\c{t}ie este ecua\c{t}ia diferen\c{t}ial\A\
cu derivate par\c{t}iale pentru func\c{t}ia de und\A\ $\psi(r,\theta,\phi)$ 
a electronului \h n 
atomul de hidrogen. \^{I}mpreun\A\ cu diferitele condi\c{t}ii pe care
$\psi(r,\theta,\phi)$
trebuie s\A\ le \h ndeplineasc\A\ 
[de exemplu, $\psi(r,\theta,\phi)$ trebuie s\A\ aib\A\ o valoare
unic\A\ pentru fiecare punct spa\c{t}ial ($r,\theta,\phi$)],
aceast\A\ ecua\c{t}ie specific\A\ de manier\A\ complet\A\ comportamentul
electronului hidrogenic.
Pentru a vedea care este acest comportament, vom rezolva ec. 20 pentru 
$\psi(r,\theta,\phi)$ \c{s}i vom interpreta rezultatele ob\c{t}inute.

%%%%%%%%%%%%%%%%%%%%%%%%%%%%%%%%%%%%%%%%%%%%%%%%%%%%%%%%%%%%%%%%%%%%
\section*{Separarea de variabile \h n coordonate sferice}
%Ecua\c{t}ia Schr\"odinger}
Ceea ce este cu adev\A rat util \h n scrierea ecua\c{t}iei Schr\"odinger 
\h n coordonate sferice pentru atomul de hidrogen 
const\A\ \h n faptul c\A\ astfel se poate realiza u\c{s}or separarea  \h n trei 
ecua\c{t}ii independente, fiecare unidimensional\A\ .  
Procedeul de separare este de a c\A uta solu\c{t}iile pentru care func\c{t}ia de 
und\A\ $\psi(r, \theta, \phi)$ are forma unui produs de trei 
func\c{t}ii, fiecare \h ntr-una din cele trei variabile sferice: $R(r)$, 
care depinde numai de $r$; 
$\Theta(\theta)$ care depinde numai de $\theta$; \c{s}i $\Phi(\phi)$ care 
depinde numai de $\phi$ \c{s}i este practic analog separ\A rii ecua\c{t}iei
Laplace. Deci
\begin{equation} %42
\psi(r, \theta, \phi) = R(r)\Theta(\theta)\Phi(\phi)~.
\end{equation}
Func\c{t}ia $R(r)$ descrie varia\c{t}ia func\c{t}iei de und\A\ 
$\psi$ a electronului de-a lungul razei vectoare dinspre nucleu, 
cu $\theta$ \c{s}i $\phi$ constante. Varia\c{t}ia lui $\psi$ cu  
unghiul zenital $\theta$ de-a lungul unui meridian al unei sfere 
centrat\A\ \h n nucleu este descris\A\ numai de c\A tre func\c{t}ia 
$\Theta(\theta)$  pentru $r$ \c{s}i $\phi$ constante. \^{I}n sf\h r\c{s}it, 
func\c{t}ia 
$\Phi(\phi)$ descrie cum variaz\A\ $\psi$ cu unghiul azimutal 
$\phi$ de-a lungul unei paralele a unei sfere centrat\A\ \h n nucleu, 
\h n condi\c{t}iile \h n care $r$ \c{s}i $\theta$ sunt men\c{t}inute constante.

Folosind $\psi=R\Theta\Phi$, vedem c\A\
\begin{equation} %43
\frac{\partial \psi}{\partial r} = \Theta \Phi \frac{d 
R}{d r}~,  
\end{equation}
\begin{equation} %44
\frac{\partial \psi}{\partial \theta} = R\Phi \frac{d
\Theta}{d \theta}~, 
\end{equation} 
\begin{equation} %45 
\frac{\partial \psi}{\partial \phi} = R\Theta \frac{d
\Phi}{d\phi }~.  
\end{equation}
Evident, acela\c{s}i tip de formule se men\c{t}ine pentru derivatele de ordin 
superior nemixte.
Subtituindu-le \h n ec. 20, dup\A\ \h mp\A r\c{t}irea  
cu $R\Theta \Phi$ se ob\c{t}ine
\begin{equation} %46
\frac{\sin^{2}\theta}{R} \frac{d}{d r}\left(r^{2} \frac{d
R}{d r}\right)+\frac{\sin\theta}{\Theta} \frac{d}{d 
\theta}\left(\sin\theta
\frac{d \Theta}{d \theta}\right)+\frac{1}{\Phi}  
\frac{d^{2} \Phi}{d \phi^{2}} +
\frac{2mr^{2}\sin^{2}\theta}{\hbar^{2}} \left(\frac{e^{2}}{4\pi 
\epsilon_{0}r} + E\right) = 0~.
\end{equation}
Al treilea termen al acestei ecua\c{t}ii este func\c{t}ie numai de unghiul 
$\phi$, \h n timp ce ceilal\c{t}i doi sunt func\c{t}ii de $r$ \c{s}i $\theta$. 
Rescriem ecua\c{t}ia anterioar\A\ \h n forma
\begin{equation} %47
\frac{\sin^{2}\theta}{R} \frac{\partial}{\partial r}\left(r^{2} \frac{\partial
R}{\partial r}\right)+\frac{\sin\theta}{\Theta} \frac{\partial}{\partial
\theta}\left(\sin\theta
\frac{\partial \Theta}{\partial \theta}\right)+
\frac{2mr^{2}\sin^{2}\theta}{\hbar^{2}} \left(\frac{e^{2}}{4\pi 
\epsilon_{0}r} +
E\right) = -\frac{1}{\Phi}\frac{\partial^{2} \Phi}{\partial \phi^{2}}~.
\end{equation}
Aceast\A\ ecua\c{t}ie poate fi corect\A\ numai dac\A\ cei doi membri sunt 
egali cu aceea\c{s}i constant\A\ , pentru c\A\ sunt func\c{t}ii de variabile 
diferite. Este convenabil s\A\ not\A m aceast\A\ constant\A\ cu $m_{l}^{2}$.  
Ecua\c{t}ia diferen\c{t}ial\A\ pentru func\c{t}ia $\Phi$ este
\begin{equation}  %48
-\frac{1}{\Phi}\frac{\partial^{2} \Phi}{\partial \phi^{2}} = m_{l}^{2}~.
\end{equation}
Dac\A\ se subtituie $m_{l}^{2}$ \h n partea dreapt\A\ a ec. 26 \c{s}i se divide 
ecua\c{t}ia rezultant\A\ cu $\sin^{2}\theta$, dup\A\ o regrupare a termenilor, 
se ob\c{t}ine
%%%%%%%%%%%%%%%%%%%%%%%%%%%%%%%%%
\begin{equation} %49
\frac{1}{R} \frac{d}{d r}\left(r^{2} \frac{d
R}{d r}\right) + 
\frac{2mr^{2}}{\hbar^{2}} \left(\frac{e^{2}}{4\pi \epsilon_{0}r} 
+ E\right) = \frac{m_{l}^{2}}{\sin^{2}\theta} - \frac{1}{\Theta \sin\theta} 
\frac{d}{d\theta}\left(\sin\theta\frac{d 
\Theta}{d \theta}\right)~.
\end{equation}
%%%%%%%%%%%%%%%%%%%%%%%%%%%%%%%%
\^{I}nc\A\ odat\A\ se prezint\A\ o ecua\c{t}ie \h n  care apar variabile
diferite \h n fiecare 
membru, ceea ce oblig\A\ la egalarea ambilor cu aceea\c{s}i constant\A\ . 
Din motive care se vor vedea mai t\h rziu, vom nota aceast\A\ constant\A\  
prin $l(l+1)$. Ecua\c{t}iile pentru func\c{t}iile $\Theta(\theta)$ \c{s}i 
$R(r)$ sunt
%%%%%%%%%%%%%%%%%%%%%%%%%%%%%
\begin{equation} %29
\frac{m_{l}^{2}}{\sin^{2}\theta} - \frac{1}{\Theta 
\sin\theta}\frac{d}{d\theta}\left(sin\theta \frac{d\Theta}{d\theta}\right) 
= l(l +1)
\end{equation}
%%%%%%%%%%%%%%%%%%%%%%%%%%%%
\c{s}i
\begin{equation} %30
\frac{1}{R}\frac{d}{dr}\left(r^{2}\frac{dR}{dr}\right) + 
\frac{2mr^{2}}{\hbar^{2}}\left(\frac{e^{2}}{4\pi \epsilon_{0}r} + 
E\right) = l(l+1)~.
\end{equation}
Ecua\c{t}iile 27, 29 \c{s}i 30 se scriu \h n mod normal \h n forma
\begin{equation} %52
\frac{d^{2}\Phi}{d\phi^{2}} + m_{l}^{2}\Phi = 0~,
\end{equation}
\begin{equation} %53
\frac{1}{\sin\theta}\frac{d}{d\theta}\left(\sin\theta 
\frac{d\Theta}{d\theta}\right) + 
\left[l(l+1)-\frac{m_{l}^{2}}{\sin^{2}\theta}\right]\Theta = 0~,
\end{equation}
\begin{equation} %54
\frac{1}{r^{2}}\frac{d}{dr}\left(r^{2}\frac{dR}{dr}\right) + 
\left[\frac{2m}{\hbar^{2}}\left(\frac{e^{2}}{4\pi \epsilon_{0}r} + 
E\right) - \frac{l(l+1)}{r^{2}}\right]R = 0~.
\end{equation}

Fiecare dintre aceste ecua\c{t}ii este o ecua\c{t}ie diferen\c{t}ial\A\
ordinar\A\ pentru o 
func\c{t}ie de o singur\A\ variabil\A\ . \^{I}n felul acesta s-a reu\c{s}it 
simplificarea ecua\c{t}iei Schr\"odinger pentru atomul de 
hidrogen care, ini\c{t}ial, era o ecua\c{t}ie diferen\c{t}ial\A\ par\c{t}ial\A\
pentru o func\c{t}ie $\psi$ de trei variabile.

%%%%%%%%%%%%%%%%%%%%%%%%%%%%%%%%%%%%%%%%%%%%%%%%%%%%%%%%%%%%%%%%%%
\section*{Interpretarea constantelor de separare: numere cuantice}

\subsection*{Solu\c{t}ia pentru partea azimutal\A\ }
Ec. 31 se rezolv\A\ u\c{s}or pentru a g\A si urm\A toarea solu\c{t}ie 
\begin{equation} %55
\Phi(\phi) = A_{\phi}e^{im_{l}\phi}~,
\end{equation}
unde $A_{\phi}$ este constanta de integrare. Una dintre condi\c{t}iile 
stabilite mai \h nainte pe care trebuie s\A\ le \h ndeplineasc\A\
o func\c{t}ie de und\A\ (\c{s}i prin urmare deasemenea 
$\Phi$, care este o component\A\ a func\c{t}iei complete $\psi$) este 
s\A\ aib\A\ o valoare unic\A\ pentru fiecare punct din spa\c{t}iu f\A r\A\
excep\c{t}ie. De exemplu, se 
observ\A\ c\A\ $\phi$ \c{s}i $\phi + 2\pi$ se identific\A\ \h n acela\c{s}i 
plan meridian. De aceea, trebuie ca $\Phi(\phi)= \Phi(\phi + 
2\pi)$, adic\A\ $Ae^{im_{l}\phi} = Ae^{im_{l}(\phi + 2\pi)}$. Aceasta se poate
\h ndeplini numai c\h nd $m_{l}$ este 0 sau un num\A r \h ntreg pozitiv sau 
negativ $(\pm 1, \pm 2, \pm 3,...)$. Acest num\A r $m_{l}$ se cunoa\c{s}te ca 
num\A rul cuantic magnetic al atomului de hidrogen \c{s}i este rela\c{t}ionat cu  
direc\c{t}ia momentului cinetic $L$ pentru c\A\ s-a putut fi asociat cu 
efectele c\h mpurilor magnetice axiale asupra electronului. Num\A rul cuantic 
magnetic $m_{l}$ este determinat de c\A tre num\A rul cuantic orbital 
$l$, care la r\h ndul s\A u determin\A\ modulul momentului cinetic al
electronului. 

Interpretarea num\A rului cuantic orbital $l$ nu este nici ea f\A r\A\ unele
probleme. S\A\ examin\A m ec. 33, care corespunde p\A r\c{t}ii 
radiale $R(r)$ 
a func\c{t}iei de und\A\ $\psi$. Aceast\A\ ecua\c{t}ie este rela\c{t}ionat\A\ 
numai cu aspectul radial al mi\c{s}c\A rii electronilor, adic\A\ , 
cu apropierea \c{s}i dep\A rtarea de nucleu (pentru elipse); 
totu\c{s}i, este prezent\A\ \c{s}i energia total\A\ a 
electronului $E$. Aceast\A\ energie include energia cinetic\A\ a 
electronului \h n mi\c{s}care orbital\A\ care nu are nimic de-a face cu 
mi\c{s}carea radial\A\ . Aceast\A\ contradic\c{t}ie se poate elimina cu
urm\A torul ra\c{t}ionament: energia cinetic\A\ $T$ a electronului are 
dou\A\ p\A r\c{t}i: $T_{radial}$ datorat\A\ mi\c{s}c\A rii
de apropiere \c{s}i dep\A rtare de nucleu, \c{s}i $T_{orbital}$ datorat\A\
mi\c{s}c\A rii \h n jurul nucleului. 
Energia poten\c{t}ial\A\ $V$ a electronului este  
energia electrostatic\A\ . Prin urmare, energia sa total\A\ este
%%%%%%%%%%%%%%%%%%%%%%%%
\begin{equation}  %35
E = T_{radial} + T_{orbital} - \frac{e^{2}}{4\pi \epsilon_{0}r}~.
\end{equation}
Substituind aceast\A\ expresie a lui $E$ \h n ec. 33 ob\c{t}inem, dup\A\ o  
regrupare a termenilor,
\begin{equation}  %57
\frac{1}{r^{2}}\frac{d}{dr}\left(r^{2}\frac{dR}{dr}\right) + 
\frac{2m}{\hbar^{2}}\left[T_{radial} + T_{orbital} - 
\frac{\hbar^{2}l(l+1)}{2mr^{2}}\right]R=0~.
\end{equation}
Dac\A\ ultimii doi termeni din paranteze se 
anuleaz\A\ \h ntre ei, ob\c{t}inem o ecua\c{t}ie 
diferen\c{t}ial\A\ pentru mi\c{s}carea pur radial\A\ . 
Impunem deci condi\c{t}ia
\begin{equation}  %58
T_{orbital} = \frac{\hbar^{2}l(l+1)}{2mr^{2}}~.
\end{equation}
Energia cinetic\A\ orbital\A\ a electronului este \h ns\A\
\begin{equation}  %38
T_{orbital} = \frac{1}{2}mv_{orbital}^{2}
\end{equation}
\c{s}i cum momentul cinetic $L$ al electronului este
\begin{equation}  %60
L = mv_{orbital}r~,
\end{equation}
putem exprima energia cinetic\A\ orbital\A\ \h n forma
\begin{equation}  %61
T_{orbital} = \frac{L^{2}}{2mr^{2}}
\end{equation}
De aceea avem
\begin{equation}  %62
\frac{L^{2}}{2mr^{2}} = \frac{\hbar^{2}l(l+1)}{2mr^{2}}
\end{equation}
\c{s}i deci
\begin{equation}  %42
L = \sqrt{l(l+1)}\hbar~.
\end{equation}
Interpretarea  acestui rezultat este c\A\ , \h ntruc\h t num\A rul 
cuantic orbital $l$ este limitat la valorile $l=0,1,2,...,(n-1)$, 
electronul poate avea numai momentele cinetice $L$ care se specific\A\ 
prin intermediul ec. 42. Ca \c{s}i \h n cazul energiei totale $E$,  
momentul cinetic se conserv\A\ \c{s}i este cuantizat, iar unitatea sa 
natural\A\ de m\A sur\A\ \h n mecanica cuantic\A\ este 
$\hbar=h/2\pi=1.054 \times 10^{-34}$ J.s. 

\^{I}n mi\c{s}carea planetar\A\ macroscopic\A\ , num\A rul 
cuantic care descrie momentul unghiular este at\h t de mare c\A\ 
separarea \h n st\A ri discrete ale momentului cinetic nu se poate 
observa experimental. De exemplu, un electron  
al c\A rui num\A r cuantic orbital este 2, are un 
moment cinetic $L=2.6 \times 10^{-34}$ J.s., \h n timp ce momentul 
cinetic al planetei noastre este $2.7 \times 10^{40}$ J.s.!

Se obi\c{s}nuie\c{s}te 
s\A\ se noteze st\A rile de moment cinetic cu litera $s$ 
pentru $l=0$, cu $p$ pentru $l=1$, $d$ pentru $l=2$, etc.
Acest cod alfabetic provine din clasificarea empiric\A\ a 
spectrelor \h n a\c{s}a numitele serii  
principal\A\ , difuz\A\ \c{s}i fundamental\A\ , care este anterioar\A\ mecanicii
cuantice.
%As\ii\, un estado 
%$s$ es el que no tiene momento angular, un estado $p$ tiene el momento 
%angular $\sqrt{2}\hbar$, etc.

Combinarea num\A rului cuantic total cu litera corespunz\A toare momentului
cinetic este o alt\A\ nota\c{t}ie frecvent folosit\A\ pentru st\A rile atomice. 
De exemplu, o stare \h n care $n=2$, $l=0$ este o stare  
$2s$, iar una \h n care $n=4$, $l=2$ este o stare $4d$.

Pe de alt\A\ parte, pentru interpretarea num\A rului cuantic magnetic, 
vom \c{t}ine cont c\A\ la fel ca pentru impulsul lineal, momentul cinetic este
un vector \c{s}i deci pentru al descrie se necesit\A\ specificarea  
direc\c{t}iei, sensului \c{s}i modulului s\A u. 
Vectorul $L$ este perpendicular 
planului \h n care are loc mi\c{s}carea de rota\c{t}ie \c{s}i direc\c{t}ia 
\c{s}i sensul s\A u sunt date de regula m\h inii drepte 
(de produs vectorial): degetul mare are 
direc\c{t}ia \c{s}i sensul lui $L$ 
c\h nd celelalte patru degete sunt \h n direc\c{t}ia de rota\c{t}ie.

Dar ce semnifica\c{t}ie se poate da unei direc\c{t}ii \c{s}i sens \h n  
spa\c{t}iul limitat al unui atom de hidrogen ? R\A spunsul este simplu dac\A\
ne g\h ndim c\A\ un electron care gireaz\A\ \h n jurul unui nucleu reprezint\A\
un circuit minuscul, care ca dipol magnetic prezint\A\ un c\h mp 
magnetic corespunz\A tor. \^{I}n consecin\c{t}\A\ , un electron atomic
cu moment 
cinetic interac\c{t}ioneaz\A\ cu un c\h mp magnetic extern $B$. Num\A rul 
cuantic magnetic $m_{l}$ specific\A\ direc\c{t}ia lui $L$, 
determinat\A\ de componenta lui $L$ \h n direc\c{t}ia c\h mpului. Acest 
fenomen se cunoa\c{s}te \h n mod comun drept cuantizare spa\c{t}ial\A\ .

Dac\A\ alegem direc\c{t}ia c\h mpului magnetic ca ax\A\ 
$z$, componenta lui $L$ \h n aceast\A\ direc\c{t}ie este
\begin{equation} %64
L_{z} = m_{l}\hbar~.
\end{equation}
Valorile posibile ale lui $m_{l}$ pentru o valoare dat\A\ a lui $l$, 
merg de la $+l$ 
p\h n\A\ la $-l$, trec\h nd prin 0, astfel c\A\ orient\A rile
posibile ale 
vectorului moment cinetic $L$ \h ntr-un c\h mp magnetic sunt $2l+1$. C\h nd 
$l=0$, $L_{z}$ poate avea numai valoarea zero; c\h nd $l=1$, $L_{z}$ 
poate fi $\hbar$, 0, sau $-\hbar$; c\h nd $l=2$, $L_{z}$ ia numai una dintre 
valorile 
$2\hbar$, $\hbar$, 0, $-\hbar$, sau $-2\hbar$, \c{s}i a\c{s}a mai departe. 
Men\c{t}ion\A m c\A\ $L$ nu poate fi exact alineat (paralel sau 
antiparalel) cu $B$, pentru c\A\ $L_{z}$ este \h ntotdeauna mai mic dec\h t 
modulul $\sqrt{l(l+1)}\hbar$ momentului unghiular total.

Cuantizarea spa\c{t}ial\A\ a momentului cinetic orbital al atomului de 
hidrogen se arat\A\ \h n fig. 6.1.

%%%%%%%%%%%%%%
\vskip 2ex
\centerline{
\epsfxsize=120pt
\epsfbox{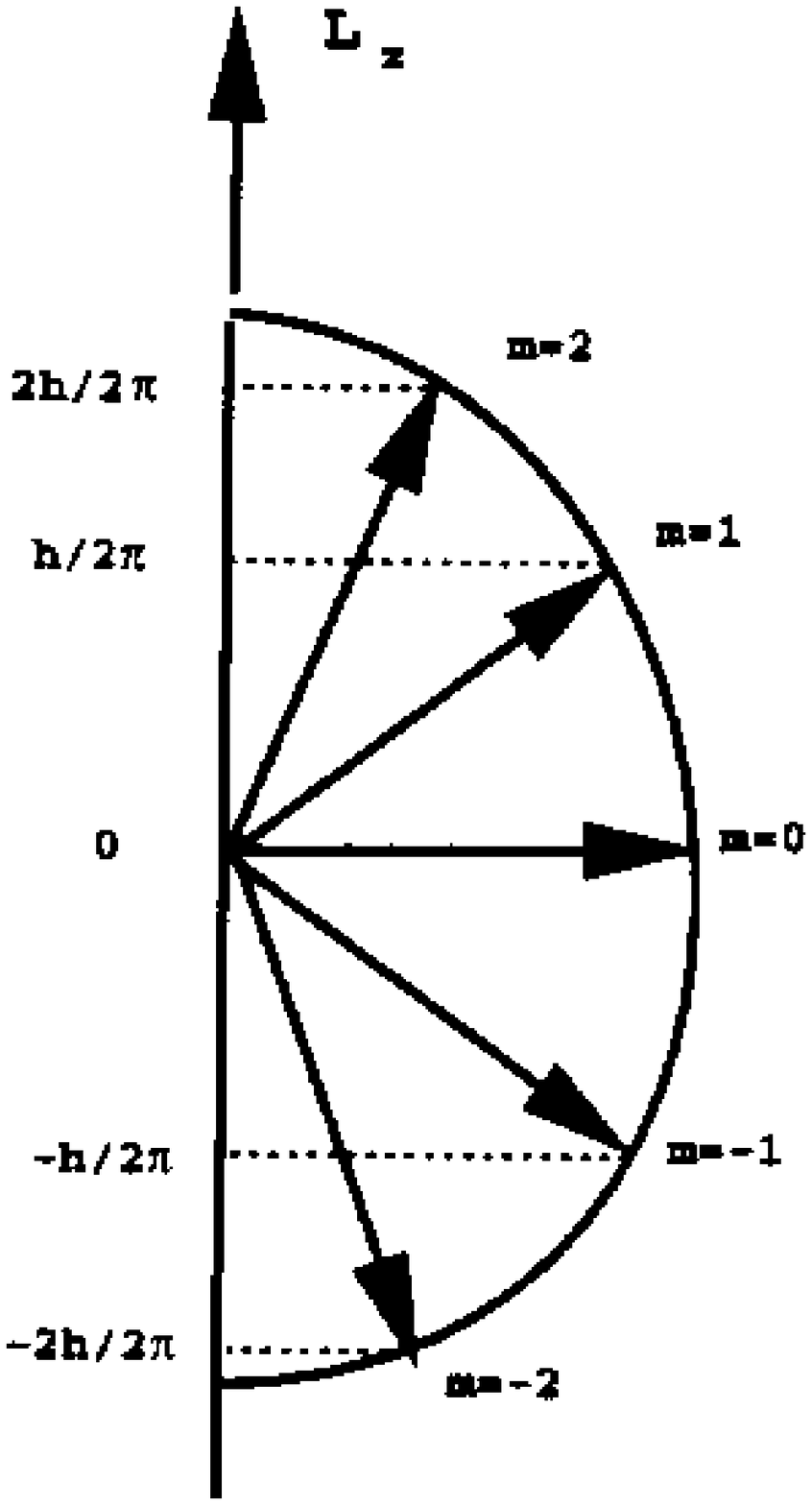}}
\vskip 4ex
\begin{center}
{\small Fig. 6.1:
Cuantizarea spa\c{t}ial\A\ a momentului cinetic pentru st\A ri $l=2$,
$L=\sqrt{6}\hbar$.}
\end{center}
%%%%%%%%%%%%%%%%

Trebuie s\A\ consider\A m electronul 
caracterizat de c\A tre un anumit  
$m_{l}$ ca av\h nd o orientare determinat\A\ a 
momentului s\A u cinetic $L$ fa\c{t}\A\ de un c\h mp magnetic 
extern \h n cazul \h n care acesta se aplic\A\ .

\^{I}n absen\c{t}a c\h mpului magnetic extern, direc\c{t}ia axei $z$ 
este complet arbitrar\A\ . De aceea, componenta 
lui $L$ \h n orice direc\c{t}ie pe care o alegem este $m_{l}\hbar$; 
c\h mpul magnetic extern ofer\A\ o 
direc\c{t}ie de referin\c{t}\A\ privilegiat\A\ din punct de vedere experimental. 
%Un c\h mp  
%magnetic nu este unica direc\c{t}ie de referin\c{t}\A posibil\A\ . De 
%exemplu, la l\ii nea entre los dos \'atomos $H$ en la mol\'ecula de 
%hidr\'ogeno $H_{2}$ tiene tanto significado experimental como la 
%direcci\'on de un campo magn\'etico y, a lo largo de esta l\ii nea, las 
%componentes de los momentos angulares de los \'atomos de $H$ est\'an 
%determinados por sus valores $m_{l}$.

Dece este cuantizat\A\ numai componenta lui $L$ ?  
R\A spunsul se rela\c{t}ioneaz\A\ cu faptul c\A\ $L$ nu poate fi direc\c{t}ionat 
de manier\A\ arbitrar\A\ ; \h ntotdeauna descrie un con 
centrat pe axa de cuantizare \h n a\c{s}a fel \h nc\h t proiec\c{t}ia 
sa $L_{z}$ este 
$m_{l}\hbar$. Motivul pentru care se produce acest fenomen este principiul de 
incertitudine: dac\A\ $L$ ar fi fix \h n spa\c{t}iu, \h n a\c{s}a fel \h nc\h t
$L_{x}$, $L_{y}$ \c{s}i $L_{z}$ ar avea valori bine definite, electronul 
ar fi confinat \h ntr-un plan bine definit. De exemplu, dac\A\ $L$ ar fi fixat
de-a lungul direc\c{t}iei $z$, electronul ar avea tendin\c{t}a de a se 
men\c{t}ine \h n planul $xy$  (fig. 6.2a).

%%%%%%%%%%%%%%
\vskip 2ex
\centerline{
\epsfxsize=180pt
\epsfbox{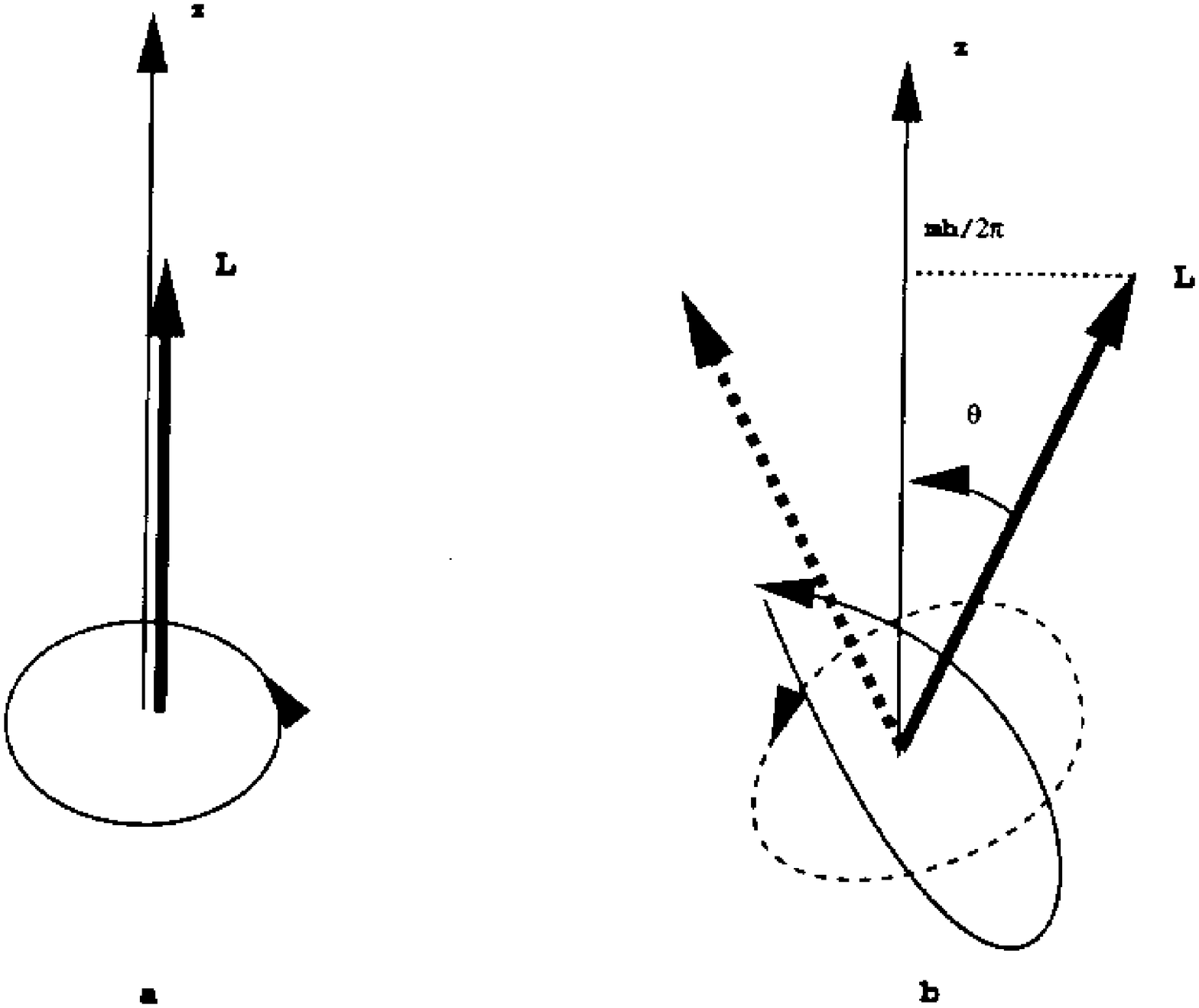}}
\vskip 4ex
\begin{center}
{\small{Fig. 6.2: Principiul de incertitudine interzice o direc\c{t}ie fix\A\
\h n spa\c{t}iu a momentului cinetic.}\\
}
\end{center}
%%%%%%%%%%%%%%%%

Acest lucru poate s\A\ aib\A\ loc numai \h n situa\c{t}ia \h n care 
componenta $p_{z}$ a
impulsului electronului \h n direc\c{t}ia $z$ este infinit de 
incert\A\ , ceea ce desigur este imposibil dac\A\ face parte din atomul de 
hidrogen. Totu\c{s}i, cum \h n realitate numai o component\A\ 
$L_{z}$ a lui $L$ \h mpreun\A\ cu $L^2$ au valori definite \c{s}i $\mid 
L \mid > \mid L_{z} \mid$, electronul nu este limitat la un plan unic 
(fig. 6.2b), iar dac\A\ ar fi a\c{s}a, ar exista o incertitudine
\h n coordonata $z$ a electronului. Direc\c{t}ia lui $L$ se schimb\A\ \h n
mod constant (fig. 6.3), astfel c\A\ valorile medii ale lui $L_{x}$ \c{s}i 
$L_{y}$ sunt 0, de\c{s}i $L_{z}$ are \h ntotdeauna valoarea $m_{l}\hbar$.

%%%%%%%%%%%%%%
\vskip 2ex
\centerline{
\epsfxsize=180pt
\epsfbox{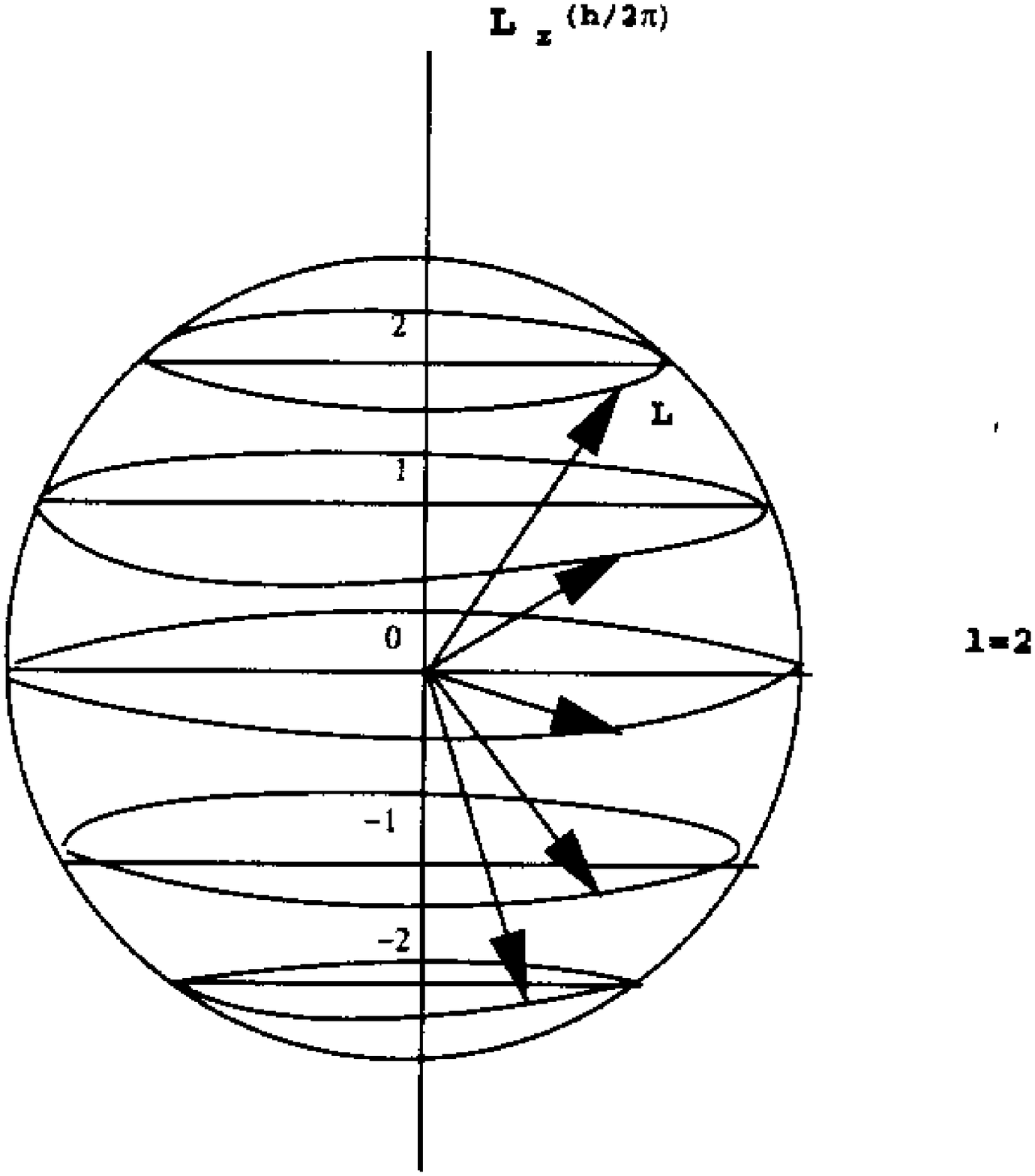}}
\vskip 4ex
\begin{center}
{\small{Fig. 6.3: Vectorul moment cinetic prezint\A\ o precesie constant\A\ 
\h n jurul axei $z$.}\\
}
\end{center}
%%%%%%%%%%%%%%%%

Solu\c{t}ia pentru $\Phi$ trebuie s\A\ satisfac\A\ deasemenea  
condi\c{t}ia de 
normalizare, care este dat\A\ de c\A tre ec. 2. Deci pentru $\Phi$ avem
\begin{equation} %65
\int_{0}^{2\pi} \mid \Phi \mid^{2}d\phi = 1
\end{equation}
\c{s}i substituind $\Phi$ se ob\c{t}ine
\begin{equation} %66
\int_{0}^{2\pi} A_{\phi}^{2}d\phi = 1~.
\end{equation}
Astfel $A_{\phi}=1/\sqrt{2\pi}$ \c{s}i deci $\Phi$  
normalizat\A\ este dat\A\ de
\begin{equation} %67
\Phi(\phi) = \frac{1}{\sqrt{2\pi}}e^{im_{l}\phi}~.
\end{equation}

%%%%%%%%%%%%%%%%%%%%%%%%%%%%%%%%%%%%%%%%%%%%%%%%%%%%%%%%%%%%%%%%%%%
\subsection*{Solu\c{t}ia pentru partea polar\A\ }
Ecua\c{t}ia diferen\c{t}ial\A\ pentru partea polar\A\ $\Theta(\theta)$ are o
solu\c{t}ie mai complicat\A\ fiind dat\A\ de polinoamele  Legendre asociate
\begin{equation} %68
P_{l}^{m_{l}}(x) = 
(-1)^{m_{l}}(1-x^{2})^{m_{l}/2} 
\frac{d^{m_{l}}}{dx^{m_{l}}}P_{l}(x) = 
(-1)^{m_{l}}\frac{(1-x^{2})^{m_{l}/2}}{2^{l}l!}\frac{d^{m_{l} + 
l}}{dx^{{m_{l} + l}}}(x^{2} - 1)^{l}~.
\end{equation}
Aceste func\c{t}ii satisfac urm\A toarea rela\c{t}ie de ortogonalitate
\begin{equation} %69
\int_{-1}^{1} [P_{l}^{m_{l}}(cos\theta)]^{2}dcos\theta = 
\frac{2}{2l+1}\frac{(l+m_{l})!}{(l-m_{l})!}~.
\end{equation}
\^{I}n cazul mecanicii cuantice, $\Theta(\theta)$ este dat\A\ 
de polinoamele Legendre normalizate, respectiv, dac\A\
\begin{equation} %70
\Theta(\theta) = A_{\theta}P_{l}^{m_{l}}(cos\theta)~,
\end{equation}
atunci condi\c{t}ia de normalizare este dat\A\ de
\begin{equation} %71
\int_{-1}^{1} A_{\theta}^{2}[P_{l}^{m_{l}}(cos\theta)]^{2}dcos\theta = 1~.
\end{equation}
Prin urmare constanta de normalizare pentru partea polar\A\ este
\begin{equation} %72
A_{\theta} = \sqrt{\frac{2l+1}{2} \frac{(l-m_{l})!}{(l+m_{l})!}}
\end{equation}
\c{s}i prin urmare, func\c{t}ia $\Theta(\theta)$ deja normalizat\A\ este
\begin{equation} %73
\Theta(\theta) = 
\sqrt{\frac{2l+1}{2}\frac{(l-m_{l})!}{(l+m_{l})!}} P_{l}^{m_{l}}(cos\theta)
\end{equation}

Pentru obiectivele noastre, cea mai important\A\ proprietate a acestor 
func\c{t}ii este c\A\ , a\c{s}a cum s-a men\c{t}ionat deja, 
exist\A\ numai c\h nd constanta $l$ 
este un num\A r \h ntreg egal sau mai mare dec\h t 
$\mid m_{l}\mid$, care este valoarea absolut\A\ a lui $m_{l}$. Aceast\A\ 
condi\c{t}ie se poate scrie sub forma setului de valori disponibile 
pentru $m_{l}$  
\begin{equation} %74
m_{l} = 0,\pm 1, \pm 2,...,\pm l~. 
\end{equation}

%%%%%%%%%%%%%%%%%%%%%%%%%%%%%%%%%%%%%%%%%
\subsection*{Unificarea p\A r\c{t}ilor azimutal\A\ \c{s}i polar\A\ : armonicele sferice}
Solu\c{t}iile pentru p\A r\c{t}ile azimutal\A\ \c{s}i polar\A\ se pot
uni pentru a 
forma armonicele sferice, care depind de $\phi$ \c{s}i 
$\theta$ \c{s}i contribuie la simplificarea manipul\A rilor algebrice 
ale func\c{t}iei de und\A\ complet\A\ $\psi(r,\theta,\phi)$. 
Armonicele sferice se introduc \h n felul urm\A tor:
\begin{equation} %75
Y_{l}^{m_{l}}(\theta,\phi) = (-1)^{m_{l}} \sqrt{\frac{2l+1}{4\pi} 
\frac{(l-m_{l})!}{(l+m_{l})!}} P_{l}^{m_{l}}(cos\theta)e^{im_{l}\phi}~.
\end{equation} 
Factorul suplimentar $(-1)^{m_{l}}$  nu produce nici o problem\A\
pentru c\A\ 
ec. Schr\"odinger este linear\A\ \c{s}i homogen\A\ \c{s}i este convenabil
pentru studiul del momentului cinetic. Se cunoa\c{s}te ca factorul de faz\A\ 
Condon-Shortley, efectul s\A u fiind de a introduce o 
alternan\c{t}\A\ a semnelor $\pm$.

%%%%%%%%%%%%%%%%%%%%%%%%%%%%%%%%%%%%%%%%%%%%%%%%%%%%%%%%%%%%%%%%%%%%
\subsection*{Solu\c{t}ia pentru partea radial\A\ }
Solu\c{t}ia pentru partea radial\A\ $R(r)$ 
a func\c{t}iei de und\A\ $\psi$ a atomului de hidrogen este ceva mai
complicat\A\ \c{s}i aici este unde apar diferen\c{t}e mai mari fa\c{t}\A\ de
ecua\c{t}ia Laplace \h n electrostatic\A\ . Rezultatul final se exprim\A\ \h n 
func\c{t}ie de polinoamele asociate Laguerre (Schr\"odinger 1926). 
Ecua\c{t}ia radial\A\ 
se poate rezolva \h n form\A\ analitic\A\ exact\A\ 
numai c\h nd E este pozitiv sau pentru una din 
urm\A toarele valori negative $E_{n}$ (\h n care caz electronul este legat
atomului)
\begin{equation} %76
E_{n} = 
-\frac{m 
e^{4}}{32\pi^{2}\epsilon_{0}^{2}\hbar^{2}}\left(\frac{1}{n^{2}}\right)~, 
\end{equation}
unde $n$ este un num\A r \h ntreg numit num\A rul cuantic 
principal \c{s}i descrie cuantizarea energiei electronului \h n 
atomul de hidrogen. Aceast spectru discret a fost ob\c{t}inut pentru prima 
dat\A\ de c\A tre Bohr cu metode empirice de cuantizare \h n 1913 \c{s}i apoi 
de c\A tre Pauli \c{s}i respectiv Schr\"odinger \h n 1926.

O alt\A\ condi\c{t}ie care trebuie s\A\ fie satisf\A cut\A\
pentru a rezolva ecua\c{t}ia radial\A\ este ca $n$ s\A\ fie \h ntotdeauna 
mai mare dec\h t $l$. Valoarea sa minim\A\ este $l+1$. 
Invers, condi\c{t}ia asupra lui $l$ este
\begin{equation} %77
l = 0,1,2,...,(n-1) 
\end{equation}

Ecua\c{t}ia radial\A\ se poate scrie \h n forma 
%%%%%%%%%%%%%%%%%%%%%%%%%%%
\begin{equation} %78
r^{2}\frac{d^{2}R}{dr^{2}} + 2r\frac{dR}{dr} + \left[\frac{2m 
E}{\hbar^{2}}r^{2} + \frac{2me^{2}}{4\pi \epsilon_{0} \hbar^{2}}r - 
l(l+1)\right]R = 0~,
\end{equation}
%%%%%%%%%%%%%%%%%%%%%%%%%%%%%
Dup\A\ \h mp\A r\c{t}irea cu $r^2$, se folose\c{s}te substitu\c{t}ia 
$\chi (r) =rR$ pentru a elimina termenul \h n $\frac{dR}{dr}$ \c{s}i a 
ob\c{t}ine forma standard a ec. Schr\"odinger radiale cu poten\c{t}ial efectiv
$U(r)=-{\rm const}/r + l(l+1)/r^2$
(poten\c{t}ial electrostatic plus barier\A\ centrifugal\A\ ). Aceast\A\ 
procedur\A\ se aplic\A\ numai pentru a discuta o nou\A\ condi\c{t}ie 
obligatorie de frontier\A\ , ob\c{ti}nerea spectrului fiind prin intermediul
ecua\c{t}iei pentru $R$.
Diferen\c{t}a \h ntre o ec. Schr\"odinger radial\A\ \c{s}i
una \h n toat\A\ linia real\A\ este c\A\ o condi\c{t}ie de frontier\A\ 
suplimentar\A\ trebuie impus\A\ \h n origine ($r=0$). Poten\c{t}ialul 
coulombian apar\c{t}ine unei clase de poten\c{t}iale care se numesc slab
singulare, pentru care ${\rm lim} _{r\rightarrow 0}=U(r)r^2=0$. Se \h ncearc\A\
solu\c{t}ii de tipul $\chi \propto r^{\nu}$, ceea ce implic\A\ 
$\nu (\nu -1)=l(l+1)$ cu solu\c{t}iile $\nu _1 =l+1$ \c{s}i $\nu _2=-l$, exact
ca \h n cazul electrostaticii. Solu\c{t}ia negativ\A\ se elimin\A\ \h n cazul
$l\neq 0$ pentru c\A\ duce la divergen\c{t}a integralei de normalizare \c{s}i
deasemenea nu respect\A\ normalizarea la func\c{t}ia delta \h n cazul spectrului
continuu, iar 
cazul $\nu _2 =0$ se elimin\A\ din condi\c{t}ia de finitudine a energiei 
cinetice medii.Concluzia final\A\ este c\A\ $\chi (0)=0$ pentru orice $l$.

%%%%%%%%%%%%%%%%%%%%%%%%
%despre care se \c{s}tie c\A\ are solu\c{t}ii polinomiale de tip 
%Laguerre asociate, care \h ndeplinesc urm\A toarea  
%condi\c{t}ie de normalizare
%\begin{equation} %79
%\int_{0}^{\infty}e^{-\rho}\rho^{2l}[L_{n+l}^{2l+1}(\rho)]^{2}\rho^{2}d\rho = 
%\frac{2n[(n+l)!]^{3}}{(n-l-1)!}~.
%\end{equation}
%%%%%%%%%%%%%%%%%%%%%%%%%%%%%

Revenind la analiza ecua\c{t}iei pentru func\c{t}ia radial\A\ $R$, se pune mai
\h nt\h i problema adimensionaliz\A rii ecua\c{t}iei.
Aceasta se face observ\h nd c\A\ se poate
forma o singur\A\ scal\A\ de spa\c{t}iu \c{s}i timp din combina\c{t}ii ale
celor trei constante fizice care intr\A\ \h n aceast\A\ problem\A\, respectiv
$e^2$, $m$ \c{s}i $\hbar$. Acestea sunt raza Bohr $a_{0}=\hbar ^2/me^2=0.529\cdot 
10 ^{-8}$ cm. \c{s}i $t_{0}=\hbar ^3/me^4=0.242 10^{-16}$ sec., care se numesc 
unit\A \c{t}i atomice. Folosind aceste unit\A \c{t}i ob\c{t}inem
%%%%%%%%%%%%%%%%%%%%%%%%%%%%
\begin{equation} %78b
\frac{d^{2}R}{dr^{2}} + \frac{2}{r}\frac{dR}{dr} + \left[2 
E + \frac{2}{r} - 
\frac{l(l+1)}{r^2}\right]R = 0~,
\end{equation}
%%%%%%%%%%%%%%%%%%%%%%%%%%%%%%%
unde ne intereseaz\A\ spectrul discret ($E<0$). Cu nota\c{t}iile 
$n=1/\sqrt{-E}$ \c{s}i $\rho=2r/n$ se ajunge la:
%%%%%%%%%%%%%%%%%%%%%%%%%%%
\begin{equation} %78c
\frac{d^{2}R}{d\rho ^{2}} + \frac{2}{\rho}\frac{dR}{d\rho} + 
\left[\frac{n}{\rho}-\frac{1}{4} - 
\frac{l(l+1)}{\rho ^2}\right]R = 0~.
\end{equation}
%%%%%%%%%%%%%%%%%%%%%%%%%
Pentru $\rho \rightarrow \infty$, ecua\c{t}ia se reduce la 
$\frac{d^{2}R}{d\rho ^{2}}=\frac{R}{4}$ cu solu\c{t}ii $R\propto e^{\pm\rho /2}$.
Se accept\A\ pe baza condi\c{t}iei de normalizare numai exponen\c{t}iala 
atenuat\A\ . Pe de alt\A\ parte asimptotica de zero, a\c{s}a cum am 
comentat deja, este $R\propto \rho ^{l}$. Prin urmare, putem substitui $R$
printr-un produs de trei func\c{t}ii radiale  $R=\rho ^{l}e^{-\rho /2}F(\rho)$, 
dintre care primele dou\A\ sunt p\A r\c{t}ile asimptotice, iar a treia este 
func\c{t}ia radial\A\ \h n regiunea intermediar\A\ , 
care ne intereseaz\A\ cel mai mult pentru c\A\
ne d\A\ spectrul energetic. Ecua\c{t}ia pentru $F$ este
%%%%%%%%%%%%%%%%%%%%%%%%%%%
\begin{equation} %78d
\rho\frac{d^{2}F}{d\rho ^{2}} + (2l+2-\rho)\frac{dF}{d\rho} + 
(n-l-1)F = 0~.
\end{equation}
%%%%%%%%%%%%%%%%%%%%%%%%%
care este un caz particular de ecua\c{t}ie hipergeometric\A\ confluent\A\ 
\h n care cei doi parametri `hiper'geometrici depind de $n,l$ \c{s}i care se
poate identifica cu ecua\c{t}ia pentru
polinoamele Laguerre asociate $L_{n+l}^{2l+1}(\rho)$ \h n fizica matematic\A\ .
Astfel, forma normalizat\A\ a lui $R$ este:
\begin{equation} %80
R_{nl}(r) = 
-\frac{2}{n^2}\sqrt{\frac{(n-l-1)!}{2n[(n+l)!]^{3}}}
e^{-\rho /2}\rho^{l} L_{n+l}^{2l+1}(\rho)~,
\end{equation}
%unde $\rho=2r/na_{0}$ \c{s}i $a_{0}=\hbar^{2}/me^{2}$.
unde s-a folosit condi\c{t}ia de normalizare a polinoamelor Laguerre:
%%%%%%%%%%%%%%%%%%%%%%%%%%%%%%%%%%%%%
\begin{equation} %79
\int_{0}^{\infty}e^{-\rho}\rho^{2l}[L_{n+l}^{2l+1}(\rho)]^{2}\rho^{2}d\rho = 
\frac{2n[(n+l)!]^{3}}{(n-l-1)!}~.
\end{equation}
%%%%%%%%%%%%%%%%%%%%%%%%%%%%%

Avem deci solu\c{t}iile fiec\A reia dintre ecua\c{t}iile care depind numai de o
singur\A\  
variabil\A\ \c{s}i prin urmare putem construi func\c{t}ia de 
und\A\ pentru fiecare stare electronic\A\ \h n atomul de hidrogen, respectiv 
dac\A\ $\psi(r,\theta,\phi)=R(r)\Theta(\theta)\Phi(\phi)$, 
atunci func\c{t}ia de und\A\ complet\A\ este
\begin{equation} %81
\psi(r,\theta,\phi)={\cal N}_{H}(\alpha r)^{l} 
e^{-\alpha r/2} L_{n+l}^{2l+1}(\alpha r) 
P_{l}^{m_{l}}(cos\theta)e^{im_{l}\phi}~,
\end{equation}
unde ${\cal N}_{H}=-\frac{2}{n^2}
\sqrt{\frac{2l+1}{4\pi}\frac{(l-m_{l})!}{(l+m_{l})!} 
\frac{(n-l-1)!}{[(n+l)!]^{3}}}$ \c{s}i $\alpha=2/na_{0}$.

Utiliz\h nd armonicele sferice, solu\c{t}ia se scrie \h n felul urm\A tor 
\begin{equation}  %82
\psi(r,\theta,\phi)=-\frac{2}{n^2}\sqrt{\frac{(n-l-1)!}{[(n+l)!]^{3}}}
(\alpha r)^{l}
e^{-\alpha r/2} L_{n+l}^{2l+1}(\alpha r)Y_{l}^{m_{l}}(\theta,\phi)~.
\end{equation}

Aceast\A\ formul\A\ se poate considera rezultatul matematic final pentru 
solu\c{t}ia ec. Schr\"odinger \h n cazul  
atomului de hidrogen pentru oricare stare sta\c{t}ionar\A\
a electronului s\A u. \^{I}ntr-adev\A r, se pot vedea \h n mod explicit at\h t
dependen\c{t}a asimptotic\A\ c\h t \c{s}i cele dou\A\ seturi ortogonale complete,
polinoamele Laguerre asociate \c{s}i respectiv armonicele sferice,
corespunz\A toare acestei ecua\c{t}ii lineare cu 
derivate par\c{t}iale de ordinul doi. Coordonatele parabolice 
[$\xi=r(1-\cos\theta)$, $\eta =r(1+\cos \theta)$, $\phi=\phi$], 
sunt un alt set de variabile \h n care ec.
Schr\"odinger pentru atomul de hidrogen este u\c{s}or de separat 
(E. Schr\"odinger, Ann. Physik {\bf 80}, 437, 1926; 
P.S. Epstein, Phys. Rev. {\bf 28}, 695, 1926; 
I. Waller, Zf. Physik {\bf 38}, 635, 1926). 
Solu\c{t}ia 
final\A\ se exprim\A\ ca produsul unor factori de natur\A\ asimptotic\u{a},
armonice azimutale \c{s}i dou\A\ seturi de
polinoame Laguerre asociate \h n $\xi$, respectiv $\eta$. Spectrul energetic 
($-1/n^2$) \c{s}i degenerarea ($n^2$) evident nu se modific\u{a}.

% Aceast\A\ func\c{t}ie de und\A\ nu are o interpretare 
%fizic\A\ imediat\A\ , dar modulul p\A trat 
%$\mid \psi (r) \mid^{2}$  
%este propor\c{t}ional cu probabilitatea de a g\A si experimental  
%electronul \h ntre $r$ \c{s}i $r+dr$ acel loc \c{s}i la orice moment .
%En la tabla 1 se dan las funciones
%de onda normalizadas del \'atomo de hidr\'ogeno para $n=1,2,3$.

%%%%%%%%%%%%%%%%%%%%%%%%%%%%%%%%%%%%%%%%%%%%%%%%%%%%%%%%%%%%%%%%%%%%%%%%%
\section*{Densitatea de probabilitate electronic\A\ }
\^{I}n modelul lui Bohr al atomului de hidrogen, electronul se rote\c{s}te \h n
jurul nucleului pe traiectorii circulare sau eliptice. Dac\A\ se realizeaz\A\
un 
experiment adecuat, s-ar putea vedea c\A\ electronul ar fi \h ntotdeauna
situat \h n limitele expeimentale la o 
distan\c{t}\A\ fa\c{t}\A\ de nucleu $r=n^{2}a_{0}$ (unde $n$ este num\A rul 
cuantic care numeroteaz\A\ orbita \c{s}i $a_{0}=0.53$ $\AA$ este raza
orbitei celei mai apropiate de nucleu, cunoscut\A\ ca raza Bohr) 
\c{s}i \h n planul ecuatorial $\theta=90^{o}$, 
\h n timp ce unghiul azimutal $\phi$ poate varia \h n timp.

Teoria cuantic\A\ a atomului de hidrogen modific\A\  
concluziile modelului lui Bohr \h n dou\A\ aspecte importante. 
\^{I}n primul r\h nd, nu se 
pot da valori exacte pentru $r,\theta,\phi$, ci numai 
probabilit\A \c{t}i relative de a g\A si electronul \h ntr-o zon\A\ 
infinitezimal\A\ dat\A\ a spa\c{t}iului. 
Aceast\A\ imprecizie este, desigur, o consecin\c{t}\A\ a naturii 
ondulatorii a electronului. \^{I}n al doilea r\h nd, nu se poate 
g\h ndi c\A\ electronul se mi\c{s}c\A\ \h n jurul nucleului \h n sensul 
conven\c{t}ional clasic, pentru c\A\ densitatea de probabilitate 
$\mid \psi \mid^{2}$ nu 
depinde de timp \c{s}i poate varia considerabil \h n func\c{t}ie de zona 
infinitezimal\A\ unde se calculeaz\A\ .   

Func\c{t}ia de und\A\ $\psi$ a electronului \h n atomul de hidrogen 
este $\psi=R\Theta\Phi$ unde $R=R_{nl}(r)$ descrie cum 
se schimb\A\ $\psi$ cu $r$ c\h nd numerele cuantice orbital \c{s}i total 
au valorile $n$ \c{s}i $l$; $\Theta=\Theta_{lm_{l}}(\theta)$ descrie la r\h ndul
lui varia\c{t}ia lui $\psi$ cu $\theta$ c\h nd numerele 
cuantice magnetic \c{s}i orbital au valorile $l$ \c{s}i $m_{l}$; 
\h n sf\h r\c{s}it, 
$\Phi=\Phi_{m_{l}}(\phi)$ d\A\ schimbarea lui $\psi$ cu 
$\phi$ c\h nd num\A rul cuantic magnetic are valoarea $m_{l}$.  
Densitatea de probabilitate $\mid \psi \mid^{2}$ se poate scrie 
\begin{equation}  %83
\mid \psi \mid^{2} = \mid R \mid^{2} \mid \Theta \mid^{2} \mid \Phi \mid^{2}~.
\end{equation}
%unde se \h n\c{t}elege c\A\ dac\A\ func\c{t}ia este complex\A\ , 
%trebuie s\A\ se \c{t}in\A\ cont c\A 
%cuenta que su cuadrado se debe sustituir por el producto de ella y su 
%conjugada compleja.
Densitatea de probabilitate $\mid \Phi \mid^{2}$, care m\A soar\A\
posibilitatea de a g\A si 
electronul la un unghi azimutal $\phi$ dat, este o 
constant\A\ care nu depinde de $\phi$. Prin urmare, densitatea 
de probabilitate electronic\A\ este simetric\A\ fa\c{t}\A\ de axa $z$, 
independent de starea cuantic\A\ ``magnetic\A\ " (at\h ta timp c\h t nu se
aplic\A\ un c\h mp magnetic extern), ceea ce face ca electronul s\A\ aib\A\ 
aceea\c{s}i probabilitate de a se g\A si
\h n orice direc\c{t}ie azimutal\A\ .  
Partea radial\A\ $R$ a func\c{t}iei de und\A\ , spre deosebire de $\Phi$, nu
numai c\A\ variaz\A\ cu $r$, ci \c{s}i o face \h n mod diferit pentru
fiecare combina\c{t}ie de numere cuantice $n$ \c{s}i $l$. Fig. 6.4 arat\A\
grafice ale lui $R$ \h n func\c{t}ie de $r$ pentru st\A rile $1s$, $2s$,
\c{s}i $2p$ ale atomului de hidrogen. $R$ este maxim 
\h n centrul nucleului ($r=0$) pentru toate st\A rile $s$, \h n timp ce
este zero \h n $r=0$ pentru toate st\A rile care au moment cinetic.

%%%%%%%%%%%%%%
\vskip 2ex
\centerline{
\epsfxsize=280pt
\epsfbox{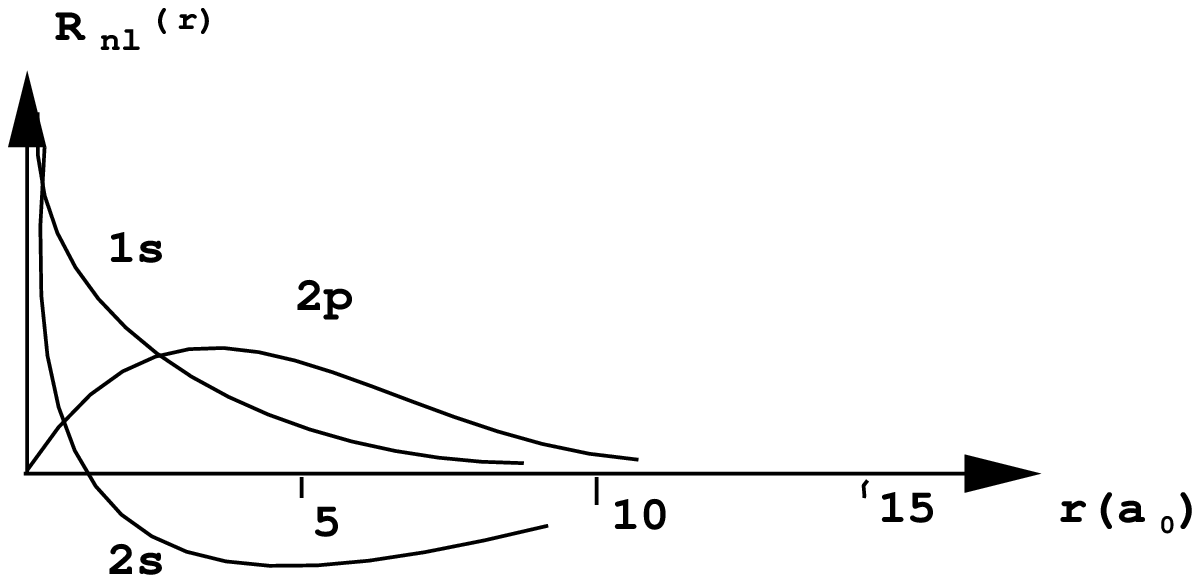}}
\vskip 4ex
\begin{center}
{\small{Fig. 6.4: Grafice aproximative ale func\c{t}iilor 
radiale $R_{1s}$, $R_{2s}$, $R_{2p}$; ($a_0=0.53$ \AA ).}\\
}
\end{center}
%%%%%%%%%%%%%%%%

%%%%%%%%%%%%%%
\vskip 2ex
\centerline{
\epsfxsize=280pt
\epsfbox{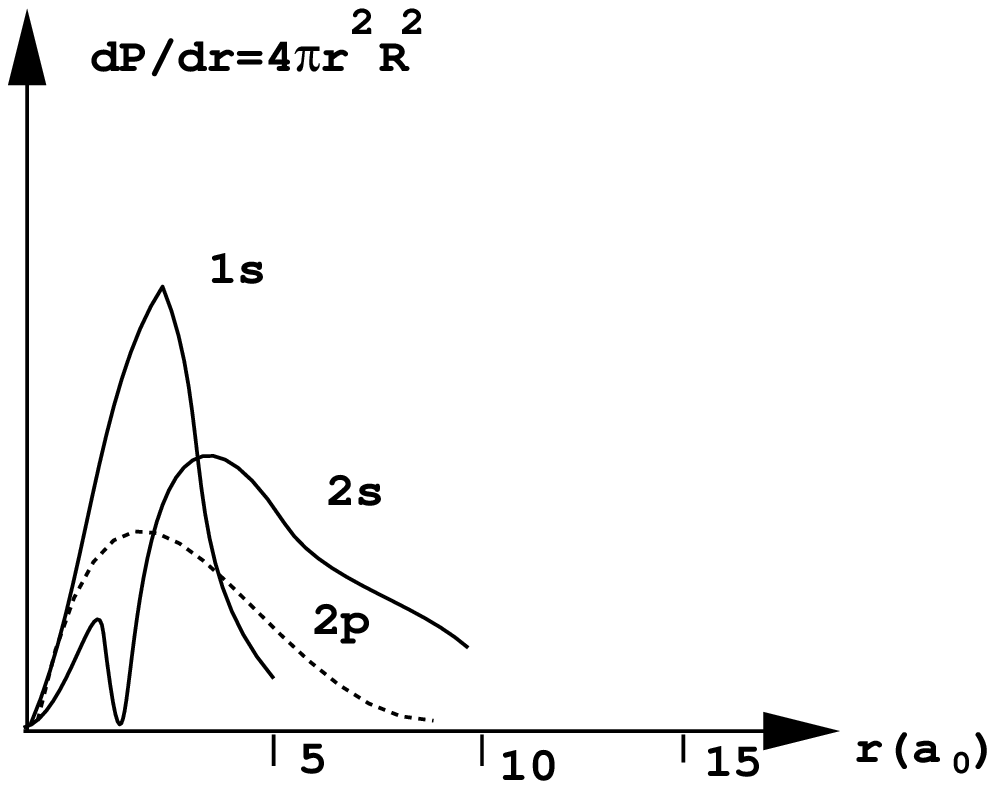}}
\vskip 4ex
\begin{center}
{\small{Fig. 6.5: Densitatea de probabilitate de a g\A si electronul atomului 
de hidrogen \h ntre $r$ \c{s}i $r+dr$ fa\c{t}\A\ de nucleu pentru st\A rile 
$1s$, $2s$, $2p$.}\\
}
\end{center}
%%%%%%%%%%%%%%%%

Densitatea de probabilitate electronic\A\ \h n punctul $r,\theta,\phi$ este 
propor\c{t}ional\A\ cu $\mid \psi \mid^{2}$, dar probabilitatea real\A\ \h n 
elementul de volum infinitezimal $dV$ este 
$\mid \psi \mid^{2}dV$. \^{I}n coordonate polare sferice
\begin{equation}  %84
dV=r^{2}\sin\theta dr d\theta d\phi~,
\end{equation}
\c{s}i cum $\Theta$ \c{s}i $\Phi$ sunt func\c{t}ii normalizate,  
probabilitatea numeric\A\ real\A\ $P(r)dr$ de a g\A si electronul la o 
distan\c{t}\A\ fa\c{t}\A\ de nucleu cuprins\A\ \h ntre $r$ \c{s}i $r+dr$ este
\begin{eqnarray}  %85
P(r)dr & = & r^{2}\mid R \mid^{2}dr \int_{0}^{\pi} 
\mid\ \Theta \mid^{2} \sin\theta d\theta \int_{0}^{2\pi} 
\mid\ \Phi \mid^{2}d\phi \nonumber\\
& = & r^{2}\mid R \mid^{2}dr
\end{eqnarray}
$P(r)$ este reprezentat\A\ \h n fig. 6.5 pentru acelea\c{s}i st\A ri 
ale c\A ror func\c{t}ii radiale $R$ apar \h n fig. 6.4; \h n principiu,
curbele sunt foarte diferite. Observ\A m imediat c\A\ $P(r)$ nu 
este maxim\A\ \h n nucleu pentru st\A rile $s$, a\c{s}a cum este $R$, 
av\h nd 
maximul la o distan\c{t}\A\ finit\A\ de acesta. Valoarea cea mai probabil\A\
a lui $r$ pentru un electron $1s$ este exact $a_{0}$, care este raza Bohr. 
Totu\c{s}i, valoarea medie a lui $r$ pentru un electron $1s$ este $1.5a_{0}$, 
ceea ce pare ciudat la prima vedere, pentru c\A\ nivelele de 
energie sunt acele\c{s}i \h n mecanica cuantic\A\ \c{s}i \h n modelul 
lui Bohr. Aceast\A\ aparent\A\ discrepan\c{t}\A\ se elimin\A\ dac\A\ se \c{t}ine 
cont de faptul c\A\ 
energia electronului depinde de $1/r$ \c{s}i nu direct  
de $r$, iar valoarea medie a lui $1/r$ pentru un electron $1s$ este 
exact $1/a_{0}$.

Func\c{t}ia $\Theta$ variaz\A\ cu unghiul polar $\theta$ pentru toate 
numerele cuantice $l$ \c{s}i $m_{l}$, excep\c{t}ie f\A c\h nd $l=m_{l}=0$, care
sunt st\A ri $s$. Densitatea de probabilitate $\mid\ \Theta \mid^{2}$ pentru 
o stare $s$ este o constant\A\ (1/2), ceea ce \h nseamn\A\ c\A\ , \h ntruc\h t 
$\mid \Phi \mid^{2}$ este deasemenea constant\A\ , densitatea de probabilitate 
electronic\A\ $\mid \psi \mid^{2}$ are aceea\c{s}i valoare pentru o valoare a lui 
$r$ dat\A\ , \h n toate direc\c{t}iile. \^{I}n alte st\A ri, 
electronii au un comportament unghiular care uneori ajunge s\A\ fie foarte
complicat.
Aceasta se poate vedea \h n fig.6.5, unde se arat\A\ densit\A \c{t}ile de 
probabilitate electronic\A\ pentru diferite st\A ri atomice
\h n func\c{t}ie de $r$ \c{s}i $\theta$. (Termenul care se reprezint\A\ este 
$\mid \psi \mid^{2}$ \c{s}i nu $\mid \psi \mid^{2}dV$). Deoarece  
$\mid \psi \mid^{2}$ este independent de $\phi$, o reprezentare 
tridimensional\A\ a lui $\mid \psi \mid^{2}$ se ob\c{t}ine prin rota\c{t}ia
unei reprezent\A ri particulare \h n jurul unei axe verticale, ceea ce poate 
ar\A ta c\A\
densit\A \c{t}ile de probabilitate pentru st\A rile $s$ au  
simetrie sferic\A\ , \h n timp ce toate celelalte nu o posed\A\ . Se ob\c{t}in
\h n acest fel loburi mai mult sau mai pu\c{t}in pronun\c{t}ate, 
care au forme caracteristice pentru fiecare stare \h n parte \c{s}i care  
\h n chimie joac\A\ un rol important \h n determinarea
modului \h n care interac\c{t}ioneaz\A\ atomii \h n interiorul moleculelor.
\\
\\
{\bf 6N. Not\u{a}}:

\noindent
1. E. Schr\"odinger a ob\c{t}inut premiul Nobel \h n 1933 
(\h mpreun\A\ cu Dirac)
pentru ``descoperirea de noi forme productive ale teoriei
atomice". Schr\"odinger a scris o remarcabil\A\ serie de patru
articole intitulat\A\  ``Quantisierung als Eigenwertproblem"  [``Cuantizarea ca
problem\A\ de autovalori"] (I-IV,
primite la redac\c{t}ia revistei Annalen der Physik \h n
27 Ianuarie, 23 Februarie, 10 Mai \c{s}i 21 Iunie 1926).

%\newpage
\section*{{\huge 6P. Probleme}}
  
{\bf Problema 6.1} - S\A\ se ob\c{t}in\A\ formulele pentru orbitele stabile
\c{s}i pentru nivelele de energie ale electronului \h n atomul de hidrogen
folosind numai argumente bazate pe lungimea de und\A\ de Broglie
asociat\A electronului \c{s}i valoarea `empiric\A\ ' $5.3 \cdot 10^{-11}$ m 
pentru raza Bohr. 

{\bf Solu\c{t}ie}: Lungimea de und\A\ a electronului este dat\A\ de
$
\lambda = \frac{h}{mv}%\nonumber
$
\h n timp ce dac\A\ egal\A m for\c{t}a electric\A\ cu  
for\c{t}a centripet\A\ , respectiv
$
\frac{mv^{2}}{r} = \frac{1}{4\pi \epsilon_{0}}\frac{e^{2}}{r^{2}}%\nonumber
$
ob\c{t}inem c\A\ viteza electronului este dat\A\ de
$
v = \frac{e}{\sqrt{4\pi \epsilon_{0} mr}}~.%\nonumber
$%\end{eqnarray}
\^{I}n aceste condi\c{t}ii, lungimea de und\A\ a electronului este
$
\lambda = \frac{h}{e}\sqrt{\frac{4\pi \epsilon_{0}r}{m}}%\nonumber
$.
Acum, dac\A\ folosim valoarea $5.3 \times 10^{-11}$m pentru raza $r$ a 
orbitei electronice, vedem c\A\ lungimea de und\A\ a electronului este 
$\lambda=33 \times 10^{-11}$ m. Aceast\A\ lungime de und\A\ are exact 
aceea\c{s}i valoare ca circumferin\c{t}a orbitei electronului, $2\pi 
r=33 \times 10^{-11}$ m. Dup\A\ cum se poate vedea, orbita electronului \h n 
atomul de hidrogen corespunde astfel unei unde ``\h nchis\A\ \h n ea 
\h ns\A \c{s}i" (adic\A\ de tip sta\c{t}ionar). 
Acest fapt se poate compara cu vibra\c{t}iile unui inel de alam\A\ . 
Dac\A\ lungimile de und\A\ sunt un submultiplu 
al circumferin\c{t}ei sale, inelul ar
putea continua starea sa vibratorie pentru foarte mult timp cu 
disipare redus\A\
(st\A ri `proprii' de vibra\c{t}ie sau unde sta\c{t}ionare). Dac\A\ \h ns\A\ 
num\A rul de lungimi de und\A\ nu este \h ntreg
se va produce o interferen\c{t}\A\ negativ\A\ pe m\A sur\A\ ce undele se
propag\A\ de-a lungul inelului \c{s}i vibra\c{t}iile vor disp\A rea 
foarte repede. 
Astfel, se poate afirma c\A\ un 
electron  se poate roti indefinit \h n jurul nucleului f\A r\A\ a radia 
energia de care dispune at\h ta timp c\h t orbita con\c{t}ine un num\A r 
\h ntreg 
de lungimi de und\A\ de Broglie. Cu acestea, avem condi\c{t}ia de stabilitate 
\begin{eqnarray}
n\lambda = 2\pi r_{n}~,\nonumber
\end{eqnarray}
unde $r_{n}$ este raza orbitei care con\c{t}ine $n$ lungimi  
de und\A\ . Substituind $\lambda$, avem
\begin{eqnarray}
\frac{nh}{e}\sqrt{\frac{4\pi \epsilon_{0}r_{n}}{m}} = 2\pi r_{n}~,\nonumber
\end{eqnarray}
\c{s}i deci orbitele stabile ale electronului sunt
\begin{eqnarray}
r_{n} = \frac{n^{2}\hbar^{2}\epsilon_{0}}{\pi me^{2}}~.\nonumber
\end{eqnarray}
%\\

Pentru nivelele de energie, avem $E=T+V$ \c{s}i prin substituirea  
energiilor poten\c{t}ial\A\ \c{s}i cinetic\A\ ob\c{t}inem
\begin{eqnarray}
E = \frac{1}{2}mv^{2} - \frac{e^{2}}{4\pi \epsilon_{0}r}~,\nonumber
\end{eqnarray}
sau echivalent
\begin{eqnarray}
E_{n} = -\frac{e^{2}}{8\pi \epsilon_{0}r_{n}}~.\nonumber
\end{eqnarray}
Substituind valoarea lui $r_{n}$ \h n ultima ecua\c{t}ie ob\c{t}inem
\begin{eqnarray}
E_{n} = 
-\frac{me^{4}}{8\epsilon_{0}^{2}\hbar^{2}} 
\left(\frac{1}{n^{2}}\right)~.\nonumber 
\end{eqnarray}
%ceea ce ne d\A\ nivelele de energie.
\\

{\bf Problema 6.2} - Teorema lui Uns\"old spune c\A\ , pentru orice valoare
a num\A rului cuantic orbital $l$, densit\A \c{t}ile de probabilitate, sumate
peste toate subst\A rile posibile, de la  $m_{l}=-l$ p\h n\A\ la $m_{l}=+l$ 
dau o 
constant\A\ independent\A\ de unghiurile $\theta$ sau $\phi$, adic\A\
\begin{eqnarray}
\sum_{m_{l}=-l}^{+l} \mid \Theta_{lm_{l}} \mid^{2} \mid \Phi_{m_{l}} 
\mid^{2} = ct.\nonumber 
\end{eqnarray}

Aceast\A\ teorem\A\ arat\A\ c\A\ orice atom sau ion cu subst\A ri \h nchise
prezint\A\ o 
distribu\c{t}ie sferic simetric\A\ de sarcin\A\ electric\A\ . S\A\ se verifice 
teorema Uns\"old pentru $l=0$, $l=1$ \c{s}i $l=2$.
%con ayuda de la tabla 1.

{\bf Solu\c{t}ie}: Avem pentru $l=0$, $\Theta_{00}=1/\sqrt{2}$ \c{s}i
$\Phi_{0}=1/\sqrt{2\pi}$, deci vedem c\A\ 
\begin{eqnarray}
\mid \Theta_{0,0} \mid^{2} \mid \Phi_{0} \mid^{2} = \frac{1}{4\pi}~.\nonumber
\end{eqnarray}

Pentru $l=1$ avem
\begin{eqnarray}
\sum_{m_{l}=-1}^{+1} \mid \Theta_{lm_{l}} \mid^{2} 
\mid \Phi_{m_{l}} \mid^{2} = \mid \Theta_{1,-1} \mid^{2} \mid \Phi_{-1} 
\mid^{2} + \mid \Theta_{1,0} \mid^{2} \mid \Phi_{0} \mid^{2} + \mid 
\Theta_{1,1} \mid^{2} \mid \Phi_{1} \mid^{2}~.\nonumber
\end{eqnarray}
Pe de alt\A\ parte func\c{t}iile de und\A\ sunt: 
$\Theta_{1,-1}=(\sqrt{3}/2)sin\theta$, 
$\Phi_{-1}=(1/\sqrt{2\pi})e^{-i\phi}$, 
$\Theta_{1,0}=(\sqrt{6}/2)cos\theta$, $\Phi_{0}=1/\sqrt{2\pi}$, 
$\Theta_{1,1}=(\sqrt{3}/2)sin\theta$, $\Phi_{1}=(1/\sqrt{2\pi})e^{i\phi}$~,
care substituite \h n ecua\c{t}ia anterioar\A\ conduc la
\begin{eqnarray}
\sum_{m_{l}=-1}^{+1} \mid \Theta_{lm_{l}} \mid^{2} 
\mid \Phi_{m_{l}} \mid^{2} = \frac{3}{8\pi}sen^{2}\theta + 
\frac{3}{4\pi}cos^{2}\theta + \frac{3}{8\pi}sen^{2}\theta = 
\frac{3}{4\pi}\nonumber 
\end{eqnarray}
\c{s}i din nou ob\c{t}inem o constant\A\ .

Pentru $l=2$ avem
\begin{eqnarray}
\sum_{m_{l}=-2}^{+2} \mid \Theta_{lm_{l}} \mid^{2}
\mid \Phi_{m_{l}} \mid^{2} =\nonumber
\end{eqnarray}
\begin{eqnarray}
\mid \Theta_{2,-2} \mid^{2} \mid \Phi_{-2} \mid^{2}
\mid \Theta_{2,-1} \mid^{2} \mid \Phi_{-1} \mid^{2} 
+ \mid \Theta_{2,0} \mid^{2} \mid \Phi_{0} \mid^{2} 
+ \mid \Theta_{2,1} \mid^{2} \mid \Phi_{1} \mid^{2}
+ \mid \Theta_{2,2} \mid^{2} \mid \Phi_{2} \mid^{2}\nonumber
\end{eqnarray}
\c{s}i func\c{t}iile de und\A\ sunt:
$\Theta_{2,-2}=(\sqrt{15}/4)sin^{2}\theta$, 
$\Phi_{-2}=(1/\sqrt{2\pi})e^{-2i\phi}$, 
$\Theta_{2,-1}=(\sqrt{15}/2)sin\theta cos\theta$,
$\Phi_{-1}=(1/\sqrt{2\pi})e^{-i\phi}$,
$\Theta_{2,0}=(\sqrt{10}/4)(3cos^{2}\theta-1)$,
$\Phi_{0}=1/\sqrt{2\pi}$,
$\Theta_{2,1}=(\sqrt{15}/2)sin\theta cos\theta$,
$\Phi_{1}=(1/\sqrt{2\pi})e^{i\phi}$,
$\Theta_{2,2}=(\sqrt{15}/4)sin^{2}\theta$,
$\Phi_{2}=(1/\sqrt{2\pi})e^{2i\phi}$,
care substituite \h n ecua\c{t}ia anterioar\A\ dau
\begin{eqnarray}
\sum_{m_{l}=-2}^{+2} \mid \Theta_{lm_{l}} \mid^{2}
\mid \Phi_{m_{l}} \mid^{2} = \frac{5}{4\pi}~,\nonumber
\end{eqnarray}
ceea ce din nou verific\A\ teorema Uns\"old.
\\

{\bf Problema 6.3} - Probabilitatea de a g\A si un electron atomic
a c\A rui
func\c{t}ie de und\A\ radial\A\ este cea de stare fundamental\A\
$R_{10}(r)$ \h n afara unei sfere de raz\A\ Bohr $a_{0}$ 
centrat\A\ \h n nucleu este
\begin{eqnarray}
\int_{a_{0}}^{\infty} \mid R_{10}(r) \mid^{2}r^{2}dr~.\nonumber 
\end{eqnarray}
%Func\c{t}ia de und\A\ $R_{10}(r)$ corespunde st\A rii fundamentale a 
%atomului de hidrogen, iar $a_{0}$ este raza Bohr. 
S\A\ se calculeze probabilitatea de a g\A si 
electronul \h n starea fundamental\A\ atomic\A\ la o 
distan\c{t}\A\ de nucleu mai mare de $a_{0}$.

{\bf Solu\c{t}ie}: Func\c{t}ia de und\A\ radial\A\ care corespunde 
st\A rii fundamentale este
\begin{eqnarray}
R_{10}(r) = \frac{2}{a_{0}^{3/2}}e^{-r/a_{0}}~.\nonumber
\end{eqnarray}
Substituind-o \h n integral\A\ ob\c{t}inem

\begin{eqnarray}
\int_{a_{0}}^{\infty} \mid R(r) \mid^{2}r^{2}dr = 
\frac{4}{a_{0}^{3}} \int_{a_{0}}^{\infty} r^{2} e^{-2r/a_{0}}dr ~,\nonumber
\end{eqnarray}
sau
\begin{eqnarray}
\int_{a_{0}}^{\infty} \mid R(r) \mid^{2}r^{2}dr =
\frac{4}{a_{0}^{3}}\left[-\frac{a_{0}}{2}r^{2}e^{-2r/a_{0}}  
-\frac{a_{0}^{2}}{2}re^{-2r/a_{0}}
-\frac{a_{0}^{3}}{4}e^{-2r/a_{0}}\right]_{a_{0}}^{\infty}~.\nonumber
\end{eqnarray}
Aceasta ne conduce la:
\begin{eqnarray}
\int_{a_{0}}^{\infty} \mid R(r) \mid^{2}r^{2}dr = \frac{5}{e^{2}}
\approx 68 \% \; !!~,\nonumber
\end{eqnarray}
care este probabilitatea cerut\A\ \h n aceast\A\ problem\A\ .

%\end{document}

\newpage
%%%%%%%%%%%%%%%%%%%%%%%%%%%%%%%%%%%%%%%%%%%%%%%%%%%%%%%%%%%%%%%
%%%%%%%%%%%%%%%%%%%%%%%%%%%%%%%%%%%%%%%%%%%%%%%%%  Ciocniri cuantice
%%%%%%%%%%%%%%%%%%%%%%%%%%%%%%%%%%%%%%%%%%%%%%%%%%%%%%%%%%%%%%%%%
%\protect
%\setcounter{equation}
\begin{center}{\huge 7. CIOCNIRI CUANTICE}
\end{center}
%\author{\it Daniel Jim\'enez Alvarez}
%\date{}
%\maketitle
%%%%%%%%%%%%%%%%%%%%%%%%%%%%%%%%%%%%%%%%%%%%%%%%%%%%%%%%%%%%%%%%%%%%

\section*{\bf Introducere}
%%%%%%%%%%%%%%%%%%%%%%%%%%%%%%
\setcounter{equation}{0}
Pentru ini\c{t}iere \h n teoria cuantic\A\ de \h mpr\A \c{s}tiere 
ne vom servi de
rezultate deja cunoscute de la \h mpr\A \c{s}tierea clasic\A\ \h n c\h mpuri
centrale \c{s}i vom presupune anumite situa\c{t}ii care vor simplifica 
calculele f\A r\A\ \h ns\A\ a ne \h ndep\A rta prea mult
de problema ``real\u{a}". \c{S}tim c\A\ \h n studiul experimental al unei
ciocniri putem ob\c{t}ine date care ne pot ajuta
s\A\ \h n\c{t}elegem distribu\c{t}ia materiei ``\c{t}int\u{a}", sau mai bine spus
interac\c{t}iunea \h ntre fasciculul incident \c{s}i ``\c{t}int\u{a}".  
Ipotezele pe care le vom presupune corecte sunt:

i) Particulele nu au spin, ceea ce 
{\em nu} \h nseamn\A\ c\A\ acesta nu este important \h n ciocniri.

ii) Ne vom ocupa numai de dispersia elastic\A\  pentru care
posibila structur\A\ intern\A\ a particulelor nu se ia \h n considerare.

iii) \c{T}inta este suficient de
sub\c{t}ire pentru a putea neglija \h mpr\A \c{s}tierile
multiple.

iv) Interac\c{t}iunile sunt descrise printr-un poten\c{t}ial care 
depinde numai de pozi\c{t}ia relativ\A\ a particulelor.

Aceste ipoteze elimin\A\ o serie de efecte cuantice \c{s}i m\A\ resc 
corectitudinea unor
rezultate bazice din teoria ciocnirilor clasice. Astfel definim:

\begin{equation}
\frac{d\sigma}{d\Omega}=\frac{I(\theta,\varphi)}{I_{0}}~,
\end{equation}

\noindent
unde $d\sigma$ este elementul de unghi solid, $I_{0}$ este
num\A rul de particule incidente pe unitate de arie \c{s}i
$I{}d\Omega$ este num\A rul de particule dispersate \h n elementul de
unghi solid $d\Omega$. 

Cu aceste concepte binecunoscute \c{s}i cu ajutorul m\A rimii asimptotice
parametru de impact $b$ asociat fiec\A rei particule clasice incidente 
ajungem la importanta formul\A\ clasic\A\

\begin{equation}
\frac{d\sigma}{d\Omega}=\frac{b}{\sin\theta}\vert 
\frac{db}{d\theta}\vert ~.
\end{equation}

Dac\A\ dorim s\A\ cunoa\c{s}tem \h n termeni cuantici fenomenologia
de ciocnire,
trebuie s\A\ studiem evolu\c{t}ia \h n timp a unui pachet de unde. 
Fie $F_{i}$ fluxul de particule al fascicolului incident, adic\A\ , 
num\A rul de particule pe unitate de timp care intersecteaz\A\ 
o suprafa\c{t}\A\ unitar\A\ transversal\A\ axei de propagare. 
Vom pozi\c{t}iona un detector departe de zona de ac\c{t}iune efectiv\A\ a
poten\c{t}ialului, care sub\h ntinde un unghi
solid $d\Omega$; cu acesta putem \h nregistra num\A rul de 
particule $dn/dt$ dispersate \h n unitatea de timp \h n $d\Omega$ \h n 
direc\c{t}ia $(\theta,\varphi)$.

%%%%%%%%%%%%%%
\vskip 2ex
\centerline{
\epsfxsize=280pt
\epsfbox{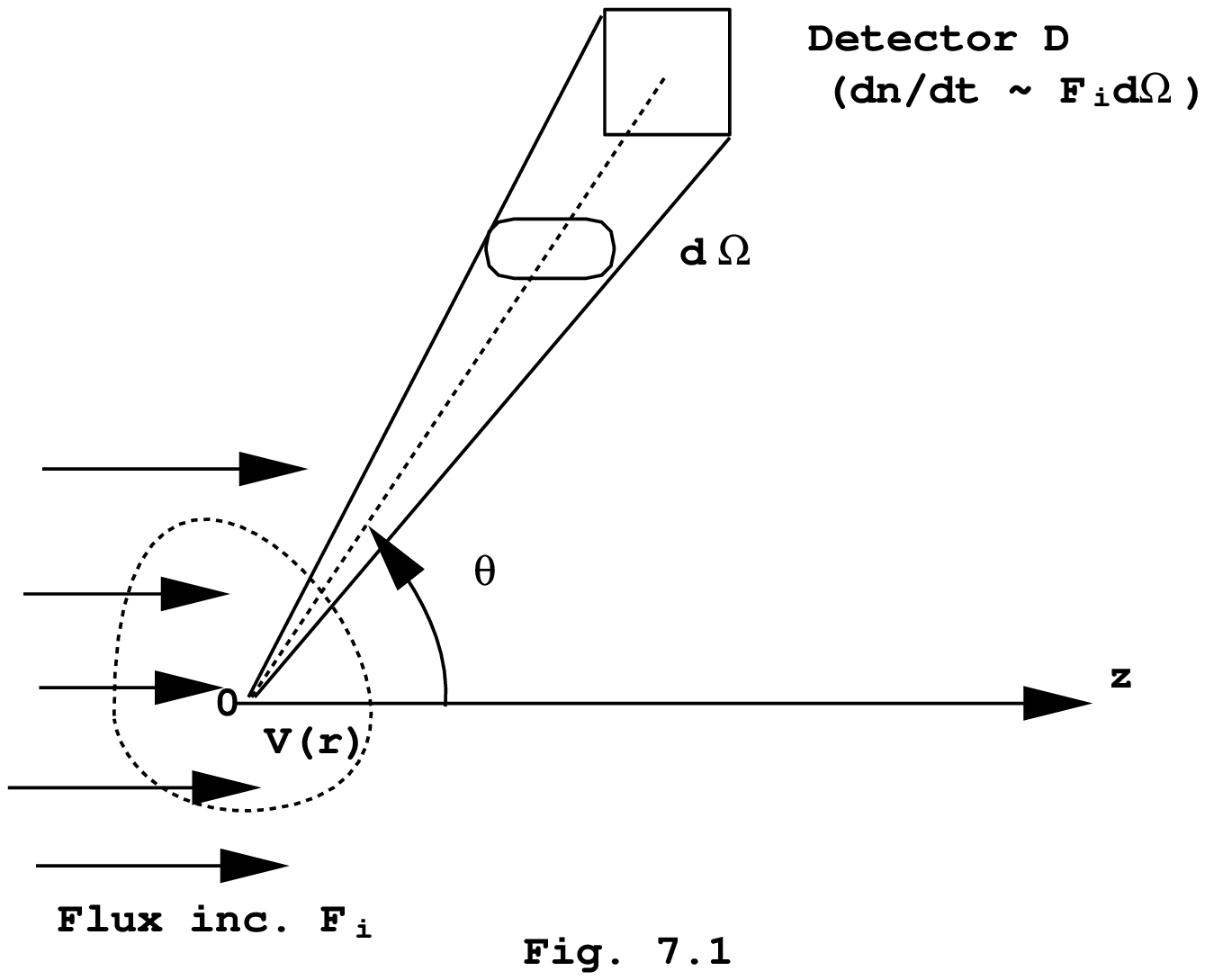}}
\vskip 4ex
%\begin{center}
%{\small{Fig. x}\\
%}
%\end{center}
%%%%%%%%%%%%%%%%

\noindent
$dn/dt$ este propor\c{t}ional cu $d\Omega$ \c{s}i 
$F_{i}$. S\A\ numim $\sigma (\theta,\varphi)$  coeficientul de 
propor\c{t}ionalitate \h ntre $dn$ \c{s}i $F_{i} d\Omega$:
\begin{equation}
 dn=\sigma (\theta,\varphi)F_{i} d\Omega~,
\end{equation}

\noindent
care este prin defini\c{t}ie sec\c{t}iunea diferen\c{t}ial\A\ transversal\A\ .

Num\A rul de particule pe unitatea de timp care ajung la 
detector este egal cu num\A rul de particule care intersecteaz\A\ o 
suprafa\c{t}\A\ $\sigma (\theta,\varphi) d\Omega$ situat\A\ perpendicular 
pe axa fasciculului. Sec\c{t}iunea total\A\ de dispersie este prin 
defini\c{t}ie:
\begin{equation}
 \sigma=\int \sigma (\theta,\varphi) d\Omega~. 
\end{equation}

Cum putem orienta axele de coordonate conform alegerii dorite, o vom face 
\h n a\c{s}a fel ca axa fasciculului incident de particule 
s\A\ coincid\A\ cu axa z (aceasta pentru simplificarea calculelor, unde vom 
folosi coordonatele sferice). \\
\^{I}n regiunea negativ\A\ a axei, pentru $t$ negativ mare, 
particula va fi practic liber\A\ : nu este afectat\A\ de 
$V({\bf r})$ \c{s}i starea sa se poate reprezenta prin unde plane. Prin 
urmare func\c{t}ia de und\A\ trebuie s\A\ con\c{t}in\A\ termeni de forma 
$e^{ikz}$, unde $k$ este constanta care apare \h n ecua\c{t}ia 
Helmholtz. Prin analogie cu optica, forma undei dispersate este:
\begin{equation}
 f(r)= \frac{e^{ikr}}{r}~. 
\end{equation}

\^{I}ntr-adev\A r:
\begin{equation}
 (\nabla ^{2} + k^{2})e^{ikr} \neq 0
\end{equation} 

\c{s}i
\begin{equation} 
(\nabla ^{2} + k^{2}) \frac{e^{ikr}}{r}=0
\end{equation}
pentru $r>r_{0}$, unde $r_{0}$ este orice num\A r pozitiv.

Presupunem c\A\ mi\c{s}carea particulei este descris\A\ de 
Hamiltonianul:
\begin{equation}
H=\frac{{\rm \bf p^2}}{2\mu}+V=H_{0}+V~.
\end{equation}

V este diferit de zero numai \h ntr-o mic\A\ vecin\A tate \h n jurul
originii. \c{S}tim c\A\ %evolu\c{t}ia 
un pachet de unde \h n $t=0$ se poate scrie:

\begin{equation}
 \psi({\bf{r}},0)=\frac{1}{(2\pi)^\frac{3}{2}}\int 
\varphi({\rm \bf k})\exp[i{\rm \bf k\cdot (r-r_{0})}]
{\rm {\bf d^{3}k}}~,
\end{equation}

\noindent
unde $\psi$ este o func\c{t}ie semnificativ nenul\A\ \h n segmentul
(l\A rgimea) 
$\Delta {\rm \bf k}$
centrat \h n jurul lui ${\rm \bf k_{0}}$. Presupunem deasemena
c\A\ ${\rm \bf k_{0}}$ este
paralel la ${\rm \bf r_{0}}$, dar de sens opus.
Pentru a vedea \h n mod cantitativ ce se \h nt\h mpl\A\ cu pachetul de 
unde
c\h nd la un moment ulterior ciocne\c{s}e \c{t}inta \c{s}i este  
dispersat de aceasta ne putem folosi de dezvoltarea lui $\psi({\rm \bf r},0)$
\h n func\c{t}iile proprii 
$\psi_{n}({\rm \bf r})$ ale lui $H$, respectiv
$\psi({\bf{r}},0)=\sum_{n}c_{n}\psi_{n}(\bf{r})$. Astfel, pachetul
de unde la timpul $t$ este:
\begin{equation}
\psi({\bf 
r},t)=\sum_{n}c_{n}\varphi _{n}({\bf r})\exp(-\frac{i}{\hbar}E_{n}t)~.
\end{equation}

Aceasta este o func\c{t}ie proprie a operatorului $H_{0}$ \c{s}i nu a lui $H$, 
dar putem 
substitui aceste func\c{t}ii proprii cu func\c{t}ii proprii 
particulare ale lui $H$, pe care le vom nota cu $\psi_{k}^{(+)}(\bf{r})$.  
Forma asimptotic\A\ a acestora din urm\A\ este de tipul:
\begin{equation}
\psi _{k}^{(+)}(\bf{r})\simeq e^{i\bf{k\cdot r}} +
f({\rm \bf r})\frac{e^{ikr}}{r}~,
\end{equation}

unde, cum este uzual,
${\rm \bf p}=\hbar {\rm \bf k}$ \c{s}i  $E=\frac{\hbar ^{2}k^{2}}{2m}$.

Aceasta corespunde unei unde plane ca fascicul incident 
\c{s}i o und\A\ sferic\A\ divergent\u{a}, despre care se poate spune c\A\
este rezultatul  
interac\c{t}iunii \h ntre fascicul \c{s}i \c{t}int\u{a}. Aceste solu\c{t}ii
ale ec.
Schr\"odinger exist\A\ \h n realitate, \c{s}i putem dezvolta 
$\psi ({\rm \bf r},0)$ \h n
unde plane \c{s}i $\psi _{k}({\rm \bf r})$:
\begin{equation}
\psi({\rm \bf r},0)=\int \varphi ({\rm \bf k})\exp(-i{\rm \bf k\cdot
r_{0}})\psi _{{\rm \bf k}}({\rm \bf r}) d^{3}k~,
\end{equation}
unde $ \hbar\omega= \frac{\hbar^{2}k^{2}}{2m}$.
%%%%%%%%%%%%%%%%%%%%%%%%%%%%%%%%%%%%%%%%%%%%%%%%%%%%%%%%%%%%%%%%%%%%%%%%%%%%%
Se poate spune deci c\A\ unda sferic\A\ divergent\A\ nu are nici o 
contribu\c{t}ie la pachetul de unde ini\c{t}ial.
%%%%%%%%%%%%%%%%%%%%%%%%%%%%%%%%%%%%%%%%%%%%%%%%%%%%%%%%%%%%%%%%%%%%%%%%%%%%%%

\section*{\bf \^{I}mpr\A \c{s}tierea unui pachet de unde}
%%%%%%%%%%%%%%%%%%%%%%%%%%%%%%%%%%%%%%%%%%%%%%%%%%%%%%%%

Orice und\A\ sufer\A\ \h n cursul propag\A rii o dispersie. De aceea nu se 
poate ignora efectul undei divergente din acest punct de vedere. 
Se poate folosi urm\A torul truc: 

\begin{equation} 
\omega= 
\frac{\hbar}{2m}k^{2}= 
\frac{\hbar}{2m}[{\bf k_{0}+(k-k_{0}})]^{2}= 
\frac{\hbar}{2m}[2{\bf k_{0}\cdot k - k_{0}^{2}+ (k-k_{0})^{2}}]~,
\end{equation}
  
\noindent pentru a neglija ultimul termen \h n paranteze. 
Substituind $\omega$ \h n $\psi$, cerem ca: 
\( \frac{\hbar}{2m}({\bf k-k_{0}})^{2}T \ll 1 \),
unde $T \simeq \frac{2mr_{0}}{\hbar k_{0}}$ \c{s}i deci:

\begin{equation}
\frac{(\Delta k)^{2}r_{0}}{k_{0}} \ll 1~. 
\end{equation}

\noindent Aceast\A\ condi\c{t}ie ne spune c\A\ pachetul de unde nu se 
disperseaz\A\ \h n mod apreciabil 
chiar \c{s}i atunci c\h nd se deplaseaz\A\ pe o distan\c{t}\A\
macroscopic\A\ $r_{0}$.

Aleg\h nd direc\c{t}ia vectorului $\bf{k}$ al undei incidente de-a lungul uneia
dintre cele trei direc\c{t}ii carteziene, putem scrie \h n coordonate sferice

\( \psi_{k}(r,\theta,\varphi) \simeq e^{ikz} + 
\frac{f(k,\theta,\varphi)e^{ikr}}{r}~. \) 

\^{I}ntruc\h t $H$, operatorul Hamiltonian (que hemos considerat p\h n\A\ acum
nu ca operator pentru c\A\ 
rezultatele sunt acelea\c{s}i) este invariant 
la rota\c{t}iile \h n axa z, putem alege condi\c{t}iile de frontier\A\ 
deasemenea invariante, astfel c\A\ :

\( \psi_{k}(r,\theta,\varphi)\simeq e^{ikz}+\frac{f(\theta)e^{ikr}}{r}~.\)

\noindent Acest tip de func\c{t}ii se cunosc ca unde de \h mpr\A \c{s}tiere.
Coeficientul $f(\theta)$ al undei divergente se cunoa\c{s}te ca 
amplitudine de \h mpr\A \c{s}tiere.\\

\section*{\bf Amplitudinea de probabilitate \h n \h mpr\A \c{s}tieri}
%%%%%%%%%%%%%%%%%%%%%%%%%%%%%%%%%%%%%%%%%%%%%%%%%%%%%%%%%%%%%%%%%%%%%%

Ecua\c{t}ia Schr\"odinger de rezolvat este:
\begin{equation}
 i\hbar \frac{\partial\psi}{\partial t}= - \frac{\hbar^{2}}{2m} 
\nabla^{2}\psi + V({\bf r},t)\psi~. 
\end{equation}

\noindent
Expresia
\begin{equation}
P({\bf r},t)= \psi^{*}({\bf r},t)\psi ({\bf r},t)=\vert \psi ({\bf 
r},t) \vert ^{2} 
\end{equation}

\noindent se interpreteaz\A\ , cf. lui Max Born, ca o densitate de 
probabilitate dac\A\ func\c{t}ia de und\A\ se normalizeaz\A\ astfel ca:
\begin{equation}
\int \vert \psi ({\rm \bf r},t) \vert ^{2}  d^{3}r = 1~.
\end{equation}

Desigur integrala de 
normalizare a lui $\psi$ trebuie s\A\ fie independent\A\ de timp. 
%dac\A\ avem \h n vedere ecua\c{t}ia Schr\"odinger. 
Acest lucru se poate nota \h n felul urm\A tor:
\begin{equation}
I= \frac{\partial}{\partial t} \int _{\Omega} P({\rm \bf r},t) d^{3}r=
\int_{\Omega} (\psi^{*}\frac{\partial\psi}{\partial t}
+\frac{\partial\psi^{*}}{\partial t}\psi) d^{3}r 
\end{equation}

\noindent \c{s}i din ec. Schr\"odinger:
\begin{equation}
\frac{\partial\psi}{\partial t}= \frac{i\hbar}{2m}
\nabla ^{2}\psi-\frac{i}{\hbar}V({\bf r},t)\psi
\end{equation} 

\noindent rezult\A\ :
$$
I=\frac{i\hbar}{2m} \int_{\Omega}
[\psi^{*}\nabla^{2}-(\nabla^{2}\psi^{*})\psi]d^{3}r = \frac{i\hbar}{2m} 
\int_{\Omega} \nabla \cdot 
[\psi^{*}\nabla\psi-(\nabla\psi^{*})\psi]d^{3}r=
$$
\begin{equation}
=\frac{i\hbar}{2m} \int_{A}[\psi^{*}\nabla\psi-(\nabla\psi^{*})\psi]_{n}
dA~, 
\end{equation}

\noindent unde s-a folosit teorema Green pentru evaluarea integralei 
de volum. 
$dA$ este elementul de suprafa\c{t}\A\ pe frontiera care 
delimiteaz\A\ regiunea de 
integrare \c{s}i 
$[\quad]_{n}$ denot\A\ componenta \h n direc\c{t}ia normal\A\ la elementul de
suprafa\c{t}\A\ $dA$.

Definind:
\begin{equation}
 {\bf 
S}({\bf r},t)=\frac{\hbar}{2im} [\psi^{*}\nabla\psi-(\nabla\psi^{*})\psi]~, 
\end{equation}

ob\c{t}inem:
\begin{equation}
 I= \frac{\partial}{\partial t} \int_{\Omega} P({\bf r},t) d^{3}r= - 
\int _{\Omega} \nabla\cdot {\bf S} d^{3}r = -\int_{A} S_{n} dA~,
\end{equation}

\noindent pentru pachete de und\A\ \h n care $\psi$ se pune zero la  
distan\c{t}e mari 
\c{s}i integrala de normalizare converge, integrala de suprafa\c{t}\A\ este 
zero c\h nd $\Omega$ este tot spa\c{t}iul. Se poate demonstra (se poate consulta
P. Dennery \& A. Krzywicki, {\it Mathematical methods for physicists}) c\A\ 
integrala de 
suprafa\c{t}\A\ este zero, astfel c\A\ integrala de normalizare este 
constant\A\ \h n timp \c{s}i deci se satisface cerin\c{t}a ini\c{t}ial\A\ . 
Din aceea\c{s}i 
ecua\c{t}ie pentru ${\bf S}$ ob\c{t}inem:
\begin{equation}
 \frac{\partial P({\bf r},t)}{\partial t} + \nabla \cdot {\bf S}({\bf 
r},t)= 0~, 
\end{equation}

\noindent care este o ecua\c{t}ie de continuitate  
cu fluxul de densitate $P$ \c{s}i curent de densitate ${\bf S}$, 
f\A r\A\ nici un fel de surse (pozitive sau negative).
% As\'{\i}, es razonable interpretar ${\bf S}$ como 
%una densidad de corriente de probabilidad. Por semejanza con la 
%electrodin\'amica, 
Dac\A\ interpret\A m $\frac{\hbar}{im}\nabla$ ca un fel de `operator'
vitez\A\  (ca \c{s}i \h n cazul timpului nu se poate vorbi de un operator 
vitez\A\ \h n sens riguros), atunci:
\begin{equation}
{\bf S}({\bf r}, t)= Re(\psi ^{*}\frac{\hbar}{im}\nabla\psi)~.
\end{equation}
Calculul efectiv al densit\A \c{t}ii de curent pentru o und\A\ de
\h mpr\A \c{s}tiere este de tip `truc' \c{s}i nu \h l consider\A m
ilustrativ. Rezultatul final este $j_{r}=\frac{hk}{mr^2}|f(\theta)|^2$,
unde direc\c{t}ia $\theta=0$ nu se include.

\section*{\bf Func\c{t}ia Green \h n teoria de \h mpr\A \c{s}tiere}
%%%%%%%%%%%%%%%%%%%%%%%%%%%%%%%%%%%%%%%%%%%%%%%%%%%%%%%%%%%%%%%%%%%

O alt\A\ form\A\ de a scrie ecua\c{t}ia Schr\"odinger de rezolvat este
$(-\frac{\hbar^{2}}{2m} \nabla^{2} + V)\psi = E\psi $ sau
$(\nabla^{2} + k^{2})\psi = U\psi $ unde:
$ k^{2}=\frac{2mE}{\hbar^{2}}$ \c{s}i $U=\frac{2mV}{\hbar^{2}}$.

Rezult\A\ mai convenabil de transformat aceast\A\ ecua\c{t}ie la o form\A\ 
integral\A\ . Aceasta se poate face dac\A\ vom considera $U\psi$ 
din partea dreapt\A\ 
a ecua\c{t}iei ca o inomogeneitate, ceea ce ne permite s\A\ construim 
solu\c{t}ia ecua\c{t}iei cu ajutorul func\c{t}iei Green (nucleu integral), 
care prin defini\c{t}ie este solu\c{t}ia lui:
\begin{equation}
\label{eq:e1}
(\nabla^{2}+k^{2})G(\bf{r,r'}) =  
\delta(\bf{r-r'})~. 
\end{equation} 

\noindent Solu\c{t}ia ecua\c{t}iei Schr\"odinger se
d\A\ ca suma solu\c{t}iei ecua\c{t}iei omogene \c{s}i a 
solu\c{t}iei inomogene de tip Green:
\begin{equation}
\psi(\bf{r})=\lambda(\bf{r})-\int
G(\bf{r,r'})U(\bf{r'})\psi(\bf{r'})d^{3}r'~.
\end{equation}

C\A ut\A m acum o func\c{t}ie $G$ care s\A\ fie un produs de fun\c{t}ii
linear independente, cum sunt de exemplu undele plane:
\begin{equation}
G({\bf r,r'}=\int A({\bf q})e^{i{\bf q\cdot (r-r')}}dq~.
\end{equation}

\noindent Folosind ecua\c{t}ia \ref{eq:e1}, avem:
\begin{equation}
\int A({\bf q})(k^{2}-q^{2})e^{i{\bf q\cdot(r-r')}}dq=
\delta{\bf(r-r')}~,
\end{equation}
 
\noindent care se transform\A\ \h ntr-o identitate dac\A\ :
\begin{equation}
A({\bf q})= (2\pi)^{-3}(k^{2}-q^{2})^{-1}~.
\end{equation}

\noindent De aici rezult\A\ :

\begin{equation}
G({\bf r,r'})=\frac{1}{(2\pi)^{3}} \int 
\frac{e^{iqR}}{k^{2}-q^{2}}d^{3}q~,
\end{equation}
cu $R=\vert {\bf r-r'} \vert$.
Dup\A\ un calcul folosind metode de variabil\A\
complex\A\ \footnote{Se poate vedea problema 7.1.}, ajungem la:
\begin{equation}
G(R)= - \frac{1}{4\pi} \frac{e^{ikR}}{R}~.
\end{equation}

Aceast\A\ func\c{t}ie nu este determinat\A\ \h n mod univoc;  
func\c{t}ia
Green poate fi oricare solu\c{t}ie a ecua\c{t}iei \ref{eq:e1}; 
Alegerea uneia particulare se face prin impunerea
condi\c{t}iilor de frontier\A\ asupra func\c{t}iilor proprii $\psi_{k}({\bf
r})$.

Func\c{t}ia Green ob\c{t}inut\A\ \h n aceste condi\c{t}ii este:
\begin{equation}
G({\bf r,r'})= -\left( \frac{e^{ik \vert {\bf r-r'} \vert}}{4\pi
\vert{\bf r-r'}\vert }\right)~.
\end{equation}

\^{I}n acest fel, ajungem la ecua\c{t}ia integral\A\ pentru func\c{t}ia de
und\A\ de ciocnire:
\begin{equation}
\psi (k,{\bf r})= \varphi (k,{\bf r}) - \frac{m}{2 \pi \hbar^{2}} \int
\frac{e^{ik \vert {\bf r-r'} \vert}}{ {\bf r-r'} } U({\bf r'}) \psi (k,{\bf 
r})d{\bf r}~,
\end{equation}
unde $\varphi$ este o solu\c{t}ie a ecua\c{t}iei Helmholtz. Not\h nd
$ \vert {\bf r-r'} \vert = R $:
\begin{equation}
(\nabla^{2}+k^{2})\psi=(\nabla^{2}+k^{2})[\varphi + \int G({\bf r,r'}) 
U({\bf r'}) \psi({\bf r'}) d^{3}r']
\end{equation}

\noindent \c{s}i presupun\h nd c\A\ putem schimba ordinea opera\c{t}iilor
\c{s}i pune operatorul $\nabla ^2$ \h n interiorul integralei: 
\begin{equation}
(\nabla^{2}+k^{2})\psi= \int (\nabla^{2}+k^{2}) G ({\bf r,r'}) U({\bf
r'}) \psi({\bf r'}) d^{3}r'= U({\bf r}) \psi ({\bf r})~,
\end{equation}
ceea ce ne arat\A\ c\A\ se verific\A\ faptul c\A\
$G(R)= \frac{1}{4\pi} \frac{e^{ikR}}{R}$ este solu\c{t}ie.

\section*{\bf Teorema optic\A\ }
%%%%%%%%%%%%%%%%%%%%%%%%%%%%%%%%
 
Sec\c{t}iunea diferen\c{t}ial\A\ total\A\ este dat\A\ de:
\begin{equation}
\sigma_{tot}(k)= \int \frac{d\sigma}{d\Omega} d\Omega~.
\end{equation}

S\A\ exprim\A m acum $f(\theta)$ ca func\c{t}ie de \c{s}iftul de faz\A\
$S_{l}(k)=e^{2i\delta_{l}(k)}$ \h n forma:
\begin{equation}
f(\theta)=  \frac{1}{k} \sum_{l=0}^{\infty} (2l+1) e^{i\delta_{i}(k)} 
\sin \delta_{l}(k) P_{l}(\cos \theta)
\end{equation}

\noindent
atunci
$$
\sigma_{tot} = \int [\frac{1}{k} \sum_{l=0}^{\infty} (2l+1)
e^{i\delta_{l}(k)}\sin \delta_{l}(k) P_{l}(\cos \theta)]
$$
\begin{equation}
[\int 
[\frac{1}{k} \sum_{l'=0}^{\infty} (2l'+1)e^{i\delta_{l'}(k)}\sin 
\delta_{l'}(k) P_{l'}(\cos \theta)]~.
\end{equation}
Folosind acum $\int P_{l}(\cos\theta)P_{l'}(\cos\theta)= \frac{4\pi}{2l+1}
\delta_{ll'}$ ob\c{t}inem
\begin{equation}
\sigma_{tot}= \frac{4\pi}{k^{2}} \sum_{l=0}^{\infty} (2l+1)\sin
\delta_{l}(k)^{2}~.
\end{equation}
Ceea ce ne intereseaz\A\ este c\A\ :
$$
{\rm Im} f(0)=\frac{1}{k} \sum_{l=0}^{\infty} (2l+1)
{\rm Im}[e^{i\delta_{l}(k)}\sin \delta_{l}(k)]P_{l}(1) =
\frac{1}{k} \sum _{l=0}^{\infty} (2l+1) \sin
\delta_{l}(k)^{2}=
$$
\begin{equation}
\frac{k}{4\pi} \sigma_{tot}~.
\end{equation}

Aceast\A\ rela\c{t}ie este cunoscut\A\ ca {\em teorema optic\A\ }. 
Semnifica\c{t}ia sa
fizic\A\ este c\A\ interferen\c{t}a undei incidente cu unda 
dispersat\A\ la unghi zero produce ``ie\c{s}irea" particulei din 
unda incident\A\ , ceea ce permite conservarea probabilit\A \c{t}ii.

\section*{\bf Aproxima\c{t}ia Born} 
%%%%%%%%%%%%%%%%%%%%%%%%%%%%%%%%%%%

S\A\ consider\A m situa\c{t}ia din Fig. 7.2:

%%%%%%%%%%%%%%
\vskip 2ex
\centerline{
\epsfxsize=120pt
\epsfbox{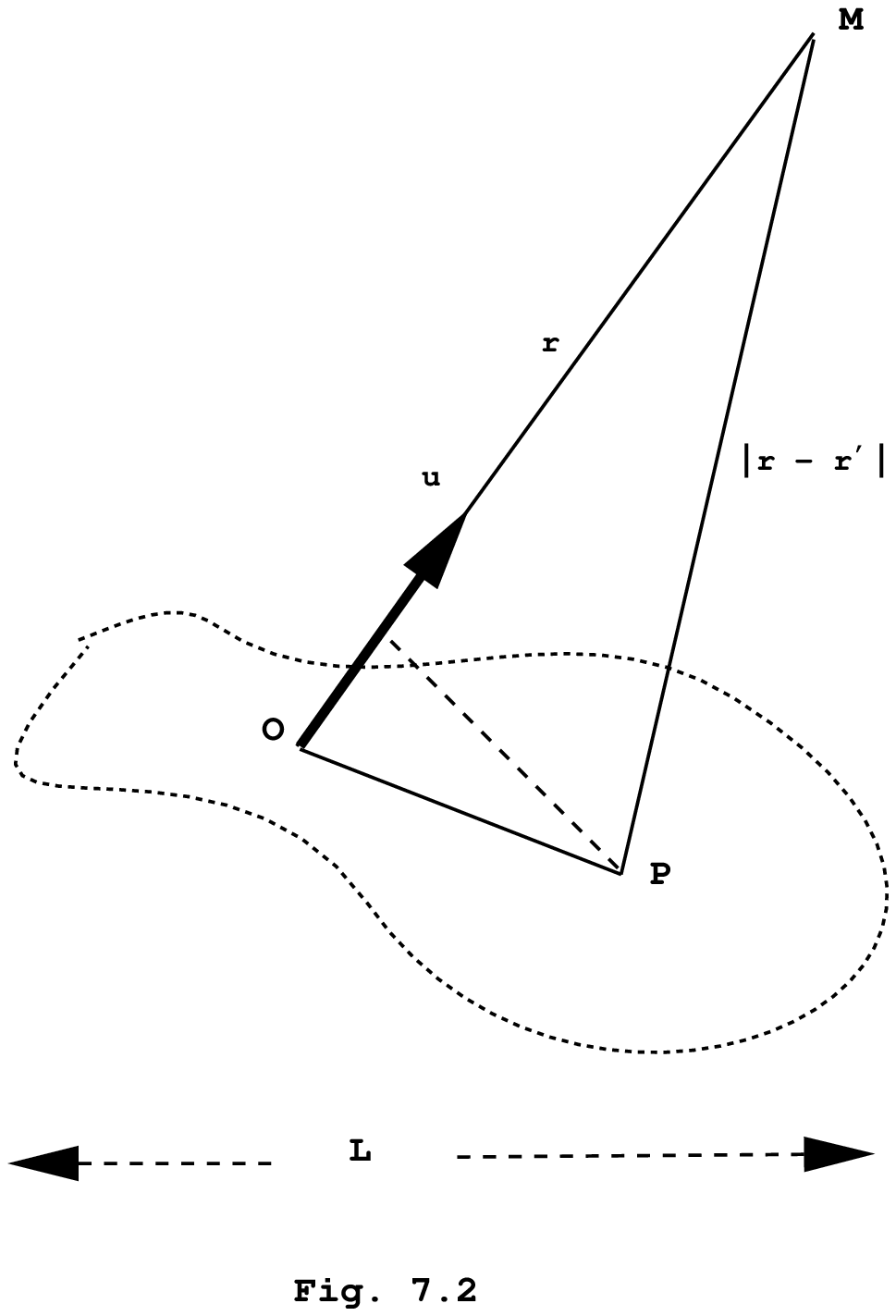}}
\vskip 2ex
%\begin{center}
%{\small{Fig. x}\\
%}
%\end{center}
%%%%%%%%%%%%%%%%
Punctul de observare M este departe de P, care se afl\A\ \h n
regiunea
de influen\c{t}\A\ a poten\c{t}ialului $U$, cu $r\gg L$, $r'\ll l$.
Segmentul MP, care corespunde la $\vert {\bf r-r'} \vert$, este \h n aceste
condi\c{t}ii geometrice
aproximativ egal cu proiec\c{t}ia lui MP pe MO:
\begin{equation}
\vert {\bf r-r'} \vert \simeq r-{\bf u \cdot r'}~,
\end{equation}
\noindent unde ${\bf u}$ este vectorul unitar (versor) \h n direc\c{t}ia 
${\bf r}$. Atunci, pentru $r$ mare:
\begin{equation}
G=- \frac{1}{4\pi} \frac{e^{ik \vert {\bf r-r'} \vert}}{\vert {\bf 
r-r'}\vert} \simeq_{r \rightarrow \infty}  -\frac{1}{4 \pi} 
\frac{e^{ikr}}{r} e^{-ik {\bf u \cdot r}}~.
\end{equation}

\noindent Substituim $G$ \h n expresia integral\A\ a func\c{t}iei de
und\A\ de ciocnire pentru a ob\c{t}ine:
\begin{equation}
\psi({\bf r})= e^{ikz} - \frac{1}{4\pi} \frac 
{e^{ikr}}{r} 
\int e^{-ik {\bf u \cdot r}}U({\bf r'})\psi ({\bf r'}) 
d^{3}r'~.
\end{equation}

\noindent Aceasta deja nu mai este o func\c{t}ie de distan\c{t}a $r=OM$, ci
numai de $\theta$ \c{s}i $\psi$; atunci:
\begin{equation}
f(\theta, \psi)= - \frac{1}{4\pi} \int e^{-ik {\bf u\ 
\cdot r}} U({\bf r'}) \psi ({\bf r'}) d^{3}r'~.
\end{equation}
Definim acum vectorul de und\A\ incident ${\bf k_{i}}$
ca un vector de modul $k$ dirijat de-a lungul 
axei polare a fasciculului astfel c\A\ :
$ e^{ikz}=e^{i {\bf k_{i} \cdot r}}$;
similar, ${\bf k_{d}}$, de modul
$k$ \c{s}i cu direc\c{t}ia fixat\A\ prin $\theta$ \c{s}i $\varphi$, se 
nume\c{s}te vector de und\A\ `deplasat' \h n direc\c{t}ia 
$(\theta, \varphi)$:
$ {\bf k_{d}}= k{\bf u} $

Vectorul de und\A\ transferat \h n direc\c{t}ia $(\theta, 
\varphi)$ se introduce prin: ${\bf K}= {\bf k_{d}-k_{i}}$.

%%%%%%%%%%%%%%
\vskip 1ex
\centerline{
\epsfxsize=80pt
\epsfbox{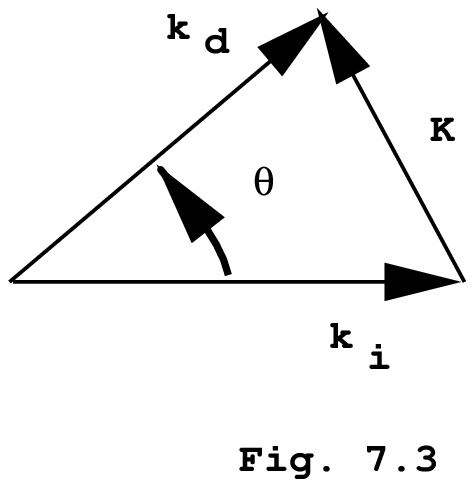}}
\vskip 2ex
%\begin{center}
%{\small{Fig. x}\\
%}
%\end{center}
%%%%%%%%%%%%%%%%

Cu aceasta, putem scrie ecua\c{t}ia integral\A\ de dispersie \h n forma:
\begin{equation}
\label{eq:e3}
\psi ({\bf r})= e^{i{\bf k_{i}\cdot r}} + \int
G({\bf r,r'}) U({\bf r'}) \psi({\bf r'}) d^{3}r'
\end{equation}

Acum putem \h ncerca rezolvarea acestei ecua\c{t}ii \h n mod iterativ. Pun\h nd 
${\bf r} \rightarrow {\bf r'}; {\bf r'} \rightarrow {\bf r''}$, putem scrie: 
\begin{equation}
\psi ({\bf r'})= e^{i{\bf k_{i}\cdot r'}} + \int G({\bf r',r''}) U({\bf
r''}) \psi({\bf r''}) d^{3}r''~.
\end{equation}

Substituind \h n \ref{eq:e3} ob\c{t}inem:
$$
\psi({\bf r})= e^{i{\bf k}_{i}\cdot r} + \int G({\bf r,r'})U({\bf
r'})e^{i{\bf k_{i} \cdot r'}}d^{3}r'
+ 
$$
\begin{equation} \label{eq:e4}
\int \int G({\bf r,r'})U({\bf
r'})G({\bf r',r''})U({\bf r''}) \psi({\bf r''})d^{3}r'' d^{3}r'~.
\end{equation}

Primii doi termeni din partea dreapt\A\ sunt 
cunoscu\c{t}i \c{s}i numai al treilea con\c{t}ine func\c{t}ia necunoscut\A\ 
$\psi({\bf r})$. Putem repeta acest procedeu: \h nlocuind ${\bf r}$ 
cu ${\bf r''}$ \c{s}i ${\bf r'}$ cu ${\bf r'''}$ ob\c{t}inem 
$\psi ({\bf r''})$~, 
pe care putem s\A\ o reintroducem \h n ec. \ref{eq:e4}:
$$
\psi({\bf r}) = e^{i {\bf k_{i} \cdot r}} + \int G({\bf r,r'})U({\bf 
r'}) e^{i {\bf k_{i} \cdot r'}}
+
$$
$$
\int \int G({\bf r,r'})U({\bf r'}) G({\bf r',r''})U({\bf r''})e^{i {\bf
k_{i} \cdot r''}}d^{3}r'd^{3}r''+$$
\begin{equation}
 \int \int \int  G({\bf r,r'})U({\bf r'}) G({\bf r',r''})U({\bf
r''})e^{i {\bf k_{i}\cdot r''}} G({\bf r'',r'''})U({\bf r'''}) \psi ({\bf 
r'''})~.
\end{equation}

\noindent Primii trei termeni sunt cunoscu\c{t}i; func\c{t}ia
necunoscut\A\ $\psi({\bf r})$ se afl\A\ \h n al patrulea termen. \^{I}n acest 
fel, prin itera\c{t}ii construim func\c{t}ia de und\A\ de dispersie 
sta\c{t}ionar\A\ . Not\A m c\A\ fiecare termen \h n dezvoltarea \h n serie 
prezint\A\ o putere superioar\A\
\h n poten\c{t}ial fa\c{t}\A\ de cel precedent. Putem continua \h n acest fel 
p\h n\A\ c\h nd ob\c{t}inem o expresie neglijabil\A\ \h n partea dreapt\A\ , 
\c{s}i 
ob\c{t}inem $\psi({\bf r})$ \h n func\c{t}ie numai de m\A rimi cunoscute.

Substituind expresia lui $\psi({\bf r})$ \h n $f(\theta, \varphi)$ 
ob\c{t}inem dezvoltarea \h n serie Born a amplitudinii de 
\h mpr\A \c{s}tiere. Limit\h ndu-ne la primul ordin \h n $U$, 
trebuie s\A\ se fac\A\ doar substituirea 
lui $\psi({\bf r'})$ cu $e^{i{\bf k_{i}\cdot r'}}$ \h n partea dreapt\A\ a
ecua\c{t}iei pentru a ob\c{t}ine:
%%%%%%%%%%%
$$
f^{(B)}(\theta, \varphi)= \frac{-1}{4\pi}  \int e^{i{\bf k_{i}\cdot
r'}} U({\bf r'}) e^{-ik {\bf u\cdot r'}} d^{3}r'=
\frac{-1}{4\pi} \int e^{-i{\bf (k_{d}-k_{i})\cdot r'}} U({\bf r'})
d^{3}r'=
$$
\begin{equation}
\frac{-1}{4\pi} \int e^{-i{\bf K \cdot r'}}U({\bf r'})d^{3}r'
\end{equation}

\noindent
${\bf K}$ este vectorul de und\A\ transferat definit mai \h naite. Vedem c\A\ 
sec\c{t}iunea de dispersie se rela\c{t}ioneaz\A\ \h n mod simplu
cu poten\c{t}ialul, dac\A\ \c{t}inem cont de
$V({\bf r})= \frac{\hbar^{2}}{2m} U({\bf r})$ \c{s}i
$\sigma (\theta,\varphi)= \vert f(\theta, \varphi) \vert^{2}$.
Rezultatul este:
\begin{equation}
\sigma^{(B)} (\theta,\varphi)=\frac{m^{2}}{4\pi^{2}\hbar^{4}} \vert
\int e^{-i{\bf K \cdot r}} V({\bf r})d^{3}r \vert^{2}
\end{equation}

Direc\c{t}ia \c{s}i modulul vectorului undei dispersate ${\bf K}$ depinde 
de modulul $k$ al lui ${\bf k_{i}}$ \c{s}i ${\bf k_{d}}$ precum \c{s}i
de direc\c{t}ia de 
\h mpr\A \c{s}tiere $(\theta,\varphi)$. Pentru $\theta$ \c{s}i $\varphi$ 
da\c{t}i, sec\c{t}iunea eficace este o func\c{t}ie de $k$, 
energia fasciculului incident. Analog, pentru 
o energie dat\A\ , $\sigma^{(B)}$ este o func\c{t}ie de
$\theta$ \c{s}i $\varphi$. Aproxima\c{t}ia Born permite ca studiind
varia\c{t}ia sec\c{t}iunii eficace diferen\c{t}iale \h n func\c{t}ie de
direc\c{t}ia de \h mpr\A \c{s}tiere \c{s}i energia incident\A\ s\A\ ob\c{t}inem  
informa\c{t}ii asupra poten\c{t}ialului $V({\bf r})$. \\

%atencion: para imprimir es: dvips blabla.dvi, para pasar a otro archivo
%que sea ps entonces es: dvips -o bla.dvi bla.ps, para escoger impresora 
%es la opcion -p %

\noindent \underline{{\bf 7N. Not\u{a}}}: Unul dintre primele articole de 
\h mpr\A \c{s}tiere cuantic\A\ este:

\noindent
M. Born, ``Quantenmechanik der Stossvorg\"ange" [``Mecanica cuantic\A\ a 
proceselor de ciocnire"],
Zf. f. Physik {\bf 37}, 863-867 (1926)

\bigskip

%%%%%%%%%%%%%%%%%%%%%%%%%%%%%%%
%\documentclass{article}
%\begin{document}
%%%%%%%%%%%%%%%%%%%%%%%%%%%%%%
%\newpage
\section*{{\huge 7P. Probleme}}

{\bf Problema 7.1}

\noindent{\bf Calculul de variabil\A\ complex\A\ a func\c{t}iei Green}

Reamintim c\A\ am ob\c{t}inut deja rezultatul:

\( G({\bf r,r'})=\frac{1}{(2\pi)^{3}} \int
\frac{e^{iqR}}{k^{2}-q^{2}}d^{3}q~, \)
cu $R=\vert {\bf r-r'} \vert$.
Cum $d^{3}q=q^{2} \sin\theta dq d\theta d\phi$, ajungem,
dup\A\ ce integr\A m \h n variabilele unghiulare , la:

\( G({\bf r,r'})= \frac{i}{4\pi^{2}R}\int_{-\infty} ^{\infty}
\frac{(e^{-iqR}-e^{iqR})}{k^{2}-q^{2}} q dq~. \)

\noindent S\A\ punem:
$C=\frac{i}{4\pi^{2}R}$; \c{s}i s\A\ separ\A m integrala \h n dou\A\ 
p\A r\c{t}i:

%%%%%%%%%%%%%%%%%%%%%%%%%%%%%%%%
\( C(\int _{-\infty} ^{\infty} \frac{e^{-iqR}}{k^{2}-q^{2}} q dq -
\int _{-\infty} ^{\infty}\frac{e^{iqR}}{k^{2}-q^{2}} q dq)~. \)
%%%%%%%%%%%%%%%%%%%%%%%%%%%

\noindent S\A\ facem acum $q \rightarrow -q$ \h n prima integral\A\ :

\(  \int _{-\infty} ^{\infty} \frac{e^{-i(-q)R}}{k^{2}-(-q)^{2}} (-q)
d(-q)= \int _{\infty} ^{-\infty} \frac{e^{iqR}}{k^{2}-q^{2}} q dq
= -\int _{-\infty} ^{\infty} \frac{e^{iqR}}{k^{2}-q^{2}} q dq \)

\noindent astfel c\A\ :

\( G({\bf r,r'})= -2C ( \int_{-\infty} ^{\infty} 
\frac{qe^{iqR}}{k^{2}-q^{2}}
dq)~.\)

\noindent Substituind $C$,
%\( C= \frac{i}{2\pi^{2}}R \) \\
%\noindent 
ob\c{t}inem:

\( G({\bf r,r'})= \frac{-i}{2\pi^{2}R}\int_{-\infty} ^{\infty}
\frac{qe^{iqR}}{k^{2}-q^{2}}dq \)

\^{I}n aceast\A\ form\A\ integrala se poate evalua cu ajutorul reziduurilor
polilor pe care \h i posed\A\ , folosind metodele de variabil\A\ 
complex\A\ . Not\A m c\A\ exist\A\ poli simpli \h n pozi\c{t}iile
$q=_{-}^{+}k$.

%%%%%%%%%%%%%%
\vskip 2ex
\centerline{
\epsfxsize=280pt
\epsfbox{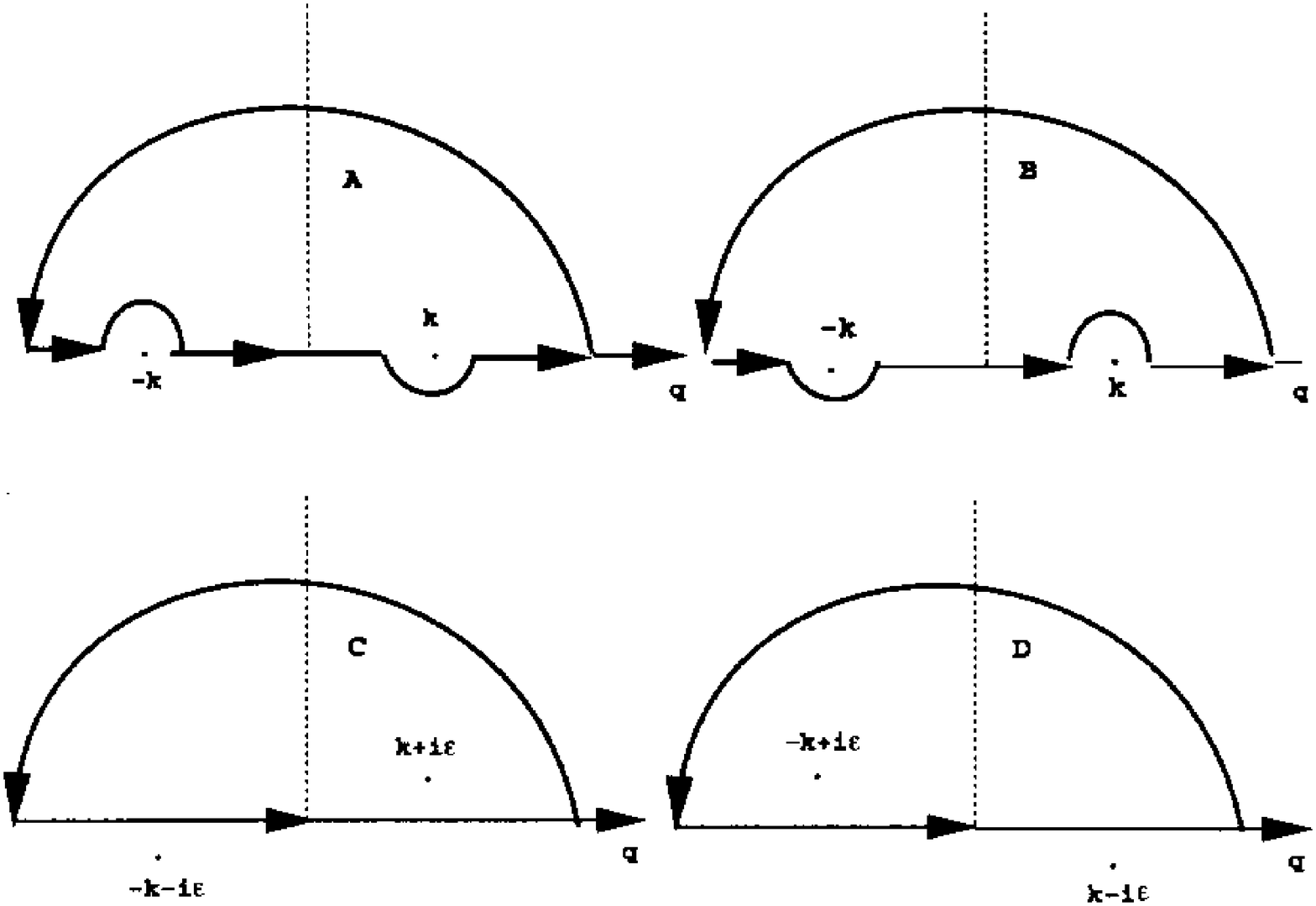}}
\vskip 4ex
\begin{center}
{\small{Fig. 7.4: Reguli de contur \h n jurul polilor pentru $G_{+}$ \c{s}i 
$G_{-}$}\\
}
\end{center}
%%%%%%%%%%%%%%%%

Folosim conturul din figura 7.4, care \h nconjoar\A\ polii 
%AQUI VA UNA IMAGEN DEL CONTORNO DE INTEGRACION 
\h n modul ar\A tat, pentru c\A\ acesta d\A\ efectul fizic
corect, pentru c\A\ de acord cu teorema reziduurilor,

\( G(r)= - \frac{1}{4\pi} \frac{e^{ikr}}{r}\quad({\rm Im} k > 0) \) ,

\(G(r)= - \frac{1}{4\pi} \frac{e^{-ikr}}{r}\quad ({\rm Im} k < 0) \)

Solu\c{t}ia care ne intereseaz\A\ este prima, pentru c\A\ d\A\  unde dispersate
{\em divergente}, \h n timp ce a doua solu\c{t}ie reprezint\A\ unde
dispersate convergente. Mai mult, combina\c{t}ia linear\A\ 

\( \frac{1}{2} \lim_{\epsilon\rightarrow 0} [G_{k+i\epsilon} +
G_{k-i\epsilon}] = - {\frac{1}{4\pi}} \frac{\cos kr}{r} \)

\noindent corespunde undelor sta\c{t}ionare.

\noindent
Evaluarea formal\A\ a integralei se poate face lu\h nd
$k^{2}-q^{2}\rightarrow k^{2}+i\epsilon-q^{2}$ , astfel c\A\ :
\(\int _{-\infty} ^{\infty}\frac{qe^{iqR}}{k^{2}-q^{2}}dq \rightarrow
\int _{-\infty} ^{\infty}\frac{qe^{iqR}}{(k^{2}+i\epsilon)-q^{2}}dq~. \)

\noindent Aceasta este posibil pentru $R>0$, de aceea conturul pentru 
calcul va fi situat \h n semiplanul complex superior. Astfel, polii
integrantului se afl\A\ \h n:
$q=_{-}^{+}\sqrt{k^{2}+i\epsilon} \simeq
^{+}_{-}(k+\frac{i\epsilon}{2k})$.
Procedeul de luare a limitei c\h nd $\epsilon \rightarrow 0$ trebuie
efectuat {\em dup\A\ } evaluarea integralei.\\

{\bf Problema 7.2}

\noindent {\bf Forma asimptotic\A\ a func\c{t}iei radiale}

Cum s-a v\A zut deja \h n capitolul {\it Atomul de hidrogen} 
partea radial\A\ a ec. Schr\"odinger
se poate scrie:

\( ( \frac {d^2}{dr^{2}} + \frac{2}{r} \frac{d}{dr} ) 
R_{nlm}(r)-\frac{2m}{\hbar^{2}}[V(r)+\frac{l(l+1) 
\hbar^{2}}{2mr^{2}}]R_{nlm}(r)+\frac{2mE}{\hbar^{2}}R_{nlm}(r)=0~. \)

\noindent $n,l,m$ sunt numerele cuantice sferice. De acum \h nainte nu se vor mai
scrie din motive de comoditate. 
$R$ este func\c{t}ia de und\A\ radial\A\ (depinde numai de 
$r$). Vom presupune c\A\ poten\c{t}ialele cad la zero mai repede 
dec\h t $1/r$, \c{s}i \h n plus c\A\ $\lim_{r \rightarrow 0} r^{2}V(r)=0$.

Folosim acum $u(r)=rR$, \c{s}i cum:
$(\frac{d^{2}}{dr^{2}} + \frac{2}{r} \frac{d}{dr})\frac{u}{r} =
\frac{1}{r} \frac{d^{2}}{dr^{2}}u$, avem

\( \frac{d^{2}}{dr^{2}}u + 
\frac{2m}{\hbar^{2}}[E-V(r)-\frac{l(l+1)\hbar^{2}}{2mr^{2}}] u=0~. \)

\noindent Not\A m c\A\ poten\c{t}ialul prezint\A\ un termen suplimentar:

\( V(r)\rightarrow V(r)+\frac{l(l+1)\hbar^{2}}{2mr^{2}}~,\)

\noindent care corespunde unei bariere centrifugale repulsive. Pentru o
particul\A\ liber\A\ $V(r)=0$ \c{s}i ecua\c{t}ia devine 

\( [\frac{d^{2}}{dr^{2}} + \frac{2}{r} 
\frac{d}{dr})-\frac{l(l+1)}{r^{2}}]R + k^{2}R=0~. \)

\noindent Introduc\h nd variabila $\rho=kr$, ob\c{t}inem

\( \frac{d^{2}R}{d\rho^{2}} + \frac{2}{\rho} \frac{dR}{d\rho} - 
\frac{l(l+1)}{\rho^{2}}R + R=0~. \)

\noindent
Solu\c{t}iile acestei ecua\c{t}ii sunt a\c{s}a numitele {\it func\c{t}ii 
Bessel sferice}. Solu\c{t}ia regular\A\ este:

\( j_{l}(\rho)=(-\rho)^{l} (\frac{1}{\rho} \frac{d}{d\rho})^{l} (\frac{\sin 
\rho}{\rho})~, \)

\noindent iar cea iregular\A\ :

\( n_{l}(\rho)= - (-\rho)^{l} (\frac{1}{\rho} \frac{d}{d\rho})^{l} 
(\frac{\cos \rho}{\rho})~. \)

Pentru $\rho$ mare, func\c{t}iile de interes sunt func\c{t}iile 
Hankel sferice:

\( h_{l}^{(1)}(\rho)=j_{l}(\rho)+ in_{l}(\rho) \)
\c{s}i
\( h_{l}^{(2)}(\rho)=[ h_{l}^{(1)}(\rho)]^{*}~. \)

De interes deosebit este comportamentul pentru $\rho \gg l$:
\begin{equation}
\label{eq:P1}
j_{l}(\rho) \simeq \frac{1}{\rho} \sin
(\rho-\frac{l\pi}{2})
\end{equation}  
\begin{equation}
\label{eq:P2}
 n_{l}(\rho) \simeq - \frac{1}{\rho} \cos(\rho-\frac{l\pi}{2})~. 
\end{equation}

\noindent \c{s}i atunci

\( h_{l}^{(1)} \simeq  -\frac{i}{\rho} e^{i(\rho - l\pi/2)}~. \)

Solu\c{t}ia regular\A\ \h n origine este:
$R_{l}(r)=j_{l}(kr)$

Forma asimtotic\A\ este (folosind ec. \ref{eq:P1})

\( R_{l}(r) \simeq \frac{1}{2ikr}[e^{-ikr-l\pi/2}-e^{ikr-l\pi/2}]~.\)

\bigskip

{\bf Problema 7.3}

\noindent {\bf Aproxima\c{t}ia Born pentru poten\c{t}iale Yukawa}

S\A\ consider\A m un poten\c{t}ial de forma:

\begin{equation} 
V({\bf r})= V_{0} \frac{e^{-\alpha r}}{r}~, 
\end{equation}

\noindent cu $V_{0}$ \c{s}i $\alpha$ constante reale \c{s}i $\alpha$ 
pozitiv\A\ . Poten\c{t}ialul 
este atractiv sau repulsiv \h n func\c{t}ie de semnul lui $V_{0}$; 
cu c\h t este mai mare $\vert V_{0} \vert$, cu at\h t este mai intens
poten\c{t}ialul.
Presupunem c\A\ $\vert V_{0} \vert$ este suficient de mic pentru ca  
aproxima\c{t}ia Born s\A\ func\c{t}ioneze. Conform formulei 
ob\c{t}inute anterior, amplitudinea de 
dispersie este dat\A\ de:\\

\( f^{(B)}(\theta, \varphi)= - \frac{1}{4\pi} \frac{2mV_{0}}{\hbar^{2}} 
\int e^{-i {\bf K \cdot r}} \frac{e^{-\alpha r}}{r} d^{3}r~. \)

Cum acest poten\c{t}ial depinde numai de $r$, integr\A rile unghiulare 
se pot face u\c{s}or, ajung\h nd astfel la forma:

\( f^{(B)}(\theta, \varphi)=  \frac{1}{4\pi} \frac{2mV_{0}}{\hbar^{2}} 
\frac{4\pi}{\vert {\bf K} \vert} \int_{0}^{\infty} \sin \vert {\bf K} 
\vert r \frac{e^{-\alpha r}}{r} r dr~. \)

A\c{s}adar, ob\c{t}inem:

\( f^{(B)}(\theta, \varphi)=  -\frac{2mV_{0}}{\hbar^{2}} \frac 
{1}{\alpha^{2} + \vert {\bf K}\vert^{2}}~.\)

%AQUI VA UNA FIGURA 

Din figur\A\ se observ\A\ c\A\ :
$\vert {\bf K} \vert = 2k \sin \frac{\theta}{2}$;
prin urmare:

\( \sigma^{(B)}(\theta)=\frac{4m^{2}V_{0}^{2}}{\hbar^{4}} 
\frac{1}{[\alpha^{2} + 4k^{2} \sin \frac{\theta}{2}^{2}]^{2}}~. \)

Sec\c{t}iunea total\A\ se ob\c{t}ine prin integrare:

\( \sigma^{(B)} = \int \sigma^{(B)}(\theta) d\Omega= 
\frac{4m^{2}V_{0}^{2}}{\hbar^{4}} \frac{4\pi}{\alpha^{2}(\alpha^{2}+4k^{2})}
~. \)

% \end{document}

\newpage
%%%%%%%%%%%%%%%%%%%%%%%%%%%%%%%%%%%%%%%%%%%%%%%%%%%%%%%%%%%%%%%%%%%%%%%%%%%%%
%%%%%%%%%%%%%%%%%%%%%%%%%%%%%%%%%%%%%%%%%%%%%%%%%%%%%%%%%%%%%%%%%%%%%%%%%%%
%%%%%%%%%%%%%%%%%%%%%%%%%%%%%%%%%%%%%%%%%%%%%%%%%%%%%%%%%%%%    88888
%\documentstyle[aps,preprint,tighten]{revtex}
%\begin{document}
%\draft
\def\bi{bigskip}
\def\noi{noindent}
%\protect
%\setcounter{equation}
%%%%%%%%%%%%%%%%%%%%%%%%%%%%%%%%%
\begin{center}
{\huge 8. UNDE PAR\c{T}IALE}
\end{center}
%\author{Pedro Basilio Espinoza Padilla}
%\address{Universidad de Guanajuato, Instituto de F\'isica, \\ Le\'on,
%Guanajuato; M\'exico.}
%\maketitle
%\begin{abstract}
%%%%%%%%%%%%%%%%%%%%%%%%%%%%%%%%%
%\begin{center}
%En el presente trabajo,
%Vom explica pe scurt \h n ce consist\A\ metoda undelor par\c{t}iale
%\h n studiul problemelor de dispersie cuantic\A\ . 
%\end{center}
%\end{abstract}
%%%%%%%%%%%%%%%%%%%%%%%%%%%%%%%%%%%%

\section*{Introducere}
\setcounter{equation}{0}
Metoda undelor par\c{t}iale se refer\A\ la particule  
care interac\c{t}ioneaz\A\ \h ntr-o regiune restr\h ns\A\ de spa\c{t}iu
cu o alta, care  prin caracteristicile sale
este cunoscut\A\ ca centru de \h mpr\A \c{s}tiere (de exemplu faptul c\A\ se 
poate considera fix\A\ ). \^{I}n afara acestei regiuni, 
interac\c{t}ia \h ntre cele dou\A\ particule se poate considera neglijabil\A\ .
\^{I}n acest fel este posibil s\A\ se descrie particula \h mpr\A \c{s}tiat\A\
cu urm\A torul Hamiltonian:
\begin{equation}
H=H_0+V~,
\end{equation}

\noindent
unde $H_0$ corespunde Hamiltonianului de particul\A\ liber\A\ .
Deci problema noastr\A\ este de a rezolva urm\A toarea ecua\c{t}ie:
\begin{equation}
(H_0+V) \mid \psi \rangle = E \mid \psi \rangle~.
\end{equation}

Este evident c\A\ spectrul va fi continuu (studiem cazul \h mpr\A \c{s}tierii
elastice). Solu\c{t}ia ecua\c{t}iei precedente este dat\A\ de: 
\begin{equation}
\mid \psi \rangle = \frac {1}{E-H_0} V\mid \psi \rangle + \mid \phi \rangle~.
\end{equation}

Cu o analiz\A\ u\c{s}oar\A\ putem s\A\ vedem c\A\ pentru $V=0$ 
ob\c{t}inem  
solu\c{t}ia $\mid \phi \rangle $, adic\A\ , solu\c{t}ia corespunz\A toare 
particulei libere. Trebuie notat c\A\ operatorul $\frac{1}{E-H_0}$
\h ntr-un anumit sens este anomal, pentru c\A\ are un continuu de poli
pe axa real\A\ care coincid cu valorile proprii ale lui $H_0$. Pentru
a `sc\A pa' de aceast\A\ problem\A\
s\A\ producem o mic\A\ deplasare \h n direc\c{t}ia imaginar\A\ ($\pm i\epsilon$)
a t\A ieturii de pe axa real\A\ :
\begin{equation}
\mid \psi^{\pm} \rangle = \frac {1}{E-H_0 \pm i\varepsilon} V\mid \psi^{\pm} 
\rangle + \mid \phi \rangle
\end{equation}

Aceast\A\ ecua\c{t}ie este cunoscut\A\ ca ecua\c{t}ia Lippmann-Schwinger.
\^{I}n final deplasarea polilor va fi \h n sens pozitiv de la axa imaginar\A\
(pentru ca principiul de cauzalitate s\A\ nu fie violat [cf. 
Feynman]). S\A\ lu\A m reprezentarea x:
\begin{equation}
\langle {\bf {x}}\mid \psi^{\pm} \rangle =\langle {\bf {x}}\mid \phi \rangle + 
\int d^{3} x^{'}\left \langle {\bf {x}} \vert \frac {1}{E-H_0 \pm i\varepsilon 
}\vert {\bf {x^{'}}} \right \rangle \langle {\bf {x^{'}}} \mid V\mid 
\psi^{\pm}\rangle~.
\end{equation}

Primul termen din partea dreapt\A\   
corespunde unei particule libere \h n timp ce al doilea termen se 
interpreteaz\A\ ca o 
und\A\ sferic\A\ care `iese' din centrul de \h mpr\A \c{s}tiere. 
Nucleul integralei anterioare
se poate asocia cu o func\c{t}ie Green (sau propagator) \c{s}i este foarte
simplu s\A\ se calculeze:
\begin{equation}
G_{\pm}({\bf {x}},{\bf {x^{'}}})=\frac{\hbar^{2}}{2m}\left \langle {\bf {x}} \vert 
\frac {1}{E-H_0 \pm i\varepsilon}\vert {\bf {x^{'}}} \right \rangle = 
-\frac{1}{4\pi} \frac{e^{\pm ik\mid {\bf {x}}-{\bf {x^{'}}}\mid}}{\mid {\bf 
{x}}-{\bf {x^{'}}}\mid}~,
\end{equation}

\noindent
unde $E={\hbar^{2}}{k^2}/2m$.
A\c{s}a cum am v\A zut mai \h nainte func\c{t}ia de und\A\ se poate scrie
ca o 
und\A\ plan\A\ plus una sferic\A\ care iese din centrul de \h mpr\A \c{s}tiere
(p\h n\A\ la un factor constant):
\begin{equation}
\langle {\bf {x}}\mid \psi^{+} \rangle =e^{{\bf {k}}\cdot {\bf {x}}} + 
\frac{e^{ikr}}{r} f({\bf {k}},{\bf {k^{'}}})~.
\end{equation}

M\A rimea $f({\bf {k}},{\bf {k^{'}}})$ care apare \h n ec. 7
se cunoa\c{s}te ca amplitudine de dispersie \c{s}i se poate scrie
explicit \h n forma:
\begin{equation}
f({\bf {k}},{\bf {k^{'}}})=-\frac{1}{4\pi} {(2\pi )^3}\frac{2m}{\hbar^2}\langle 
{\bf {k^{'}}}\mid V \mid \psi^{+} \rangle~. 
\end{equation}

S\A\ definim acum un operator T astfel c\A\ :
\begin{equation}
T\mid \phi \rangle = V\mid \psi^{+} \rangle
\end{equation}

Dac\A\  multiplic\A m ecua\c{t}ia Lippmann-Schwinger cu V \c{s}i folosim 
defini\c{t}ia anterioar\A\ ob\c{t}inem:
\begin{equation}
T\mid \phi \rangle = V\mid \phi \rangle + V\frac{1}{E-H_0+
i\varepsilon}T\mid \phi 
\rangle ~. 
\end{equation}
Iter\h nd ecua\c{t}ia anterioar\A\ (ca \h n teoria de
perturba\c{t}ii)
putem ob\c{t}ine aproxima\c{t}ia Born \c{s}i corec\c{t}iile sale de ordin 
superior.

\section*{Metoda undelor par\c{t}iale}
%%%%%%%%%%%%%%%%%%%%%%%%%%%%%%%%%%%%%%%%%%

S\A\ consider\A m acum cazul unui poten\c{t}ial central nenul. \^{I}n acest
caz, pe baza defini\c{t}iei (9) se deduce c\A\ operatorul $T$ comut\A\ cu
$\vec {L}^{2} $ \c{s}i $\vec {L}$; de aici se spune c\A\ 
$T$ este un operator scalar. \^{I}n acest fel 
pentru a u\c{s}ura calculele este convenabil s\A\ se foloseasc\A\ 
coordonate sferice, 
pentru c\A\ dat\A\ simetria problemei, operatorul $T$ va fi diagonal. Acum, 
s\A\ vedem ce form\A\ ia expresia (8) pentru amplitudinea de dispersie:
\begin{equation}
f({\bf {k}},{\bf {k^{'}}})={\rm const.}\sum_{lml^{'}m^{'}} 
%\sum_{m} \sum_{l^{'}} \sum_{m^{'}} 
\int dE\int 
dE^{'}\langle {\bf {k^{'}}}\mid E^{'} l^{'} m^{'} \rangle \langle E^{'} 
l^{'} m^{'}\mid T\mid Elm\rangle \langle Elm\mid \bf {k} \rangle~,
\end{equation}
unde ${\rm const.}=-\frac{1}{4\pi}\frac{2m}{\hbar^2} {(2\pi)^3}$.
Dup\A\ c\h teva calcule se ob\c{t}ine:
\begin{equation}
f({\bf {k}},{\bf {k^{'}}})=-\frac{4\pi^2}{k}\sum_{l}\sum_{m} T_{l} (E) 
Y^{m}_{l} ({\bf {k^{'}}})Y^{m^{*}}_{l}(\bf {k})~.
\end{equation}

Aleg\h nd sistemul de coordonate astfel ca vectorul $\bf {k}$ s\A\ aib\A\ 
aceea\c{s}i direc\c{t}ie cu axa  
orientat\A\ z, se ajunge la concluzia c\A\ la amplitudinea de dispersie 
vor contribui numai armonicele sferice cu m egal cu zero; dac\A\  
definim $\theta$ ca unghiul \h ntre ${\bf {k}}$ \c{s}i ${\bf {k^{'}}}$ 
vom avea:
\begin{equation}
Y^{0}_{l} ({\bf {k^{'}}})=\sqrt {\frac{2l+1}{4\pi}} P_{l}(cos\theta)~.
\end{equation}

Cu urm\A toarea defini\c{t}ie:
\begin{equation}
f_{l}(k)\equiv-\frac{\pi T_{l} (E)}{k}~,
\end{equation}

ec. (12) se poate scrie \h n forma urm\A toare:
\begin{equation}
f({\bf {k}},{\bf {k^{'}}})=f(\theta)=\sum^{\infty}_{l=0} 
(2l+1)f_{l}(k)P_{l}(cos\theta)~.
\end{equation}

Pentru $f_{l}(k)$ se poate da o interpretare simpl\A\ pe baza dezvolt\A rii
unei unde plane \h n unde sferice. Astfel putem scrie 
func\c{t}ia $\langle {\bf {x}}\mid \psi^{+} \rangle$ pentru
valori mari ale lui $r$ \h n forma
%tener la forma:
$$%\begin{displaymath}
\langle {\bf {x}}\mid \psi^{+} \rangle = \frac{1}{{(2\pi )^{3/2}}}\left[ 
{e^{ikz}}+f(\theta ) \frac{{e^{ikr}}}{r}\right] =
$$%\end{displaymath}
$$%\begin{displaymath}
\frac{1}{{(2\pi)^{3/2}}}\left[ \sum_{l} (2l+1)P_{l}(\cos\theta 
)\left(\frac{{e^{ikr}}-{e^{i(kr-l\pi )}}}{2ikr} \right) 
+\sum_{l}(2l+1)f_{l}(k)P_{l}(\cos\theta )\frac{{e^{ikr}}}{r}\right]%=
$$%\end{displaymath}
\begin{equation}
=\frac{1}{{(2\pi )^{3/2}}}\sum_{l} 
(2l+1)\frac {P_{l}(\cos\theta )}{2ik}\left[ \left[ 
1+2ikf_{l}(k)\right]\frac{{e^{ikr}}}{r}-\frac{{e^{i(kr-l\pi 
)}}}{r} \right]~.
\end{equation}

Aceast\A\ expresie se poate interpreta dup\A\ cum urmeaz\A\ . 
Cei doi termeni exponen\c{t}iali corespund unor unde sferice, primul unei unde
emergente \c{s}i al doilea uneia convergente; \h n plus efectul de 
\h mpr\A \c{s}tiere
se vede convenabil \h n  coeficientul undei emergente, care este egal cu unu 
c\h nd nu exist\A\ dispersor.

\section*{Deplas\A ri (\c{s}ifturi) de faz\A\ }
%%%%%%%%%%%%%%%%%%%%%%%%%%%%%%%%%%%%%%%%%%%%%%%

S\A\ ne imagin\A m acum o suprafa\c{t}\A\ \h nchis\A\ centrat\A\ \h n
dispersor. Dac\A\ presupunem c\A\ nu exist\A\ creare \c{s}i nici anihilare de  
particule se verific\A\ :
\begin{equation}
\int {\bf {j}}\cdot d{\bf {S}}=0~,
\end{equation}

\noindent
unde regiunea de integrare este suprafa\c{t} definit\A\ mai \h nainte
\c{s}i ${\bf {j}}$ este densitatea de curent de 
probabilitate. \^{I}n plus, datorit\A\ conserv\A rii  momentului 
cinetic ecua\c{t}ia anterioar\A\ trebuie s\A\ se verifice pentru fiecare
und\A\ par\c{t}ial\A\ 
(cu alte cuvinte, toate undele par\c{t}iale au diferite valori
ale proiec\c{t}iilor momentului cinetic, ceea ce le face diferite. 
Formularea teoretic\A\ ar fi echivalent\A\ dac\A\ se consider\A\ pachetul de 
unde ca un flux de 
particule care nu interac\c{t}ioneaz\A\ \h ntre ele; mai mult, pentru c\A\
poten\c{t}ialul problemei noastre este central, momentul cinetic al fiec\A rei
``particule'' se va conserva ceea ce ne permite s\A\ spunem c\A\ 
particulele continu\A\ s\A\ fie acelea\c{s}i). Cu aceste considera\c{t}ii,
putem s\A\ afirm\A m c\A\ at\h t unda divergent\A\ c\h t \c{s}i cea emergent\A\ 
difer\A\ cel mult printr-un factor de faz\A\ . Deci, dac\A\ definim:
\begin{equation}
S_{l}(k)\equiv 1+ 2ikf_{l}(k)
\end{equation}

\noindent trebuie s\A\ avem
\begin{equation}
\mid S_{l}(k)\mid =1~.
\end{equation}

Rezultatele anterioare se pot interpreta cu ajutorul 
conserv\A rii probabilit\A \c{t}ilor \c{s}i erau de `a\c{s}teptat'
pentru c\A\ nu am presupus c\A\ exist\A\ creare \c{s}i anihilare de
particule, astfel c\A\ 
influen\c{t}a centrului dispersor consist\A\ pur \c{s}i simplu \h n 
a ad\A uga un factor de faz\A\ \h n  
componentele undelor emergente \c{s}i \h n virtutea unitarit\A \c{t}ii
factorului de faz\A\ \h l putem scrie \h n forma:
\begin{equation}
S_{l}=e^{2i\delta_{l}}~,
\end{equation}

\noindent 
unde $\delta_{l}$ este real \c{s}i este func\c{t}ie de k. 
Pe baza defini\c{t}iei (18) putem s\A\ scriem:
\begin{equation}
f_{l}=\frac{{e^{2i\delta_{l}}}-1}{2ik}=\frac{{e^{i\delta_{l}}}\sin 
(\delta_{l})}{k}=\frac{1}{k\cot (\delta_{l})-ik}~. 
\end{equation}

Sec\c{t}iunea total de \h mpr\A \c{s}tiere ia forma urm\A toare:
$$%\begin{displaymath}
\sigma_{total}=\int \mid f(\theta){\mid ^2}d\Omega =
$$%\end{displaymath}
$$%\begin{displaymath}
\frac{1}{{k^2}}{\int _{0} ^{2\pi}}d\phi {\int _{-1} ^{1}}d(\cos (\theta 
))\sum_{l} \sum_{{l^{'}}}(2l+1)(2{l^{'}}+1){e^{i\delta_{l}}}\sin 
(\delta_{l}){e^{i\delta_{{l^{'}}}}} \sin (\delta_{{l^{'}}})P_{l}P_{{l^{'}}}
$$%\end{displaymath}
\begin{equation}
=\frac{4\pi }{{k^2}}\sum_{l} (2l+1)\sin {^2}(\delta_{{l^{'}}})~.
\end{equation}

\section*{Determinarea \c{s}ifturilor de faz\A\ }
%%%%%%%%%%%%%%%%%%%%%%%%%%%%%%%%%%%%%%%%%%%%%%%%%%%%%%%%
S\A\ consider\A m acum un poten\c{t}ial V astfel c\A\ se anul\A\ pentru $r>R$, 
unde parametrul R se cunoa\c{s}te ca ``raz\A\ de ac\c{t}iune a 
poten\c{t}ialului'', astfel c\A\  
regiunea $r>R$ evident trebuie s\A\ corespund\A\ unei unde sferice
liber\A\ (neperturbat\A\ ). Pe de alt\A\ parte, forma cea mai general\A\ 
de dezvoltare a unei unde plane \h n unde sferice este:
\begin{equation}
\langle {\bf {x}}\mid \psi^{+} \rangle =\frac{1}{{(2\pi )^{3/2}}}\sum_{l} 
{i^{l}} (2l+1)A_{l}(r)P_{l}(\cos \theta ) \quad (r>R)~,
\end{equation}

unde coeficientul $A_{l}$ este prin defini\c{t}ie:
\begin{equation}
A_{l}={c_{l} ^{(1)}}{h_{l} ^{(1)}}(kr)+{c_{l} ^{(2)}}{h_{l} ^{(2)}}(kr)~,
\end{equation}

\c{s}i unde ${h_{l} ^{(1)}}$ \c{s}i ${h_{l} ^{(2)}}$ sunt func\c{t}iile
Hankel sferice ale c\A ror forme asimptotice sunt:
%\begin{displaymath}
$$ 
{h_{l} ^{(1)}} \sim \frac{{e^{i(kr-l\pi /2)}}}{ikr}
$$
%\end{displaymath}
$$%\begin{displaymath}
{h_{l} ^{(2)}} \sim - \frac{{e^{-i(kr-l\pi /2)}}}{ikr}~.
$$%\end{displaymath}

Examin\h nd forma asimptotic\A\ a expresiei (23) care este:
\begin{equation}
\frac{1}{{(2\pi )^{3/2}}}\sum_{l}(2l+1)P_{l}\left[ \frac{{e^{ikr}}}{2ikr}-
\frac{{e^{-i(kr-l\pi)}}}{2ikr} \right]~,
\end{equation}

se poate vedea c\A\ :
\begin{equation}
{c_{l} ^{(1)}}=\frac{1}{2} e^{2i\delta_{l}} \qquad {c_{l} ^{(2)}}=\frac{1}{2}~.
\end{equation}

Aceasta permite scrierea func\c{t}iei de und\A\ radial\A\ pentru $r>R$ 
\h n forma:
\begin{equation}
A_{l}=e^{2i\delta _{l}}\left[ \cos \delta _{l} j_{l} (kr)
- \sin \delta _{l}n_{l} 
(kr)\right]~. \end{equation}

Folosind ecua\c{t}ia anterioar\A\ putem evalua evalua derivata sa 
logaritmic\A\
\h n r=R, i.e., exact la frontiera zonei de ac\c{t}iune a poten\c{t}ialului:
\begin{equation}
\beta _{l}\equiv \left( \frac{r}{A_{l}}\frac{dA_{l}}{dr}\right)_{r=R}=
kR\left[
\frac{{j_{l}^{'}}\cos \delta _{l}-{n_{l}^{'}}(kR)\sin \delta _{l}}{j_{l}\cos
\delta _{l}-{n_{l}}(kR)\sin \delta_{l}}\right]~.
\end{equation}

\noindent
$j_{l}^{'}$ este derivata lui $j_{l}$ \h n raport cu $kr$ \c{s}i evaluat\A\ \h n
$r=R$. Alt rezultat important pe care \h l putem ob\c{t}ine cunosc\h nd
resultatul anterior este \c{s}iftul de faz\A\ :
\begin{equation}
\tan \delta _{l}=\frac{kR{j_{l}^{'}}(kR)-\beta _{l}
j_{l}(kR)}{kR{n_{l}^{'}}(kR)-
\beta_{l} n_{l}(kR)}~.
\end{equation}

Pentru a ob\c{t}ine solu\c{t}ia completa a problemei \h n acest caz
este necesar s\A\ se fac\A\ calculele pentru
$r<R$, adic\A\ , \h n interiorul razei de ac\c{t}iune al poten\c{t}ialului.
Pentru cazul unui poten\c{t}ial central, ecua\c{t}ia 
Schr\"odinger \h n trei dimensiuni este:
\begin{equation}
\frac{{d^{2}}u_{l}}{d{r^{2}}}+\left( {k^{2}}-\frac{2m}{{\hbar ^{2}}} 
V-\frac{l(l+1)}{{r^{2}}} \right) u_{l}=0~,
\end{equation}

unde $u_{l}=rA_{l}(r)$ este supus\A\ condi\c{t}iei de 
frontier\A\ $u_{l}\mid _{r=0} \quad =0$. Astfel, putem calcula 
derivata logaritmic\A , care \h n virtutea propriet\A \c{t}ii de
continuitate a derivatei logaritmice (care este
echivalent\A\ cu continuitatea derivatei \h ntr-un punct de discontinuitate) ne
conduce la:
\begin{equation}
\beta_{l} \mid_{interior}=\beta_{l}\mid_{exterior}~.
\end{equation}

\section*{Un exemplu: \h mpr\A \c{s}tierea pe o sfer\A\ solid\A\ }
%%%%%%%%%%%%%%%%%%%%%%%%%%%%%%%%%%%%%%%%%%%%%%%%%%%%%%%%%%%%%%%%%%%
S\A\ trat\A m acum un caz specific.
Fie un poten\c{t}ial definit prin:
\begin{equation}
V=\left\{ 
\begin{array} {ll}
\infty & \mbox{ $r<R$} \\
0      & \mbox {$r>R~.$} 
\end{array}
 \right.
\end{equation}

Se \c{s}tie c\A\ o particul\A\ nu poate penetra \h ntr-o regiune unde 
poten\c{t}ialul este infinit, astfel c\A\ func\c{t}ia de und\A\ 
trebuie s\A\ se anuleze \h n $r=R$; din faptul c\A\ sfera este impenetrabil\A\ 
rezult\A\ deasemenea c\A\ :
\begin{equation}
A_{l}(r)\mid_{r=R} =0~.
\end{equation}

Astfel, din ec. (27) avem:
\begin{equation}
\tan \delta_{l} = \frac{j_{l} (kR)}{n_{l} (kR)}~.
\end{equation}

Se vede c\A\ se poate calcula u\c{s}or \c{s}iftul de faz\A\  pentru orice
$l$.
S\A\ consider\A m acum cazul $l=0$ (\h mpr\A \c{s}tiere de und\A\ s) pentru care
avem:
$$%\begin{displaymath}
\delta_{l} = -kR
$$%\end{displaymath}

\c{s}i din ec. (27):

\begin{equation}
A_{l=0}(r)\sim \frac{\sin kr}{kr}\cos\delta_{0}+\frac{\cos 
kr}{kr}\sin\delta_{0}=\frac{1}{kr}\sin (kr+\delta_{0})~.
\end{equation}

\noindent
Vedem c\A\ fa\c{t}\A\ de mi\c{s}carea liber\A\ exist\A\ o contribu\c{t}ie 
adi\c{t}ional\A\ de o faz\A\ . 
Este clar c\A\ \h ntr-un caz mai general diferitele unde vor avea diferite 
\c{s}ifturi de faz\A\ ceea ce provoac\A\ o distorsie tranzitorie
\h n pachetul de unde dispersat.
S\A\ studiem acum cazul energiilor mici,
i.e., $kR<<1$. \^{I}n acest caz, 
expresiile pentru func\c{t}iile Bessel (folosite pentru a descrie func\c{t}iile 
Hankel sferice) sunt urm\A toarele:
\begin{equation}
j_{l} (kr)\sim \frac{(kr)^{l}}{(2l+1)!!}
\end{equation}
\begin{equation}
n_{l} (kr)\sim -\frac{(2l-1)!!}{(kr)^{l+1}}
\end{equation}

care ne conduc la:
\begin{equation}
\tan\delta_{l} = \frac{-(kR)^{2l+1}}{(2l+1)[(2l-1)!!]^{2}}~.
\end{equation}

Din aceast\A\ formul\A\ putem s\A\ vedem c\A\ o contribu\c{t}ie
apreciabil\A\ la \c{s}iftul de faz\A\ este dat de undele cu $l=0$ \c{s}i cum 
$\delta_{0}=-kR$ ob\c{t}inem pentru sec\c{t}iunea eficace:
\begin{equation}
\sigma_{total}=\int\frac{d\sigma}{d\Omega}d\Omega=4\pi R^{2}~.
\end{equation}

De aici se ajunge la concluzia c\A\ sec\c{t}iunea eficace de \h mpr\A \c{s}iere
cuantic\A\ este de patru ori mai mare dec\h t
sec\c{t}iunea eficace clasic\A\ \c{s}i coincide cu aria
total\A\ a sferei dure. Pentru valori mari ale  energiei pachetului incident
se poate lucra cu ipoteza c\A\ toate valorile lui $l$ p\h n\A\ la o valoare 
maxim\A\ $l_{max}\sim kR$ contribuie la sec\c{t}iunea eficace total\A\ :
\begin{equation}
\sigma_{total}
=\frac{4\pi}{k^{2}}{\sum_{l=0} ^{l\sim kR}}(2l+1){\sin}^{2}\delta_{l}~.
\end{equation}

\^{I}n acest fel, pe baza ec. (34) avem:
\begin{equation}
{\sin}^{2}\delta_{l}=\frac{\tan^{2}\delta_{l}}{1+\tan^{2}\delta_{l}}=
\frac{[j_{l} (kR)]^{2}}{[j_{l} (kR)]^{2}+[n_{l} (kR)]^{2}}\sim\sin^{2}\left( 
kR-\frac{l\pi}{2}\right)~,
\end{equation}

unde am folosit expresiile:
$$%\begin{displaymath}
j_{l} (kr)\sim\frac{1}{kr}\sin\left( kr-\frac{l\pi}{2}\right)
$$%\end{displaymath}
$$%\begin{displaymath}
n_{l} (kr)\sim -\frac{1}{kr}\cos\left( kr-\frac{l\pi}{2}\right)~.
$$%\end{displaymath}

Vedem c\A\ $\delta_{l}$ descre\c{s}te cu $\frac{\pi}{2}$ de fiecare dat\A\ c\A\
 $l$ se increment\A\ cu o unitate, \c{s}i deci  
este evident c\A\ se \h ndepline\c{s}te ${\sin}^{2}\delta_{l}+
{\sin}^{2}\delta_{l+1}=1$. Aproxim\h nd ${\sin}^{2}\delta_{l}$ cu valoarea sa 
medie $\frac{1}{2}$, este simplu de ob\c{t}inut rezultatul 
pe baza sumei de numere impare:
\begin{equation}
\sigma_{total}=\frac{4\pi}{k^{2}}(kR)^{2}\frac{1}{2}=2\pi R^{2}~.
\end{equation}

Odat\A\ \h n plus rezultatul calculului bazat pe metodele de mecanic\A\ 
cuantic\A\ , de\c{s}i asem\A n\A tor,
difer\A\ totu\c{s}i de rezultatul clasic. S\A\ vedem care este originea
factorului 2; 
mai \h nt\h i vom separa ec. (15) \h n dou\A\ p\A r\c{t}i:
\begin{equation}
f(\theta )=\frac{1}{2ik}{\sum_{l=0} ^{l=kR}}(2l+1){e^{2i\delta_{l}}}P_{l}
\cos (\theta )+\frac{i}{2k}{\sum_{l=0} ^{l=
kR}}(2l+1) P_{l}\cos (\theta )=f_{\mbox{refl}}+f_{\mbox{umbr\A\ }}~.
\end{equation}

Evalu\h nd $\int |f_{\mbox{ refl}}|^{2}d\Omega$:
\begin{equation}
\int |f_{\mbox{ refl}}|^{2}d\Omega=\frac{2\pi}{4k^2}{\sum_{l=0} 
^{l_{max}}}{{\int_{-1}}^{1}}(2l+1)^{2}[P_{l}\cos (\theta )]^{2} d(\cos \theta 
)=\frac{\pi {l_{max}}^{2}}{k^{2}}=\pi R^{2}~.
\end{equation}

Analiz\h nd acum $f_{\mbox{umbr\A\ }}$ pentru unghiuri mici avem:
\begin{equation}
f_{\mbox{umbr\A\ }}\sim\frac{i}{2k}\sum (2l+1)J_{0}(l\theta )\sim ik{\int_{0} 
^{R}}bJ_{0}(kb\theta )db=\frac{iRJ_{1}(kR\theta )}{\theta}~.
\end{equation}

Aceast\A\ formul\A\ este destul de cunoscut\A\ \h n optic\A\ , 
fiind formula pentru 
difrac\c{t}ia Fraunhofer; cu ajutorul schimbului de variabil\A\ 
$z=kR\theta$ putem s\A\  
evalu\A m integrala $\int |f_{\mbox{ umbr\A\ }}|^{2}d\Omega$:
\begin{equation}
\int |f_{\mbox{ umbr\A\ }}|^{2}d\Omega \sim 2\pi R^{2}{\int_{0} 
^{\infty}}\frac{[J_{1}(z)]^{2}}{z} dz\sim\pi R^{2}~.
\end{equation}

\^{I}n sf\h r\c{s}it, neglij\h nd interferen\c{t}a \h ntre $f_{\mbox{refl}}$ 
\c{s}i $f_{\mbox{ umbr\A\ }}$ 
(pentru c\A\ faza oscileaz\A\ \h ntre $2\delta_{l+1}=2\delta_{l}-\pi$). 
Se ob\c{t}ine astfel rezultatul (42). Am etichetat unul dintre termeni cu
titlul de umbr\A\ , pentru c\A\ originea sa se  
explic\A\ u\c{s}or dac\A\ se apeleaz\A\ la 
comportamentul ondulatoriu al particulei dispersate (din punct de vedere `fizic'
nu exist\A\ nici o diferen\c{t}\A\ \h ntre un pachet de und\A\ 
\c{s}i o particul\A\ \h n acest caz). Originea sa const\A\  
\h n componentele pachetului de unde \h mpr\A \c{s}tiate \h napoi ceea ce 
produce o diferen\c{t}\A\ de faz\A\ fa\c{t}\A\ de undele incidente 
duc\h nd la o interferen\c{t}\A\ distructiv\A\ .

\section*{\^{I}mpr\A \c{s}tiere \h n c\h mp coulombian}
%%%%%%%%%%%%%%%%%%%%%%%%%%%%%%%%%%%%%%%%%%%%%%%%%%%%%%%%
S\A\ consider\A m acum un exemplu clasic \c{s}i ceva mai complicat: 
\h mpr\A \c{s}tierea de
particule \h ntr-un c\h mp coulombian. Pentru acest caz ecua\c{t}ia
Schr\"odinger este:
\begin{equation}
\left( -\frac{\hbar ^{2}}{2m}\nabla ^{2} - \frac{Z_{1}Z_{2}e^{2}}{r}\right)\psi 
({\bf {r}})=E\psi ({\bf {r}}), \qquad E>0~,
\end{equation}

\noindent
unde $m$ este masa redus\A\ a sistemului \h n interac\c{t}ie
\c{s}i evident $E>0$ deoarece trat\A m cazul dispersiei f\A r\A\ producere
de nici un fel de st\A ri legate. Ecua\c{t}ia anterioar\A\ este echivalent\A\ 
urm\A toarei expresii (pentru valori adecuate ale constantelor $k$ \c{s}i 
$\gamma$) :
\begin{equation}
\left( \nabla ^{2} +{k^{2}} +\frac{2\gamma k}{r}\right)\psi ({\bf {r}})=0~.
\end{equation}

Dac\A\ nu consider\A m bariera centrifugal\A\ a poten\c{t}ialului efectiv 
(unde $s$) ne g\A sim \h n condi\c{t}iile unei
interac\c{t}iuni coulombiene pure \c{s}i putem propune o solu\c{t}ie de forma:
\begin{equation}
\psi ({\bf {r}})={e^{i{\bf {k\cdot r}}}}\chi (u)~,
\end{equation}

\noindent
cu
$$%\begin{displaymath}
u=ikr(1-\cos\theta )=ik(r-z)=ikw~,
$$%\end{displaymath}
$$%\begin{displaymath}
{\bf {k\cdot r}}=kz~.
$$%\end{displaymath}

\noindent
$\psi ({\bf {r}})$ este solu\c{t}ia complet\A\ a ecua\c{t}iei
Schr\"odinger \c{s}i se poate a\c{s}tepta
un comportament asimptotic format din dou\A\ p\A r\c{t}i, respectiv de 
und\A\ plan\A\ ${e^{i{\bf {k\cdot r}}}}$ \c{s}i und\A\ sferic\A\ 
${r^{-1}e^{ikr}}$. Definind noi variabile:
\begin{displaymath}
z=z \qquad w=r-z \qquad \lambda =\phi~,
\end{displaymath}

\noindent
cu ajutorul rela\c{t}iilor anterioare, ec. (48) ia forma:
\begin{equation}
\left[ u \frac{d^{2}}{du^{2}}+(1-u)\frac{d}{du}-i\gamma\right]\chi (u)=0~.
\end{equation}

Pentru a rezolva aceast\A\ ecua\c{t}ie, trebuie studiat mai \h nt\h i
comportamentul s\A u asimptotic, dar cum acesta a fost deja prezentat, 
func\c{t}ia 
de und\A\ asimptotic\A\ normalizat\A\ care se ob\c{t}ine \h n final 
ca rezultat al tuturor calculelor anterioare este:
\begin{equation}
\psi_{\bf k} ({\bf {r}})=\frac{1}{(2\pi )^{3/2}}\left( {e^{i[{\bf {k\cdot 
r}}-\gamma ln(kr-{\bf {k\cdot r}})]}}+
\frac{f_{c}(k,\theta){e^{i[kr+\gamma 
ln2kr]}}}{r}\right)~.
\end{equation}

Dup\A\ cum vedem, func\c{t}ia de und\A\ anterioar\A\ prezint\A\ termeni 
care o fac s\A\ difere apreciabil de ec. (7). Acest lucru se datoreaz\A\
faptului c\A\ for\c{t}a 
coulombian\A\ este de raz\A\ infinit\A\ de ac\c{t}iune. 
Efectuarea calculului exact pentru 
amplitudinea de \h mpr\A \c{t}iere coulombian\A\ este destul de dificil de 
realizat. %(de hecho casi todos los c\'alculos de este problema). 
Aici vom da numai rezultatul final 
pentru func\c{t}ia de und\A\ normalizat\A\ :
\begin{equation}
\psi_{\bf k} ({\bf {r}})=\frac{1}{(2\pi )^{3/2}}\left( {e^{i[{\bf {k\cdot r}}-
\gamma ln(kr-{\bf {k\cdot 
r}})]}}+\frac{g_{1}^{*}(\gamma )}{g_{1}(\gamma )}\frac{\gamma}{2k\sin 
(\theta /2) ^{2}}\frac{e^{i[kr+\gamma ln2kr]}}{r}\right)~,
\end{equation}

unde $g_{1}(\gamma )=\frac{1}{\Gamma (1-i\gamma )}$.

\^{I}n ceea ce prive\c{s}te analiza de unde par\c{t}iale o vom reduce la
prezentarea rezultatelor deja  
discutate \h ntr-un mod c\h t mai clar posibil.
Mai \h nt\h i scriem func\c{t}ia de und\A\ (49) $\psi ({\bf {r}})$ \h n 
urm\A toarea form\A\ :
\begin{equation}
\psi ({\bf {r}})={e^{i{\bf {k\cdot r}}}}\chi (u)=A{e^{i{\bf {k\cdot 
r}}}}\int_{C}{e^{ut}}{t^{i\gamma -1}}(1-t)^{-i\gamma}dt~,
\end{equation}

\noindent
unde $A$ este o constant\A\ de normalizare \c{s}i toat\A\ partea integral\A\
este  
transformata Laplace invers\A\ a transformatei directe a ecua\c{t}iei 
(50). O form\A\ convenabil\A\ a ecua\c{t}iei anterioare este:
\begin{equation}
\psi ({\bf {r}})=A\int_{C}{e^{i{\bf {k\cdot r}}}(1-t)}{e^{ikrt}}(1-t)
d(t,\gamma )dt
\end{equation}

\noindent
cu
\begin{equation}
d(t,\gamma )={t^{i\gamma -1}}(1-t)^{-i\gamma -1}~.
\end{equation}

\^{I}n cadrul analizei de unde par\c{t}iale proced\A m la a scrie:
\begin{equation}
\psi ({\bf {r}})={\sum_{l=0} ^{\infty}}(2l+1){i^{l}}P_{l}(\cos\theta )A_{l}(kr)~,
\end{equation}

\noindent
unde
\begin{equation}
A_{l}(kr)=A\int_{C}{e^{ikrt}}j_{l}[kr(1-t)](1-t)d(t,\gamma )~.
\end{equation}

Aplic\h nd rela\c{t}iile \h ntre func\c{t}iile Bessel sferice \c{s}i 
func\c{t}iile Hankel sferice avem:
\begin{equation}
A_{l}(kr)=A_{l}^{(1)}(kr)+A_{l}^{(2)}(kr)~.
\end{equation}

Evaluarea acestor coeficien\c{t}i nu o vom prezenta aici (fiind destul de 
complicat\A\ ). Rezult\A\ c\A\ :
\begin{equation}
A_{l}^{(1)}(kr)=0
\end{equation}
\begin{equation}
A_{l}^{(2)}(kr)\sim -\frac{Ae^{\pi\gamma /2}}{2ikr}[2\pi ig_{1}(\gamma)]
\left( 
e^{-i[kr-(l\pi /2)+\gamma \ln 2kr]}-{e^{2i\eta_{l} (k)}}
e^{i[kr-(l\pi /2)+\gamma 
\ln 2kr]}\right)
\end{equation}

\noindent
unde
\begin{equation}
{e^{2i\eta_{l} (k)}}=\frac{\Gamma (1+l-i\gamma )}{\Gamma (1+l+i\gamma )}~.
\end{equation}

\section*{Calculul amplitudinii de \h mpr\A \c{s}tiere coulombian\A\ }
%%%%%%%%%%%%%%%%%%%%%%%%%%%%%%%%%%%%%%%%%%%%%%%%%%%%%%%%%%%%%%%%%%%%%%%%

Dac\A\ efectu\A m  transformata Laplace a ec. (50) ob\c{t}inem:
\begin{equation}
\chi (u)=A\int_{C} e^{ut}t^{i\gamma-1}(1-t)^{-i\gamma}dt~.
\end{equation}

Conturul $C$ merge de la $-\infty $ la $\infty$ \c{s}i se \h nchide
pe deasupra axei reale. \^{I}n aceste condi\c{t}ii, vedem c\A\ exist\A\
doi poli: c\h nd $t=0$ \c{s}i $t=1$. Cu schimbul de variabil\A\ $s=ut$ 
ob\c{t}inem:
\begin{equation}
\chi (u)=A\int_{C_{1}}e^{s}s^{i\gamma -1}(u-s)^{-i\gamma}~.
\end{equation}

$\chi (u)$ trebuie s\A\ fie regular\A\ \h n zero \c{s}i \h ntr-adev\A r:
\begin{equation}
\chi (0)=(-1)^{-i\gamma}A\int_{C_{1}}\frac{e^{s}}{s}ds~.
=(-1)^{-i\gamma}A2\pi i
\end{equation}

Lu\h nd acum limita $u\to\infty$, s\A\ facem o deplasare infinitezimal\A\ 
(pentru a elimina faptul c\A\ polii sunt pe contur) \c{s}i cu un schimb de 
variabil\A\  astfel c\A\ $\frac{s}{u}=
-\frac{(s_{0}\pm i\varepsilon)}{i\kappa}$, 
vedem c\A\ aceast\A\ expresie tinde la zero c\h nd $u\to\infty$. Deci, 
putem s\A\ dezvolt\A m 
$(u-s)$ \h n serie de puteri de $\frac{s}{u}$ pentru polul cu $s=0$. 
Dar aceast\A\ dezvoltare  nu este bun\A\ \h n $s=1$, pentru c\A\ \h n acest caz
$s=-s_{0}+i(\kappa\pm\varepsilon)$ \c{s}i de aici rezult\A\ c\A\ $\frac{s}{u}=
1-\frac{(s_{0}\pm 
i\varepsilon )}{\kappa}$ tinde la $1$ c\h nd $\kappa\to\infty$; dar dac\A\
facem schimbul de variabil\A\ $s^{'}=s-u$ aceast\A\ problem\A\ se elimin\A\ :
\begin{equation}
\chi (u)=A\int_{\rm C_{2}}\left([e^{s}s^{i\gamma
-1}(u-s)^{-i\gamma}]ds+[e^{s^{'}+u}(-s^{'})^{i\gamma}(u+s^{'})^{i\gamma-1}]
ds^{'}\right)~.
\end{equation}

Dezvolt\h nd seriile de puteri este u\c{s}or de calculat integralele 
precedente, dar \h n rezultat trebuie s\A\ se ia 
limita $\frac{s}{u}\to 0$ pentru a ob\c{t}ine formele asimptotice corecte
pentru \h mpr\A \c{s}tierea coulombian\A\ :
$$%\begin{displaymath}
\chi (u)\sim 2\pi iA\left[u^{-i\gamma}g_{1}(\gamma )-(-u)^{i\gamma 
-1}e^{u}g_{2}(\gamma )\right]
$$%\end{displaymath}
$$%\begin{displaymath}
2\pi g_{1}(\gamma )=i\int_{\rm C_{2}}e^{s}s^{i\gamma -1}ds
$$%\end{displaymath}
\begin{equation}
2\pi g_{2}(\gamma )=i\int_{\rm C_{2}}e^{s}s^{-i\gamma}ds~.
\end{equation}

Dup\A\ acest \c{s}ir de schimb\A ri de variabile, ne \h ntoarcem la 
$s$-ul original pentru a ob\c{t}ine:
$$%\begin{displaymath}
(u^{*})^{i\gamma}=(-i)^{i\gamma}[k(r-z)]^{i\gamma}
=e^{\gamma\pi /2}e^{i\gamma \ln k(r-z)}
$$%\end{displaymath}
\begin{equation}
(u)^{-i\gamma}=(i)^{-i\gamma}[k(r-z)]^{-i\gamma}=e^{\gamma\pi /2}e^{-i\gamma 
\ln k(r-z)}~.
\end{equation}

Calculul lui $\chi$ odat\A\ efectuat, este echivalent cu a avea 
$\psi_{\bf k} ({\bf {r}})$ pornind de la (49).

\section*{Aproxima\c{t}ia eikonal\A\ }
%%%%%%%%%%%%%%%%%%%%%%%%%%%%%%%%%%%%%%

Vom face o scurt\A\ expozi\c{t}ie a aproxima\c{t}iei
eikonale a c\A rei filosofie este aceea\c{s}i cu cea care 
se face
c\h nd se trece de la optica ondulatorie la optica
geometric\A\ \c{s}i de aceea este corect\A\ c\h nd
poten\c{t}ialul variaz\A\ pu\c{t}in pe distan\c{t}e comparabile cu 
lungimea de und\A\ a pachetului de unde dispersat, adic\A\ , pentru cazul 
$E>>|V|$. Astfel aceast\A\ aproxima\c{t}ie poate fi considerat\A\ ca o 
aproxima\c{t}ie cuasiclasic\A\ . Mai \h nt\h i propunem c\A\ func\c{t}ia de 
und\A\ cuasiclasic\A\ are forma binecunoscut\A\ :
\begin{equation}
\psi\sim e^{iS({\bf r})/\hbar}~,
\end{equation}

\noindent
unde S satisface ecua\c{t}ia Hamilton-Jacobi, cu solu\c{t}ia:
\begin{equation}
\frac{S}{\hbar}=\int_{-\infty}^{z}\left[ k^{2}-\frac{2m}{\hbar ^{2}}V\left( \sqrt 
{b^{2}+z'^{2}}\right)\right]^{1/2}dz'+ {\mbox{ constant\u{a}}}~.
\end{equation}

Constanta aditiv\A\ se alege \h n a\c{s}a fel \h nc\h t:
\begin{equation}
\frac{S}{\hbar}\to kz\qquad {\rm pentru} \qquad V\to 0~.
\end{equation}

Termenul care multiplic\A\ poten\c{t}ialul se poate interpreta ca un schimb de
faz\A\ al pachetului de unde, av\h nd urm\A toarea form\A\ explicit\A\
\begin{equation}
\Delta (b)\equiv \frac{-m}{2k\hbar^{2}}\int_{-\infty}^{\infty} V\left( 
\sqrt{b^{2}+z^{2}}\right)dz~.
\end{equation}

\^{I}n cadrul metodei de unde par\c{t}iale aceast\A\ 
aproxima\c{t}ie are urm\A toarea aplica\c{t}ie. \c{S}tim c\A\ 
aproxima\c{t}ia eikonal\A\ este corect\A\ la energii \h nalte, 
unde exist\A\ multe unde par\c{t}iale care 
contribuie la dispersie. Astfel putem considera $l$ ca o variabil\A\
continu\A\ \c{s}i prin analogie cu mecanica clasic\A\ punem $l=bk$.
\^{I}n plus, cum deja am men\c{t}ionat $l_{max}=kR$, care substituit \h n
expresia (15) conduce la:
\begin{equation}
f(\theta )=-ik\int bJ_{0}(kb\theta )[e^{2i\Delta (b)}-1]db~.
\end{equation}

\bigskip

\section*{{\huge 8P. Probleme}}

{\em Problema 8.1}

\bigskip

\noindent
S\A\ se ob\c{t}in\A\ deplasarea de faz\A\ (\c{s}iftul) \c{s}i sec\c{t}iunea 
diferen\c{t}ial\A\ de \h mpr\A \c{s}tiere la unghiuri mici pentru un centru
de \h mpr\A \c{s}tiere de poten\c{t}ial $U(r)=\frac{\alpha}{r^2}$. S\A\ se 
\c{t}in\A\ cont de faptul c\A\ \h n \h mpr\A \c{s}tierile de unghiuri mici 
principala contribu\c{t}ie o dau undele par\c{t}iale cu $l$ mari.

\bigskip

{\bf Solu\c{t}ie}:

Rezolv\h nd ecua\c{t}ia 
$$
R_{l}^{''}+\Bigg[k^2-\frac{l(l+1)}{r^2}-\frac{2m\alpha}{\hbar ^2 r^2}\Bigg]=0
$$
cu condi\c{t}iile de frontier\A\ $R_{l}(0)=0$, $R_{l}(\infty)=N$, unde $N$ este
un num\A r finit, ob\c{t}inem
$$
R_{l}(r)=A\sqrt{r}I_{\lambda}(kr)~,
$$
unde $\lambda=\Bigg[(l+\frac{1}{2})^2+\frac{2m\alpha}{\hbar ^2}\Bigg]^{1/2}$ 
\c{s}i $I$ este prima func\c{t}ia Bessel modificat\A\ .

Pentru determinarea lui $\delta _{l}$ se folose\c{s}te expresia asimptotic\A\
a lui $I_{\lambda}$:
$$
I_{\lambda}(kr)\propto \left(\frac{2}{\pi kr}\right)^{1/2}\sin (kr-
\frac{\lambda\pi}{2}+\frac{\pi}{4})~.
$$
Prin urmare
$$
\delta _{l}=-\frac{\pi}{2}\left(\lambda-l-\frac{1}{2}\right)=
-\frac{\pi}{2}\left(\Bigg[(l+\frac{1}{2})^2
+\frac{2m\alpha}{\hbar ^2}\Bigg]^{1/2}-\left(l+\frac{1}{2}\right)\right)~.
$$
Condi\c{t}ia $l$ mari de care se vorbe\c{s}te \h n problem\A\ ne conduce la:
$$
\delta _{l}=-\frac{\pi m \alpha}{(2l+1)\hbar ^2}~,
$$
de unde se vede c\A\ $|\delta _{l}|\ll 1$ pentru $l$ mari.

Din expresia general\A\ a amplitudinii de \h mpr\A \c{s}tiere
$$
f(\theta)=\frac{1}{2ik}\sum _{l=0}^{\infty}(2l+1)P_{l}(\cos \theta)
(e^{2i\delta _{l}}-1)~,
$$
la unghiuri mici $e^{2i\delta _{l}}\approx 1+2i\delta _{l}$ \c{s}i
$$
\sum _{l=0}^{\infty}P_{l}(\cos \theta)=\frac{1}{2\sin \frac{\theta}{2}}~.
$$
Astfel:
$$
f(\theta)=-\frac{\pi \alpha m}{k \hbar ^2}\frac{1}{2 \sin \frac{\theta}{2}}~.
$$
Prin urmare:
$$
\frac{d\sigma}{d\theta}=\frac{\pi ^3 \alpha ^2 m}{2\hbar ^2 E}{\mbox ctg}
\frac{\theta}{2}~.
$$

%}

%\end{document}

\end{document}